%% file: memoirthesis.tex
\RequirePackage[l2tabu]{nag}		% Warns for incorrect (obsolete) LaTeX usage

\documentclass[a4paper,12pt,openbib,oldfontcommands]{memoir} %add 'draft' to turn draft option on (see below)
\usepackage[T1]{fontenc}
\usepackage[font=small,labelfont=bf,justification=justified,format=plain]{subcaption}
\usepackage{enumerate,amsthm,amssymb,color}
\usepackage[table]{xcolor}
\usepackage{tcolorbox}
\usepackage{afterpage}
\usepackage{ifpdf}
\usepackage{chemist}
\usepackage{appendix}
\usepackage{textpos}
\usepackage{pst-all}
\usepackage{exscale}
\usepackage{mathptmx}
\usepackage[scaled=.90]{helvet}
\usepackage{courier}

\usepackage{tkz-euclide}
\usepackage{import}
\usepackage{algorithm}

\usepackage{algcompatible}

\usepackage{amsfonts} 					%Calls Amer. Math. Soc. (AMS) fonts
\usepackage[centertags]{amsmath}			%Writes maths centred down
\usepackage{newlfont}	%Helpful package for fonts and symbols
\usepackage{textgreek}
\usepackage{physics}
\usepackage{graphicx}					%Calls figure environment
\usepackage{longtable,rotating}		%Long tab environments including rotation. 
\usepackage[utf8]{inputenc}			%Needed to encode non-english characters 
\usepackage{multicol} %directly for mac
\usepackage{multirow}
\usepackage{float}	
\usepackage{etoolbox}
\usepackage{verbatim}					%Permits pre-formated text insertion
\usepackage{latexsym}					%Extra symbols
\usepackage{url}						%Supports url commands

\usepackage[spanish,english]{babel}		%For languages characters and hyphenation
\usepackage{siunitx}                 %Creates coloured text and background
\usepackage[colorlinks=true,citecolor=blue,linkcolor=magenta]{hyperref}   %Creates hyperlinks in cross references
\usepackage{memhfixc}					%Must be used on memoir document 
\usepackage{enumerate}					%For enumeration counter
\usepackage{footnote}					%For footnotes
\usepackage{microtype}					%Makes pdf look better.
\usepackage{rotfloat}					%For rotating and float environments 
\usepackage{alltt}						%LaTeX commands are not disabled in 

\usepackage{mathtools}
\usepackage{dsfont}
\usepackage{bbm}
\usepackage{bm}
\usepackage{synttree}
\usepackage{bbold}
\usepackage[normalem]{ulem}
\usepackage{geometry}
\usepackage{thmtools}
\usepackage{framed}
\usepackage[noend]{algpseudocode}
\usepackage{chngcntr}

\usepackage{booktabs}
\usepackage{eurosym}
\DeclareMathAlphabet{\softcal}{OMS}{cmsy}{m}{n}
\usepackage{enumitem}

\usepackage{stmaryrd}

\definecolor{quantumviolet}{HTML}{53257F}
\definecolor{navy}{RGB}{47,60,126}
\definecolor{darkviolet}{RGB}{99,56,142}
\definecolor{darkgreen}{RGB}{39,174,96}

\newcommand{\xmark}{%
\textcolor{red}{
\tikz[scale=0.23] {
    \draw[line width=0.7,line cap=round] (0,0) to [bend left=6] (1,1);
    \draw[line width=0.7,line cap=round] (0.2,0.95) to [bend right=3] (0.8,0.05);
}}}

\newcommand{\cmark}{%
\textcolor{darkgreen}{
\tikz[scale=0.23] {
    \draw[line width=0.7,line cap=round] (0.25,0) to [bend left=10] (1,1);
    \draw[line width=0.8,line cap=round] (0,0.35) to [bend right=1] (0.23,0);
}}}

\usepackage[nameinlink]{cleveref} % 'nameinlink' to mimic \autoref's behavior
\crefname{section}{Section}{Sections}
\crefname{equation}{Equation}{Equations}
\crefname{figure}{Figure}{Figures}
\crefname{table}{Table}{Tables}
\crefname{appendix}{Appendix}{Appendices}
\crefname{theorem}{Theorem}{Theorems}
\crefname{thm}{Theorem}{Theorems}
\crefname{cor}{Corollary}{Corollaries}
\crefname{lemma}{Lemma}{Lemmas}
\crefname{proposition}{Proposition}{Propositions}
\crefname{definition}{Definition}{Definitions}
\crefname{algorithm}{Algorithm}{Algorithms}

\let\autoref\cref

\hypersetup{
    colorlinks=true,
    linkcolor=navy,
    citecolor=quantumviolet,
    urlcolor=quantumviolet,
    breaklinks=true,
}
\let\mathcal\relax % <- Just in case (This is usually unnecessary).
\usepackage[cal=cm]{mathalfa}

\usepackage{tabularx}
    \newcolumntype{L}{>{\raggedright\arraybackslash}X}

\makeatletter
\newcommand{\IfRestatedTF}[2]{\ifthmt@thisistheone #2\else #1\fi}
\makeatother

\usepackage{lettrine}
 \newcommand{\initial}[1]{%
 	\lettrine[lines=3,lhang=0.33,nindent=0em]{
		\color{gray}
     		{\textsc{#1}}}{}}
\ifpdf
\fi
\ifdraftdoc 
	\usepackage{draftwatermark}				%Sets watermarks up.
	\SetWatermarkScale{0.3}
	\SetWatermarkText{\bf Draft: \today}
\fi
\settrimmedsize{297mm}{210mm}{*}
\settypeblocksize{634pt}{448.13pt}{*} 
\setulmargins{4cm}{*}{*} 
\setlrmargins{*}{*}{1.5} 
\setmarginnotes{17pt}{51pt}{\onelineskip} 
\setheadfoot{\onelineskip}{2\onelineskip} 
\setheaderspaces{*}{2\onelineskip}{*} 
\checkandfixthelayout
\OnehalfSpacing 
\setsecnumdepth{subsection} 
\maxsecnumdepth{subsubsection}
\usepackage{calc,soul}
\makeatletter 
\newlength\dlf@normtxtw 
\newsavebox{\feline@chapter} 
\newcommand\feline@chapter@marker[1][4cm]{%
	\sbox\feline@chapter{% 
		\resizebox{!}{#1}{\fboxsep=1pt%
			\colorbox{gray}{\color{white}\thechapter}% 
		}}%
		\rotatebox{90}{% 
			\resizebox{%
				\heightof{\usebox{\feline@chapter}}+\depthof{\usebox{\feline@chapter}}}% 
			{!}{\scshape\so\@chapapp}}\quad%
		\raisebox{\depthof{\usebox{\feline@chapter}}}{\usebox{\feline@chapter}}%
} 
\newcommand\feline@chm[1][4cm]{%
	\sbox\feline@chapter{\feline@chapter@marker[#1]}% 
	\makebox[0pt][c]{% aka \rlap
		\makebox[1cm][r]{\usebox\feline@chapter}%
	}}
\makechapterstyle{daleifmodif}{

	\renewcommand\printchapternum{\null\hfill\feline@chm[2.5cm]\par}

} 
\makeatother 
\chapterstyle{daleifmodif}
\makepagestyle{myvf} 
\makeoddfoot{myvf}{}{\thepage}{} 
\makeevenfoot{myvf}{}{\thepage}{} 
\makeheadrule{myvf}{\textwidth}{\normalrulethickness} 
\makeevenhead{myvf}{\small\textsc{\leftmark}}{}{} 
\makeoddhead{myvf}{}{}{\small\textsc{\rightmark}}
\newcommand{\clearemptydoublepage}{\clearpage{\pagestyle{empty}\cleardoublepage}}

\makeindex
\patchcmd{\subequations}{}%
{}{}{}

\newcommand{\pgftextcircled}[1]{                                                                    %Defines encircled text
    \setbox0=\hbox{#1}%
    \dimen0\wd0%
    \divide\dimen0 by 2%
    \begin{tikzpicture}[baseline=(a.base)]%
        \useasboundingbox (-\the\dimen0,0pt) rectangle (\the\dimen0,1pt);
        \node[circle,draw,outer sep=0pt,inner sep=0.1ex] (a) {#1};
    \end{tikzpicture}
}
\newcommand{\socal}[1]{\textrm{$\softcal{#1}$}}

\newcommand{\E}{\socal{E}}

\let\oldsqrt\sqrt
\def\sqrt{\mathpalette\DHLhksqrt}
\def\DHLhksqrt#1#2{%
\setbox0=\hbox{$#1\oldsqrt{#2\,}$}\dimen0=\ht0
\advance\dimen0-0.2\ht0
\setbox2=\hbox{\vrule height\ht0 depth -\dimen0}%
{\box0\lower0.4pt\box2}}
\newcommand{\mycaption}[2][\@empty]{
	\captionnamefont{\scshape} 
	\changecaptionwidth
	\captionwidth{0.9\linewidth}
	\captiondelim{.\:} 
	\indentcaption{0.75cm}
	\captionstyle[\centering]{}
	\setlength{\belowcaptionskip}{10pt}
	\ifx \@empty#1 \caption{#2}\else \caption[#1]{#2}
}

\declaretheoremstyle[
    name=Theorem,
    mdframed={
  skipabove=8pt,
  skipbelow=6pt,
  hidealllines=true,
  backgroundcolor={myblue},
  innerleftmargin=8pt,
  innerrightmargin=8pt}
]{thmsty}
\theoremstyle{plain}
\newtheorem{thm}{Theorem}[chapter]
\theoremstyle{plain}
\newtheorem{conj}{Conjecture}[chapter]
\theoremstyle{plain}

\theoremstyle{plain}
\newtheorem{lemma}{Lemma}[chapter]
\theoremstyle{plain}
\newtheorem{cor}{Corollary}[chapter]
\theoremstyle{plain}

\newtheorem{definition}{Definition}[chapter]

\newcounter{prot}[chapter]

\definecolor{gray}{rgb}{0.5,0.5,0.5}

\newcommand{\Var}{\text{Var}}
\newcommand{\sumrw}{\sum_{\substack{ i_1,i'_1\in R(\omega) \\j_1,j'_1\in R(\omega)}}}

\newcommand{\mmstate}[1]{\mathbb{1}\kern-1pt /\kern-1pt #1 }

\newcommand{\per}[1]{\mathop{\mathsf{Per}\left(#1\right)}}
\newcommand{\pr}[2][]{
	\mathop{
		\ifx &#1&
		\mathrm{Pr}
		\else
            \mathrm{Pr}_{#1}
		\fi
		\left[#2\right]}
}

\DeclareMathSymbol{\qm}{\mathalpha}{operators}{"3F}
\DeclareMathAlphabet{\mathbbold}{U}{bbold}{m}{n}

\begin{document}
\frontmatter
\pagenumbering{gobble}
\input{frontmatter/titleSU} 
%
\input{frontmatter/title}% Jury et al.
%
% \input{frontmatter/dedication}
%
\input{frontmatter/Acknowledgements.tex}

%
\input{frontmatter/abstract}

\clearemptydoublepage

\input{frontmatter/Resume.tex}

\clearemptydoublepage
%

%\clearemptydoublepage
%
\renewcommand{\contentsname}{Table of Contents}
\maxtocdepth{subsection}
\addtocontents{toc}{\protect\thispagestyle{empty}}
\tableofcontents*
\addtocontents{toc}{\par\nobreak \mbox{}\hfill{\bf Page}\par\nobreak}
\clearemptydoublepage

\mainmatter
\import{chapters/01_Introduction/}{Introduction.tex}

\clearemptydoublepage

\import{chapters/02_Preliminaries/}{Preliminaries.tex}

\clearemptydoublepage

%\import{chapters/03_Subspace_Preserving_VQCs}{Subspace_Preserving_VQCs.tex}
%\clearemptydoublepage

\import{chapters/04_HW_Preserving_Methods/}{HW_Preserving_Methods.tex}
\clearemptydoublepage

\import{chapters/05_Photonic_Sub_Optimals_Models}{Photonic_Suboptimal_models.tex}
\clearemptydoublepage

\import{chapters/06_Subspace_Preserving_Algorithms}{Subspace_Preserving_Algorithms.tex}
\clearemptydoublepage

\import{chapters/07_VQC_as_Fourier_Models}{VQC_as_Fourier_Models.tex}
\clearemptydoublepage

\import{chapters/08_Fourier_Surrogates}{Fourier_Surrogates.tex}
\clearemptydoublepage

\import{chapters/09_Conclusion_and_Outlook}{Conclusion_and_Outlook.tex}
\clearemptydoublepage
%
%
% And the appendix goes here
\appendix

\import{Appendix}{Proof_HW_Preserving_Trainability.tex}
\clearemptydoublepage

\import{Appendix}{Experimental_Details_PQCNN.tex}
\clearemptydoublepage

\import{Appendix}{Proof_Fourier_Surrogate.tex}

\clearemptydoublepage

\bibliographystyle{ieeetr}

\refstepcounter{chapter}
\bibliography{References}

%
%\clearemptydoublepage 
%\thispagestyle{empty}

\ % The empty page

% Add index

%   
\end{document}

%% file: frontmatter/titleSU.tex
 \definecolor{marine}{RGB}{29,39,105}
 \setlength{\columnseprule}{0pt}
 \setlength\columnsep{10pt}

\label{form}
%%%%%%%%%%%%%%%%%%%%%%%%%%%%%%%%%%%%%%%%%%%%%%%%%%%%%%%%%%%%%%%%%%%%%%%%%%%%%%%%%%%%%%%%%%%%%%%%%%%%%%%%%%%%%%%%%%%%%%%%%%%%%%%%%%%%%%%%%%%%%%%%%%%%%%%%%%%%%%%%%%%%%%%
%%%%%%%%%%%%%%%%%%%%%%%%%%%%%%%%%%%%%%%%%%%%%%%%%%%%%%%%%%%%%%%%%%%%%%%%%%%%%%%%%%%%%%%%%%%%%%%%%%%%%%%%%%%%%%%%%%%%%%%%%%%%%%%%%%%%%%%%%%%%%%%%%%%%%%%%%%%%%%%%%%%%%%%
%%% Formulaire / Form
%%% Remplacer les paramètres des \newcommand par les informations demandées / Replace \newcommand parameters by asked informations
%%%%%%%%%%%%%%%%%%%%%%%%%%%%%%%%%%%%%%%%%%%%%%%%%%%%%%%%%%%%%%%%%%%%%%%%%%%%%%%%%%%%%%%%%%%%%%%%%%%%%%%%%%%%%%%%%%%%%%%%%%%%%%%%%%%%%%%%%%%%%%%%%%%%%%%%%%%%%%%%%%%%%%%
%%%%%%%%%%%%%%%%%%%%%%%%%%%%%%%%%%%%%%%%%%%%%%%%%%%%%%%%%%%%%%%%%%%%%%%%%%%%%%%%%%%%%%%%%%%%%%%%%%%%%%%%%%%%%%%%%%%%%%%%%%%%%%%%%%%%%%%%%%%%%%%%%%%%%%%%%%%%%%%%%%%%%%%

\newcommand{\PhDTitle}{Quantum Machine Learning for Industrial Applications} 	%% Titre de la thèse / Thesis title
\newcommand{\PhDname}{Léo Monbroussou} 															%% Civilité, nom et prénom /  Civility, first name and name 
\newcommand{\NNT}{20XXSACLXXXX} 															%% Numéro National de Thèse (donnée par la bibliothèque à la suite du 1er dépôt)/ National Thesis Number (given by the Library after the first deposit)

\newcommand{\ecodoctitle}{Dénomination} 													%% Nom de l'ED. Voir site de l'Université Paris-Saclay / Full name of Doctoral School. See Université Paris-Saclay website
\newcommand{\ecodocacro}{Sigle}																%% Sigle de l'ED. Voir site de l'Université Paris-Saclay / Acronym of the Doctoral School. See Université Paris-Saclay website
\newcommand{\ecodocnum}{000} 																%% Numéro de l'école doctorale / Doctoral School number
\newcommand{\PhDspeciality}{voir spécialités par l'ED} 										%% Spécialité de doctorat / Speciality 
\newcommand{\PhDworkingplace}{Nom de l'établissement} 										%% Établissement de préparation / PhD working place : l'Université Paris-Sud, l'Université de Versailles-Saint-Quentin-en-Yvelines, l'Université d'Evry-Val-d'Essonne, l'Institut des sciences et industries du vivant et de l'environnement (AgroParisTech), CentraleSupélec,l'Ecole normale supérieure de Cachan, l'Ecole Polytechnique, l'Ecole nationale supérieure de techniques avancées, l'Ecole nationale de la statistique et de l’administration économique, HEC Paris, l'Institut d'optique théorique et appliquée, Télécom ParisTech, Télécom SudParis   
\newcommand{\defenseplace}{Ville de soutenance} 											%% Ville de soutenance / Place of defense
\newcommand{\defensedate}{Date} 															%% Date de soutenance / Date of defense

%%% Établissement / Institution
%%% Si la thèse a été produite dans le cadre d'une co-tutelle, commenter la partie "Pas de co-tutelle" et décommenter la partie "Co-tutelle" / If the thesis has been prepared in guardianship, comment the part "Pas de co-tutelle" and uncomment the part "Co-tutelle"

	%%%%%%%%%%%%%%%%%%%%%%%%%
	%%% Pas de co-tutelle %%%
	%%%%%%%%%%%%%%%%%%%%%%%%%

\newcommand{\logoEtt}{blank}																%% NE PAS MODIFIER / DO NOT MODIFY
\newcommand{\vpostt}{0.1} 																	%% NE PAS MODIFIER / DO NOT MODIFY
\newcommand{\hpostt}{6}																		%% NE PAS MODIFIER / DO NOT MODIFY
\newcommand{\logoEt}{etab} 																	%% Logo de l'établissement de soutenance. Indiquer le sigle / Institution logo. Indicate the acronym : AGRO, CENTSUP, ENS, ENSAE, ENSTA, HEC, IOGS, TPT, TSP, UEVE, UPSUD, UVSQ, X 
\newcommand{\vpos}{0.1}																		%% À modifier au besoin pour aligner le logo verticalement / If needed, modify to align logo vertilcally
\newcommand{\hpos}{11}																		%% À modifier au besoin pour aligner le logo horizontalement / If needed, modify to align logo horizontaly

\newcommand{\jurynameA}{Prénom Nom}
\newcommand{\juryadressA}{Statut, Établissement (Unité de recherche)}
\newcommand{\juryroleA}{Président}

%%% Membre n°2 (Rapporteur) / Member n°2 (Reviewer)
\newcommand{\jurynameB}{Prénom Nom}
\newcommand{\juryadressB}{Statut, Établissement (Unité de recherche)}
\newcommand{\juryroleB}{Rapporteur}

%%% Membre n°3 (Rapporteur) / Member n°3 (Reviewer)
\newcommand{\jurynameC}{Prénom Nom}
\newcommand{\juryadressC}{Statut, Établissement (Unité de recherche)}
\newcommand{\juryroleC}{Rapporteur}

%%% Membre n°4 (Examinateur) / Member n°4 (Examiner)
\newcommand{\jurynameD}{Prénom Nom}
\newcommand{\juryadressD}{Statut, Établissement (Unité de recherche)}
\newcommand{\juryroleD}{Examinateur}

%%% Membre n°5 (Directeur de thèse) / Member n°5 (Thesis supervisor)
\newcommand{\jurynameE}{Prénom Nom}
\newcommand{\juryadressE}{Statut, Établissement (Unité de recherche)}
\newcommand{\juryroleE}{Directeur de thèse}

%%% Membre n°6 (Co-directeur de thèse) / Member n°6 (Thesis co-supervisor)
\newcommand{\jurynameF}{Prénom Nom}
\newcommand{\juryadressF}{Statut, Établissement (Unité de recherche)}
\newcommand{\juryroleF}{Co-directeur de thèse}

%%% Membre n°7 (Invité) / Member n°7 (Guest)
\newcommand{\jurynameG}{Prénom Nom}
\newcommand{\juryadressG}{Statut, Établissement (Unité de recherche)}
\newcommand{\juryroleG}{Invité}

%%% Membre n°8 (Invité) / Member n°8 (Guest)
\newcommand{\jurynameH}{Prénom Nom}
\newcommand{\juryadressH}{Statut, Établissement (Unité de recherche)}
\newcommand{\juryroleH}{Invité}

%% Il est possible d'ajouter des membres supplémentaires selon le même modèle / More jury members can be added according to the same model

\label{layout}
%%%%%%%%%%%%%%%%%%%%%%%%%%%%%%%%%%%%%%%%%%%%%%%%%%%%%%%%%%%%%%%%%%%%%%%%%%%%%%%%%%%%%%%%%%%%%%%%%%%%%%%%%%%%%%%%%%%%%%%%%%%%%%%%%%%%%%%%%%%%%%%%%%%%%%%%%%%%%%%%%%%%%%%
%%%%%%%%%%%%%%%%%%%%%%%%%%%%%%%%%%%%%%%%%%%%%%%%%%%%%%%%%%%%%%%%%%%%%%%%%%%%%%%%%%%%%%%%%%%%%%%%%%%%%%%%%%%%%%%%%%%%%%%%%%%%%%%%%%%%%%%%%%%%%%%%%%%%%%%%%%%%%%%%%%%%%%%
%%% Mise en page / Page layout      
%%% NE RIEN MODIFIER EXCEPTÉ LA PARTIE CONCERNANT LE JURY (voir \label{jury}) SI BESOIN / DO NOT MODIFY EXCEPT SECTION CONCERNING JURY (see \label{jury}) IF NEEDED
%%%%%%%%%%%%%%%%%%%%%%%%%%%%%%%%%%%%%%%%%%%%%%%%%%%%%%%%%%%%%%%%%%%%%%%%%%%%%%%%%%%%%%%%%%%%%%%%%%%%%%%%%%%%%%%%%%%%%%%%%%%%%%%%%%%%%%%%%%%%%%%%%%%%%%%%%%%%%%%%%%%%%%%
%%%%%%%%%%%%%%%%%%%%%%%%%%%%%%%%%%%%%%%%%%%%%%%%%%%%%%%%%%%%%%%%%%%%%%%%%%%%%%%%%%%%%%%%%%%%%%%%%%%%%%%%%%%%%%%%%%%%%%%%%%%%%%%%%%%%%%%%%%%%%%%%%%%%%%%%%%%%%%%%%%%%%%%

% Méta-données du PDF / PDF meta-datas
\hypersetup{
	pdfauthor={\PhDname},
	pdfsubject={Manuscrit de thèse de doctorat},
	pdftitle={\PhDTitle},
}

\begin{titlingpage}

\thispagestyle{empty}

%\color{bordeau} \hfill \vfill \tiny \ecodocnum
\begin{textblock}{5}(-3.7,-4.5)
	%\vspace{10mm}
	\includegraphics[scale=0.7]{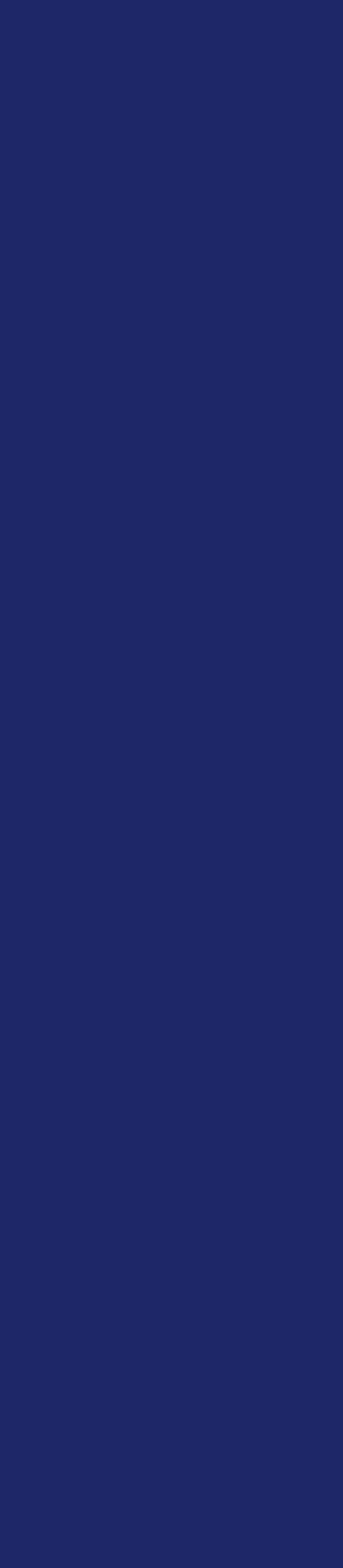}%width=3cm, height=1\paperheight
\end{textblock}

\begin{textblock}{5}(-2,-1.7)
	%\vspace{10mm}
	\includegraphics[scale=0.35]{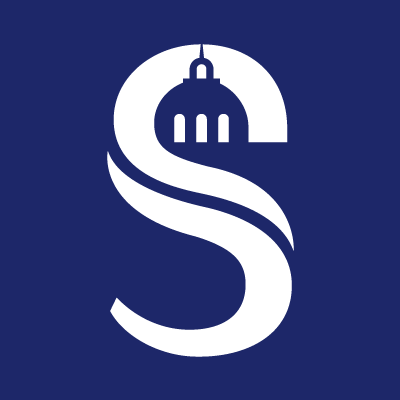}
\end{textblock}

\begin{textblock}{1}(-0.5,4.9)
	\huge{\rotatebox{90}{\color{white}{\fontsize{38}{54}\selectfont Sorbonne Université}}}
\end{textblock}

%\vspace{6cm}
%% Texte
\begin{textblock}{11}(2.8,0)
	\textblockcolour{white}
	
	%\begin{center}  
	\begin{center}
		\fontsize{1cm}{1.1cm}\selectfont{\textcolor{marine}{\PhDTitle}} \bigskip %% Titre de la thèse 
		\color{black} %% Couleur noire du reste du texte
        \vfill
		\vspace{0.6cm}
		\Large{\textsc{\PhDname}\\ \vspace{5.5mm}} %% Nom du docteur
		\vfill
        \normalsize{PhD Thesis - Sorbonne Université.\\
        CIFRE Funding - Collaboration between LIP6 and Naval Group \\
        Ecole Doctorale Informatique, Télécommunications et Electronique (n°\:572).\\ Computer Science.\\
        %%%%%%%%%%%%%%%%%%%%%%
        %%% Naval Group IP %%%
        %\vspace{5mm}
        %{\tiny ©Naval Group SA All rights reserved 2025. Both the content and the form of this document/software are the property of Naval Group SA and/or of third party. It is formally prohibited to use, copy, modify, translate, disclose or perform all or part of this document/software without obtaining Naval Group SA's prior written consent or authorization. Any such unauthorized use, copying, modification, translation, disclosure or performance by any means whatsoever shall constitute an infringement punishable by criminal or civil law and, more generally, a breach of Naval Group SA's rights.}  \\
        %%%%%%%%%%%%%%%%%%%%%%
        \vspace{11mm}
        \normalsize PhD defence held publicly on November 26th, 2025, \\ with the following thesis committee:\\
        }
        
    \vspace{2.0cm}

    \begin{minipage}{1\textwidth}  
        \begin{multicols}{2}
        \small
        \textbf{Oleksandr Kyriienko}: Rapporteur,\\
        \textit{Professor in Quantum Technologies, University of Sheffield, United-Kingdom}\\
        \textbf{Marco Cerezo}: Examinateur,\\
        \textit{Staff Scientist, Los Alamos National Laboratory, United States of America}\\
        \textbf{William Clements}: Examinateur,\\
        \textit{Head Of Machine Learning, ORCA Computing, United Kingdom}\\
        \textbf{Alex B. Grilo}: Co-Directeur de Thèse,\\
        \textit{Chargé de Recherche CNRS, LIP6, Sorbonne Université, France}\\
        \textbf{Romain Kukla:} Invité\\
        \textit{Ingénieur de Recherche, Naval Group, France}
        
        \bigskip
        
        \bigskip
        
        \bigskip
        \bigskip
        \textbf{Iordanis Kerenidis}: Rapporteur,\\
        \textit{Directeur de Recherche CNRS, IRIF, Université Paris Cité, France}\\
        \textbf{Mehrnoosh Sadrzadeh}: Examinatrice,\\
        \textit{Professor of Computer Science, University College London, United Kingdom}\\
        \textbf{Elham Kashefi}: Directrice de Thèse,\\
        \textit{Directrice de Recherche CNRS, LIP6, Sorbonne Université, France, and Professor, University of Edinburgh, United Kingdom}\\
        \textbf{Mathilde Portais}: Co-Encadrante de Thèse,\\
        \textit{Ingénieure de Recherche, Naval Group, France}\\
        \end{multicols}
    \end{minipage}
		\bigskip
        \bigskip
        \bigskip
        \bigskip

    \begin{figure*}[htbp]
      \centering
      \setlength{\tabcolsep}{1em} % spacing between images
      \begin{tabular}{c c c}
        \raisebox{-0.5\height}{\includegraphics[height=4cm]{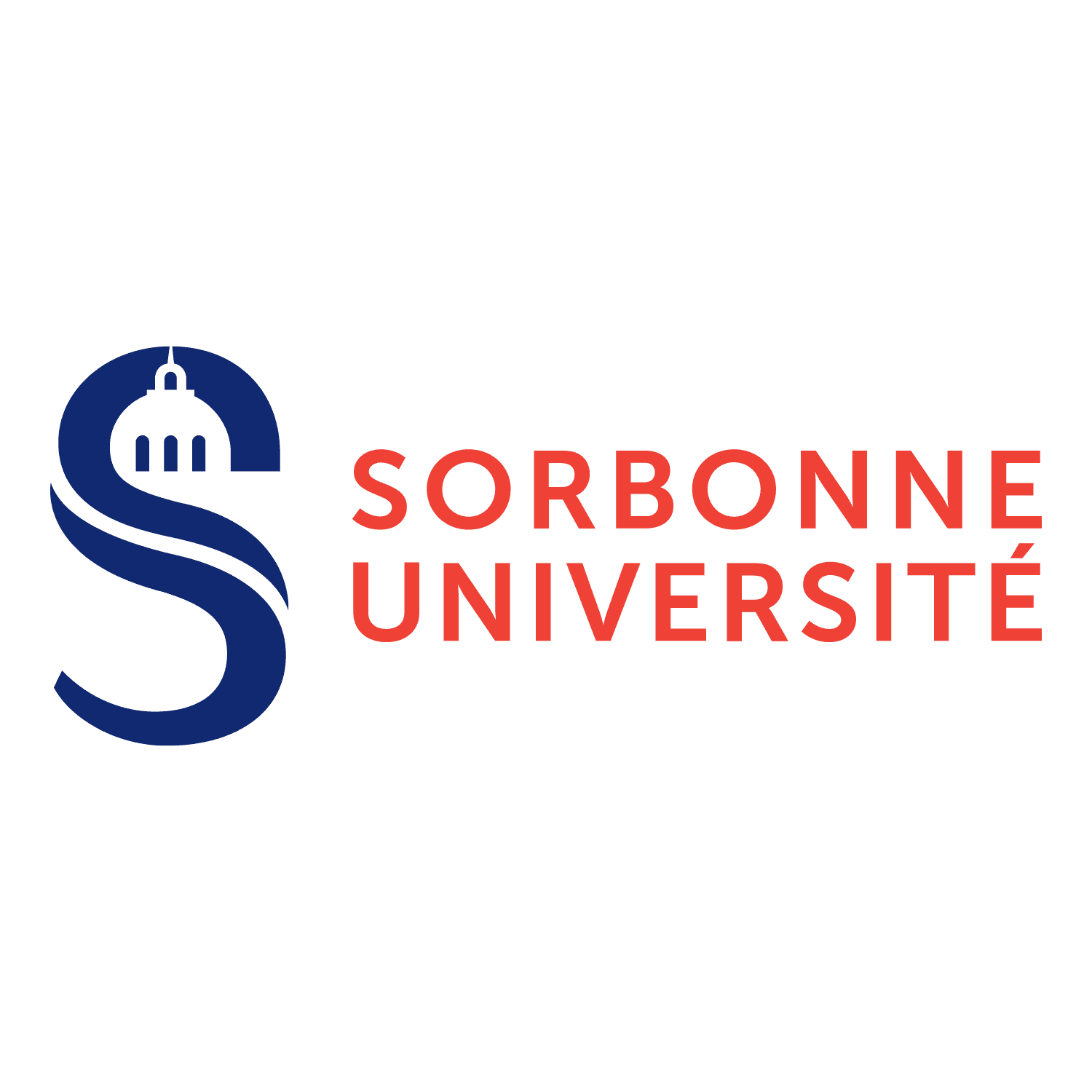}} 
        \hspace{1cm}
        \raisebox{-0.5\height}{\includegraphics[height=2cm]{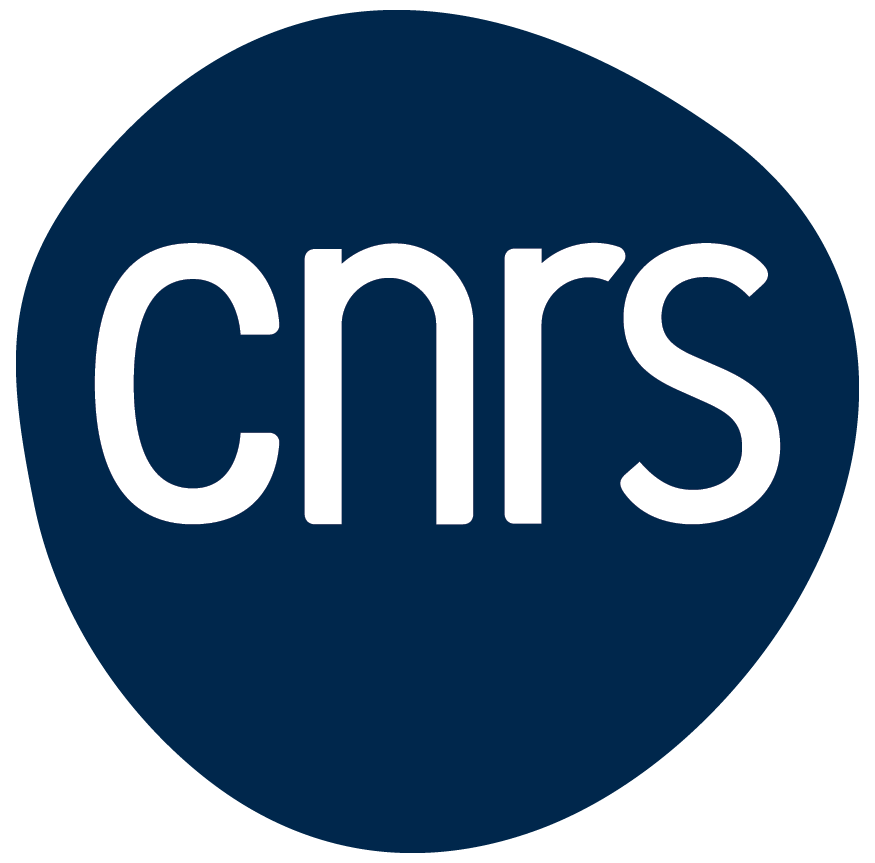}}
        \hspace{1cm}
        \raisebox{-0.5\height}{\includegraphics[height=1.2cm]{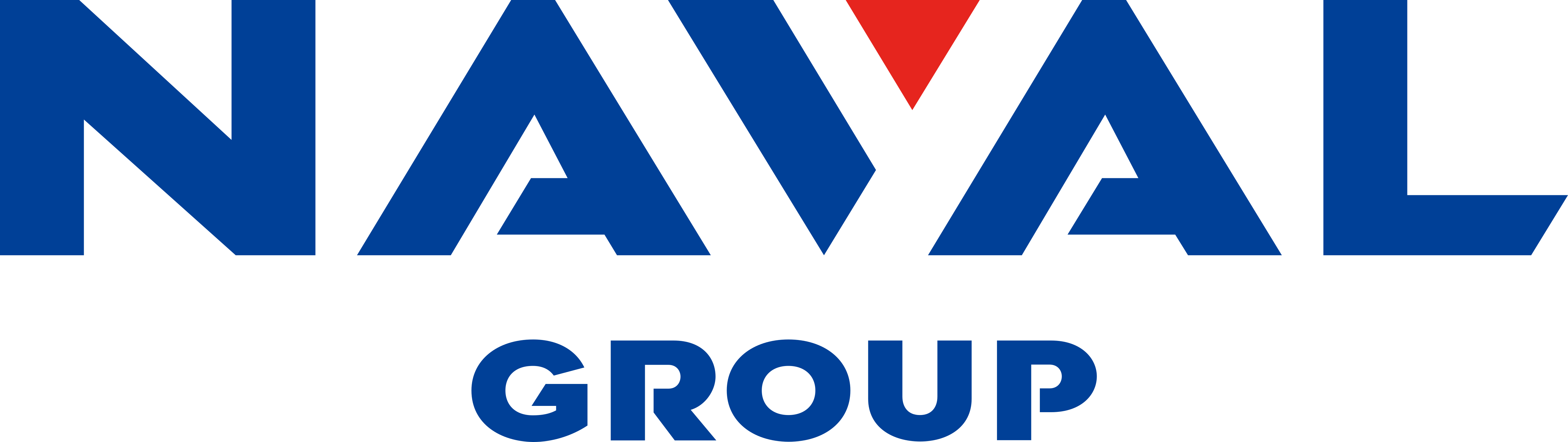}}
      \end{tabular}
      %\caption{}
    \end{figure*}

    \end{center}
\end{textblock}
\end{titlingpage}

%% file: frontmatter/title.tex
%
% File: Title.tex
% Author: V?ctor Bre?a-Medina
% Description: Contains the title page
%
% UoB guidelines:
% 
% At the top of the title page, within the margins, the dissertation should give the title and, if 
% necessary, sub-title and volume number. If the dissertation is in a language other than English, the 
% title must be given in that language and in English. The full name of the author should be in the 
% centre of the page. At the bottom centre should be the words ?A dissertation submitted to the 
% University of Bristol in accordance with the requirements for award of the degree of ? in the 
% Faculty of ...?, with the name of the school and month and year of submission. The word count of 
% the dissertation (text only) should be entered at the bottom right-hand side of the page.
%
%
\begin{titlingpage}
\begin{SingleSpace}
\calccentering{\unitlength} 
\begin{adjustwidth*}{\unitlength}{-\unitlength}
\begin{center}
\rule[0.5ex]{\linewidth}{2pt}\vspace*{-\baselineskip}\vspace*{3.2pt}
\rule[0.5ex]{\linewidth}{1pt}\\[\baselineskip]
{\Huge \PhDTitle }\\[6mm] %\\ \vspace{0.2cm} \\ \vspace{0.4cm} 
\rule[0.5ex]{\linewidth}{1pt}\vspace*{-\baselineskip}\vspace{3.2pt}
\rule[0.5ex]{\linewidth}{2pt}\\
\vspace{7.5mm}
{\large\textsc{Léo Monbroussou}}\\
\vspace{7.5mm}
{Thèse de Doctorat de Sorbonne Université.\\
Ecole Doctorale Informatique, Télécommunications et Electronique (n°\:572).\\ Specialité Informatique.\\
\vspace{11mm}
\normalsize Thèse présentée et soutenue à Paris le 26 Novembre 2025,\\
en présence du jury suivant:\\
}
\vspace{8mm}
% \begin{minipage}{10cm}
% A dissertation submitted to Télécom Paristech in accordance with the requirements of the degree of \textsc{Doctor of Quantum Information} in the INFRES Department.
% \end{minipage}\\
    \begin{minipage}{1\textwidth}  
        \begin{multicols}{2}
        \small
        \textbf{Oleksandr Kyriienko}: Rapporteur,\\
        \textit{Professeur des Technologies Quantiques, Université de Sheffield, Royaume-Uni}\\
        \textbf{Marco Cerezo}: Examinateur,\\
        \textit{Chercheur, Laboratoire National de Los Alamos, États-Unis d'Amérique}\\
        \textbf{William Clements}: Examinateur,\\
        \textit{Directeur du Machine Learning, ORCA Computing, Royaume-Uni}\\
        \textbf{Alex B. Grilo}: Co-Directeur de Thèse,\\
        \textit{Chargé de Recherche CNRS, LIP6, Sorbonne Université, France}\\
        \textbf{Romain Kukla:} Invité\\
        \textit{Ingénieur de Recherche, Naval Group, France}
        
        \bigskip
        
        \bigskip
        
        \bigskip
        \bigskip
        \textbf{Iordanis Kerenidis}: Rapporteur,\\
        \textit{Directeur de Recherche CNRS, IRIF, Université Paris Cité, France}\\
        \textbf{Mehrnoosh Sadrzadeh}: Examinatrice,\\
        \textit{Professeur en Informatique, University College London, Royaume-Uni}\\
        \textbf{Elham Kashefi}: Directrice de Thèse,\\
        \textit{Directrice de Recherche CNRS, LIP6, Sorbonne Université, France, et Professeur, Université d'Edimbourg, Royaume-Uni}\\
        \textbf{Mathilde Portais}: Co-Encadrante de Thèse,\\
        \textit{Ingénieure de Recherche, Naval Group, France}\\
        \end{multicols}
    \end{minipage}

% \textbf{M. Nicolas Fabre}: Examinateur,\\
% \textit{Associate Professor, Telecom Paris, France.}\\
%\vspace{13mm}
%{ }
%\vspace{12mm}
\end{center}
%\begin{flushright}
%{\small Word count: ten thousand and four}
%\end{flushright}
\end{adjustwidth*}
\end{SingleSpace}
\global\let\clearpagegood\clearpage
\global\let\clearpage\relax
\end{titlingpage}
\global\let\clearpage\clearpagegood

%% file: frontmatter/Acknowledgements.tex
%
% file: dedication.tex
% author: V?ctor Bre?a-Medina
% description: Contains the text for thesis dedication
%

\chapter*{Acknowledgements}
\thispagestyle{empty}
\begin{SingleSpace}
\initial{I} would like to start by thanking the jury for agreeing to review my work, and for the thoughtful feedback and comments they provided. I am proud to have your names on my thesis, each of you has been a genuine source of inspiration to me as researchers in this field. This thesis would never have been possible without my supervisors. Alex, you have been a constant source of support; your kindness and dedication made these years unforgettable. Elham, no words feel adequate. Your endless optimism, your vision for our research, your care for everyone around you, your team spirit, every effort I noticed, and all those I did not. I am endlessly grateful, and I hope to keep learning from you. Jonas Landman also deserves his place here: your supervision and leadership shaped these years profoundly. I would also like to thank Mathilde and Naval Group for the industrial supervision and their broader commitment to the quantum ecosystem. \\

Throughout this thesis, I had the great honour of collaborating with many outstanding researchers. In particular, I want to acknowledge the LIP6 QML team. Slimane, working with you was both joyful and formative. Eliott, your hard work, humour, and dedication are irreplaceable. I am confident machines won't be able to replace you. Hela, it was a privilege to work alongside you: your scientific rigour and integrity are something I deeply admire. Hugo, you are the other star of the team; I cherish our conversations at the lab. Snehal, you are a phenomenal co-worker. You fight for your ideas with passion, and you are super creative.  Verena, I think everyone who has met you would agree you are kindness itself. You are a passionate, hard worker, super fun researcher, and you have been a constant inspiration. You clearly are the main reason why photonic QML is so cool. I also warmly thank Tigran, Armando, Constantin, Yidong, and Letao, whose internship was a pleasure to supervised. I was lucky to meet the Scottish branch of the team on several occasions, and I heartily thank Raul, Craig, Brian, Caitlin, James, Abbas, Marine, Stuart, Chirag, Sean, Ioannis. A very special thanks to Mina and Ramin for your guidance, kindness and generosity throughout these years.\\

I am deeply grateful to the entire LIP6 quantum information group,  past and present members, for fostering such a stimulating and welcoming environment. A special thanks to the permanent members Marco, Yoann, Fred, Jessica, Damian, and Eleni. \\

Collaborative projects have been among the most enriching parts of this thesis. I am particularly grateful to Ulysse Chabaud, whose joyful generosity towards the whole community has meant so much to me. The PHOQUSING project brought me into contact with amazing italian teams: the brilliant Beatrice Polacchi, Fabio Sciarrino, Taira Giordani, Eugenio Caruccio, Giovanni Rodari, Francesco Hoch, Gonzalo Carvacho, Nicolò Spagnolo, Mattia Bossi, Abhiram Rajan, Niki Di Giano, Riccardo Albiero, Francesco Ceccarelli, and Roberto Osellame. Visiting part of the team at La Sapienza was an honour, and I learned enormously from all of you. \\

These years also brought me into contact with many talented researchers I wish to acknowledge: Naomie Chmielewski, Ulysse Rémond, Joseph Michael, Gerard Milburn, Raj Patel, Danijela Marković, Pierre-Emmanuel Emeriau, Daphne Wang, Ariane Soret, Nicolas Heurtel, Zoë Holmes, Vincent Danos, Bo Yang, Dominik Leichtle, Ross Grassie, André Ferreira-Martins, Renato Farias, and the wonderful Natansh Mathur.\\

J'aimerais maintenant remercier mes proches, dont le soutien a compté énormément. Je suis profondément reconnaissant envers Blandine et sa famille, à qui je dois tant. Merci à Sylvie et William Leduc pour leur générosité et leur soutien tout au long de mes études. Ce paragraphe ne peut rendre justice à toutes les amitiés précieuses de ces années, je m'excuse d'avance pour les oublis inévitables. Merci à Manon, Max, Cléments, Rachel, Juliette, Maureen, Charly, Uruk. À mes musiciens préférés, Olivier et Julien. À Lise et Antoine pour leur amitié et leur curiosité. À Maxime et Auriane, je vous dédie respectivement les sections 5.2 et 2.2.1. À Bastien, si bon compagnon de route. À Marie, qui j'espère sait à quel point je lui suis reconnaissant pour son amitié et sa gentillesse. À mon frère Emmanuel, à ma sœur Manon, et à Ysaline. \\

Enfin, et surtout, merci beaucoup Myriam.

\end{SingleSpace}
% \clearpage

%% file: frontmatter/abstract.tex
%
% File: abstract.tex
% Author: V?ctor Bre?a-Medina
% Description: Contains the text for thesis abstract
%
% UoB guidelines:
%
% Each copy must include an abstract or summary of the dissertation in not
% more than 300 words, on one side of A4, which should be single-spaced in a
% font size in the range 10 to 12. If the dissertation is in a language other
% than English, an abstract in that language and an abstract in English must
% be included.
\chapter*{Abstract (English)}
\begin{SingleSpace}
\thispagestyle{empty}

\initial{R}ecent advances in Machine Learning have transformed numerous industrial sectors, yet classical paradigms face fundamental limitations: rapidly growing data volumes, rising computational costs, significant energy consumption, and the physical scaling limits of conventional hardware architectures. Quantum computing has emerged as a promising computational paradigm to address these challenges, giving rise to the field of Quantum Machine Learning (QML). In this thesis, the theoretical foundations of QML are investigated, with a focus on near-term and future practical applications. Three central challenges are addressed: the trainability of variational quantum circuits, their expressivity, and their resistance to efficient classical simulation. The trainability of Hamming-weight preserving variational quantum circuits is first studied, and theoretical guarantees are established that resolve an open conjecture on the absence of barren plateaus for this circuit family. Subspace-preserving QML algorithms are then introduced, including photonic circuits and quantum convolutional neural networks, and are designed to mimic classical ML subroutines while offering polynomial quantum advantage. Finally, variational quantum circuits are analyzed as quantum Fourier models, and a framework is derived to jointly characterize expressivity and trainability, from which conditions are obtained under which quantum models provably separate from their classical counterparts. These contributions are intended to advance the theoretical roadmap for harnessing near-term and future quantum technologies in real-world applications.
\end{SingleSpace}

% \global\let\clearpagegood\clearpage
\global\let\clearpage\relax
\newpage
\chapter*{Abstract (Français)}
\global\let\clearpage\clearpagegood
\thispagestyle{empty}
\begin{SingleSpace}

\initial{L}es récentes avancées en apprentissage automatique ont transformé de nombreux secteurs industriels, mais les paradigmes classiques se heurtent à des limites fondamentales~: volumes de données croissants, coûts de calcul élevés, consommation énergétique importante, et contraintes physiques sur le passage à l'échelle des architectures matérielles conventionnelles. L'informatique quantique a émergé comme un paradigme computationnel prometteur pour dépasser ces limites, donnant naissance au domaine de l'apprentissage automatique quantique (QML). Dans cette thèse, les fondements théoriques du QML sont étudiés, avec un accent sur les applications pratiques à court et moyen terme. Trois défis centraux sont abordés~: la capacité d'entraînement des circuits quantiques variationnels, leur expressivité, et leur résistance à aux méthodes de simulation et d'approximation classiques. La capacité d'entraînement des circuits quantiques variationnels préservant le poids de Hamming est tout d'abord étudiée, et des garanties théoriques sont établies, résolvant une conjecture ouverte sur l'absence de Barren Plateaux pour cette famille de circuits. Des algorithmes QML préservant les sous-espaces sont ensuite introduits, incluant des circuits photoniques et des réseaux de neurones convolutifs quantiques, conçus pour reproduire des sous-routines d'apprentissage classiques tout en offrant un avantage quantique polynomial. Enfin, les circuits quantiques variationnels sont analysés comme des modèles de Fourier quantiques, et un cadre théorique est dérivé pour caractériser conjointement expressivité et entraînabilité, à partir duquel des conditions sont obtenues sous lesquelles les modèles quantiques se séparent provablement de leurs homologues classiques. Ces contributions visent à faire progresser la communauté scientifique pour l'exploitation des technologies quantiques à court et moyen terme dans des applications industrielles.
\end{SingleSpace}

%% file: frontmatter/Resume.tex
\chapter*{Résumé en Français}

\initial{L}es récentes avancées en apprentissage automatique ont transformé de nombreux secteurs industriels~: santé, finance, logistique, fabrication, découverte de matériaux, en permettant la modélisation prédictive, l'optimisation et l'automatisation à une échelle sans précédent. Cependant, le paradigme classique de l'apprentissage automatique se heurte à des limites fondamentales~: croissance rapide des volumes de données, coûts de calcul croissants, consommation énergétique considérable des grands modèles, et limites physiques des architectures matérielles conventionnelles. Ces défis motivent l'exploration de nouveaux paradigmes computationnels capables d'étendre les capacités de l'apprentissage automatique classique et de fournir un avantage concurrentiel dans des applications réelles. \\

L'informatique quantique a émergé comme un candidat prometteur pour dépasser ces limites. Bien que Richard Feynman ait initialement imaginé son usage dans les années 1980 pour accélérer la simulation de systèmes physiques quantiques, des propositions plus récentes ont démontré des algorithmes promettant des accélérations exponentielles pour des problèmes pertinents pour l'industrie. Dans le contexte de l'apprentissage automatique quantique (QML), l'algorithme HHL pour la résolution de systèmes d'équations linéaires en est un exemple célèbre. Cependant, cet algorithme repose sur des hypothèses mathématiques très restrictives et des exigences matérielles difficiles à garantir en pratique, illustrant l'importance de lire attentivement les conditions d'application de tels résultats théoriques. Bien que des algorithmes QML pour ordinateurs quantiques tolérants aux fautes aient été proposés pour plusieurs applications
importantes, des efforts considérables restent nécessaires pour concevoir des méthodes compatibles avec le matériel actuel.

Outre les avancées technologiques côté hardware nécessaires à l'émergence de l'usage de l'informatique quantique, les méthodes computationnelles proposées sont source de débat quant à leurs capacité à dépasser les capacité du calcul classique. Des travaux ont montré que des modèles classiques peuvent parfois égaler les performances d'algorithmes quantiques qui ont initialement été décrits comme plus performants.  \\

Les algorithmes quantiques variationnels ont émergé comme l'une des approches les plus prometteuses pour le court terme, en employant  une approche hybride et des optimiseurs classiques pour entraîner des circuits quantiques paramétrés, les rendant naturellement adaptés aux contraintes des dispositifs NISQ. Malgré leur promesse, la communauté QML fait face à trois défis majeurs interdépendants~: garantir la convergence des méthodes entraînant des circuits quantiques variationnels, contrainte par le phénomène de gradient évanescent dit \textit{Barren Plateau}~; garantir l'expressivité, c'est-à-dire donner des garanties fortes quant à l'intéret des méthodes lors d'un passage à l'échelle, pour laquelle des métriques ont été proposées mais dont une valeur élevée conduit souvent à des problèmes d'entraînabilité~; et la résistance à la simulation classique efficace, qui conditionne l'existence d'un véritable avantage quantique. En l'absence d'une recette générale permettant de concevoir des méthodes quantiques provablement utiles à l'échelle industrielle, cette thèse propose d'aborder plusieurs de ces défis théoriques. \\
 
Le chapitre des préliminaires présente les notions mathématiques fondamentales nécessaires à la compréhension des résultats. Une mesure d'expressivité couramment utilisée pour un circuit quantique variationnel est la proximité de la distribution des matrices unitaires générées à la mesure de Haar, caractérisée par la notion de $2$-\textit{design}. Sous cette hypothèse, le modèle quantique présente un phénomène de concentration exponentielle~: sa variance est inversement proportionnelle à la dimension de l'espace mathématique qui définit le circuit, conduisant aux Barren Plateaux. Les circuits préservant les sous-espaces sont ensuite présentés, ainsi que leurs propriétés d'entraînabilité et de simulation classique. En particulier, l'algèbre de Lie dynamique (DLA) caractérise le groupe de Lie associé aux matrices unitaires atteignables~: si la DLA est de dimension polynomiale, le circuit peut être simulé classiquement en temps polynomial. Une conjecture préexistante dans la littérature proposait que la variance du gradient de la fonction de coût soit inversement proportionnelle à la dimension de la DLA~; cette conjecture est invalidée dans le chapitre suivant. \\
 
Le troisième chapitre est consacré à l'étude de l'entraînabilité et de la contrôlabilité de circuits quantiques variationnels préservant le poids de Hamming (HW). Ces circuits utilisent des portes à deux qubits, en particulier la porte \textit{Reconfigurable Beam Splitter} (RBS) et la porte \textit{Fermionic Beam Splitter} (FBS), qui préservent les sous-espaces engendrés par les états de base à poids de Hamming fixé. La porte FBS présente une contrôlabilité inférieure à la porte RBS~: le comportement du circuit dans un sous-espace à HW~$k > 1$ est entièrement déterminé par celui dans le sous espace HW~$1$, la dimension maximale de la DLA étant bornée par $n(n-1)/2$. La notion d'encodeur de données quantique préservant le HW est introduite, réalisant un encodage d'amplitude dans le sous-espace de HW~$k$. La contrôlabilité du circuit est étudiée à l'aide de la Matrice d'Information de Fisher Quantique (QFIM), dont le rang maximal est une métrique de contrôlabilité dans l'espace des états. Deux algorithmes sont proposés pour concevoir de tels encodeurs~: l'un procède par ajout itératif de portes jusqu'à saturation du rang de la QFIM, l'autre par réduction depuis un circuit surparamétrisé. Le résultat central de ce chapitre établit, sans aucune hypothèse de $2$-design, une expression analytique exacte de la variance du gradient de la fonction de coût pour les circuits RBS et FBS. Il est démontré que si les portes sont organisées en une structure périodique connectée (\textit{Connected Periodic Structure Ansatz}), alors après un nombre polynomial de répétitions, la variance du gradient décroît avec la dimension de l'espace de Hilbert associé à un HW fixé. Ce résultat démontre l'absence de Barren Plateau pour ces circuits lorsque le HW est fixé, et invalide la conjecture mentionnée~: la variance du gradient dépend de la dimension du sous-espace de Hilbert et non de la dimension de la DLA. Par ailleurs, bien que le régime où ces garanties théoriques sont obtenues soit classiquement simulable en temps polynomial, un avantage polynomial reste envisageable, notamment pour des plateformes photoniques à haut taux de répétition. \\
 
Le quatrième chapitre s'intéresse aux circuits d'optique quantique linéaire, plateformes particulièrement prometteuses pour le calcul quantique à court terme en raison de leur taux de répétition élevé et de leurs propriétés naturelles de préservation du nombre de photons. Un réseau optique linéaire sur $m$ modes est caractérisé par l'homomorphisme photonique $\varphi : SU(m) \to SU(d_n)$, qui relie l'unitaire d'évolution à un photon $W^1 \in SU(m)$ à l'unitaire d'évolution à $n$ photons $W^n = \varphi(W^1)$ via des permanents de matrices. Dans ce chapitre, une attention particulière est accordée au contrôle de ces circuits photoniques~: la dimension de l'ensemble des matrices unitaires accessibles par un circuit d'optique linéaire est bornée par $\min(p, m^2-1)$, où $p$ est le nombre de portes paramétrées. Cette limitation de contrôlabilité est une contrainte sur l'expressivité du modèle de sortie. Pour dépasser cette limite, un nouveau schéma d'\textit{injection d'état} (SI) est proposé~: il consiste à mesurer un ou plusieurs modes entre deux blocs d'optique linéaire et à ré-injecter des photons selon le résultat de la mesure, sans nécessiter de reconfiguration en temps réel du circuit quantique. Ce schéma se distingue du schéma d'optique linéaire adaptative par l'absence de reconfiguration des unitaires en temps réel et par un délai optique considérablement réduit dans le cas d'utilisation de commutateur électronique ou piezo-électrique en comparaison à la reconfiguration de portes quantiques basés sur des effets thermiques. Est ensuite étudié la  pureté de l'état quantique de sortie après plusieurs couches d'injection. Des bornes inférieures sont offertes pour l'évolution de cette pureté en utilisant différentes hypothèses, notamment le régime de non-collision. Le lien entre la pureté et la capacité à distinguer deux états de sorties différents est présenté, avec la notion introduite de distinguabilité des états de sortie. Enfin, il est montré que l'optique linéaire augmentée d'injection d'état permet de générer des probabilités de sortie qui sont conjecturées comme difficiles à calculer classiquement, en échappant aux algorithmes classiques permettant une estimation efficacaes de ces probabilités. \\
 
Le cinquième chapitre présente de nouveaux algorithmes QML basés sur les circuits préservant les sous-espaces, conçus pour remplacer des sous-routines d'apprentissage automatique classiques tout en offrant un avantage quantique polynomial. Une architecture de réseau de neurones convolutif quantique préservant le poids de Hamming est proposée, analogue au CNN classique~: une couche de convolution par filtre RBS, une couche de pooling par portes CNOT et mesures réalisant une non-linéarité similaire à un average pooling tout en préservant la structure tensorielle de l'état, et une couche dense orthogonale réalisée par un réseau de neurones orthogonal via des portes RBS. Des simulations sur les jeux de données MNIST, Fashion-MNIST et CIFAR-10 montrent que l'architecture QCNN atteint des performances comparables au CNN classique avec un nombre réduit de paramètres (755 contre 990), et dépasse même le CNN classique sur MNIST et Fashion-MNIST. Une version photonique de cette architecture exploite l'encodage tensoriel sur des états de Fock~: une image classique est encodée dans une superposition d'états à un photon par registre, la couche de convolution est réalisée à l'aide de BeamSplitters, et la couche de pooling est réalisée grâce au protocole d'injection d'état présenté dans le chapitre précédent. Cette architecture a été validée expérimentalement sur une plateforme photonique hybride, comprenant une source de photons uniques à boîte quantique semi-conductrice, deux circuits intégrés programmables à 8 et 12 modes, et des détecteurs supraconducteurs à nanofils. Les résultats de classification sur des jeux d'images de barres et de rayures ($4 \times 4$ pixels) et MNIST ($8 \times 8$ pixels) donnent des précisions comparables aux résultats de référence pour les architectures QCNN tolérantes aux fautes. \\
 
Le sixième chapitre étudie l'expressivité des circuits quantiques variationnels à travers le prisme des modèles de Fourier. Il est établi que tout circuit variationnel qui encode des données classiques en paramétrant des hamiltoniens avec ces données peut être décrit comme une série de Fourier dans l'entrée classique~: $f(x, \theta) = \sum_{\omega \in \Omega} c_\omega(\theta) e^{i\omega^T x}$, où le spectre $\Omega$ est déterminé par les valeurs propres des hamiltoniens d'encodage et les coefficients de Fourier dépendent principalement des couches entraînables. Une notion cruciale est introduite~: la redondance $|R(\omega)|$ d'une fréquence $\omega$, définie comme le nombre de paires de chemins dans l'arbre spectral quantique générant cette fréquence. Sous l'hypothèse que chaque couche entraînable forme un $2$-design exact, la variance du coefficient de Fourier $c_\omega$ est proportionnelle à la redondance normalisée $\widetilde{|R(\omega)|} := |R(\omega)|/d^2$~: les fréquences à haute redondance présentent une variance relativement plus grande, tandis que celles à faible redondance sont exponentiellement concentrées. Ce résultat est étendu au cas de l'hypothèse d'approximate $2$-design~: la variance reste bornée par un polynôme en $\widetilde{|R(\omega)|}$ et en $\varepsilon$, démontrant que le phénomène d'expressivité évanescente, situation où la variance de certains coefficients de Fourier décroît exponentiellement, peut persister au-delà de l'hypothèse exacte de $2$-design. Le cas d'une architecture constituée de blocs locaux est également traité, permettant de capturer l'interaction entre un observable local et le circuit. Il est démontré que l'expressivité évanescente est conceptuellement distincte de la concentration du modèle entier~: des régimes existent où les coefficients de Fourier sont exponentiellement concentrés sans que le modèle global le soit. Un résultat complémentaire établit la borne $\sum_{\omega} |c_\omega(\theta)|^2 \leq \|O\|_\infty^2$ pour tout paramètre $\theta$. Des simulations numériques confirment que les fréquences à faible redondance sont plus difficiles à atteindre lors de l'entraînement, étayant l'intuition que l'expressivité évanescente affecte la convergence pratique du modèle. \\
 
Le septième chapitre étudie les conditions sous lesquelles les modèles QML peuvent éviter l'approximation classique basée sur une méthode de machine learning classique appelée "Random Fourier Features". Les VQCs sont des modèles linéaires dans un espace de caractéristiques~: $f_Q(x) = \beta_Q^\top \phi(x)$. Dans le régime surparamétré ($p > M$), la descente de gradient classique converge vers l'estimateur de norme minimale (\textit{Minimum Norm Least Squares}, MNLS) parmi les solutions qui minimisent le risque empirique. Contrairement au cas classique, le vecteur de poids d'un VQC ne reste pas dans l'espace engendré par les données d'entraînement et ne converge pas vers le MNLS~; c'est cette distinction qui constitue la condition nécessaire pour un avantage quantique. Un modèle quantique interpolant $f_Q$ présente un avantage quantique potentiel si son vecteur de poids est de norme très supérieure à celle du MNLS. Cette condition est analysée pour plusieurs cas d'étude, dans le cadre de données continues comme discrètes. Sont notamment considérés des modèles quantiques associées à des fonctions difficiles à réaliser classiquement et d'importance pour des problèmes de cryptographie, comme le logarithme discret. La dequantisation est impossible en cohérence avec la complexité supposée du problème. Les liens entre la norme du vecteur de poids et la concentration du modèle sont également discutés. Les exemples développés avec des entrées continues réalisent la condition nécessaire pour éviter la méthode d'approximation classique, mais au prix d'une concentration des modèles qui empêche leurs correctes entraînements. Un résultat constructif établit l'existence de familles de modèles de Fourier avec un grand vecteur de poids, non concentrés et bornés, dont la réalisation par un circuit quantique constitue une question ouverte centrale. \\
 
Cette thèse adopte une approche pragmatique face au paradoxe fondamental du QML~: les régimes où des garanties théoriques rigoureuses peuvent être obtenues (absence de Barren Plateau, entraînabilité prouvée) sont souvent classiquement simulables en temps polynomial, remettant en cause l'avantage exponentiel initialement recherché. Plutôt que de viser un avantage exponentiel, la thèse s'intéresse à exploiter un avantage polynomial important, combiné à un matériel approprié comme des plateformes photoniques à haut taux de répétition, afin de rendre l'informatique quantique pratiquement utile, en particulier pour accélérer des blocs de construction fondamentaux de l'intelligence artificielle. Ces travaux, majoritairement théoriques, contribuent à faire progresser la communauté scientifique pour l'exploitation des technologies quantiques à court et moyen terme dans des applications industrielles réelles.

%% file: chapters/01_Introduction/Introduction.tex
\let\textcircled=\pgftextcircled
\chapter{Introduction}

%%%%%%%%%%%%%%%%%%%%%%%    
    \section{QML for Industrial Applications}

Over the past decades, advances in Machine Learning (ML) have transformed a wide range of industries, enabling predictive modeling, optimization, and automation at unprecedented scales. From healthcare and finance to logistics, manufacturing, and materials discovery, data-driven methods are increasingly central to decision-making and innovation. However, the classical paradigm of ML faces fundamental challenges that vary depending on the use case: the rapid growth of data volumes, the rising computational cost of training complex models, concerns over security and privacy, the significant energy consumption of large-scale ML models, and the physical limits of conventional hardware architectures in scaling to ever larger problem sizes. These challenges motivate the exploration of new computational paradigms that can extend the capabilities of classical ML and provide a competitive advantage in real-world applications.

Quantum computing has emerged as a promising candidate to address these challenges. While Richard Feynman originally advocated its use in the 1980s for accelerating the simulation of quantum physics and chemistry, more recent proposals have demonstrated algorithms that promise exponential speedups over classical methods for problems highly relevant to industry. In the context of Quantum Machine Learning (QML), one of the most famous examples is the HHL algorithm (named after Aram Harrow, Avinatan Hassidim, and Seth Lloyd) for solving systems of linear equations in logarithmic time \cite{harrow_quantum_2009}. However, this algorithm relies on stringent resource requirements and mathematical assumptions that are difficult to guarantee in practice, highlighting the importance of carefully reading the \emph{fine print} of such theoretical results \cite{aaronson_read_2015}. Although QML algorithms for fault-tolerant quantum computers (FTQCs) have been proposed for several impactful applications \cite{lloyd_quantum_2014, biamonte_quantum_2017, kerenidis_q-means_2019}, significant effort is still required to design methods that are compatible with near-term hardware \cite{preskill_quantum_2018}. In addition with strong hardware requirements, the community is still proposing new methods, and debating of the impact of existing proposals. This debate on "dequantization" of quantum methods can be illustrated by \cite{tang_quantum_2021}, where the author proposed a classical model that matches the performance of previously proposed quantum algorithms for principal component analysis \cite{lloyd_quantum_2014} and nearest-centroid clustering \cite{lloyd_quantum_2013}. These efforts by the community are crucial not only to enable the early adoption of quantum computers in industry, but also to support hardware providers by demonstrating applications that require fewer resources.

Variational quantum algorithms (VQAs) \cite{cerezo_variational_2021} have emerged as one of the most promising approaches for developing near-term useful methods. These algorithms employ classical optimizers to train parametrized quantum circuits, making them naturally suited to the constraints of noisy intermediate-scale quantum (NISQ) devices. Despite their promise, the integration of VQAs into industry still faces significant challenges, including issues of trainability and the difficulty of rigorously identifying and formalizing quantum advantage. As a result, directly mapping existing VQA-based methods to industrial use cases is, at present, premature. Instead, the community must focus on developing new algorithms and theoretical tools that can bridge the gap between today’s hardware and the demands of industrial utility.

This thesis investigates the intersection of quantum machine learning and its potential for practical applications, with a focus on theoretical foundations and mathematical analysis. As there is currently no general recipe to design quantum methods that are provably useful at scale for solving industrial problem, this thesis propose to address several of the theoretical challenges that have emerged in recent years. 

In particular, it explores what forms of plausible quantum advantage can be expected, identifies theoretical conditions under which algorithmic guarantees may be established, and characterizes regimes in which quantum models resist efficient classical simulation. Through these contributions, this work aims to advance the road-map for harnessing near-term and future quantum technologies for practical impact.

    \section{Objectives and QML Challenges}

Machine Learning has made it possible to solve many industrial problems that traditional computational methods either struggled with or could only address at a higher cost. Intuitively, quantum computing seems like a good candidate to improve some machine learning methods, as it offers large computational speed up on some problems. In particular, the output model of variational quantum circuits can be considered as a linear algebraic computation in a very large Hilbert space. 

However, the Quantum Machine Learning (QML) community is facing important challenges in the design of useful near-term applications. First, training unitary matrices of exponential size seems to cause a vanishing gradient phenomena called Barren Plateau \cite{mcclean_barren_2018,larocca_barren_2025}, limiting the \textbf{trainability} of variational quantum circuits. Secondly, the QML community has tried to offer figures of merit to characterize the quality of quantum models, that can be called \textbf{expressivity}. While the pertinence of such figures of merit will be discussed in this Thesis, it has been shown that having high expressivity metrics often leads to trainability issues. Finally, studies have shown that approximating or simulating quantum circuits, and quantum models can sometime be done using \textbf{surrogate models}.

\begin{figure}
    \centering
    \includegraphics[width=1\linewidth]{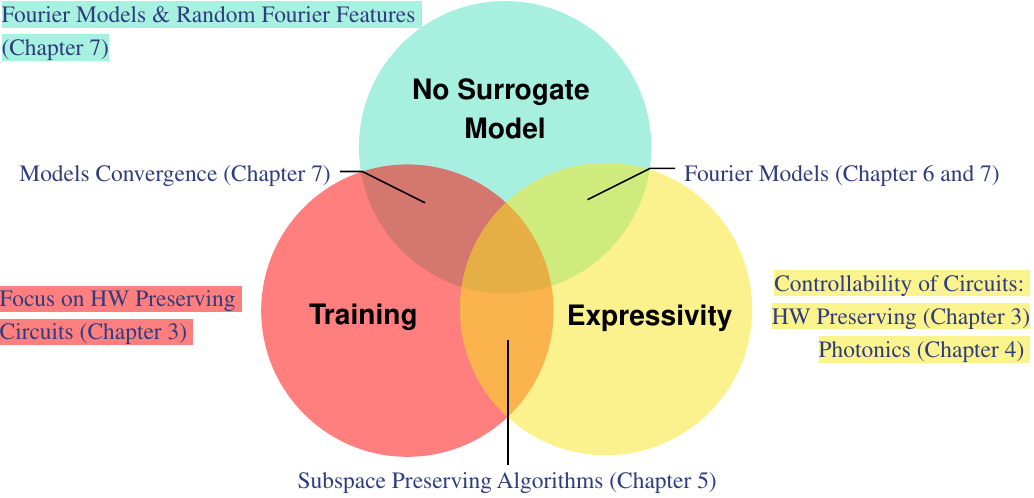}
    \caption{Quantum Machine Learning Challenges, and corresponding part of the Thesis.}
    \label{fig:QML_Challenges}
\end{figure}

In this Thesis, each of theses challenges will be discussed and compared to the recent advancements of the community. First, the trainability of subspace preserving variational quantum circuits is studied, as those circuits are believed to avoid Barren Plateau under an uniform distribution hypothesis (discussed in \autoref{chap:Preliminaries}) over a particular group of interest. In \autoref{chap:HW_Preserving_Methods}, a focus on Hamming Weight (HW) preserving circuits is presented, with theoretical guarantees on the trainability of such circuits with a different set of hypothesis, answering a prior conjecture on the trainability of such circuit. As with other studies from the community, these results tend to show that regimes where theoretical guarantees on the training can be offered also turn out to be "classically simulable" regimes, i.e., cases wher the computation can be done on a classical computer in polynomial time.

In later part of the thesis, such regime is handled as polynomial advantage can still be valuable for quantum industrial utility, especially with high repetition rate such as photonic ones.

Figures of merit for expressivity are discussed in the Thesis. In particular, the controllability of quantum circuits is studied in \autoref{chap:HW_Preserving_Methods} and \autoref{chap:Photonic_Suboptimal}, and motivation for its importance is presented through the lens of the need for a quantum data loading procedure. However, theoretical figures of merit are not always convincing to ensure the utility of a quantum algorithm for industrial scales. In \autoref{chap:Subspace_Preserving_Algorithms}, the choice of mimicking classical machine learning subroutines is made, in order to motivate the good scaling of the newly designed algorithms.

Finally, the expressivity of variational quantum circuits is studied through the lens of Fourier quantum models in \autoref{chap:VQC_as_Fourier_Models}. This framework, not restricted to subspace preserving quantum circuits, is quite general and allows to compare quantum circuit outputs and classical learning methods. In particular, it allows to compare their convergence, and to discuss the idea of approximation techniques as developed in \autoref{chap:Fourier_Surrogates}.

%%%%%%%%%%%%%%%%%%%%%%%
    \section{Thesis Outline}

In \autoref{chap:Preliminaries}, important mathematical notions, tools, and frameworks are introduced. Figures of merit and important concepts for expressivity, trainability, classical simulation and approximation that are useful for each part of the thesis are presented. In particular, the theory of subspace preserving variational quantum circuits is explored as an introductive part both of Hamming-weight preserving quantum circuits and photonic algorithms later discussed. This chapter recalls the state of the art results and explains why the following studies contributed to the quantum machine learning field, and to the theoretical development of applications. \\

In \autoref{chap:HW_Preserving_Methods}, a study of specific Hamming weight preserving variational quantum circuits is proposed. In particular, the control of such circuits is studied through the lens of the design of a quantum data loader, and theoretical results on its trainability are offered and compared with recent results.\\

In \autoref{chap:Photonic_Suboptimal}, a study of linear optical quantum circuits, that are near term, suboptimal, and particle number preserving is proposed. Limitations on the controllability capacities are highlighted, and a new scheme for near term applications is offered.\\

In \autoref{chap:Subspace_Preserving_Algorithms}, new quantum machine learning building blocks are presented, using Hamming weight preserving circuits, particle number preserving circuits, and the scheme previously introduced. This chapter offers a change in perspective: using a suboptimal model to offer a polynomial advantage while mimicking classical machine learning building blocks to ensure utility, along with theoretical guarantees on the training and good scaling of the performances.\\

In \autoref{chap:VQC_as_Fourier_Models}, variational quantum circuits are studied as Fourier models. Expressivity and its connections with trainability are explored, offering theoretical tools to design such circuits.\\

In \autoref{chap:Fourier_Surrogates}, the study of variational quantum circuits Fourier models is used to design surrogate models and to compare the convergence of classical and quantum models. Such results allow for guidelines to ensure a separation between quantum and classical learning models.

%%%%%%%%%%%%%%%%%%%%%%%
    \section{Contributions}

\begin{enumerate}
    \item The results of \autoref{chap:HW_Preserving_Methods} were published in Quantum as:
    \begin{itemize}
    \item \cite{monbroussou_trainability_2025} \textit{Trainability and Expressivity of Hamming-Weight Preserving Quantum Circuits for Machine Learning}, with Eliott Z. Mamon, Jonas Landman, Alex B. Grilo, Romain Kukla, and Elham Kashefi.
    \end{itemize}
    \textit{Contribution to the project:} I initiated the project, proposed the space-efficient encoding, expressivity study, the study of trainability, and contributed on all aspects. \\

    \item The results of \autoref{chap:Photonic_Suboptimal} were published in Physical Review Research as:
    \begin{itemize}
    \item \cite{monbroussou_toward_2025} \textit{Toward quantum advantage with photonic state injection}, with Eliott Z. Mamon, Hugo Thomas, Verena Yacoub, Ulysse Chabaud and Elham Kashefi.
    \end{itemize}
    \textit{Contribution to the project:} I initiated the project, proposed the state injection method, and focused on the link with the purity of the state and its controllability. I also participated in the proof of the probability estimation using state injection.\\

    \item The results of \autoref{chap:Subspace_Preserving_Algorithms} on the Hamming weight preserving QCNN were published in Quantum Science and Technology as:
    \begin{itemize}
    \item \cite{monbroussou_subspace_2025} \textit{Subspace preserving quantum convolutional neural network architectures}, with Jonas Landman, Letao Wang, Alex B. Grilo, and Elham Kashefi.
    \end{itemize}
    \textit{Contribution to the project:} I initiated the project, proposed each part of the algorithms, and designed the preliminary version of the \href{https://github.com/ptitbroussou/HW_QCNN}{simulation library}.\\
     
    \item The results of \autoref{chap:Subspace_Preserving_Algorithms} on the Photonic QCNN were published in Advanced Photonics as:
    \begin{itemize}
    \item \cite{monbroussou_photonic_2025} \textit{Photonic Quantum Convolutional Neural Networks with Adaptive State Injection}, with Beatrice Polacchi, Verena Yacoub, Eugenio Caruccio, Giovanni Rodari, Francesco Hoch, Gonzalo Carvacho, Nicolò Spagnolo, Taira Giordani,  Mattia Bossi, Abhiram Rajan, Niki Di Giano, Riccardo Albiero, Francesco Ceccarelli, Roberto Osellame, Elham Kashefi, and Fabio Sciarrino.
    \end{itemize}
    \textit{Contribution to the project:} I initiated the project with Elham Kashefi and Fabio Sciarrino, proposed the photonic algorithm, and contributed in every theoretical aspects of the study. I also help designing the experiment by adapting the algorithm to the photonic chip and post-selection, and I designed the \href{https://github.com/ptitbroussou/Photonic_Subspace_QML_Toolkit}{Photonic Subspace QML toolkit}. Parts of the software that led to the experiments were performed with Perceval \cite{heurtel_perceval_2023} by Fabio Sciarrino's team, in order to include noise in the simulations.\\

    \item The results of \autoref{chap:VQC_as_Fourier_Models} were published in Quantum as:
    \begin{itemize}
    \item \cite{mhiri_constrained_2024} \textit{Constrained and Vanishing Expressivity of Quantum Fourier Models}, with Hela Mhiri, Mario Herrero-Gonzalez, Slimane Thabet, Elham Kashefi, and Jonas Landman.
    \end{itemize}
    \textit{Contribution to the project:} This project was initiated and led by Hela Mhiri, who I help with the theorems proofs and interpretation of the results.\\

    \item The results of \autoref{chap:Fourier_Surrogates} were published in NPJ QI as:
    \begin{itemize}
    \item \cite{thabet_when_2025} \textit{When Quantum and Classical Models Disagree: Learning Beyond Minimum Norm Least Square}, with Slimane Thabet, Eliott Z. Mamon, and Jonas Landman.
    \end{itemize}
    \textit{Contribution to the project:} This project was initiated and led by Slimane Thabet. I focused on the examples of separation from classical to quantum models based on results from \cite{mhiri_constrained_2024}, and helped with the interpretation of the results.\\
\end{enumerate}

The results in this thesis were also presented as talks at \href{https://qtml2025.cqt.sg/accepted-papers/}{QTML 2025} (Singapore), \href{https://icoqc2025.sciencesconf.org/}{International Conference on Quantum Computing} (Institut Henry Poincarré), \href{https://briefingsource.edst.ibm.com/register/SZGRQECH}{IBM workshop on advancing quantum computing} (IBM London), Albert Fert Laboratory Seminar (invited by Danijela Markovic), \href{https://indico.cern.ch/event/1433194/timetable/#20250123.detailed}{International Conference on Quantum Technologies for High-Energy Physics 2025} (CERN), \href{https://qei2025.sciencesconf.org/resource/page/id/6}{Quantum Energy Initiative 2025} (Grenoble), \href{https://initiative-hpc-maths.gitlab.labos.polytechnique.fr/site/pages/2024-03-01-monbroussou.html}{CMAP Laboratory Seminar} (Ecole Polytechnique). The results ere also presented as poster presentations at national and international conferences such as QCTIP 2025 (Berlin), QCTIP 2024 (Edinburgh), QTML 2023 (CERN), and ECML PKDD 2022. I also presented my work in LMS Research School \href{https://icms.ac.uk/activities/workshop/lms-research-school-quantum-machine-learning-and-hamiltonian-simulation/}{Quantum Machine Learning and Hamiltonian Simulation} (organized by the International Centre for Mathematical Sciences).

%% file: chapters/02_Preliminaries/Preliminaries.tex
\let\textcircled=\pgftextcircled
 \chapter{Preliminaries} \label{chap:Preliminaries}

\initial{T}\textit{his chapter presents the mathematical notions that are important to understand the results of the Thesis the status of the QML community at the time that these projects were carried out. Basic knowledge of linear algebra, quantum computing, and variational quantum circuits is assumed. First, the main metric of expressivity for quantum models used by the community is introduced, which corresponds to studying the distribution of achievable unitary matrices. Importantly, a distribution that is uniform over an exponentially large group is generally interpreted as highly expressive and enables the use of powerful mathematical tools such as the Weingarten calculus. These tools were used to show that a uniform distribution over an exponentially large group implies a link between model concentration and expressivity. This can lead to vanishing gradient phenomena during the training of quantum circuits, an important problem known as the Barren Plateau (BP) phenomenon, which is recalled here. Subspace-preserving variational quantum circuits are also introduced, along with existing results from the literature on their trainability and classical simulation. A prior conjecture on trainability is recalled, which is later invalidated in \autoref{chap:HW_Preserving_Methods}. These results help to understand what is often referred to as a \textbf{curse of dimensionality} in variational quantum circuits. The general intuition supported by this Thesis and other works is that circuits that are exponentially hard to simulate classically also tend to exhibit Barren Plateaus, at least in settings where theoretical guarantees can be provided. This Thesis focuses on the use of subspace-preserving quantum circuits acting on polynomial-sized subspaces, aiming for quantum utility based on a polynomial advantage. This is particularly relevant for suboptimal photonic hardware (based on linear optics and adaptivity techniques as introduced in \autoref{chap:Photonic_Suboptimal}) that offers a high repetition rate. In addition, other figures of merit for expressivity are discussed, such as the controllability of quantum circuits and their corresponding Fourier models.}

    \section{Expressivity Measures of Quantum Models}\label{sec:Expressivity_Measures_Q_Models}

This Section discusses the different metrics of expressivity that are relevant to this study.

        \subsection{Distribution of Trainable Unitary Matrices and Model Concentration}\label{subsec:Distribution_Unitary_Model_Concentration}
        
A common expressivity measure for a variational quantum circuit (VQC) that has been extensively used in the literature \cite{holmes_connecting_2022,sim_expressibility_2019} is how uniformly the ensemble of the generated unitaries explores the unitary group. When learning a function generated by a VQC, it is usually sufficient to characterize the distance to the Haar measure up to the second moment. Hence, a VQC that forms a 2-design is defined as follows:

\begin{definition}[2-design]\label{def:2-design}
    A VQC $U(\Theta)$ is said to form a 2-design if the ensemble of unitaries it generates $\{U(\theta)\}_{\theta \in \Theta}$ (understood with a chosen distribution on the parameter space $\Theta$) produces first and second moments equal to those of the Haar measure over the space of all unitaries. The $2^{\text{nd}}$ moment superoperator of the distribution generated by $U(\Theta)$ is defined as:
    \begin{equation}
        M^{(2)}_{U(\Theta)} = \int_{\Theta} dU(\theta) U(\theta)^{\otimes 2} \otimes (U(\theta)^\ast)^{\otimes 2} 
    \end{equation}
    Let $U_H$ be the Haar ensemble of unitary matrices. The (vectorized) superoperator is then given by:
    \begin{equation}\label{Eq:Superoperator_A_2-design}
        \mathcal{A}^{(2)}_{U(\Theta)} = M^{(2)}_{U_H} - M^{(2)}_{U(\Theta)}
    \end{equation}
      such that $U(\Theta)$ forms an exact 2-design if $\mathcal{A}^{(2)}_{U(\Theta)} = 0$.
\end{definition}

To characterize the landscape of a quantum model, its variance with respect to the distribution of trainable parameters (the VQC distribution) is computed, and the Chebyshev inequality is used to quantify its \emph{concentration} around its average value. In this context, the mathematical concept of a 2-design allows to ease the calculation of the $2^{\text{nd}}$ moment superoperator $M^{(2)}_{U(\Theta)}$ \cite{miszczak_symbolic_2017}. 
However, such an assumption on the parameterized circuit leads to the quantum \emph{model exponential concentration} phenomena where the variance of the model vanishes with the dimension of the considered exponentially big Hilbert space \cite{arrasmith_equivalence_2022,mcclean_barren_2018,holmes_connecting_2022}.
Hence, the model exponential concentration is formally defined as follows. \\

\textit{Quantum models} on \textit{n} qubits can be defined as the family of parameterized functions $f : \mathcal{X} \times \Theta \rightarrow \mathbb{R}$ obtained by measuring the expectation value of some Hermitian observable $O$, such that:
 \begin{equation}\label{Eq:Quantum_Model}
    f(x,\theta) = \bra{0} U(x,\theta)^\dagger O U(x,\theta) \ket{0} 
 \end{equation}
  where $U(x,\theta)$ is a $2^n$-dimensional unitary , $\theta \in \Theta$ is the vector of trainable parameters and $x= (x_1,\dots,x_D) \in \mathcal{X} \subset \mathbb{R}^D$ is the classical data vector, with $\mathcal{X}$ the input space.

\begin{definition}[Model Exponential Concentration]\label{def:Model_Concentration}
    Consider a quantum model $f(x,\theta)$ such as defined in Eq. \eqref{Eq:Quantum_Model}. The model is said to exhibit a concentration phenomenon with respect to the set of trainable parameters $\theta$ when:
    \begin{equation}
        \Var_{\theta}[f(x,\theta)] = \mathcal{O}\left(\frac{1}{b^n}\right)
    \end{equation}
    for some constant $b > 1$.
\end{definition}

Let $\mathcal{Y}$ be a label space corresponding to the input space $\mathcal{X}$, and $\mathcal{D}$ the probability distribution over $\mathcal{X} \times \mathcal{Y}$ given by the considered dataset. A \textit{loss function} is a measurable map
\begin{equation}
    \ell : \mathbb{R} \times \mathcal{Y} \longrightarrow \mathbb{R},
\end{equation}
which quantifies the discrepancy between the prediction $f(x,\theta)$ of the quantum model defined in~\eqref{Eq:Quantum_Model} and the target label $y \in \mathcal{Y}$. The associated \emph{expected risk} is then defined as
\begin{equation}
    \mathcal{C}(\theta) := \mathbb{E}_{(x,y)\sim\mathcal{D}}\!\left[\ell\!\left(f(x,\theta),\, y\right)\right].
\end{equation}

It has been shown in \cite{holmes_connecting_2022} that model concentration can be connected to Barren Plateaus \cite{mcclean_barren_2018, larocca_barren_2025}, defined as:

\begin{definition}[Barren Plateau]\label{def:Barren_Plateau}
    The loss function landscape of a $n$-qubit VQC is said to exhibit a Barren Plateau (BP) if for all $\lambda$:
    \begin{equation}
        \mathbb{E}_{\theta}[\partial_{\theta_\lambda} \mathcal{C}(\theta)] = 0, \;\;\; \mathrm{Var}_{\theta}[\partial_{\theta_\lambda} \mathcal{C}(\theta)] = O\left(\frac{1}{b^n}\right),
    \end{equation}
    with $b > 1$.
\end{definition}

The $\varepsilon$\emph{-distance} to a 2-design can also be considered, where $\varepsilon$ is the (vectorized) superoperator $\mathcal{A}^{(2)}_{U(\Theta)}$ norm\footnote{Different norms can be used to characterize the $\varepsilon$-distance to a 2-design. They are indeed equivalent up to exponential factors in the number of qubits \cite{low_pseudo-randomness_2010}.}.  

\begin{definition}\label{def:monomial_norm}[Monomial definition of $\varepsilon$-approximate 2-design]
    An ansatz $U(\Theta)$ forms a monomial $\varepsilon$-approximate 2-design if:
    \begin{equation}
        \max_{p,q,r,s \in [d]} \; |(\mathcal{A}^{(2)}_{U(\Theta)})_{p,q,r,s}| \leq \frac{\varepsilon}{d^2}
    \end{equation}
    where $(\mathcal{A}^{(2)}_{U(\Theta)})_{p,q,r,s}$ is a coefficient of the $d^4$-dimensional matrix $\mathcal{A}^{(2)}_{U(\Theta)}$.
\end{definition}

The Monomial definition of an $\varepsilon$-approximate 2-design has been used in the literature for studying anti-concentration of random circuits. However, previous works \cite{larocca_diagnosing_2022,holmes_connecting_2022} have considered $\varepsilon$ to be the infinite norm or diamond norm of the superoperator $\mathcal{A}^{(2)}_{U(\Theta)}$. 
All these measures are equivalent up to exponential factors in the number of qubits, and in the above definition $\varepsilon$ takes value between 0 and $d^2$.

        \subsection{Fourier Expressivity of Quantum Models}\label{subsec:Fourier_Expressivity_Quantum_Models}

Along with the expressivity characterization of the parameterized part in a quantum model by its $\varepsilon$-distance to a 2-design, the expressivity of a quantum model can also be examined through its Fourier representation, i.e. the signature of the specific Hamiltonian encoding strategy. 

In \autoref{chap:VQC_as_Fourier_Models} and \autoref{chap:Fourier_Surrogates}, we study those quantum models in the setting where we can express them as Fourier models.  In a recent work \cite{xiong_fundamental_2025}, authors have proposed to define the \textit{Fourier expressivity} as the smallest set of functions such that the quantum model defined in Eq. \eqref{Eq:Quantum_Model} could be expressed as a linear combination of those functions. According to this definition and the Fourier decomposition of the quantum model, the Fourier expressivity is bounded by the spectrum size.

In \autoref{chap:VQC_as_Fourier_Models}, the focus is on characterizing the expressivity of a Quantum model through Fourier lens. Indeed, it is explained how individual Fourier coefficients may suffer from exponential concentration depending on the spectrum distribution. Therefore, the term \textit{vanishing expressivity} is introduced, to describe cases where some or all Fourier coefficients are exponentially concentrated around their mean.

%%%%%%%%%%%%%%%%%%%%%%%%%%%%%%%%%%%%%%%%%

    \section{Susbpace Preserving Variational Quantum Circuits}
    \label{sec:Subspace_Preserving_VQCs}

This Section presents results and definitions for subspace preserving quantum circuits that were introduced by the QML community before or during the writing of the Thesis. Those results offer perspectives on the studies conducted in the following chapters, where subspace preserving methods are deployed for polynomial advantage and to avoid BPs, especially for non-universal photonic devices.

        \subsection{Subspace Preserving VQCs and Lie Algebra}

First, a particular case of interest for subspace preserving variational quantum circuits is the Lie Algebraic Supported Ansatz defined in \cite{fontana_characterizing_2024}. By choosing a specific set of Hamiltonian matrices to define a variational quantum circuit, its action can be limited to a set of unitary matrices of smaller dimension, called a Lie Group.   

            \subsubsection{Formal Definitions}\label{subsec:formal_def_DLA}
            
A general formulation of a parametrized unitary matrix is given by:
\begin{equation}\label{eq:PQC_Unitary_Formulation}
    U(\theta) := \prod_{i=1}^D e^{i\theta_i H_i}\,,
\end{equation}
with $\theta = (\theta_1, \dots, \theta_D)$ the set of variational parameters, and $\{H_i\}_{i=1}^D$ hermitian traceless operators. For a qubit based circuit, this set of matrices corresponds to the choice of the quantum gates. 

\begin{definition}[Dynamical Lie Algebra (DLA)]\label{def:DLA}
    Consider a circuit constructed from controllable Hamiltonian matrices as defined in \autoref{eq:PQC_Unitary_Formulation}. The Dynamical Lie algebra is defined as:
    \begin{equation}\label{eq:DLA}
        \mathfrak{g} = \text{span} \langle iH_0, \dots, iH_K\rangle_{Lie} \subseteq \mathfrak{su}(d)\, ,
    \end{equation}
    with $\left< S \right>_{Lie}$ the Lie closure, i.e., the set of all nested Lie commutators between the elements in $S$, $\mathfrak{su}(d)$ the special unitary algebra of degree $d$ (i.e. the space of skew-Hermitian traceless $d \times d$ matrices), and $\{H_0, \dots, H_K\}$ the set of distinct hamiltonian matrices used in the VQC, called the set of generators for the quantum system.  
\end{definition}

\begin{figure}
    \centering
    \includegraphics[width=0.95\linewidth]{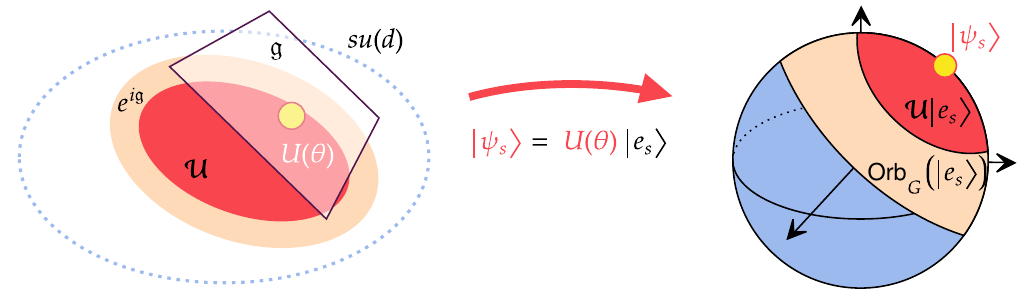}
    \caption{Representation of the unitary (on the left) and output state (on the right) spaces. The Lie algebra is the tangent space of the unitary space.}
    \label{fig:mathematical_space_DLA_Orbit}
\end{figure}

The maximal set of reachable unitary matrices $\mathcal{U} = \{U(\theta) | \theta \in \Theta\}$, with $\Theta$ the parameter space, is the Lie Group $G = e^\mathfrak{g}$. Its dimension is equal to the one of its corresponding Lie algebra. As a result, the maximal set of achievable output states is also constrained by the Dynamical Lie Algebra, as it is included in the initial state orbit defined as:
\begin{equation}\label{eq:State_Orbit}
    \text{Orb}_{G}(\rho_0) = \{ U \rho_0 U^{\dagger} | U \in G = e^{i\mathfrak{g}} \} \, ,
\end{equation}
with $\rho_0$ an initial state. The state orbit can be equivalently be defined for pure state as $\text{Orb}_G(\ket{\psi_0}) = \{U \ket{\psi_0} | U \in G \}$, with $\ket{\psi_0}$ an initial state. A graphical representation of the Lie algebra, Lie group, and state orbit are proposed in \autoref{fig:mathematical_space_DLA_Orbit} for an initial pure state $\ket{e_s}$.

        \subsubsection{Lie Algebra Supported Ansatz}

If the generators share common symmetries, i.e., there exists a hermitian operator that commutes with every generator, then every element in $\mathfrak{g}$ is block diagonal in the eigenbasis of the operator. As a consequence, every $H \in \mathfrak{g}$ and every $U \in e^{i\mathfrak{g}}$ are block diagonal in this particular basis that represents the symmetry. The state space can thus be considered as a direct sum of subspaces that are invariant under the action of $\mathfrak{g}$. 

In this Thesis, two particular symmetries that can be used to design Lie algebra supported ansatz are discussed:
\begin{itemize}
    \item Hamming weight preserving operators in \autoref{chap:HW_Preserving_Methods}, where the Hamming weight (number of qubits in state $\ket{1}$) is unchanged during the computation;
    \item Particle number preserving operators in \autoref{chap:Photonic_Suboptimal}, where the number of photons is unchanged.
\end{itemize}

\begin{figure}
    \centering
    \includegraphics[width=0.95\linewidth]{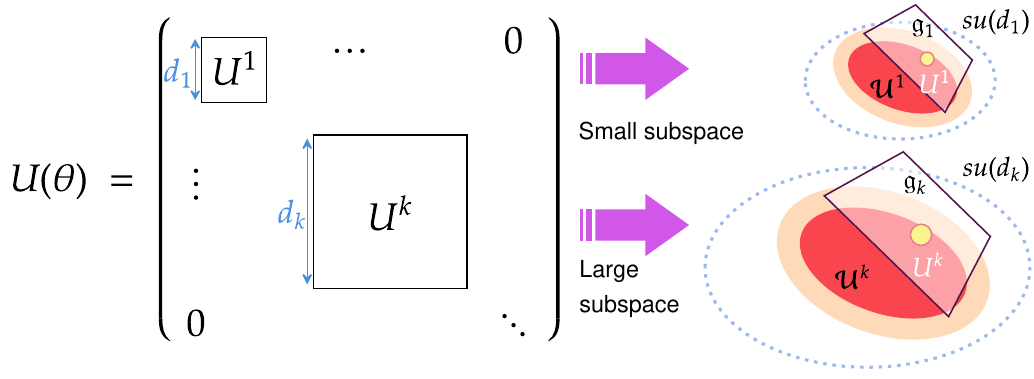}
    \caption{Schematic of a subspace parametrized unitary. The number of subspaces and their size depends on the symmetry considered. The Lie group is included into the subspace, and the set of reachable matrices is bounded by the Lie group.}
    \label{fig:enter-label}
\end{figure}

In the case where the dynamical Lie algebra is a direct sum $\mathfrak{g} = \bigoplus_k \mathfrak{g}_k$, the Hilbert space can be decomposed as well into subspaces $\mathcal{H} = \bigoplus_k \mathcal{H}_k$, with each $\mathcal{H}_k$ invariant under the action of the dynamical Lie Group $e^{i\mathfrak{g}}$. Then, if the initial state $\rho_k$ is supported entirely in $i\mathfrak{g}_k$, only the subspace $\mathcal{H}_k$ is of interest during the computation. If the subspace is of polynomial size with respect to the number of qubits, then one can classically simulate every computation in a polynomial time \cite{goh_lie-algebraic_2025}.

In \cite{larocca_diagnosing_2022}, it was shown that such ansatz can be used to avoid Barren Plateaus in VQCs. Similar to \cite{holmes_connecting_2022}, their proof is based on several hypothesis. First, the authors consider the case of subspaces preserved by the action of the Lie group. They then assume that within a subspace $\mathcal{H}_k$ of dimension $d_k$, the distribution of unitary matrices $\mathcal{U}^k$ forms a 2-design in $\mathcal{H}_k$. This requires the Lie algebra to be sufficiently large on $\mathcal{H}_k$, meaning $\mathfrak{g}_k = \mathfrak{su}(d_k)$ or $\mathfrak{g}_k = \mathfrak{u}(d_k)$. As a result, they show that the variance of the cost scales inversely with $d_k$, demonstrating that a polynomial-sized subspace can avoid Barren Plateaus. However, this setting is classically simulable in polynomial time as explained earlier.

One important question that remains from this work is whether the variance of the gradient depends on the dimension of the subspace $\mathcal{H}_k$ or on the dimension of the Lie algebra $\mathfrak{g}_k$. A conjecture asserting the dependence on the dimension of the Lie algebra was proposed in \cite{larocca_diagnosing_2022}.

\begin{conj}[From Conjecture 1 in \cite{larocca_diagnosing_2022}]\label{conj:Larocca_Var_DLA}
    Let the state $\rho$ belong to a subspace $\mathcal{H}_k$ associated with a subspace dynamical Lie algebra $\mathfrak{g}_k$ (or sub-DLA the subrepresentation in $\mathfrak{g}$ where $\rho$ has support on). Then, the scaling of the variance of the cost function partial derivative is inversely proportional to the scaling of the dimension of the DLA, i.e.
    \begin{equation}\label{eq:Conjecture_Larocca_Gradient_DLA}
        \text{Var}[\partial_\mu C(\theta)] \in \mathcal{O}\left( \frac{1}{\text{poly}(\dim(\mathfrak{g}_k))} \right)\,,
    \end{equation}
    with $C(\theta) = \text{Tr}(O U(\theta) \rho U^\dagger(\theta))$, and $\partial_\mu C(\theta) = \partial_{\theta_{\mu}} C(\theta) / \partial \theta_{\mu}$ for some $\theta_{\mu} \in \theta$.
\end{conj}

A counterexample is presented in \autoref{sec:trainability_RBS_circuit} using a Hamming weight preserving ansatz. In this example, where the initial and target states are normalized real superpositions of states with a fixed Hamming weight and the cost function is the squared Euclidean distance between the output and target states, the variance of the cost gradient depends on the dimension of the subspace $\mathcal{H}_k$.

During the writing of this thesis, it was shown in \cite{fontana_characterizing_2024, ragone_lie_2024} that:

\begin{restatable}[Adapted from Theorem 1 in \cite{ragone_lie_2024}]{thm}{RagoneLASAVarGrad}
\label{thm:RagoneLASAVarGrad}
    Suppose that $O \in i\mathfrak{g}$, or $\rho \in i\mathfrak{g}$, where the DLA $\mathfrak{g}$ can be decomposed as $\mathfrak{g} = \bigoplus_k g_k$, the variance of the loss function $l_\theta(\rho,O) = \Tr[U(\theta) \rho U^\dagger(\theta)O]$ is given by:
    \begin{equation}
        \mathrm{Var}_\theta[l_\theta(\rho,O)] = \sum_{j=1}^{k-1} \frac{\mathcal{P}_{\mathfrak{g}_j}(\rho) \mathcal{P}_{\mathfrak{g}_j}(O)}{\dim(\mathfrak{g}_j)}
    \end{equation}    
    where $\mathcal{P}_{\mathfrak{g}}(H) = \Tr[H_\mathfrak{g}^2]$ is the purity of the Hermitian operator $H \in i \mathfrak{u}(2^n)$ with respect to the operator subalgebra $\mathfrak{g} \subseteq i\mathfrak{u}(2^n)$, and with $H_\mathfrak{g}$ the orthogonal projection of $H$ onto the Lie algebra $\mathfrak{g}$.
\end{restatable}

As a result, if the initial state $\rho$, and observable $O$ lie in a polynomial large Lie algebra, the variance of the loss function avoids a Barren plateau. 

    \subsubsection{Classical Simulation of Subspace Preserving Quantum Circuits}\label{subsubsec:Classical_Simulation_Subspace_Preserving_QC}

Quantum circuits that preserve symmetries during computation may offer advantages for classical simulation. As previously stated, if the initial state $\rho_k$ is supported only on the subspace $\mathcal{H}_k$, then this will be the only subspace of interest throughout the computation.

If the subspace is of polynomial size with respect to the number of qubits, then every computation within it can be classically simulated in polynomial time. If the subspace is of exponential size but the Lie algebra is of polynomial dimension, then according to \autoref{thm:RagoneLASAVarGrad}, Barren Plateaus can be avoided provided the initial state or the observable lie in the Lie algebra. However, it was shown in \cite{goh_lie-algebraic_2025} that in such a setting, the circuit can be simulated classically in polynomial time using a basis of the Lie algebra\footnote{Note that obtaining such a Lie algebra basis is non-trivial in general, even if the basis is of polynomial size. A general algorithm for retrieving a dynamical Lie algebra basis from a set of Hamiltonian generators is provided in the supplementary materials of \cite{monbroussou_trainability_2025}; its time complexity is polynomial in the size of the Hamiltonian's subspace and the dimension of the Lie algebra.}.

%% file: chapters/04_HW_Preserving_Methods/HW_Preserving_Methods.tex
\let\textcircled=\pgftextcircled
\chapter[HW Preserving Methods]{Hamming-Weight Preserving Quantum Circuits for Machine Learning}
\label{chap:HW_Preserving_Methods}
\begin{textblock}{5.3}(0,-4)
	%\textit{`Un scientifique dans son laboratoire est un enfant placé devant des phénomènes naturels qui l'impressionnent comme des contes de fées.'\\}

%\hspace{0.5cm}--- Marie Skłodowska-Curie.
\end{textblock}

\initial{W}\textit{e previously discussed some properties of subspace preserving quantum circuits that tend to indicate a trade-off between the expressivity and trainability of such methods, in the sense that circuits which avoid Barren Plateaus due to a polynomially large Lie algebra can be simulated in polynomial time. However, those results were established under certain mathematical hypotheses that can be restrictive and were partially published during the writing of this Thesis. In this context, this chapter is dedicated to the study of the trainability and controllability of specific Hamming weight (HW) preserving variational quantum circuits. These circuits use qubit gates that preserve subspaces of the Hilbert space, spanned by basis states with fixed Hamming weight. In this study, the role of controllability is highlighted through the lens of quantum data loading. The trainability of such circuits is explored, not using any 2-design hypothesis but rather arguments on stochastic matrices, and highlights a setting where \autoref{conj:Larocca_Var_DLA} on the link between controllability and trainability of VQCs does not apply.}

\section{Hamming Weight Preserving Gates: RBS and FBS}
\label{sec:HW_gates_RBS_FBS}

This section presents the most commonly used gates in HW-preserving quantum circuits. First, the Reconfigurable Beam Splitter (RBS) gate is presented, which is easy to implement or native on many quantum devices. It is widely used in the quantum machine learning community to design variational algorithms \cite{johri_nearest_2021, cherrat_quantum_2024, jain_quantum_2024, coyle_training-efficient_2025, mathur_bayesian_2025} and quantum inspired classical methods \cite{raj_hyper_2025}.

\begin{definition}[Reconfigurable Beam Splitter gate]
    The Reconfigurable Beam Splitter (RBS) gate is a 2-qubit gate that corresponds to a $\theta$-planar rotation between the states $\ket{01}$ and $\ket{10}$:
    \begin{equation}\label{eq:RBS_2_qubit_gate}
    RBS(\theta) = e^{i \theta H_{RBS}} = \begin{pmatrix}
        1 & 0 & 0 & 0 \\
        0 & \cos(\theta) & \sin(\theta) & 0 \\
        0 & -\sin(\theta) & \cos(\theta) & 0 \\
        0 & 0 & 0 & 1 \\
        \end{pmatrix} \, , \quad \text{with} \quad H_{RBS} = \begin{pmatrix} 
        0 & 0 & 0 & 0 \\
        0 & 0 & -i & 0 \\
        0 & i & 0 & 0 \\
        0 & 0 & 0 & 0
        \end{pmatrix}\, . 
    \end{equation}
\end{definition}

Another HW-preserving gate is the Fermionic Beam Splitter (FBS) also sometimes used for QML algorithm \cite{kerenidis_quantum_2022}. This gate is not strictly a 2-qubit gate, as its action on qubits $i$ and $j$ depends on all the qubits between them.

\begin{definition}[Fermionic Beam Splitter]\label{def:FBS}
    Let $i,j \in [n]$ be qubits and $S = s_1 \dots s_n \in \{0,1\}^n$ a binary word corresponding to a basis state of fixed HW $\ket{S}$ with $n$ the total number of qubits. Then the Fermionic Beam Splitter (FBS) acts on the qubits $i$ and $j$ as the following unitary:
    \begin{equation}\label{eq:RBF_2_qubit_gate}
        \begin{pmatrix}
            1 & 0 & 0 & 0 \\
            0 & \cos(\theta) & (-1)^{f}\sin(\theta) & 0 \\
            0 & (-1)^{f+1}\sin(\theta) & \cos(\theta) & 0 \\
            0 & 0 & 0 & 1 \\
        \end{pmatrix}\, , \quad \text{with} \quad f = f_{i,j,S} = \sum_{i<l<j} s_l .
    \end{equation}
\end{definition}

Notice that those two gates have the same definition for a nearest neighbor connectivity, but have very different properties outside this setting. Both circuit preserve the same subspaces of fixed HW, that can be defined through the basis state of fixed HW.

\begin{definition}[Fixed HW state basis]\label{def:HW_basis}
    The basis of n-qubit states with fixed Hamming weight $k$ is defined as:
    \begin{equation}\label{eq:Basis_HW_k}
        B_k^n = \left\{ \ket{e} \middle| e \in \{0,1\}^n \text{ and } \text{HW} (e) = k \right\} \, ,
    \end{equation}
    with $d_k = |B_k^n| = \binom{n}{k}$, and $\text{HW}(e)$ the Hamming weight of the bit string $e$.
\end{definition}

For example, when considering $n = 3$ qubits and a HW $k=2$, the basis states are:
\begin{equation*}
    B_2^3 = \left\{ \ket{110}, \ket{101}, \ket{011} \right\} \,.
\end{equation*}

\begin{figure}
    \centering
    \includegraphics[width=0.95\textwidth]{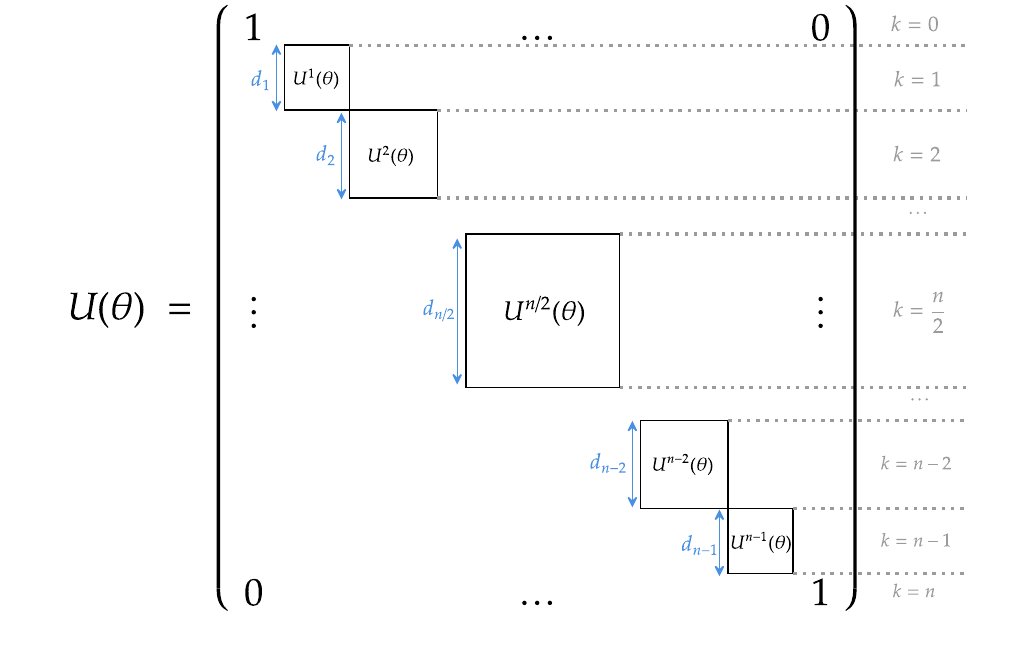}
    \caption{Block representation of the HW-preserving unitary matrices. $W$ is the $2^n \times 2^n$ unitary corresponding to a n-qubit HW-preserving quantum circuit. Each block $k$ is the unitary matrix corresponding to the preserved subspace of HW $k$, and the state basis $B_k^n$. Their size are $d_k\times d_k$ where $d_k = \binom{n}{k}$.}
    \label{fig:HW_circuit_block_unitary}
\end{figure}

As a result, circuits composed of RBS or FBS gates have an equivalent unitary matrix that is block diagonal when the states are ordered according to their HW, as illustrated in \autoref{fig:HW_circuit_block_unitary}. Each block corresponds to a fixed HW $k$, parametrized by the same set of variational parameters $\theta$, and are orthogonal matrices of size $d_k = |B_k^n| = \binom{n}{k}$. As a result, each subspace is of dimension $d_k(d_k-1)/2$. However, their DLAs (see \autoref{def:DLA}), denoted $\mathfrak{g}_k$, and their structures differ based on the choice of gate.

\begin{definition}[Compound matrix]\label{def:compound_matrix}
    Given a matrix $A \in \mathbb{R}^{n \times n}$, the compound matrix $\mathcal{A}^k$ for $k \in [n]$ is the $\binom{n}{k}$ dimensional matrix with entries $\mathcal{A}^k_{I J} = \det(A_{I J})$, where $I$ and $J$ are subsets of rows and columns of $A$ of size $k$.  
\end{definition}

Although similar to the RBS gate, the FBS is less controllable. Each block $U^k$ for $k>1$ is completely determined by $U^1$, as it is the $k$-compound matrix of $U^1$ (see \autoref{def:compound_matrix}). Therefore, the controllability of the FBS unitary matrices is upper-bounded by the controllability of the first block that represents the effect of the gate in the subspace of unary states, i.e., states of HW 1. The maximal dimension of the DLA for the FBS is $n(n-1)/2$.

\begin{figure}
  \centering
  \begin{subfigure}[b]{0.325\textwidth}
    \includegraphics[width=\textwidth]{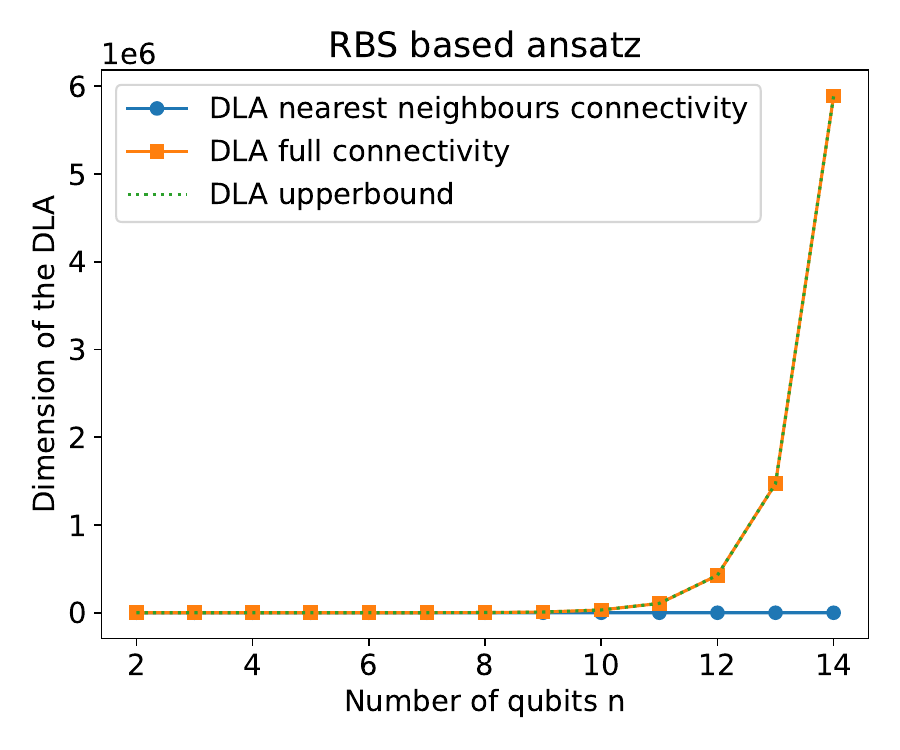}
  \end{subfigure}
  \hfill
  \begin{subfigure}[b]{0.325\textwidth}
    \includegraphics[width=\textwidth]{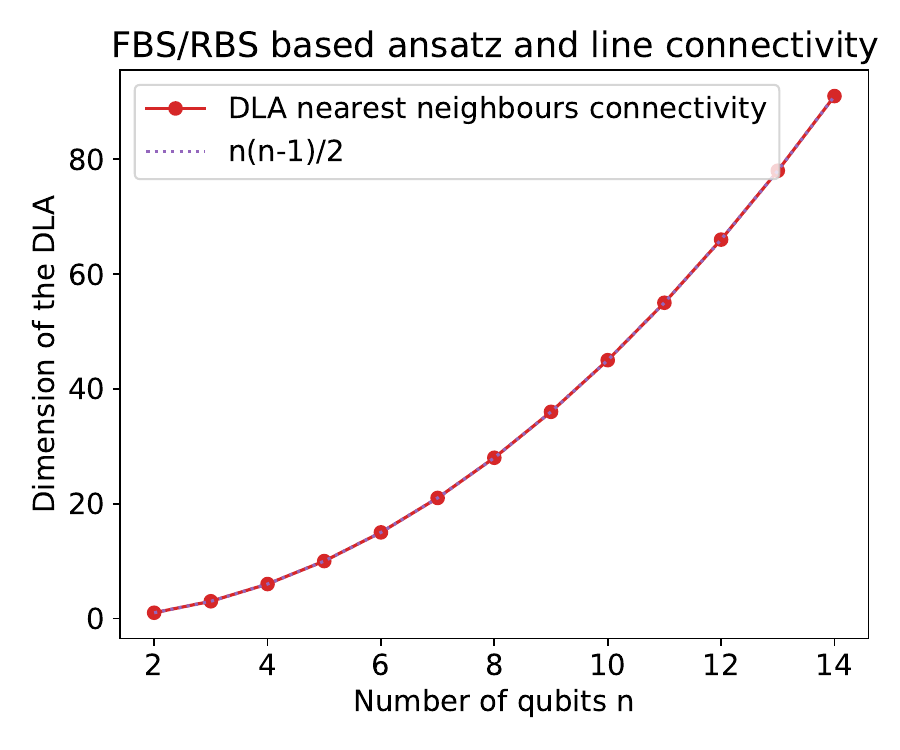}
  \end{subfigure}
    \hfill
  \begin{subfigure}[b]{0.325\textwidth}
    \includegraphics[width=\textwidth]{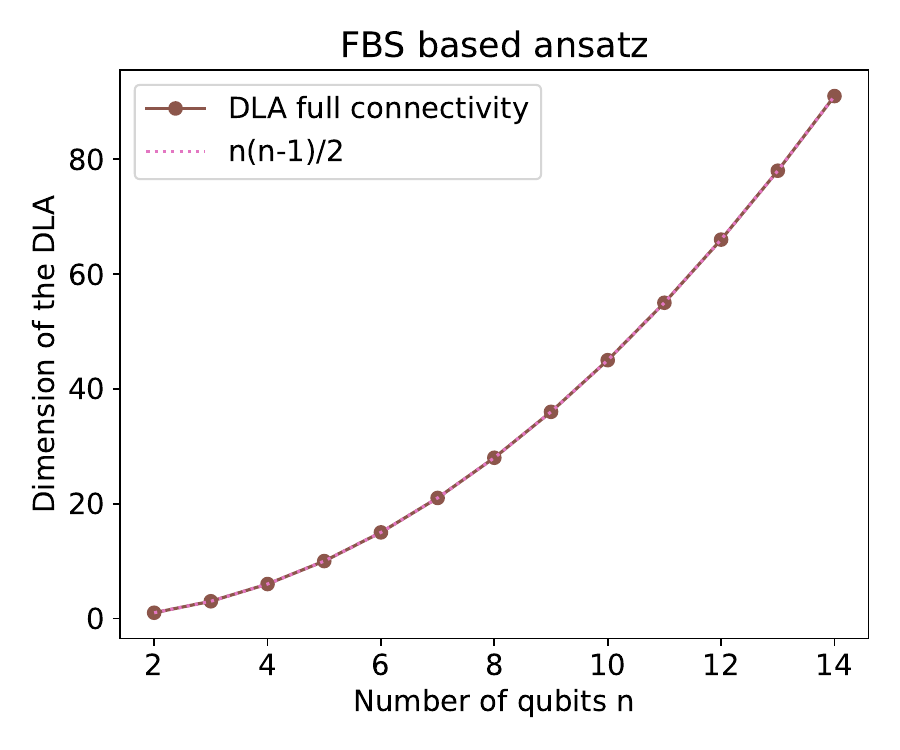}
  \end{subfigure}
  \caption{Evolution of the dimension of the DLA in the subspace of HW $k = \lfloor \frac{n}{2} \rfloor$ for: the use of RBS gates (left); the use of FBS gates with nearest neighbors connectivity (center); the use of FBS gates with full connectivity (right). This plot highlights the difference of controllability potential between RBS and FBS based quantum circuit.}
  \label{fig:DLA_RBS_FBS_evolution}
\end{figure}

The limitation in its controllability is illustrated in \autoref{fig:DLA_RBS_FBS_evolution}, and prevents the use of FBS for amplitude encoding on a subspace of HW $k$ with $k > 1$. However, FBS can be used to perform Clifford loader \cite{kerenidis_quantum_2022}, a loader determined on the entire Hilbert space and restricted to the direct sum of subspace produce by the FBS. In \cite{kerenidis_quantum_2022}, algorithms for quantum determinant sampling, singular value estimation for compound matrices, and for topological data analysis are presented using FBS gates and Clifford loaders.

\section{Space-Efficient Amplitude Encoding}
\label{sec:HW_Encoding}

This section discusses the design of HW-preserving quantum data loaders, which illustrates the importance of controllability. First, Amplitude Encoding and HW-preserving quantum data loaders are defined; these concepts will be refined in \autoref{chap:Subspace_Preserving_Algorithms}. Then, a method to achieve a data loader efficiently using HW-preserving gates and their subspace-preserving properties is proposed. Since such data loaders are classically simulable, the potential quantum speed-up for training is limited to a polynomial advantage. Nevertheless, the controllability of HW-preserving quantum circuit, analyzed through their DLA dimension and Quantum Fisher Information Matrix (QFIM) rank, is of independent interest.

        \subsection{Hamming Weight Preserving Quantum Data Loaders}\label{chap:HW_QDL}

This study starts with the definition of an Amplitude Encoding scheme and HW-preserving quantum data loaders.

\begin{definition}[Amplitude Encoding]\label{def:amplitude_encoding}
    An amplitude encoding data loader is a parametrized $n$-qubit quantum circuit that, given a classical vector $x = (x_1, \dots, x_{d}) \in \mathbb{R}^d$ (of a certain fixed length $d \leq 2^n$), prepares the quantum state:
    \begin{equation}
        \ket{x} = \frac{1}{||x||} \sum_{i=1}^{d} x_i \ket{e_i},
    \end{equation}
    where $\left\{\ket{e_1}, \dots, \ket{e_d}\right\}$ is a fixed family of $d$ orthonormal quantum states, and $||\cdot||$ denotes the $2$-norm of $\mathbb{R}^d$.  
\end{definition}

One could use the Reconfigurable Beam Splitter (RBS) gate to perform such an encoding on the basis of fixed HW (see \autoref{def:HW_basis}). This HW-preserving gate is easy to implement or native on many quantum devices. Note that the results in this chapter also hold for another HW-preserving gate named the Fermionic Beam Splitter (FBS), which was used for QML applications in \cite{kerenidis_quantum_2022} but has less favorable properties in terms of controllability, as explained previously.

The data loading scheme is explained as follows. First, the quantum state must be initialized to $\ket{e_s}$, a basis state of HW $k$. This state is then split onto the states in $B_k^n$ using RBS gates. In \cite{johri_nearest_2021}, the authors used a similar method on the unary basis $B_1^n$. Notice that achieving an amplitude encoding with such a basis would allow us to encode many more parameters, namely $\binom{n}{k} \gg n$, in an $n$ qubit state.
To design this quantum data loader, it must be ensured that any $\binom{n}{k}$-dimensional real vector $x$ can be encoded; that is, a corresponding set of RBS gate parameters $\theta = \{\theta_1, \dots, \theta_D\}$ (depending on the data point $x$) must exist such that:
\begin{equation}\label{eq:encoding_particular_equivalent_eq}
    W^{k}(\theta)\ket{e_{s}} - \frac{1}{||x||}\sum_{i=1}^{\binom{n}{k}} x_i \ket{e_i} = 0 \, .
\end{equation}
Finding the corresponding set of variational parameters or proving their existence is generally very hard when $k > 1$. Here, the focus is on the existence of an \textit{approximate} solution to the following related optimization problem:
\begin{equation}\label{eq:encoding_optimization_pb}
    \theta^\ast = \arg \min_{\theta} ||\frac{1}{||x||}\sum_{i=1}^{\binom{n}{k}} x_i \ket{e_i} - W^k(\theta)\ket{e_{s}}||_2^2\, ,
\end{equation}
which can be addressed using gradient-based optimizers. Theoretical arguments on the amenability of the loss of Eq.~\eqref{eq:encoding_optimization_pb} to gradient-based optimizers are provided later in \autoref{sec:trainability_RBS_circuit}. Namely, this cost function does not induce Barren Plateaus even though it is a type of global cost function \cite{cerezo_cost_2021}, as the Hilbert space corresponding to states of HW $k$ is not exponentially large for small $k$. Subspace preserving quantum circuits are easier to simulate in small subspaces than random quantum circuits over the entire Hilbert space \cite{anschuetz_efficient_2023}. In the case of a HW-preserving VQC, the speedup of using a quantum computer grows exponentially with $k$. Classical simulability of the encoding part itself is not an issue, if it is later combined with a trainable layer that is hard to simulate\footnote{Simple examples can be constructed in which a quantum circuit is first composed of a state preparation layer with a classically simulable encoding layer, followed by a trainable layer that does not respect the same symmetry. For example, one can 
use a unary encoding layer followed by a circuit made of $X$ rotation gates and RBS gates. The classical data would be naturally encoded in the unary basis, while the final state does not belong to the same basis.}.

        \subsection{Finding the Quantum Data Loader} \label{chap:Finding_QDL}

As explained earlier, the DLA gives an upper-bound on the controlability of a variational quantum circuit in the unitary space. Having a DLA dimension high enough is a necessary condition to design a quantum data loader. A remaining question is how to design the quantum data loader from a given subspace and the connectivity that induces the existence of such a circuit. This Section presents two algorithms to design the quantum data loader based on the study of controllability in the state space.

\begin{figure}[h!]
    \centering
    \includegraphics[width=1\textwidth]{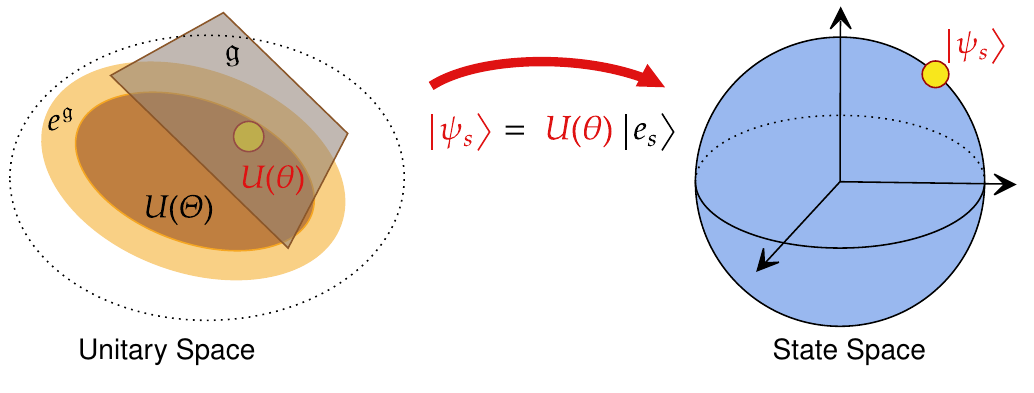}
    \caption{Representation of the unitary and output state spaces. The DLA is the tangent space of the unitary space. The possible directions for the evolution of the output state are given by the Quantum Fisher Information Matrix eigenvectors.}
    \label{fig:relevant_mathematical_spaces_for_QML}
\end{figure}

The ability of a subspace-preserving circuit to achieve amplitude encoding on one of its preserved subspaces is equivalent to the circuit perfectly controlling the state space created by its output. In particular, an RBS based VQC would achieve perfect amplitude encoding (see \autoref{def:amplitude_encoding}) on the subspace of HW $k$ if its output state could be any (real, normalized) superposition of states in $B_k^n$. That is, if the space of kets that its output explores were the entirety of a certain sphere of (real) dimension $d_k - 1$, noted $S^{d_k - 1}$ and illustrated in \autoref{fig:relevant_mathematical_spaces_for_QML}. An essential tool for studying the controllability of a quantum circuit in the state space is defined below.

\begin{definition}[Quantum Fisher Information Matrix]
\label{def:QFIM}  
The \textit{Quantum Fisher Information Matrix} (QFIM) associated to any parametrized pure state $\ket{\psi(\theta)}$ that uses $p$ continuous parameters $\theta=(\theta_1,\dots,\theta_p)$, is the following $p\times p$ real matrix assigned to each parameter vector $\theta$:
    \begin{equation}\label{eq:QFIM}
    \IfRestatedTF{
        [\mathrm{QFIM}(\theta)]_{i,j} = 4\,\mathrm{Re}\big[\, \big\langle \partial_{\theta_i} \psi(\theta) \big| \partial_{\theta_j} \psi(\theta) \big\rangle\,-\,\big\langle \partial_{\theta_i} \psi(\theta) \big| \psi(\theta) \big\rangle \big\langle \psi(\theta) \big|  \partial_{\theta_j}\psi(\theta) \big\rangle \,\big]\,.}
        {
        \begin{aligned}
           [\mathrm{QFIM}(\theta)]_{i,j} &= 4\,\mathrm{Re}\big[\, \big\langle \partial_{\theta_i} \psi(\theta) \big| \partial_{\theta_j} \psi(\theta) \big\rangle\,-\\
           & \big\langle \partial_{\theta_i} \psi(\theta) \big| \psi(\theta) \big\rangle \big\langle \psi(\theta) \big|  \partial_{\theta_j}\psi(\theta) \big\rangle \,\big]\,.
        \end{aligned}
        }
    \end{equation}
\end{definition}

For a basis state $\ket{e_s}$ of HW $k$ used as the input state to an RBS/FBS circuit $U(\theta)$, the computational cost of calculating the QFIM of the output state $\ket{\psi_s(\theta)} = U(\theta) \ket{e_s}$ depends on the subspace dimension $d_k$. Each state $\ket{\psi_s(\theta)}$ and $\ket{\partial_{\theta_i} \psi_s(\theta)}$ can be simulated as a vector of dimension $d_k$, and the overall computational cost of calculating the matrix $\mathrm{QFIM}_s(\theta)$ is $\mathcal{O}(p^2 \, d_k^2)$.

The maximal rank (over parameter space) of the QFIM is a metric of controllability in the state space \cite{liu_quantum_2019}, as it gives us the number of independent directions that can be taken by the state when tuning the gate parameters $\theta$. For this study encoding method in the subspace of HW $k$, a consequence of the fact that the kets are constrained to belong to $S^{d_k - 1}$ is that the QFIM ranks are upper-bounded by $d_k -1$ (for any parameter values):
\begin{equation}
    \max_{\theta} \; \rank [\mathrm{QFIM}_s(\theta)] \leq  d_k-1 .  
\end{equation}

\begin{figure}
  \centering
  \begin{subfigure}[b]{0.49\textwidth}
    \includegraphics[height=5.5cm]{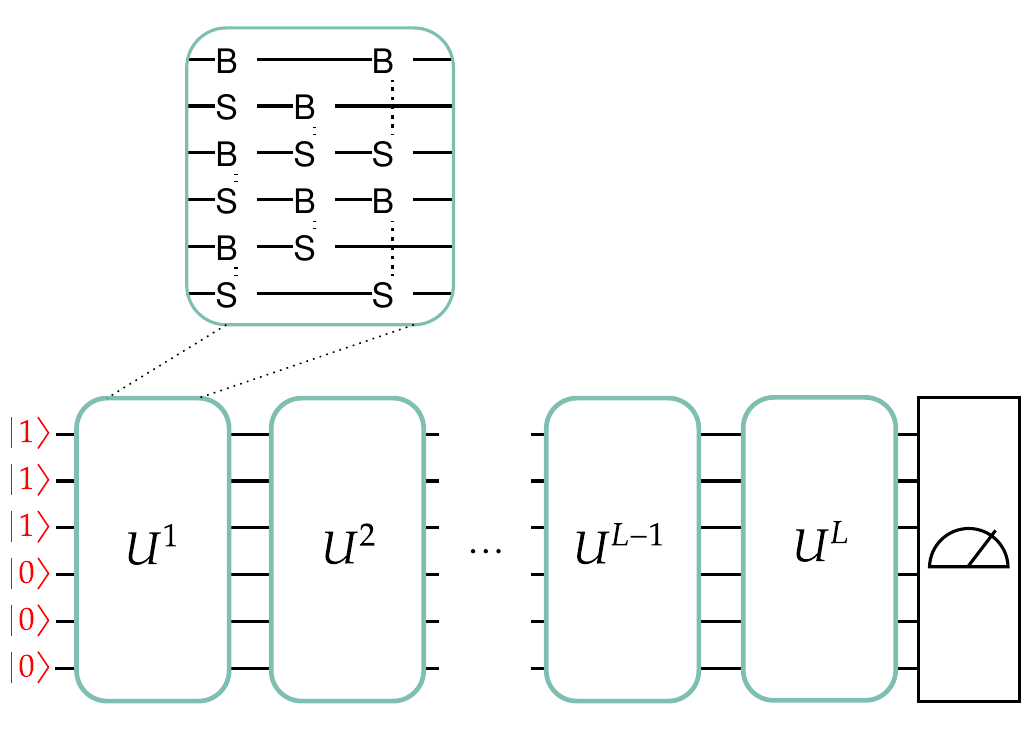}
  \end{subfigure}
  \hfill
  \begin{subfigure}[b]{0.49\textwidth}
    \includegraphics[height=5.5cm]{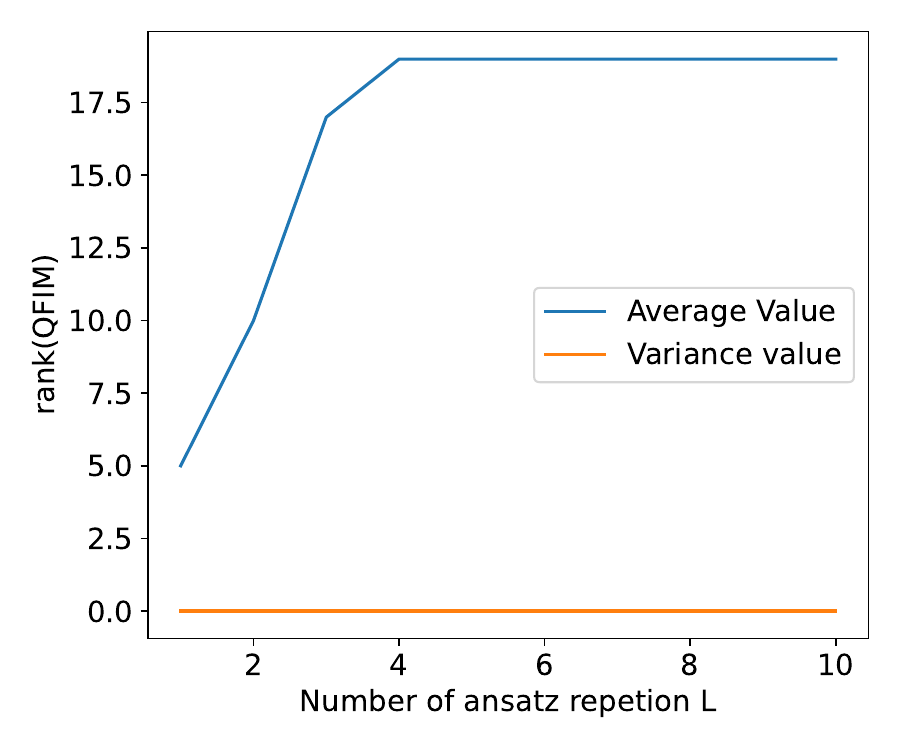}
  \end{subfigure}
  \caption{Evolution of the rank of the QFIM for a periodic structure ansatz. RBS gates are illustrated by the letters B and S separated by dotted lines. The evolution of the rank of the corresponding QFIM is given by the plot on the left right side. The derivation of the QFIM rank is done in the largest subspace ($k=n/2$ in this case $k=3$).}
  \label{fig:QFIM_rank_evolution}
\end{figure}

As in \cite{haug_capacity_2021}, one can find numerical evidence (see \autoref{fig:QFIM_rank_evolution}) that upon randomly sampling parameter values $\theta \in [0,2 \pi]^p$, the value of $\rank [\mathrm{QFIM}_s(\theta)]$ is independent of $\theta$.
In fact, this property can be justified theoretically for any standard VQC, which is stated here as \autoref{thm:QFIMrankthm}.

\begin{restatable}[Almost-constant property of QFIM rank]{thm}{QFIMrankthm}
\label{thm:QFIMrankthm}
A VQC's output state $\ket{\psi(\theta)}$ always has the following property: \textbf{almost everywhere} on the considered parameter space $\Theta$, the rank of their QFIM is constant, equal to $r_{\mathrm{max}}:=\max\limits_{\theta \in \Theta}r(\theta)$. Consequently, drawing a point $\theta \in \Theta$ uniformly at random and calculating its QFIM rank value $r(\theta)$ yields $r(\theta)=r_{\mathrm{max}}$ with probability $1$.
\end{restatable}

\begin{proof}
	This follows from the fact that the Jacobians of analytic maps have constant rank almost everywhere \cite[Prop. B.4]{bamber_how_1985}, combined with the fact that the QFIM is the Gram matrix of the Jacobian's columns (for a certain inner product).
\end{proof}

It would be easy to find specific points in the parameter space where the QFIM rank changes, for example at the poles of the Bloch sphere when considering Pauli  rotation gates on a single-qubit state. In \autoref{thm:QFIMrankthm}, "almost everywhere" means with probability one when a parameter is sampled from the uniform  distribution on the parameter space. \\  

Using the QFIM on a given subspace of HW $k$, a first algorithm is proposed to design a quantum data loader in this subspace from an initial state, created using bit-flips, and the possible generators $\mathcal{G}$ given by the qubit connectivity and the RBS gate Hamiltonian. Reaching the maximal rank (over parameter space) of the QFIM of a quantum data loader circuit is equal to $\dim(S^{d_{k}-1}) = d_k - 1$, is taken as evidence that it may achieve any state in $S^{d_{k}-1}$, i.e., achieve the amplitude encoding on the subspace of HW $k$. Following this idea, \autoref{alg:QDL_1} creates iteratively a circuit by adding RBS gates one at a time, while making sure that each added gate has actually incremented the QFIM rank, and only stops when the QFIM rank attains the dimension of the sphere $S^{d_{k}-1}$, suggesting that the data loader capability has been obtained.

\begin{algorithm}[H]
\caption{to design a HW-preserving quantum data loader}\label{alg:QDL_1}
\begin{algorithmic}[1]
\Require $\mathcal{G}$ the generators, $\ket{e_s}$ the initial state 
\State circuit = $\emptyset$
\While{$(\max_{\theta} \rank[\mathrm{QFIM}(\text{circuit}, \theta)] < d_k - 1)$}
    \For{$RBS \in \mathcal{G}$}
        \State $\text{circuit'} = \text{circuit} + RBS$
        \If{($\max_{\theta'} \rank [\mathrm{QFIM}(\text{circuit'},\theta')] > \max_{\theta} \rank [\mathrm{QFIM}(\text{circuit},\theta)]$)}
            \State $\text{circuit} = \text{circuit'}$
        \EndIf
    \EndFor
\EndWhile\label{euclidendwhile}
\State \textbf{return} \text{circuit}
\end{algorithmic}
\end{algorithm}
Using \autoref{thm:QFIMrankthm}, it suffices to calculate the QFIM rank at just one randomly sampled $\theta$ to obtain the maximum rank over parameter space. Another more heuristic approach is given by \autoref{alg:QDL_2}, using the concept of overparametrization introduced in \cite{larocca_theory_2023}:

\begin{definition}[Overparametrization]\label{def:Overparametrization}
    A VQC is overparametrized if the number of parameters $D$ is such that the QFIM, for all the states in the training set, simultaneously saturates its rank $r_{\mathrm{max}}$:
    \begin{equation}
        \max_{D \geq D_{c}, \theta} \rank[\mathrm{QFIM}_s(\theta)] = r_{\mathrm{max}}.
    \end{equation}
\end{definition}

The authors showed that for a general type of periodic-structured VQCs, we have:
\begin{equation}\label{eq:critical_threshold_nbr_parameters}
    D_c \sim \dim(DLA) \, .
\end{equation}

A quantum circuit that achieves the full rank of the QFIM can be easily constructed through overparametrization according to Eq.~\eqref{eq:critical_threshold_nbr_parameters}. Another algorithm based on the overparametrization phenomenon can be proposed, where the idea is to remove gates to reduce the circuit depth while preserving the controllability of the output state.

\begin{algorithm}[H]
\caption{to design a HW-preserving quantum data loader}\label{alg:QDL_2}
\begin{algorithmic}[1]
\Require circuit, flag = True
\While{flag}
    \State flag = False
    \For{$RBS \in \text{circuit}$}
        \State $\text{circuit'} = \text{circuit} - RBS$
        \If{($\max_{\theta'} \rank [\mathrm{QFIM}(\text{circuit'},\theta')] = \max_{\theta} \rank [\mathrm{QFIM}(\text{circuit},\theta)]$)}
            \State circuit, flag $=$ circuit', True
            \State \textbf{break for}
        \EndIf
    \EndFor
\EndWhile
\State \textbf{return} \text{circuit}
\end{algorithmic}
\end{algorithm}

The reason the rank of the QFIM can be increased to its maximum in \autoref{alg:QDL_1} is based on the results from \cite{larocca_theory_2023} on the theory of overparametrization, recalled in \autoref{def:Overparametrization}. \autoref{alg:QDL_2} must be initialized by considering a quantum circuit made of a large number of gates, as determined by the dimension of its DLA in Eq.~\eqref{eq:critical_threshold_nbr_parameters}. Those gates can be chosen randomly or in such a way to reduce the circuit depth. Using \autoref{alg:QDL_2} allows to first design a circuit with a number of gate slightly larger than the optimal, and then to reduce the circuit by removing some gates according to the QFIM. This method can be useful to first choose a circuit that corresponds to other figure of merits (favoring the use of qubits of better quality, reducing the depth) and then to avoid to derive too many time the rank of the QFIM which can be costly.

In practice, for both algorithms, particular attention must be paid to the order in which generators are tested, with regard to the circuit depth. The computational cost of both algorithms depends on the cost of calculating the rank of the QFIM, which has a computational complexity of $\mathcal{O}(p^2 \, d_k^2 + p^3)$ for classical simulation,where $p$ is the number of parameters and $d_k$ is the dimension of the chosen subspace (the $p^3$ term corresponds to the complexity of calculating the matrix rank). \autoref{alg:QDL_1} and \autoref{alg:QDL_2} are thus efficient to run for small subspaces, which correspond to the case where HW-preserving quantum circuits are trainable (see the following Section). Since such data loaders are classically simulable, the potential quantum speed-up for training is limited to a polynomial advantage. However, one could train them classically to represent classical data and then associate them with a quantum circuit that is harder to simulate—for example, by increasing the number of qubits, the Hamming weight, or incorporating gates that do not preserve Hamming weight.

\section{Trainability of HW-Preserving Quantum Circuits}
\label{sec:trainability_RBS_circuit}

It is known that some QML proposals suffer from unfavorable optimization landscape properties \cite{mcclean_barren_2018} that lead to strong limitations in their trainability. This Section presents strong results on the gradient of the cost function for VQCs composed of RBS or FBS gates. These results are not based on a Haar-random distribution of the unitary matrices, such as the 2-design hypothesis discussed in \autoref{sec:Expressivity_Measures_Q_Models}. First, the backpropagation formalism applied to RBS and FBS based VQCs is presented in \autoref{subsec:backpropagation}.Then, the resulting theorems on the variance and expectation value of the cost function gradient are presented in \autoref{subsec:Avoid_BP}.

     \subsection{Backpropagation for Gradient Calculus}\label{subsec:backpropagation}

A HW-preserving quantum circuit composed solely of RBS gates or solely of FBS gates is described. We decompose the quantum circuit as a series of such gates, for which we denote their unitary matrices in the basis $B_k^n$ by $w^{\lambda}(\theta_\lambda)$, for $\lambda=1,\dots,\lambda_{\mathrm{max}}$, with $\theta_\lambda$ denoting the a gate's angle parameter. In the subspace of HW $k$, we denote respectively the initial, intermediate, and final quantum states by $\zeta^0$, $\zeta^\lambda$ (for $\lambda=1,\dots,\lambda_{\mathrm{max}}$), and $z$ --- they are all normalized vectors in $\mathbb{R}^{d_k}$. We also denote the inner error associated to the state $\lambda$ by $\delta^{\lambda} := \partial \mathcal{C} \slash \partial \zeta^{\lambda}$.
The cost function we consider in this work is the \textit{squared Euclidean distance} between the output state $z$ of the circuit and fixed target output $y$: 
\begin{equation}\label{eq:square-euclidean-loss-function}
    \mathcal{C}(\theta) = ||z(\theta) - y||^2_2\,.
\end{equation}
We focus on this cost function because it is ubiquitous in classical machine learning tasks (where it is usually termed the \textit{$l_2$ loss}). The reader might notice that the cost function in \autoref{eq:square-euclidean-loss-function} depends on the phase on the output state $z$ (i.e. changing $z$ to $-z$ generally changes the cost function value), and thus it may not be written as an expectation value $\bra{z}O\ket{z}$ of some hermitian $O$ (as those latter functions are phase invariant). While this is true, it is possible to extend the system with a single ancillary qubit such that $\mathcal{C}(\theta)$ may be estimated through a quantum observable on the extended system (see the second tomography procedure described in \cite{landman_quantum_2022} for more details).\footnote{This in essence switches the status of the $\pm$ sign from a global phase to a local one, making it is physically observable on the extended system.}

\begin{figure}[H]
    \centering
    \includegraphics[width=0.9\textwidth]{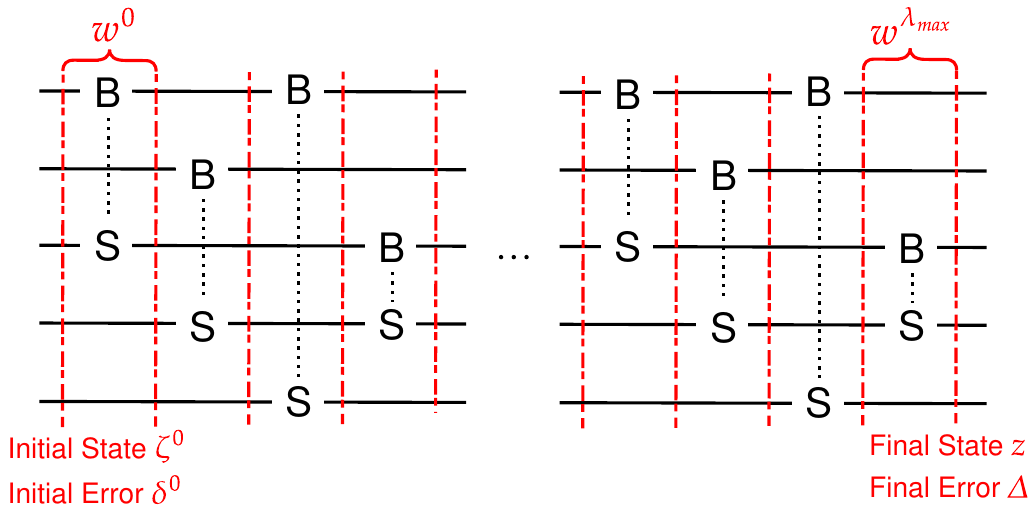}
    \caption{Decomposition of the HW-preserving quantum circuit for the backpropagation method. RBS gates are illustrated by the letters B and S separated by dotted lines.}
    \label{fig:HW_preserving_backprop}
\end{figure}

The equivalent weight matrix of our VQC is $W^k = w^{\lambda_{\mathrm{max}}} \dots w^1 w^0$. To train the circuit, each RBS parameter $\theta_\lambda$ is updated with respect to the gradient of the cost function $\mathcal{C}$. Using the chain rule, we may decompose the cost function's derivative with respect to a gate's parameter $\theta_\lambda$ in terms of the components of the subsequent quantum state $\zeta^{\lambda+1}$:  
\begin{equation}
    \frac{\partial \mathcal{C}}{\partial \theta_\lambda} = \sum_p \frac{\partial \mathcal{C}}{\partial \zeta_p^{\lambda +1}} \frac{\partial \zeta_p^{\lambda +1}}{\partial \theta_\lambda} = \sum_p \delta_p^{\lambda + 1} \frac{\partial (w_p^{\lambda} \cdot \zeta^{\lambda})}{\partial \theta_\lambda} \, .
\end{equation}

Each parameter $\theta_\lambda$ corresponds to applying a $\theta_\lambda$-planar rotation between two qubits. Such a rotation between two qubits corresponds, at the level of the subspace of HW $k$, to multiple pairs of basis directions $(l,j)$ that are undergoing a rotations in the subspace. For a circuit made of RBS gates, we have:

\begin{equation}\label{eq:backpropagation_equation}
    \frac{\partial \mathcal{C}}{\partial \theta_\lambda} =  \sum_{(l,j)} \delta_l^{\lambda} (-\sin(\theta_\lambda) \zeta_l^{\lambda} + \cos(\theta_\lambda)\zeta_j^{\lambda}) +\delta_j^{\lambda} (-\cos(\theta_\lambda) \zeta_l^{\lambda} - \sin(\theta_\lambda)\zeta_j^{\lambda}),
\end{equation}
where the sum is over all pairs $(l,j)$ of basis state indices which are undergoing a planar rotation of angle $\theta$ by the layer $\lambda$.

Similarly, for a circuit made of FBS gates, we have:
\begin{equation}\label{eq:backpropagation_equation_FBS}
    \begin{split}
    \frac{\partial \mathcal{C}}{\partial \theta_\lambda} = \sum_{(l,j)} &\delta_l^{\lambda} (-\sin(\theta_\lambda) \zeta_l^{\lambda} + (-1)^{f(a,b,\zeta_j^{\lambda})} \cos(\theta_\lambda)\zeta_j^{\lambda}) + \\
    &\delta_j^{\lambda} ((-1)^{f(a,b,\zeta_l^{\lambda})+1}\cos(\theta_\lambda) \zeta_l^{\lambda} - \sin(\theta_\lambda)\zeta_j^{\lambda}),
    \end{split}
\end{equation}
with $f(a,b,\zeta_\lambda^{\lambda}) = \sum_{a<p<b} s_p$, where $s \in \{0,1\}^{n}$ is the binary word corresponding to the state given by the index $\lambda$: $\ket{\zeta_\lambda} = \ket{s_1 \cdots s_n}$ ($a$ and $b$ are the qubits affected by the FBS).

    \subsection{Avoiding Barren Plateaus}\label{subsec:Avoid_BP}

The analytic definition of the cost function gradient provided by backpropagation (\autoref{eq:backpropagation_equation} and \autoref{eq:backpropagation_equation_FBS}) can be used to study the phenomenon of \textbf{Barren Plateaus} (BPs) defined in \autoref{def:Barren_Plateau}, a detrimental situation in which cost function gradients are exponentially suppressed.

It is possible to determine the existence of BPs under the assumption that the ensemble of parametrized unitary matrices forms an approximate 2-design \cite{holmes_connecting_2022}. In that case the quantity  $\mathrm{Var}_{\theta}[\partial_{\theta_\lambda} \mathcal{C}(\theta)]$ may be evaluated using the standard Weingarten calculus integration formulas (see e.g. \cite{miszczak_symbolic_2017}) and found to be inversely proportional to the dimension of the Hilbert space. In \cite{larocca_diagnosing_2022}, it was shown that if a subspace-preserving VQC satisfies the assumption of full controllability of the subspace (meaning the dimension of the DLA is maximal, i.e., equal to the dimension of all unitary matrices on that subspace), as well as a 2-design assumption on that subspace, then the variance of the cost gradient scales inversely with the dimension of the subspace. As a result, one could avoid BPs using a subspace invariant quantum circuit with a subspace of small dimension.

Here, it is shown that BPs can indeed be avoided for subspace-invariant quantum circuits based on RBS or FBS gates, \emph{without} making a 2-design assumption or any assumption on controllability, provided the circuit is employed only in a given HW $k$ subspace with fixed $k$ and under certain assumptions on the qubit connectivity.
The central result which enables this claim is the following \autoref{lemma:VarianceHWPreserving}, which by leveraging the specific form of RBS/FBS circuits, provides an exact analytic expression for the the variance of the cost gradient (given an initial state and a target state), for our cost function of interest. From this result, it is then possible to prove the absence of BP results depending on the situation of interest. We propose two such applications in that regard, presented as \autoref{thm:NoBPPSA} and \autoref{thm:NoBPgeneralcase}, that deal respectively with the case of a periodic connected ansatz with any input/target states, and an arbitrary circuit with randomly sampled input/target states accordng to a family of distributions. 

\begin{restatable}[Variance of RBS and FBS based VQCs]{lemma}{VarianceHWPreserving}
\label{lemma:VarianceHWPreserving}
Let us consider an $n$-qubit HW-preserving VQC made of $D\geq1$ RBS or FBS gates only, that is employed in the subspace of HW $k$ (i.e. both the initial state $\zeta^0$ and the target state $y$ are normalized real superpositions of the basis $B_k^n$), along with the cost function $\mathcal{C}(\theta)$ taken as the squared Euclidean distance between the final state $\zeta^{\lambda_{\mathrm{max}}}$ and the target state $y$. If $\theta$ is distributed uniformly in $\theta:=[0,2\pi]^D$, then we have for all $\lambda \in \llbracket1, \lambda_{\mathrm{max}}\rrbracket$:
\begin{equation}
    \mathbb{E}_{\theta}[\partial_{\theta_{\lambda}} \mathcal{C}(\theta)] = 0 \,
\end{equation}
\begin{equation}\label{eq:VarianceHWPreserving-Var}
    \begin{split}
        \mathrm{Var}_{\theta}[\partial_{\theta_{\lambda}} \mathcal{C}(\theta)] = 2 & \sum_{l,j} \left( \frac{1}{\left(2 \pi\right)^D} \int_{\theta} (\zeta_l^{\lambda})^2 + (\zeta_j^{\lambda})^2 d\theta \right) \\
        & \cdot \left( \frac{1}{\left(2 \pi\right)^D} \int_{\theta} (\tilde{y}_l^{\lambda})^2 + (\tilde{y}_j^{\lambda})^2 d\theta \right)\,,
    \end{split}
\end{equation}
with $\zeta^{\lambda} = \omega^{\lambda - 1} \dots \omega^{1} \cdot \zeta^0$ the intermediate state (before inner layer $\lambda$), $\tilde{y}^{\lambda} = (\omega^{\lambda+1})^\intercal \dots (\omega^{\lambda_{\mathrm{max}}})^\intercal \cdot y$ the back-propagated target state, and where the sum is over all pairs $(l,j)$ of basis state indices which are undergoing a planar rotation of angle $\theta$ by the layer $\lambda$.
\end{restatable}

The proof of this Lemma is based on the fact that the gradient of the cost according to one parameter $\theta_{\lambda}$ can be expressed as in \autoref{eq:backpropagation_equation}. A detailed proof can be found in \autoref{chap:proof_Lemma_Var}, first by considering state of HW $1$ in \autoref{chap:proof_lemma_unary}, then by generalizing the results to any HW in \autoref{chap:proof_lemma_general_RBS}.

To state the next theorem, we introduce the following assumption on our circuits.

\begin{definition}[CPSA]\label{def:CPSA}
We say that an RBS/FBS circuit is a \emph{Connected Periodic Structure Ansatz} (CPSA) if it is composed of $L\geq1$ parametrized repetitions of a pattern $U_0$ of RBS or FBS gates, i.e. a circuit where the equivalent unitary matrix $U(\theta)$ in the subspace of fixed HW $k$ is of the form:
\begin{equation}\label{eq:PeriodicStructureAnsatz}
    U(\theta) = \prod_{l=1}^{L} U_0(\theta_l), \quad U_0(\theta_l) = \prod_{j=1}^J e^{-i \theta_{l,j} H^j_{RBS/FBS}},
\end{equation}
and if furthermore the repeated pattern $U_0(\cdot)$ \emph{connects all qubits}, i.e. a path between any two qubits may be traced on $U_0(\cdot)$'s RBS/FBS circuit diagram.
\end{definition}

Of course, CPSA's are only possible on quantum architectures that have a \emph{connected} qubit connectivity graph. The simplest example of such a $U_0$ that connected all qubits is the diagonal line of $n-1$ RBS/FBS gates connecting qubits $1$ and $2$, $2$ and $3$, and so on. The following holds:

\begin{restatable}[Absence of Barren Plateaus, informal]{thm}{NoBPPSA}
\label{thm:NoBPPSA}
Under the same assumptions as \autoref{lemma:VarianceHWPreserving}, if additionally the gates are arranged in a CPSA (\autoref{def:CPSA}),
then there exists an integer $q\geq1$ such that if the number of repetitions $L$ grows at least as fast as $n^q$, then for all $j$, and for any $0<\alpha<1$, setting $l=\lfloor \alpha \, L \rfloor$ implies
\begin{equation}
        \begin{split}
            &\mathrm{Var}_{\theta}[\partial_{\theta_{l,j}} \mathcal{C}(\theta)] = \frac{k(n-k)}{n(n-1)}\frac{8}{d_k} + \varepsilon\,,
        \end{split}
    \end{equation}
where $\varepsilon$ decays exponentially with $n$.
Thus, after some polynomial amount of repetitions, and for angles located at any constant fraction of the depth, there is an absence of Barren Plateaus for CPSA ansatz.
\end{restatable}

The fact that this statement concerns gates located at constant ratios of the circuit depth may be interpreted as the condition that the gates are not too close to either extremities of the circuit. We concede that the proof of this theorem depends on the validity of a small conjecture that we make about spectral gaps of certain stochastic matrices (\autoref{conj:spectral-gap}), for which we present numerical evidence in \autoref{subsec:numerical-evidence-spectral-gap}.
Since \autoref{thm:NoBPPSA} states a requirement of a polynomial number of repetitions $L$ to reach its conclusion of absence of Barren Plateaus, one may wonder if it can generally be subsumed by the usual approximate $2$-design argument, since usually polynomial repetitions produce approximate $2$-designs (see Theorem 1 in \cite{larocca_diagnosing_2022}).
A discussion on the subtilities behind this is given in \autoref{subchap:design-discussion}.

The next theorem does not require any lower limit on the circuit depth:

\begin{restatable}[Evolution of the variance for RBS and FBS based quantum circuits]{thm}{NoBPgeneralcase}
\label{thm:NoBPgeneralcase}
Under the same assumptions as \autoref{lemma:VarianceHWPreserving}, if additionally the initial state $\zeta^0$ and the target state $y$ are each independently distributed on the sphere $S^{d_{k}-1}$ such that:
\begin{equation}\label{eq:hyp_distribution_NoBPGeneral}
    \forall r \in [d_k], \quad \begin{cases} \mathbb{E}[\zeta_r^0] = \mathbb{E}[y_r] = 0\,, \\ \mathbb{E}[(\zeta_r^0)^2] = \mathbb{E}[(y_r)^2] = \frac{1}{d_k}\,, \end{cases}
\end{equation}
then we have for all $\lambda \in \llbracket1, \lambda_{\mathrm{max}}\rrbracket$:
    \begin{equation}\label{eq:var_NoBPgeneralcase}
        \begin{aligned}
            \mathbb{E}_{\zeta^0,y} \mathrm{Var}_{\theta}[\partial_{\theta_{\lambda}} \mathcal{C}(\theta)] &= \frac{k(n-k)}{n(n-1)}\frac{8}{d_k}\,.
        \end{aligned}
    \end{equation}
\end{restatable}

Note that the assumption of \autoref{eq:hyp_distribution_NoBPGeneral} on the distributions lies in between the assumptions of spherical $t$-designs of $t=1$ and $t=2$ (the first line of \autoref{eq:hyp_distribution_NoBPGeneral} imposes a spherical 1-design, while the second line only enforces values of the \emph{homogeneous} $2^{\text{nd}}$ order moments, but leaves \emph{correlations} between components like $\mathbb{E}[y_1 y_2]$ unconstrained).

It can be concluded from these results that Barren Plateaus are absent for subspace-invariant RBS and FBS based quantum circuits for a fixed subspace $k$.
It is emphasized that, unlike recent related works \cite{larocca_diagnosing_2022, ragone_lie_2024, fontana_characterizing_2024} (which apply to quite general circuits), \autoref{thm:NoBPPSA} for RBS/FBS circuits does not rely on a 2-design assumption for the global unitary of the circuit.
It is also emphasized that while \autoref{thm:NoBPgeneralcase} considered particular assumptions on the input/target state distributions and cost function that do not encompass all possible learning tasks, the proof could be adapted to other tasks.

%% file: chapters/05_Photonic_Sub_Optimals_Models/Photonic_Suboptimal_models.tex
\let\textcircled=\pgftextcircled
\chapter[Photonic Sub-Optimal Models]{Photonic Sub-Optimal Models}
\label{chap:Photonic_Suboptimal}
\begin{textblock}{5.3}(0,-4)
	%\textit{`Un scientifique dans son laboratoire est un enfant placé devant des phénomènes naturels qui l'impressionnent comme des contes de fées.'\\}

%\hspace{0.5cm}--- Marie Skłodowska-Curie.
\end{textblock}

\initial{A}l\textit{though quantum computers promise large advantages over classical computing, fault-tolerant universal quantum computers are still far from being available. In particular, while photonics is one of the promising platforms for quantum computing, the technological requirements for such photonic quantum devices are huge. They often rely on the capacity to achieve adaptive measurement-based operations \cite{knill_scheme_2001, knill_quantum_2002}, and to have access to a large number of modes and initial coherent photons. In the meantime, sub-universal models have been proposed to achieve near-term quantum advantage. Those models are believed to have an intermediate computational advantage even without being able to achieve every operation that a fault-tolerant quantum computer could do. Boson Sampling \cite{gard_introduction_2015}, Gaussian Boson Sampling \cite{zhong_quantum_2020}, or IQP circuits sampling \cite{bremner_classical_2010} are good candidates, but the range of problems that one can solve using such approaches seems very limited. Finding an architecture able to offer a quantum utility to real life use case for quantum photonic device in the era of Noisy Intermediate-Scale Quantum \cite{preskill_quantum_2018} devices is an important field of research. Previous work also used linear optic circuits with post-processing strategy to simulate universal quantum computing \cite{polino_photonic_2024}, at the cost of strongly increasing the running time of the produced algorithms.}

\textit{In the previous chapters, it was explained how theoretical guarantees can be found regarding the training of subspace preserving quantum algorithms. The focus is proposed to be on QML algorithms based on such theoretical guarantees, and that focus on polynomial advantage. Linear optical circuits are good candidates for those applications, thank to their native subspace preserving properties, and their large repetition rate. However, QML applications for photonic platforms require to go beyond linear optic architectures to increase the expressivity of the resulting models, but that does not necessarily depend on whether the model is universal for quantum computation. Previous works came up with alternative solutions to incorporate non-linearity --- an important element for neural network architectures --- such as global measurements and classical activation functions between linear optical layers \cite{steinbrecher_quantum_2019}, physical non-linear blocs \cite{fu_photonic_2023}, or adaptive gates to perform learning tasks \cite{chabaud_quantum_2021}. In this Chapter, a new scheme of adaptivity is presented that allows to increase the controllability and maintain the subspace preserving properties of the circuits.}

    \section{Linear Optical Quantum Circuits}\label{sec:LO_Circuits}
        
        \subsection{Structure of Linear Quantum Optics}\label{subsec:Structure_LO}

Linear-optical networks are considered with $m$ modes and set of simple optical elements (beam-splitters and phase-shifters). In general, those circuit are used while considering input states made with $n$ identical photons that pass through the modes and optical elements and then measured to determine their locations. Here, adaptive photon-number measurements are not considered.

Each element of the linear-optical networks/circuits may either be regarded as \textit{fixed}, or \textit{parameterized}, meaning that the gate has a tunable parameter $\theta \in [0,2\pi]$ that can be freely varied, corresponding to a beam-splitter's \textit{angle} or a phase-shifter's \textit{phase}. If there are $p$ parameterized gates in the circuit, we denote by $\theta \in \Theta:=[0,2\pi]^p$ the tuple of all the parameter values. Into the circuit are sent $n$ (indistinguishable) photons in some pure quantum state. A pure quantum state of $n$ photons is a normalized vector in the \textit{$n$-photon Fock space}, which is the Hilbert space of all complex superpositions of the \textit{basis Fock states} $\ket{\bm s}$ (for all $\bm{s} = (s_1\dots,s_m)\in \mathbb N^m$ such that $s_1 + \dots + s_m = n$). We denote the set of basis Fock states by $\Phi_{m,n}$, of which there are $d_n := |\Phi_{m,n}| = \binom{m+n+1}{n}$ many. The direct sum of all the $n$-photon Fock spaces is known as \textit{the Fock space}, which is infinite-dimensional. We will use the notations $|\bm{s}| = \sum_{i=1}^m s_i$ and $\bm{s}! = \prod_{i=1}^m s_i !$.

\begin{figure}[h!]
    \centering
    \includegraphics[width=0.95\linewidth]{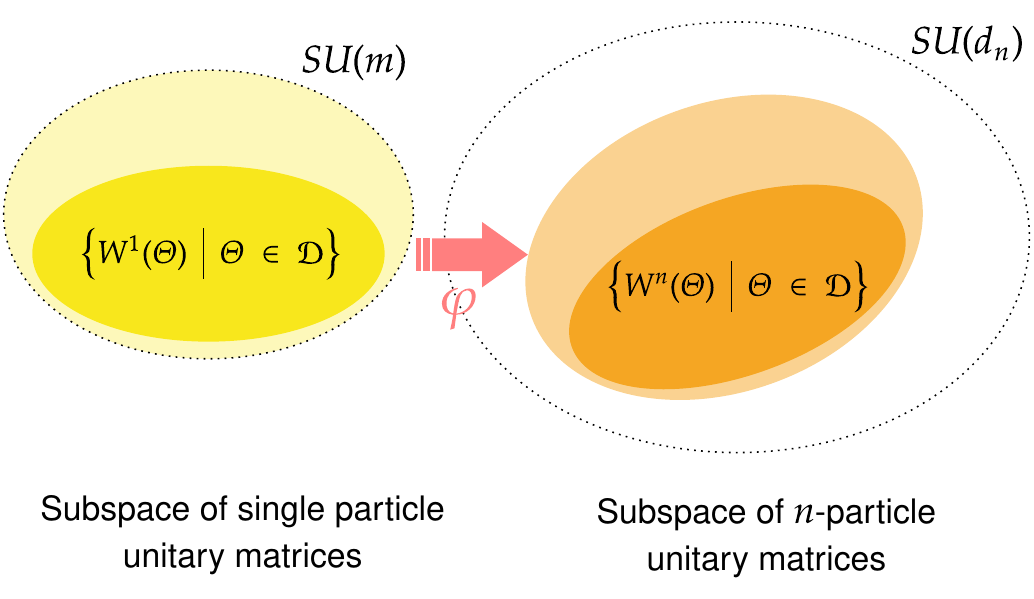}
    \caption{Representation of the spaces of reachable unitaries using linear quantum optics over $m$ modes, for $1$ photon (left) and $n$ photons (right). The homomorphism $\varphi$ is described in \autoref{thm:BS_Model_Bloc}.}
    \label{fig:Subspace}
\end{figure}

An arrangement of beam-splitters and phase-shifters over $m$ modes specifies a given unitary $W^1 \in SU(m)$\footnote{For simplicity, in this work we only ever consider \textit{special}-unitary matrices, i.e. unitary matrices $U$ with $\det(U)=1$. This is without loss of generality, as it just amounts to a convention choice in how one writes the $m \times m$ unitary matrices representing beam-splitters and phase-shifters on the $m$-mode system.}, which dictates the evolution of a single photon in the circuit. Conversely, all unitaries in $SU(m)$ may be realized as some arrangement of beam-splitters and phase-shifters \cite{reck_experimental_1994}. A single-photon unitary $W^1 \in SU(m)$ determines an $n$-photon
unitary $W^n \in SU(d_n)$, through the so-called \textit{($n$-photon) photonic homomorphism}: $W^n := \varphi(W^1)$ (see \autoref{thm:BS_Model_Bloc}). Since quantum linear optics preserves the photon number $n$, the global unitary $W$ (over the whole Fock space) of the circuit corresponds to an inifitly-sized block-diagonal matrix, where the blocks are the evolutions for a fixed photon number. Lastly, a fixed architecture of beam-splitters and phase-shifters will give rise to a parametrization $\theta\mapsto W^1(\theta)$, and so the spaces of accessible unitaries (for all possible parameter values $\theta$) will generally be smaller. We summarize the different unitary matrices introduced, the spaces they live in, and the homomorphism relation $\varphi$, in \autoref{fig:Subspace}.

\begin{figure}
    \centering
    \includegraphics[width=0.95\linewidth]{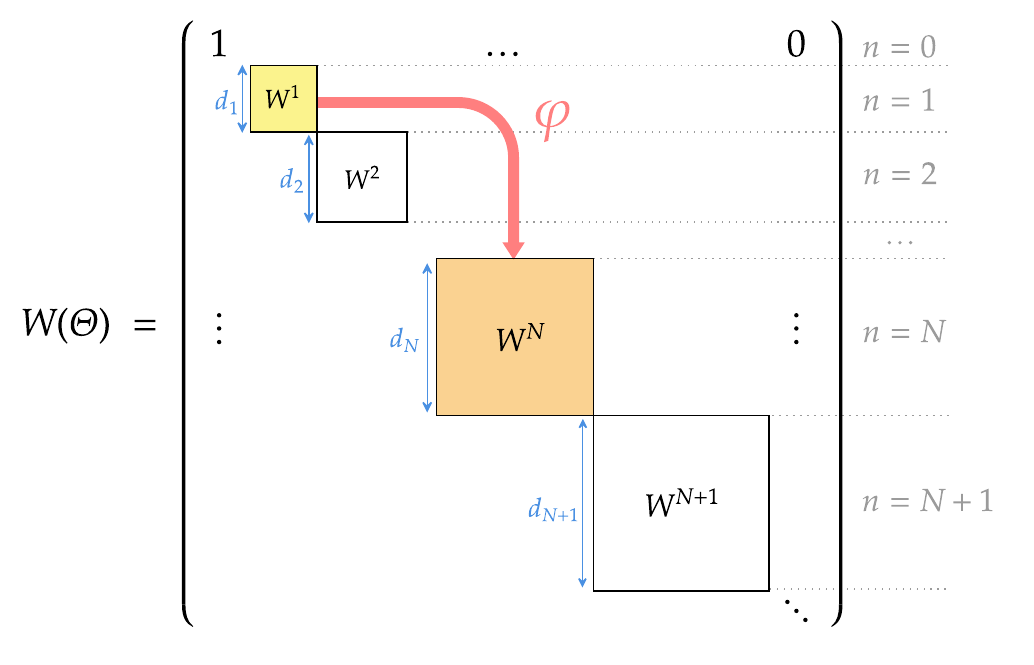}
    \caption{Representation of linear optic quantum circuit equivalent unitary matrix as a bloc diagonal matrix. Each bloc $W^n$ corresponds to a subspace of $n$ particles. The homomorphism $\varphi$ is described in \autoref{thm:BS_Model_Bloc}. Each bloc $W^n$ is a parametrized unitary matrix of dimension $d_n = \binom{m+n+1}{n}$.}
    \label{fig:Unitary_Bosonic_Circuits}
\end{figure}

The homomorphism $\varphi:SU(m)\to SU(d_n)$ describes the way in which second-quantization enforces the evolution of $n$ indistinguishable bosons, given the evolution unitary for a single boson. There are different equivalent ways to describe $\varphi$. One approach builds an expression for $\varphi(W^1)$ in terms of the matrix \textit{permanents} \cite{marcus_permanents_1965} of certain matrices related to $W^1$ \cite{aaronson_computational_2011}; another approach is to leverage the fact that $\varphi$ is an injective homomorphism between two Lie groups, and so can be understood in terms of its derivative action on the Lie Algebra $\mathfrak{su}(m)$ \cite{parellada_no-go_2023}.
We recall in \autoref{thm:BS_Model_Bloc} the first approach's expression.

\begin{restatable}[{Photonic homomorphism in terms of matrix permanents, from \cite[Section 3]{aaronson_computational_2011}}]{thm}{BSModel}\label{thm:BS_Model_Bloc}
    Given a unitary $W^1(\theta) \in SU(m)$ describing an $m$-mode linear optical circuit (possibly parametrized by $\theta$), the corresponding unitary $W^n(\theta):=\varphi(W^1 (\theta))$ describing the $n$-photon evolution is given, for all $\bm{s}, \bm{t} \in \Phi_{m,n}$, by
    \begin{equation}\label{eq:bloc_unitary_expression_BS}
        \bra{\bm{s}} W^n(\theta) \ket{\bm{t}} = \frac{\per{ W^1_{\bm{s},\bm{t}}(\theta) }}{\sqrt{\bm{s}! \bm{t} !}} \, ,
    \end{equation}
    where $W^1_{\bm{s},\bm{t}}(\theta)$ is an $n \times n$ matrix built from $W^1(\theta)$, by first taking $s_j$ copies of its $j^{\mathrm{th}}$ column and then $t_i$ copies of the $i^{\mathrm{th}}$ row of the resulting $m \times n$ matrix.
\end{restatable}

        \subsection{Connection between Reconfigurable Beam Splitters and Photonic Beam Splitters}\label{subsec:Connection_RBS_BS}

The Reconfigurable Beam Splitter (RBS) gate is a 2-qubit gate that corresponds to a $\theta$-planar rotation between the states $\ket{01}$ and $\ket{10}$:
\begin{equation}
    W_{RBS}(\theta) = \begin{pmatrix}
        1 & 0 & 0 & 0 \\
        0 & \cos(\theta) & \sin(\theta) & 0 \\
        0 & -\sin(\theta) & \cos(\theta) & 0 \\
        0 & 0 & 0 & 1 \\
        \end{pmatrix} \, \textrm{.}
    \end{equation}    
One of the main elements of Linear Optical circuits is the Beam Splitter (BS), which is a two modes gate that acts on the amplitude of input photons. It is well known that lossless two modes BS (with two input and output modes) in quantum optics is described by the unitary matrix $W_{BS}$ which has the form~\cite{makarov_theory_2022}:
\begin{equation}
    \binom{\hat{b}_1}{\hat{b}_2} = W_{BS}(T,R,\phi) \binom{\hat{a}_1}{\hat{a}_2}, \quad W_{BS}(T,R,\phi) = \begin{pmatrix} \sqrt{T} & e^{i\phi} \sqrt{R} \\
    -e^{-i\phi}\sqrt{R} & \sqrt{T} \end{pmatrix} \, ,
\end{equation}
with $\hat{b}_1$, $\hat{b}_2$ the ouptut mode annihilation operators, $\hat{a}_1$, $\hat{a}_2$ the input mode annihilation operators, $T$ and $R$ are the transmittance and reflectance ($R+T=1$), and $\phi$ is the phase shift. In this work, we will focus on the ideal model with the phase shift $\phi=0$. Using a change of variable, it comes that:
\begin{equation}
    W_{BS}(\theta)= \begin{pmatrix} \cos(\theta) & \sin(\theta) \\
    -\sin(\theta) & \cos(\theta)\\
    \end{pmatrix} \quad \quad \text{with} \quad \cos(\theta) = \sqrt{T} \quad \text{and} \quad \sin(\theta) = \sqrt{R} \,\textrm{.}
\end{equation}
Both BS and RBS perform a $\theta$-planar rotation between state $\ket{01}$ and $\ket{10}$, but those states are Fock states in the photonic case, and qubit states in the HW preserving case. However, BS and RBS are subspace-preserving gates as RBS gates preserve the HW and BS gates preserve the number of particles. As a result, one can express the equivalent unitary of both BS and RBS based quantum circuits as block-diagonal as explained in \cite{monbroussou_trainability_2025} and  \cite{monbroussou_toward_2025}. We illustrate those block diagonal equivalent unitary matrices in \autoref{fig:Unitary_Bosonic_Circuits} for $m$ modes, and in \autoref{chap:HW_Preserving_Methods}. In these figures, each block represents the equivalent unitary when considering a fixed HW $k$ or a fixed number of particles $k$. For the photonic case, the number of particles is unbounded, and the equivalent unitary is of infinite dimension. Both gates have the same impact on the initial state $\ket{00}$, $\ket{10}$, and $\ket{01}$ but they act differently on the state $\ket{11}$. In addition, other initial Fock state can be considered for the photonic BS that does not map to any state for the RBS.

Consider two quantum circuits, the first is a photonic circuit of $m$ modes, and the second is a $m$-qubit quantum circuit. When considering the subspace of HW $1$ (unitary subspace) for the RBS and the subspace of a single particle for the BS, a BS applied between modes $i$ and $j$ in the first circuit, and a RBS applied between qubit $i$ and $j$ will have the same effect. This is the reason why one can easily adapt the convolutional layer presented in \cite{monbroussou_subspace_2025}, as the tensor encoding described in Eq.~(1) of the main text ensures that, for each register, there is only one particle. 

For larger subspaces, the equivalent unitary matrices will differ. First, the size of the subspace for $m$ qubits and HW $k$ corresponds to the number of bitstring of $m$ bits and HW $k$ and is $\binom{m}{k}$, while the size of the subspace for $m$ modes and $k$ particles is $\binom{m+k-1}{k}$. In addition, authors in \cite{aaronson_computational_2011} explain that the homomorphism $\varphi$ describes the way in which second-quantization enforces the evolution of $k$ indistinguishable bosons, given the unitary evolution for a single boson. This relationship and its impact on linear optical quantum circuit controllability are discussed in \cite{monbroussou_toward_2025}. For HW preserving quantum circuits, a similar relationship between the block can exists. For example, when considering only a line connectivity, RBS acts as a Fermionic BeamSplitter gates and each block $W^k$ is the $k$-compound matrix of $W^1$ \cite{kerenidis_quantum_2022}. For a greater connectivity, such relationship disapears and even if all the blocs are highly correlated, each subspace can be perfectly controled and are not always determined by the first one \cite{monbroussou_trainability_2025}.

    \section{Limited Controllability of Linear Quantum Optics}\label{sec:Limitation_BS}

In the following we will refer to the \textit{dimension} of various subsets of unitary matrices, or of density matrices.

Consider a parametrized linear optical circuit, $W^1(\theta)$, with $p$ parametrized gates, i.e., $\theta \in \Theta=[0,2\pi]^p$. The space of $m \times m$ unitaries that are accessible, as all parameters are explored, is by definition included in $SU(m)$. Therefore,
\begin{align}
    \quad \dim(\{ W^1(\theta) \; | \; \theta \in \Theta\}) \leq \min(p,\ m^2 - 1).
\end{align}

For the corresponding $n$-photon unitaries, the existence of the injective homomorphism $\varphi$ implies that the dimension of the set of the $n$-photon unitaries reached is equal to that of the single-photon unitaries, and therefore it obeys the same limitations:
\begin{equation}
    \begin{split}
        \dim(\{ W^n(\theta) \; | \; \theta \in \Theta\}) &= \dim(\{ W^1(\theta) \; | \; \theta \in \Theta\}) \\
        &\leq \min(p,\ m^2 - 1).
    \end{split}
\end{equation}

This limitation in the set of achievable unitary matrices dimension is a constraint on the expressivity of the model output. Other figures of merit for the expressivity of quantum models exist, including the distance to a 2-design \cite{holmes_connecting_2022} that characterizes the distribution of the unitary matrices, or the Fourier expressivity \cite{xiong_fundamental_2025, mhiri_constrained_2024}. In this work, we focus on the notion of \textit{controllability} of the output state of the quantum circuit, i.e., the number of independent directions it can locally explore in the space of density matrices.

Recent works have highlighted the impact of the controllability of quantum circuits, especially in QML, by e.g. studying the rank of the Quantum Fisher Information Matrix (QFIM) \cite{larocca_theory_2023}, or the dimension of the Dynamical Lie Algebra \cite{larocca_diagnosing_2022, ragone_lie_2024, fontana_characterizing_2024} generated by the Hamiltonians in the circuit. Measurement based techniques are introduced to increase the controllability of photonic circuits. As they are not always CPTP maps, the controllability is studied at the level of the output state's density matrix.  

Studying circuits made of linear optics and adaptivity parts requires to consider a tool to characterize the controllability that is not only defined for unitary transformations. Accordingly, a new measure of controllability of the output state of a parametrized quantum circuit is introduced, using its corresponding Jacobian rank.

\begin{definition}[Number of degrees of freedom of a state]\label{def:DoF}
    The \textbf{number of degrees of freedom} of an $n$-photon state $\rho(\theta)$ at a point $\theta$ in the parameter space is defined as the rank of Jacobian matrix of the map $\rho:\Theta\to\mathbb{C}^{d_n \times d_n}$ calculated at point $\theta$:
    \begin{equation}
        \mathrm{DoF}(\rho(\theta)) = \rank[J \rho(\theta)]\,.
    \end{equation}
    The Jacobian matrix considered is actually the one of the map $\tilde{\rho}:\Theta\to \mathbb{R}^{2d_n^2}$ that results from viewing a complex matrix $\rho(\theta) \in \mathbb{C}^{d_n \times d_n}$ as a real vector $\tilde{\rho}(\theta)$ of length $2d_n^2$ (through concatenation of all the columns of the matrix, and splitting of each complex scalar into its real and imaginary parts). We recall that the Jacobian matrix of a differentiable function $f:\mathbb{R}^a \to \mathbb{R}^b$ at point $\bm{x}\in\mathbb{R}^a$ is the $b \times a$ real matrix defined by $\big[J f (\bm{x})\big]_{ij}:=\frac{\partial f_i}{\partial x_j}$.
\end{definition}

Since the considered circuits may apply successively operations that are standard parameterized optical gates, fixed optical gates, and state injections, overall the map $\theta \mapsto \rho(\theta)$ that sends a parameter tuple to the final state's density matrix is \emph{analytic}\footnote{A function is \textit{analytic} if it is smooth and if it agrees locally with its Taylor series around each point in the domain. Sines, cosines, as well as matrix products, sums, and exponentials, are all analytic; and compositions of analytic functions are still analytic. Hence, all the maps $\theta \mapsto \rho_{\mathrm{out}}(\theta)$ considered in this work --- even those including state injections --- are analytic, being only compositions of sine, cosines, and products and sums of matrices.}, and consequently, the number of degrees of freedom $\theta \mapsto \mathrm{DoF}(\rho(\theta))$ is constant \textit{almost-everywhere} on the parameter space, due to \cite[Prop. B.4]{bamber_how_1985} (a more precise writing would be exactly the same as \cite[Lemma 4]{monbroussou_trainability_2025}, with rank of density matrix's Jacobian in place of rank of the pure state's Quantum Fisher Information Matrix (QFIM)).

\begin{restatable}[Almost-constant property of number of degrees of freedom]{thm}{DOFrankthm}
\label{thm:DoFthm}
Let $\rho(\theta)$ be the density matrix of the output state of a linear optical circuit, with or without state injections. Then, its number of degrees of freedom is, \emph{almost everywhere} on the considered parameter space $\Theta$, constant and equal to
\begin{equation}\label{eq:DofThmEq}
    \mathrm{DoF}_{\mathrm{max}}(\rho):=\max\limits_{\theta \in \Theta}\mathrm{DoF}(\rho(\theta)).
\end{equation}

\end{restatable}
The practical consequence of \autoref{thm:DoFthm} is that drawing a point $\theta \in \Theta$ uniformly at random and calculating the state's number of degrees of freedom at that point yields $\mathrm{DoF}(\rho(\theta))=\mathrm{DoF}_{\mathrm{max}}(\rho)$ with probability $1$.

Hence in this work, we numerically evaluate the quantity $\mathrm{DoF}_{\mathrm{max}}(\rho)$ as follows: given an optical circuit over $m$ modes and an $n$-photon input state $\rho_{\mathrm{in}}$, we classically simulate the $d_n \times d_n$ output density matrix $\rho_{\mathrm{out}}(\theta)$ through successive applications of beam-splitters and phase-shifters unitary channels and state injection CPTP maps, using Python's library \textit{PyTorch} \cite{paszke_pytorch_2019}, which, by relying on automatic differentiation, enables us to access the Jacobian $J \rho_{\mathrm{out}}(\theta)$ of the output state. Then, we draw uniformly at random $\theta$, and calculate $\rank[J \rho(\theta)]$. With probability 1, this number is equal to $\mathrm{DoF}_{\mathrm{max}}(\rho)$ (\autoref{eq:DofThmEq}).

Since the controllability of the output state $\rho_{\mathrm{out}}(\theta)$ lies entirely in the controllability of the single-photon unitary $W^1(\theta)$ of the whole circuit, we immediately have the following limitation.

\begin{restatable}[Controllability limitation of linear quantum optics]{thm}{ControlLimitationBS}
\label{thm:Control_Limits_BS}
    Consider an $n$-photon pure state ${\rho_{\mathrm{in}} := \ketbra{\psi_{in}}}$ entering an $m$-mode linear optical circuit $W^1(\theta)$ with $p$ parametrized gates, and without state injections.
    Then, the controllability of the output density matrix $\rho_{\mathrm{out}}(\theta) := W^n(\theta) \rho_{\mathrm{in}} W^n(\theta)^\dagger$ is bounded by
    \begin{equation}
        \mathrm{DoF}_{\mathrm{max}}(\rho_{\mathrm{out}}) \leq \dim(\{ W^n(\theta) \; | \; \theta \in \Theta\}) \leq m^2 - 1.
    \end{equation}
\end{restatable}

In fact, note that in the case where the input state consists of all photons in one same mode, i.e., ${\rho_{\mathrm{in}} = \ketbra{n,0,\dots,0}}$, an even tighter bound of $O(m)$ instead of $O(m^2)$ may be shown to hold.\footnote{This bound can be obtained by an explicit calculation of the rank of the action of the derivative of the photonic homomorphism onto this initial state.}

There exists subspace preserving quantum circuits that do not suffer from such controllability limitations, e.g., RBS-based Hamming Weight preserving quantum circuits \cite{monbroussou_trainability_2025}. Those ansatz are particularly useful as they are likely to avoid vanishing gradient phenomena, the so-called \emph{barren plateau} \cite{mcclean_barren_2018}, when considering subspaces of polynomial dimension with respect to the number of qubits \cite{larocca_diagnosing_2022, ragone_lie_2024, fontana_characterizing_2024}. They can often be classically simulated \cite{cerezo_does_2024, goh_lie-algebraic_2025}, meaning that such quantum circuits would offer only polynomial advantage, if they are to offer any advantage at all. However, recent works have proposed quantum machine learning algorithms based on such ansatz \cite{jain_quantum_2024, monbroussou_subspace_2025, cherrat_quantum_2024, kerenidis_quantum_2022}. In the following, we explain how SI can increase the controllability of the quantum circuit while maintaining -- if needed -- the subspace preserving properties of the circuit.

    \section{Adaptivity Schemes}\label{sec:Adaptivity_Schemes}

        \subsection{Adaptive Linear Optic and State Injection Schemes}\label{subsec:ALO_SI_schemes}

In this Section, the State Injection (SI) scheme is introduced. This method allows to increase the expressivity of the photonic, and to perform tasks that are believed hard to do classically with fewer experimental constraints in comparison with the Adaptive Linear Optics (ALO) scheme~\cite{chabaud_quantum_2021}, as explained in \autoref{subsec:Proba_Estimation}. In the following, the SI scheme is defined in \autoref{subsubsec:scheme_def} and its experimental implementation is motivated in \autoref{subsubsec:general_exp_framework}.

            \subsubsection{Scheme Definition}\label{subsubsec:scheme_def}

\begin{figure}[h!]
  \begin{subfigure}{0.47\columnwidth}
    \includegraphics[width=\linewidth]{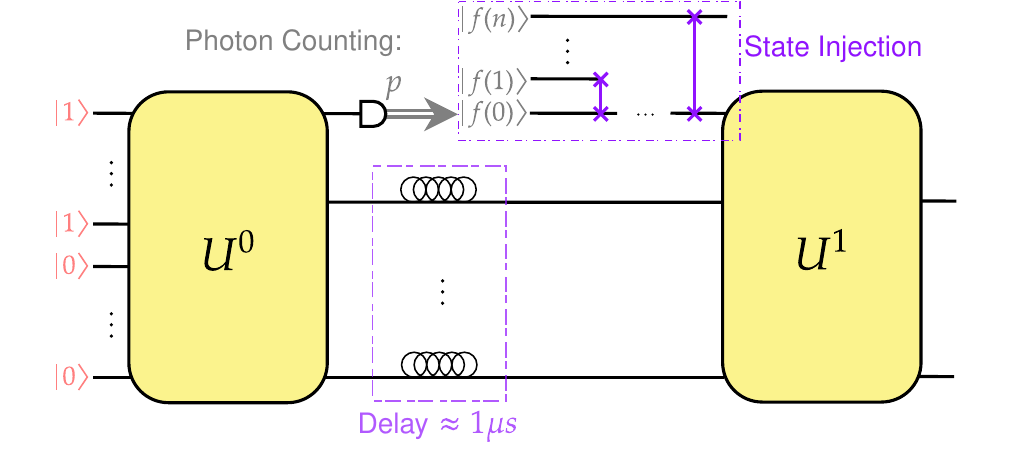}
    \caption{State Injection channel for a single mode measured. The purple SWAPs in the state injection part represent switches. The symbol $f(p)$ represents the adaptivity function that connects the measurement outcome $p$ and the injected state.}
    \label{fig:Experimental_Comparison_StateInjection}
  \end{subfigure}
  \hfill
  \begin{subfigure}{0.47\columnwidth}
    \includegraphics[width=\linewidth]{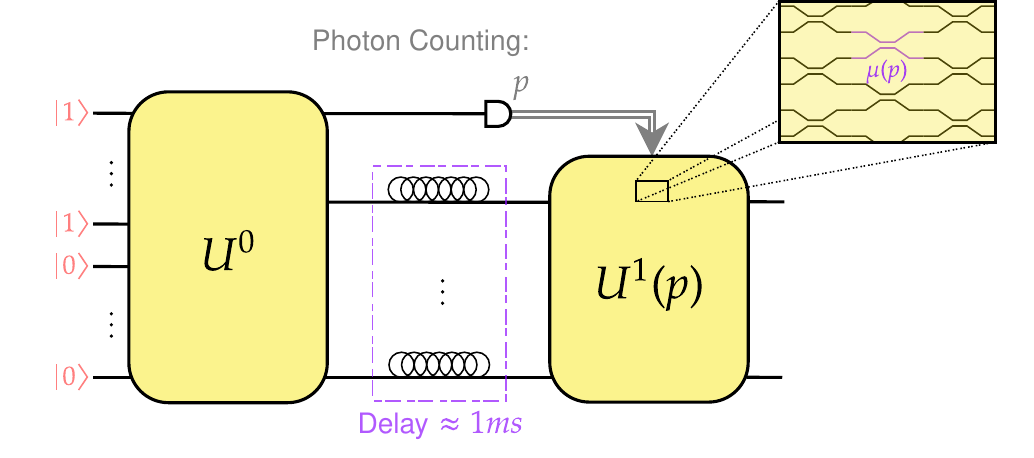}
    \caption{Feed-Forward adaptivity for a single mode measured. The symbol $\mu(p)$ represents the adaptivity function that connects the measurement outcome $p$ and the new value of the linear optical parameters.}
    \label{fig:Experimental_Comparison_AdaptiveFeedForward}
  \end{subfigure}
  \caption{Comparison of experimental requirements for the State Injection scheme (in \autoref{fig:Experimental_Comparison_StateInjection}) that we introduce, and the Feed-Forward scheme (in \autoref{fig:Experimental_Comparison_AdaptiveFeedForward}) proposed in \cite{chabaud_quantum_2021}.}
  \label{fig:Experimental_Comparison}
\end{figure}

The ALO scheme, that we illustrate in \autoref{fig:Experimental_Comparison_AdaptiveFeedForward}, was proposed by~\cite{chabaud_quantum_2021} where the authors built a \emph{feed-forward} scheme for linear optical quantum computation. Their setup is composed of an input Fock state, with $n$ photons spread across $m$ modes, and $k$ adaptive measurements. 
An adaptive measurement consists in measuring one mode using photon number resolving detector and configure the following $(m-1)\times(m-1)$ unitary according to the measurement result.

\begin{figure*}[h!]
    \includegraphics[width= 1\linewidth]{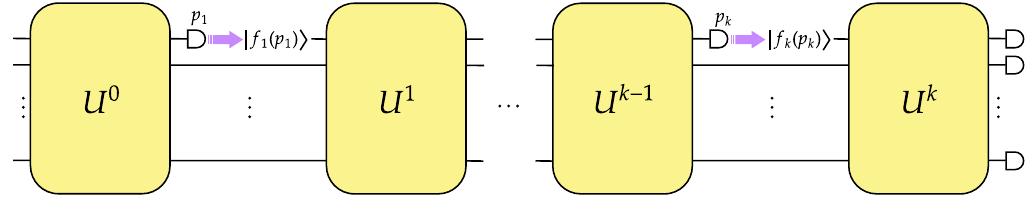}
    \caption{Quantum circuit made of linear optical blocks separated by state injections.}
    \label{fig:Introduction_framework}
\end{figure*}    

We illustrate the SI scheme in \autoref{fig:Introduction_framework}. It differs from the ALO scheme by the following. Firstly, no real time reconfiguration of the quantum circuit is required to be done within the run of the experiment as all unitary matrices are preset for each run. Secondly, our adaptive part comes from choosing what is the new Fock state to be injected in the circuit after a measurement. Lastly, the unitary does not shrink in size as the number of modes is preserved throughout the computation.

\begin{definition}[State Injection]\label{def:StateInjection}
    We call \textbf{State Injection} (SI) any operation on an $m$-mode photonic platform that performs photon-counting measurements in one or several modes, and, depending on the outcomes obtained, re-injects some photons back in one or several modes.
    Overall (since no single outcome is post-selected on), this operation is described by a CPTP map on the relevant Hilbert space.
    Different SI operations hence correspond to different choices of modes that undergo measurements and/or re-injections, and different choices of rules that map measurement outcomes to the corresponding re-injections that should be performed. We refer to the latter as a choice of \textbf{injection functions}.
\end{definition}

In the special case where the same $k$ modes (here written as adjacent modes for simplicity) are being subject to both measurements and re-injections, and the total photon count is preserved, SI operations may be detailed as follows. 
If the photon count outcomes obtained when measuring those modes are $n_1,\dots,n_k$, then the re-injection process consists of injecting, \emph{in those modes only}, the new state $\ket{f_1(n_1,\dots,n_k),\cdots,f_k(n_1,\dots,n_k)}$. The injection functions are maps of the form ${f_i:[\![0,n]\!]^k \to [\![0,n]\!]^k}$, where $[\![a, b]\!]$ denotes the set of integers between two nonnegative integers $a$ and $b$. Such functions may be chosen arbitrarily among those that respect the photon-number conservation constraint
\begin{equation}\label{eq:StateInjection-SpecialCase-photon-conservation-constraint}
    f_1(n_1,\dots,n_k) + \cdots + f_k(n_1,\dots,n_k) = n_1 + \cdots + n_k\,.
\end{equation}

Preserving the number of particles is particularly interesting as it allows to perform subspace preserving computation. Recent works have shown theoretical guarantees on the training of such variational circuits on qubits \cite{fontana_characterizing_2024, ragone_lie_2024, monbroussou_trainability_2025}. Therefore, to ensure such theoretical guarantees, the subspace must be of size polynomial in the number of modes, imposing a constant number of initial photons.

SI is a new tool that allows to go beyond the standard linear quantum optical circuits such as those used in Boson Sampling schemes. The diversity of possible encoding functions and measurement operators  offers significant flexibility in the design of quantum models, but we will mainly focus, in this paper, on the case where we count the photons in one mode and inject the same number of photons. This choice is motivated by experimental consideration (clarified in \autoref{subsubsec:general_exp_framework}), and by the preservation of the subspaces defined by a fixed number of particles. We use this example in \autoref{subsubsec:Purity}, and in \autoref{subsec:Proba_Estimation}. In  \autoref{subsubsec:general_exp_framework} we start by analyzing the general experimental framework that is common for both scheme, building on that we will highlight the differences between the ALO and SI schemes.

It is important to keep in mind that the challenges of any scheme depend on the exact task and corresponding quantum circuit. In \autoref{subsec:Proba_Estimation}, we give an instance of a learning problem where the SI scheme can require less resources. In a more "near-term" perspective, one can adapt the schemes to implement a sub-universal model that matches the experimental capacity of a platform. For example, one can choose injection functions or the adaptive function in the feed-forward scheme to not depend on the measured number of photons to avoid photon counting.

            \subsubsection{General Experimental Framework} \label{subsubsec:general_exp_framework}

In this part, the general experimental framework needed for both ALO and SI schemes is discussed. We can see that both proposals rely on a general scheme consisting of: State preparation, unitary, measurement, and adding a new unitary. In the following, we will expand the general experimental requirements for each step: 

\textbf{State Preparation.} Both schemes start with multiple single photons Fock states which requires a photon source that could emit indistinguishable photons on many parallel modes synchronously. The most common options for this are Quantum dots (QD) and time multiplexing~\cite{maring_versatile_2024,wang_high-efficiency_2017} or Spontaneous Parametric Down Conversion (SPDC) sources~\cite{zhong_12-photon_2018}.  

\textbf{Unitary.} The unitaries should be fully programmable which is the case for the photonic chips. Photonic chips, or processors, are made of waveguides shaped in beam splitters, the reflectivities of the latters are controlled by phase shifters. The waveguides are mostly engraved in glass or in silicon nitride. The phase shifters, which are the programmable parts,could be controlled by thermo-optical effects which can be quite slow taking from hundreds of microseconds to milliseconds for each reconfiguration~\cite{calvarese_strategies_2022,smith_universal_2022} or piezo-electrical (optomechanical) effects that are usually faster, tuned in hundreds of microseconds, but they have challenges for scalability~\cite{tian_piezoelectric_2024}.    

\textbf{Measurements.} Both schemes rely on single photon measurements between the unitaries. ALO requires Photons number resolving (i.e. photon counting) while SI can accept threshold detectors as well (i.e. detecting only the presence and the absence of photons) depending on the application.  

\textbf{Addition of unitary.} To be able to use the single photon measurements in real time, one should mostly consider independent chips, which implies that the spatial modes of both chips would be linked with optical fibers which imposes the implementation of a good temporal synchronization of all the modes. Noting that when we mention synchronization between the photons in state preparation or the modes between the unitaries we mean matching the temporal delays in order to maximize Hong-Ou-Mandel effect~\cite{hong_measurement_1987}.

In the case of the ALO scheme, the main challenge that needs to be adressed is the waiting time to reprogram the chips after each measurements. For example, if we consider an implementation with the most common setup relying on thermo-optical effects, we will have to put delays (or quantum memories) between the unitaries that are in the order of milliseconds, equivalent to few hundreds of kilometers of fibers which will have more losses and harder synchronization.

In the new SI scheme, we relieved the need of the real time adaptability of the consequent unitaries, we choose preset parameters for all of them instead. We rely on active real-time Fock state preparation to replace the few measured modes depending on their measurement outcome and synchronizing the new states with the unmeasured modes of each unitary, using mainly external (to the unitary) fast optical switches which would reduce the delay required for the non measured modes. Average commercial switches relying on micro or nano electro-optical technologies~\cite{thomas_noise_2025, memeo_micro-opto-mechanical_2024} can work in microseconds range which reduces the optical delay to the order of few hundreds meters.
For some applications, as in \autoref{subsec:Proba_Estimation}, this scheme will also allow the reduction of the number of photons needed at the beginning of the experiment.

        \subsection{Properties of State Injection}\label{subsec:properties_SI}

            \subsubsection{Controllability Improvement with State Injection}\label{subsubsec:Controllability_Improvement}

Since circuits that include the operations of state injection proposed in this work go beyond  linear optical circuits, they are not subject to \autoref{thm:Control_Limits_BS}, and therefore their output states $\rho_{\mathrm{out}}(\theta)$ may a priori enjoy controllability $\mathrm{DoF}(\rho_{\mathrm{out}})$ that goes beyond the $m^2 -1$ upper bound.

\begin{figure}
\centering
\includegraphics[width=0.9\linewidth]{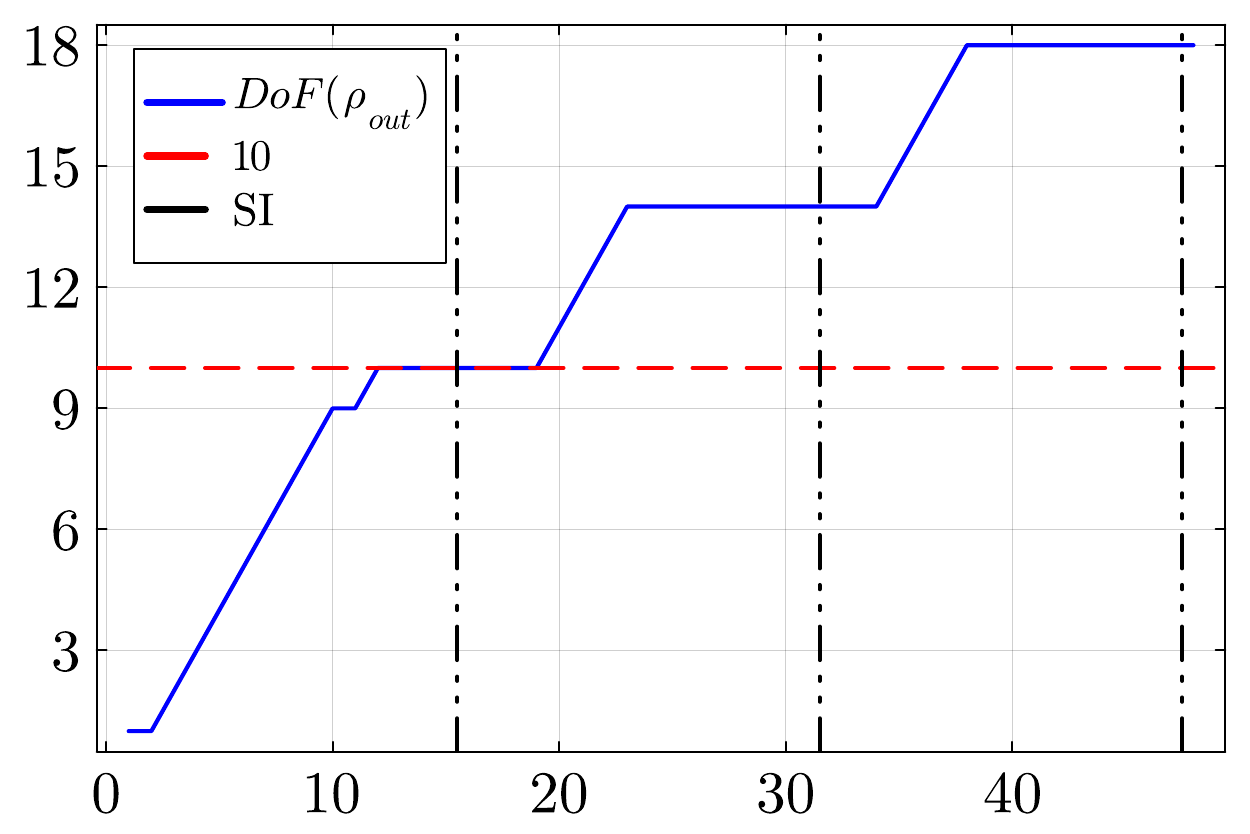}
\caption{Evolution of $DoF(\rho_{out})$ after state injections, where the injection operations consists in injecting the number of photons measured. With no state injection, the theoretical limit on the degrees of freedom is $DoF(\rho_{out})\leq \frac 1 2 m(m-1) = 10$ (red dashed line) with $m=5$ and $n=2$ photons.}
\label{fig:DoF_Evolv}
\end{figure}

In \autoref{fig:DoF_Evolv}, we explore the behavior of the  controllability one an example circuit with state injections, for $m=5$ modes, and $n=2$ photons, all input into the first mode. The type of state injection chosen here is the simplest one can consider: it is the special case of \autoref{eq:StateInjection-SpecialCase-photon-conservation-constraint} in \autoref{def:StateInjection} with $k=1$ and $f_1(n_1):=n_1$.
For a circuit with these state injections, we evaluate numerically the number of degrees of freedom $\mathrm{DoF}(\rho^{(i)})$ (\autoref{def:DoF}) of each intermediate state $\rho^{(i)}(\theta^{(i)})$ of the circuit (\autoref{fig:DoF_Evolv}). See \autoref{sec:Limitation_BS} for more details about how these numbers are calculated. The numerical results suggest a clear tradeoff between controllability and purity.

The circuit considered only consist of beam-splitters, distributed over the $m=5$ modes. 

We observe in \autoref{fig:DoF_Evolv} that indeed, the use of state injections has enabled breaking the limitation of controllability of output states of plain beam-splitters (see each little plateau just before each dashed line).

            \subsubsection{Purity Evolution with State Injections}\label{subsubsec:Purity}

Using non-unitary channels such as state injection, or other measurement-based methods, allows one to increase the controllability of the final state but decreases its purity $\Tr[\rho_{\mathrm{out}}^2]$. Many algorithms relying on non-linear channels \cite{coyle_training-efficient_2025, cong_quantum_2019} to increase the expressivity of their model do not consider the cost of reducing the purity of the final state. As controllability can be increased via SI, care should be taken not to reach the maximally mixed state. We therefore need to address 
at what rate the purity decreases when using SI, and how this decrease in purity affects the model.

For the sake of simplicity, we consider the special case in which the SI layer merely consists in measuring a single mode occupancy.
In this case, we can state the following \autoref{thm:PurityEvolution}.

\begin{restatable}{thm}{PurityEvolution}\label{thm:PurityEvolution}
    Consider a quantum circuit made of $m$ modes with an initial pure state with $n$ photons. If there is a single SI layer in the circuit, which consists in measuring the number of photons $p \in [\![ 0, n ]\!]$ in one of the modes and re-injecting the state $\ket{p}$ in that same mode, then the purity of the final state is given by
    \begin{equation}
        \gamma(\rho_{\mathrm{out}}) = \Tr[\rho_{\mathrm{out}}^2] = \sum_{i=0}^n \Pr[i]^2\, \textrm{,}
    \end{equation}
    with $\Pr[i]$ the probability of measuring $i$ photons on intermediate state just preceding the state injection. 
\end{restatable}

\begin{proof}
    We consider an initial pure state $\ketbra{\psi_0}{\psi_0}$, which is Fock state of $n$ photons over $m$ modes. We call $\rho$ the state after the state injection and $\gamma(\rho)$ its purity. Then,
    \begin{equation}
        \gamma(\rho) = \Tr[\rho^2] = \Tr[\left(\sum_{i = 0}^n \Pr[i] \ketbra{\psi^i}{\psi^i}\right)^2],
    \end{equation}
    with $\ket{\psi^i}$ the pure state post-injection corresponding to the measurement outcome $p=i$. 
    \begin{equation}
        \begin{aligned}
            \gamma(\rho) 
                &= \Tr[\sum_{i = 0}^n \Pr[i]^2 (\ketbra{\psi^i}{\psi^i})^2]  + \Tr[\sum_{0 \leq i, j \neq i \leq n} \Pr[i] \Pr[j] \ketbra{\psi^i}{\psi^i} \ketbra{\psi^j}{\psi^j}] \\
            &= \sum_{i = 0}^n\Pr[i]^2\Tr[(\ketbra{\psi^i}{\psi^i})^2]  + \sum_{0 \leq i, j \neq i \leq n} \Pr[i] \Pr[j] \Tr[\ketbra{\psi^i}{\psi^i} \ketbra{\psi^j}{\psi^j}] \\
            & = \sum_{i = 0}^n\Pr[i]^2 + \sum_{0 \leq i, j \neq i \leq n} \Pr[i] \Pr[j] \Tr[\ketbra{\psi^i}{\psi^i} \ketbra{\psi^j}{\psi^j}],
        \end{aligned}
    \end{equation}
    as for any $i$, $\ketbra{\psi^i}{\psi^i}$ is a pure state. Considering the specific injection function, we have that for all $0 \leq i \leq n$
    \begin{equation}
        \Pr[i] \ketbra{\psi^i}{\psi^i} = \Pi_i \ketbra{\psi_0}{\psi_0} \Pi_i^{\dagger},
    \end{equation}
    with $\Pi_i$ the projector of the state over all the Fock basis where there are $i$ photons on the measured modes. By definition, for all $0 \leq i,j\neq i \leq n, \Pi_i^\dagger \Pi_j = 0$.
    Therefore, it follows that
    \begin{equation}
        \gamma(\rho) = \sum_{i = 0}^n\Pr[i]^2.
    \end{equation}
\end{proof}

The result of this theorem also stands for any injection function such that two different measurements implies the injection of orthogonal states. From this result, we derive in \autoref{cor:PurityEvolutionBound} a lower bound of the purity of a linear optical circuit based on the number of SI layers it contains.

\begin{restatable}{cor}{PurityEvolutionBound}\label{cor:PurityEvolutionBound}
     We consider a quantum circuit made of $m$ modes with an initial pure state with $n$ photons, and with $L$ layers of SI separated by linear optical blocs. If each SI layer consists in measuring the number of photons $p \in [\![ 0, n ]\!]$ in one of the modes and re-injecting the state $\ket{p}$ in that same mode, then the purity of the final state is such that:
     \begin{equation}
         \gamma(\rho_{\mathrm{out}}) \geq \frac{1}{(n+1)^L}\, \textrm{.}
     \end{equation}
\end{restatable}

\begin{proof}
    We consider a circuit with $L$ state injection layers separated by linear optical circuits. We call $\rho_l$ the state produced after $l$ state injection layer. We call $\ketbra{\psi^I_l}{\psi^I_l}$ the pure state after the $l$ state injection layer when considering the set of outcome measurements $I = (i_1, \dots, i_l)$, so that $\rho_l$ is a statistical set of such states. Notice that we have: 
    \begin{equation}\label{eq:PurityInductiveRelation}
    \begin{split}
        \rho_l &= \sum_{I} U(\theta_{l-1}) \ketbra{\psi^I_{l-1}}{\psi^I_{l-1}} U^\dagger( \theta_{l-1}) \\
        &=  \sum_{i_l=0}^n U(\theta_{l-1}) \Pi_i \rho_{l-1} \Pi_i^\dagger U^\dagger( \theta_{l-1}) \, \textrm{,}
    \end{split}
    \end{equation}
    with $U(\theta_{l-1})$ the unitary corresponding to the linear optical circuit that separated the state injection layers and $\Pi_i$ the projector of the state onto all the Fock basis where there are $i_l$ photons on the measured modes for the considered layer of state injection. By definition, for all $i,j\neq i \in [n], \Pi_i^\dagger \Pi_j = 0$.
    We can thus state that
    \begin{equation}
        \gamma(\rho) = \Tr[\rho^2] = \Tr[\left(\sum_{I} \Pr[I] \ketbra{\psi^I_L}{\psi^I_L}\right)^2],
    \end{equation}
    with $I = (i_1, \dots, i_L)$, and $\Pr[I]$ the probability of measuring $i_l$ photons in the $l^{th}$ SI layer. In a similar way as in the proof of \autoref{thm:PurityEvolution}, we can show that
    \begin{equation}\label{eq:gammaRhoFromSum}
        \gamma(\rho) = \sum_I \Pr[I]^2 \Tr[\left(\ketbra{\psi^I_L}{\psi^I_L}\right)^2] = \sum_I \Pr[I]^2
    \end{equation}
    Notice that $\sum_I \Pr[I] = 1$, and that we can construct $(n+1)^L$ different vectors $I$  (for each measurement $i_l \in [\![ 0, n ]\!]$). Therefore, the sum of the square of the probability is lower bounded by
    \begin{equation}\label{eq:sumPrSquaredLB}
        \sum_I \Pr[I]^2 \geq \sum_I \left(\frac{1}{n+1}\right)^{2 L} = \frac{1}{(n+1)^L}.
    \end{equation}
    Finally,we have
    \begin{equation}
        \gamma(\rho) \geq \frac{1}{(n+1)^L} \, \textrm{.}
    \end{equation}
\end{proof}

In \autoref{cor:PurityEvolutionBound}, we consider the worst case scenario where each measurement outcome is as likely, resulting in this inverse-exponential lower bound in the purity. But if one has prior knowledge about which outcomes are more likely to occur, this lower bound may be tightened. In particular, when considering a number of modes much greater than the number of photons, one enters the so-called \textit{no-collision regime}, where the probability of measuring more than one photon is negligible. This phenomenon may be quantified using the Boson Birthday Bound introduced in \cite{aaronson_computational_2011}, and doing so, we get the following.

\begin{restatable}{cor}{PurityNoCollision}\label{cor:PurityNoCollision}
     We consider a quantum circuit made of $m$ modes with an initial pure state with $n$ photons, and with $L$ layers of SI separated by linear optical blocks. We again consider that each SI layer consists in measuring the number of photons $p \in [\![ 0, n ]\!]$ in one of the modes and re-injecting the state $\ket{p}$ in that same mode.
     
     If $m > 2n^2$, and if the linear optical blocks are considered to each be Haar distributed in the single-photon subspace, then the purity of the final state is such that:
     \begin{equation}
         \mathbb{E}_{U \in \mathcal{H}_{m,m}} [\gamma(\rho_{\mathrm{out}})] \geq \left(\frac{m-2n^2}{\sqrt{2}m}\right)^{2 L}\,.
     \end{equation}
\end{restatable}

\begin{proof}
    We recall the expression of the Boson Birthday Bound introduced in \cite{aaronson_computational_2011}:

    \begin{restatable}{thm}{BosonBirthdayBound}[{Boson Birthday Bound, adapted from \cite[Theorem 72]{aaronson_computational_2011}}]\label{thm:BosonBirthdayBound}
        Recalling that $\mathcal{H}_{m,m}$ is the Haar measure over $m \times m$ unitary matrices,
        \begin{equation}
            \mathbb{E}_{U \in \mathcal{H}_{m,m}}[\Pr[{\bm s} \in B_{m,n}]] < \frac{2n^2}{m} \, \textrm{.}
        \end{equation}
        With $\bm s$ a Fock state, and $B_{m,n}$ the set of $m$-modes and $n$-particle states where more than one photon can be per mode.  
    \end{restatable}

    We consider a circuit with $L$ state injection layers separated by linear optical circuits. We consider the case where each unitary matrix corresponding to the Linear Optical layers are Haar random matrices. We call $\ketbra{\psi_l}{\psi_l}$ the state after the $l$ state injection layer with $l \in [L]$. As in the proof of \autoref{cor:PurityEvolutionBound}, we consider the same inductive relation for the purity of the states within the circuit given by \autoref{eq:PurityInductiveRelation}. As in this previous proof, we can use \autoref{eq:gammaRhoFromSum} that we recall here:
    \begin{equation}\label{eq:defGammaRhoPr}
        \gamma(\rho) = \prod_{l=1}^L \left(\sum^n_{i=0} \Pr[l,i]^2\right).
    \end{equation}

    Using \autoref{thm:BosonBirthdayBound}, have that for any $l \in [L]$:
    \begin{equation}
        \mathbb{E}_{U \in \mathcal{H}_{m,m}}[\Pr[l,0] + \Pr[l,1]] \geq \frac{m - 2n^2}{m} \, \textrm{.}
    \end{equation}
    Therefore, we have for any $l \in [L]$:
    \begin{equation}
    \begin{aligned}
    \mathbb{E}_{U \in \mathcal{H}_{m,m}}\left[\sum_{i=0}^n \Pr[l,i]^2\right] 
         \geq \mathbb{E}_{U \in \mathcal{H}_{m,m}}\left[\sum_{i=0}^1 \Pr[l,i]^2\right] \geq 2 \left(\frac{m - 2n^2}{2m}\right)^2 \, \textrm{.} 
    \end{aligned}
    \end{equation}

    We can conclude, using \autoref{eq:defGammaRhoPr}
    \begin{equation}
        \mathbb{E}_{U \in \mathcal{H}_{m,m}}[\gamma(\rho)]  \geq  \left(\frac{m - 2n^2}{\sqrt{2} m}\right)^{2 L} \, \textrm{.}
    \end{equation}  
\end{proof}

In this setting, we notice that the purity lower bound decreases at a lower rate than the one presented in \autoref{cor:PurityEvolutionBound}. Those results could easily be adapted to more complex injection functions.

Using the results on the controllability and the purity of the final state, one can choose a particular number of SI layers and particular injection functions according to the desired specifications. In the following Section, we will discuss the impact of the state purity on the final quantum output model. 

            \subsubsection{Purity and Distinguishability}\label{subsubsec:PurityDistinguishability}

We define the concept of distinguishability of the quantum model. In what follows, we denote by $\mathcal{S}(d)$ the set of $d \times d$ density matrices.

\begin{definition}[Quantum models Distinguishability]
    Given two density matrices $\rho,\sigma \in \mathcal{S}(d)$, we introduce the following measure of distinguishability between the two states:
    \begin{equation}\label{eq:def-distinguishability-of-pair}
        \mathcal{D}(\rho, \sigma) := \max_{O} \big|\Tr[O \rho] - \Tr[O \sigma]\big|\,,
    \end{equation}
    where the maximum is taken over all observables $O \in \mathrm{Herm}(d)$ such that $||O||_{\infty} \leq 1$. Note that the quantity $\frac{1}{2}\mathcal{D}(\rho,\sigma)$ is equal to the \emph{trace-distance} $D_{\mathrm{tr}}(\rho,\sigma):=\frac{1}{2}\norm{\rho - \sigma}_1$ .

    Given now an arbitrary subset $S \subseteq \mathcal{S}(d)$ of density matrices, we define the associated distinguishability measure over $S$:
    \begin{equation}\label{eq:def-distinguishability-of-subset}
        \mathcal{D}(S) := \max_{\rho, \sigma \in S} \mathcal{D}(\rho,\sigma)\,.
    \end{equation}

    Lastly, given a \textit{quantum model} consisting of the output $\rho_{\mathrm{out}}(\theta)$ of a parametrized quantum circuit, we define the \textit{distinguishability of the quantum model} as:
    \begin{equation}\label{eq:def-distinguishability-of-model}
        \mathcal{D}( \rho_{\mathrm{out}}(\theta) ) := \mathcal{D}\Big( S\!=\!\left\{ 
\rho_{\mathrm{out}}(\theta) \; | \; \theta \in \Theta \right\} \Big)\,.
    \end{equation}
\end{definition}

A quantum model's distinguishability is an important metric to consider, as a low value would indicate a need for a high number of shots of the whole quantum experiment in order to resolve to sufficient precision the value of the observable that one wishes to estimate.
%We connect this metric with the \emph{purity} of a state $\rho \in \mathcal{S}(d)$, $\gamma(\rho)$, defined by
% \begin{equation}\label{eq:Purity_def}
%     \gamma(\rho) \hat = \Tr[\rho^2] \in [\frac{1}{d}, 1].
% \end{equation}

Intuitively, the connection between the distinguishability of two states $\mathcal{D}(\rho,\sigma)$ and their purity is clear: If two states are both too impure, they must both be relatively close to the maximally-mixed state, and therefore they should be relatively close to each other in some measure of distinguishability. This intuition is quantified in the following result:  

% in the following \autoref{thm:Distinguishability}.
% % Theorem \ref{thm:Distinguishability} offers a clear connection between the purity of the final state and the distinguishability of the output models.  

\begin{restatable}{thm}{Distinguishability}
\label{thm:Distinguishability}
    If $S \subseteq \mathcal{S}(d)$
    is a subset of $d \times d$ density matrices that are bounded in purity by a constant ${\gamma \in [1/d,\ 1]}$, i.e., for all $\rho \in S$,
    \begin{equation}
        \Tr[\rho^2] \leq \gamma,
    \end{equation}
     then the distinguishability $\mathcal{D}(S)$ of this subset (see \autoref{eq:def-distinguishability-of-subset}) satisfies
     \begin{equation}
         \mathcal{D}(S) \leq 2\sqrt{d}\,\,\sqrt{ \gamma - \frac{1}{d} }\,.
     \end{equation}
\end{restatable}

\begin{proof}
Let us begin with some notation.
We denote by $\mathrm{Herm}(d)$ the space of $d \times d$ Hermitian matrices, and given two matrices $A,B \in \mathrm{Herm}(d)$, we introduce the notation $\langle A, B \rangle := \Tr[A B]$ for the Hilbert-Schmidt inner-product in this real vector space. Denote also by $\mathcal{S}(d) \subset \mathrm{Herm}(d)$ the subset of density matrices.
We recall the definition of the Schatten $p$-norms (as they will occur in the proof, for $p=$ $1$, $2$ and $\infty$): given a matrix ${A \in \mathbb{C}^{d \times d}}$, ${\norm{A}_{p}:=\norm{\sigma}_{p}:=(\sigma_1^p + \cdots + \sigma_d^p)^{1/p}}$, where $\sigma \in \mathbb{R}^{d}$ denotes the vector of the singular values of $A$, and with the convention that $\norm{\sigma}_{\infty}:=\max_{i=1,\dots,d}(\sigma_i)$. In the case where $A \in \mathrm{Herm}(d)$, denoting by $\lambda \in \mathbb{R}^{d}$ the vector of eigenvalues of $A$, it holds that
$\norm{A}_{1} = |\lambda_1| + \cdots + |\lambda_d|$, 
$\norm{A}_{2} = (\lambda_1^2 + \cdots + \lambda_d^2)^{1/2} = \sqrt{\langle A, A \rangle}$, and
$\norm{A}_{\infty} = \max_{i=1,\dots,d} |\lambda_i|$. Note that given $A \in \mathrm{Herm}(d)$, its purity $\Tr[A^2]$ is by definition just $\norm{A}_2^2$.

Given $\rho,\rho' \in \mathcal{S}(d)$ such that
\begin{equation}\label{eq:ProofTheoremIndiscernability-assumption}
\norm{\rho}_2^2 \leq \gamma\text{\ and\ }\norm{\rho'}_2^2 \leq \gamma\,, 
\end{equation}
we have:
\begin{align}
&\mathcal{D}(\rho,\rho')\nonumber\\[5pt]
&= \max_{\norm{O}_{\infty}=1} \big| \langle \rho - \rho', O \rangle\big|\label{eq:ProofTheoremIndiscernability-eq-distinguish}\\[5pt]
&= \norm{\rho - \rho'}_1\label{eq:ProofTheoremIndiscernability-eqtoexplain1}\\[5pt]
&\leq \sqrt{d}\, \norm{\rho - \rho'}_2\\[5pt]
&\leq \sqrt{d}\, \big( \norm{\rho - \mmstate{d}}_2 + \norm{\rho' - \mmstate{d}}_2 \big)\\[5pt]
&= \sqrt{d}\,\left( \sqrt{ \norm{\rho - 0}_2^2 - \norm{\mmstate{d} - 0}_2^2 }  +  \sqrt{ \norm{\rho' - 0}_2^2 - \norm{\mmstate{d} - 0}_2^2 }  \right)\label{eq:ProofTheoremIndiscernability-eqtoexplain2}\\[5pt]
&= \sqrt{d}\,\left( \sqrt{ \norm{\rho}_2^2 - 1/d}  +   \sqrt{ \norm{\rho'}_2^2 - 1/d}  \right) \label{eq:ProofTheoremIndiscernability-eqtominorexplain3}\\[5pt]
&\leq 2\sqrt{d}\,\,\sqrt{ \gamma - 1/d}\,,\label{eq:ProofTheoremIndiscernability-eqtominorexplain4}
\end{align}
which gives the desired result.
In the above, \autoref{eq:ProofTheoremIndiscernability-eq-distinguish} is by the definition in \autoref{eq:def-distinguishability-of-pair}, \autoref{eq:ProofTheoremIndiscernability-eqtominorexplain3} uses the fact that $\norm{\mmstate{d}}_2^2 = 1/d$, and \autoref{eq:ProofTheoremIndiscernability-eqtominorexplain4} is by the assumption of \autoref{eq:ProofTheoremIndiscernability-assumption}.
We end by giving more explanations for \autoref{eq:ProofTheoremIndiscernability-eqtoexplain1,eq:ProofTheoremIndiscernability-eqtoexplain2}.

\autoref{eq:ProofTheoremIndiscernability-eqtoexplain1} is a standard equality --- that gives an operational meaning of the \textit{trace distance} of two states $\rho,\rho'$ ($\frac{1}{2}\norm{\rho - \rho'}_1$) in terms of their distinguishability ($\mathcal{D}(\rho,\rho')$). It holds because of the following two observations. First, by the $(1,\infty)$-Hölder inequality, one has that for all $O\in\mathrm{Herm}(d)$:
\begin{align}
    \big| \langle \rho - \rho', O \rangle\big| &\leq \norm{\rho - \rho'}_1 \norm{O}_\infty\,,
    \intertext{implying (\autoref{eq:ProofTheoremIndiscernability-eq-distinguish}) that}
    \mathcal{D}(\rho,\rho') &\leq \norm{\rho - \rho'}_1\,.
\end{align}
Second, by choosing the specific observable $O\in\mathrm{Herm}(d)$ given by $O:=\sum_{i=1}^d \mathrm{sign}(\lambda_i) \ketbra{e_i}$, with ${\rho - \rho' = \sum_{i=1}^d \lambda_i \ketbra{e_i}}$ the eigendecomposition of the Hermitian matrix $\rho - \rho'$, one has $\norm{O}_\infty = 1$ and
\begin{align}
    \langle \rho - \rho', O \rangle &= \norm{\rho - \rho'}_1\,,
    \intertext{implying, using \autoref{eq:ProofTheoremIndiscernability-eq-distinguish}, that}
    \mathcal{D}(\rho,\rho') &\geq \norm{\rho - \rho'}_1\,.
\end{align}

In \autoref{eq:ProofTheoremIndiscernability-eqtoexplain2}, we apply the Pythagorean theorem (in the real vector space $\mathrm{Herm}(d)$ with the geometry given by the Hilbert-Schmidt inner-product) to two right triangles. These are respectively the triangles made of the vertices $(0, \mmstate{d}, \rho)$ and $(0, \mmstate{d}, \rho')$ -- where $0$ denotes the zero vector/matrix in $\mathrm{Herm}(d)$. They are both right triangles with right angle located at point $\mmstate{d}$, since 
${\mmstate{d} \in (\mathcal{S}(d) - \mmstate{d})^\perp}$ --- that is, for any $\tau \in \mathcal{S}(d)$ one has $ \langle \mmstate{d},\ \tau - \mmstate{d} \rangle = 0$, because:
\begin{equation}
    % \begin{split}
        \langle \mmstate{d},\ \tau - \mmstate{d} \rangle = \langle \mmstate{d}, \tau \rangle - \langle \mmstate{d}, \mmstate{d} \rangle = \Tr[(\mmstate{d}) \tau] - \Tr[(\mmstate{d})^2] = \Tr[\tau]/d - 1/d = 0\,. 
    % \end{split}
\end{equation}
\end{proof}

        \subsection{Probability Estimation Problem}\label{subsec:Proba_Estimation}

In this Section, we argue on how linear optics boosted with photonic SI allows one to generate output probabilities that are believed hard to solve classically. Similar task has been tackled using ALO in \cite{chabaud_quantum_2021}. Here, we show how our proposal, which is less experimentally challenging, does offer a similar kind of advantage.

\begin{figure}
  \begin{subfigure}[t]{1\columnwidth}
    \includegraphics[width=\linewidth]{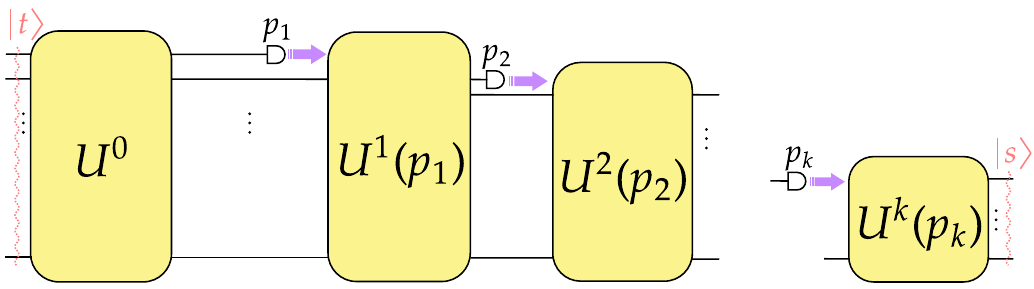}
    \caption{A cascade of $k$ linear optical interferometer, where each of the $U^l$, $1 \leq l \leq k$ is parameterized by the outcome of the measurement of the first mode of the previous interferometer $U^{l-1}$. 
    }
    \label{fig:aM}
  \end{subfigure}
  \begin{subfigure}[t]{1\columnwidth}
    \includegraphics[width=\linewidth]{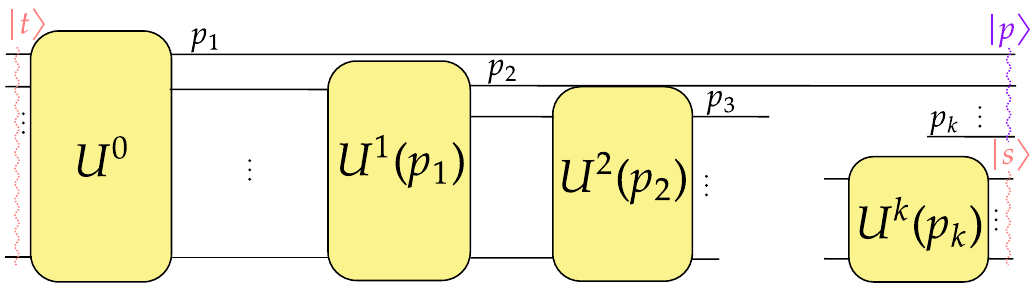}
    \caption{\emph{Equivalent model}, where no feedforward is required. We consider the same unitaries, and we post-select measurement of the modes previously used for adaptivity.}
    \label{fig:aMEM}
  \end{subfigure}
  \caption{Feedforward architecture (\autoref{fig:aM}) and its equivalent model (\autoref{fig:aMEM}). }
\label{fig:aSI}
\end{figure}

While Boson Sampling is -- under widely believed complexity theoretic assumptions --  a task intractable for a classical computer \cite{aaronson_computational_2011}, estimating a single output probability up to an inverse polynomial additive error is doable in polynomial time (in the number of involved photons) via Gurvits' algorithm \cite{aaronson_generalizing_2014}. This requires QML algorithms relying on Boson Sampling probabilities a bit of work to escape this. It was shown in many ways how a Boson Sampling-like architecture can be made universal for quantum computation \cite{knill_scheme_2001,bartolucci_fusion-based_2023,chabaud_quantum_2021}. We propose an architecture similar to \cite{chabaud_quantum_2021}, where Boson Sampling was boosted with \emph{feedforward} of the measurement result. This architecture escapes Gurvits' algorithm for probability estimation, depending on both the number of feedforwards and the number of photon measured throughout the computation (see \autoref{fig:aM}). Incidentally, this scheme becomes universal when both quantities are large (in fact, at least linear) compared to the number of input modes. However, this model comes with strong technical requirements when it comes to physical implementation. Indeed, parameterizing unitaries depending on previous measurement result, i.e., on the fly with respect to the ongoing computation, is out of reach of current technologies, due, in particular, to the heavy requirements for high speed electronics and information processing as explained in \autoref{sec:Adaptivity_Schemes}. This comes at the cost of the universality of the scheme, as we leave the question of the computational universality of the scheme open.

For a computational model, we refer to its \emph{equivalent model} (see in \autoref{fig:aSI}) as the one describing the transition amplitudes of the original computational model by a (or possibly a sum of) transition amplitude arising from an experiment when post-selection is used instead of state-injection. Accordingly, the \emph{equivalent unitary} is the unitary matrix describing that particular experiment state. Indeed, real-life experiments are to be conducted by implementing the actual computational model; the equivalent model is of great help to grasp the hardness of its classical simulation.

\begin{figure}[h!]
  \begin{subfigure}{1\columnwidth}
    \includegraphics[width=\linewidth]{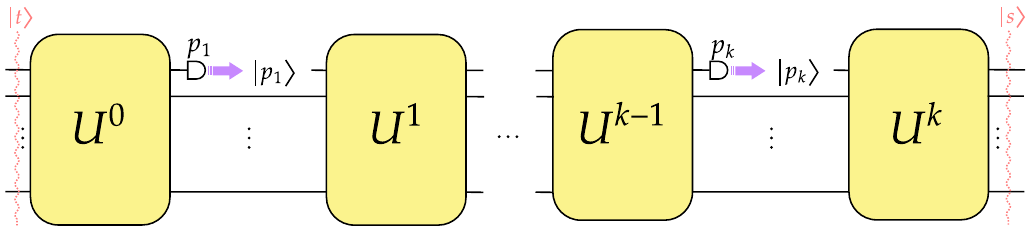}
    \caption{Probability estimation using state injection, with specific injection functions. 
    }
    \label{fig:ProbaStateInjection}
  \end{subfigure}
  \begin{subfigure}{1\columnwidth}
    \includegraphics[width=\linewidth]{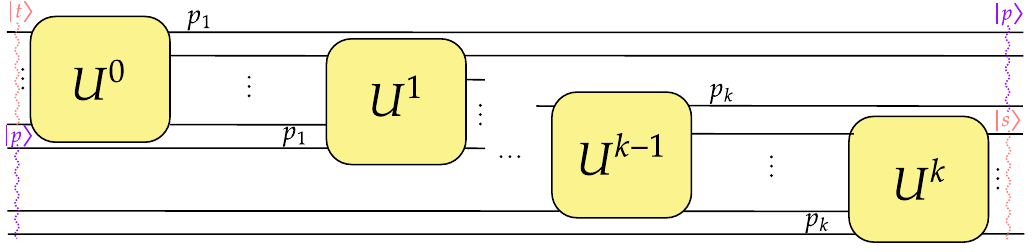}
    \caption{Equivalent model of SI presented in \autoref{fig:aM}.}
    \label{fig:EqStateInj}
  \end{subfigure}
  \caption{Feedforward architecture (\autoref{fig:aM}) and its equivalent model (\autoref{fig:aMEM}). }
\label{fig:aSIEM}
\end{figure}

Writing $\mathds{1}_l$ the identity on $SU(l)$, the equivalent unitary -- the one described in \autoref{fig:aSIEM} -- of a computation with $k$ state injections is of the form
\begin{equation}
    \tilde U = \prod_{l=0}^k(\mathds{1}_l \oplus U^l \oplus \mathds{1}_{k-l}),
\end{equation}
where $U^l \in SU(m)$ for all $0\leq l\leq k$ and ${\tilde U \in SU(m+k)}$. Thus, the output state reads
\begin{equation}
    \rho_{\mathrm{out}} = \Tr_k \left[ \hat U \left(\ket{t}\bra{t} \otimes \ket{p}\bra{p}\right)\hat U^{\dagger} \left(\mathds{1}_{m-k}\otimes \ket{p}\bra{p}\right) \right],
\end{equation}
where the $k$ last modes are traced out and $\hat U$ is the unitary action of $\tilde U$ on the multimode Fock basis.
We write $\pr[t]{p, s}$, for $p \in \Phi_{k,r} = \{p_1 \dots p_k | p_i \in \llbracket 0, r \rrbracket \text{ and } \sum_{i=1}^k p_i = r \}$, where $r\in\mathbb N$ is the total number of measured photons and $\bm s,\bm t \in \Phi_{m,n} = \{p_1 \dots p_m | p_i \in \llbracket 0, n \rrbracket  \text{ and } \sum_{i=1}^m p_i = n \}$ the probability of measuring output occupancy $s$ upon obtaining the measurement pattern $p$; indeed we get from the equivalent model and using \autoref{eq:bloc_unitary_expression_BS}, that
\begin{equation}
\begin{aligned}
        \pr[t]{p, s} 
        & = \bra{(s, p)} \varphi(\tilde U) \ket{(p, t)} \\
        & = \frac{1}{(p!)^2s!t!}\left| 
        % \per{I_{(s,p)}^{\dagger}\tilde U I_{(p,t)}} } \right|^2.  
        \per{ \tilde{U}_{(\bm{s},\bm{p}),(\bm{p},\bm{t)}} } \right|^2.
\end{aligned}
\end{equation}%W^1_{\bm{s},\bm{t}}(\theta)
Thus, the probability of measuring $s$ at the output of the interferometer is obtained by summing over all patterns, namely
\begin{equation}
    \begin{split}
        \pr[t]{s} &= \sum_{p \in \Phi_{k,r}} \pr[t]{p, s} \\
        &= \frac{1}{s!t!} \sum_{p \in \Phi_{k,r}}  \frac{1}{(p!)^2}\left| %\per{I_{(s,p)}^{\dagger}\tilde U I_{(p,t)}}\right|^2,
        \per{ \tilde{U}_{(\bm{s},\bm{p}),(\bm{p},\bm{t)}} } \right|^2,
    \end{split}
\end{equation}
where $\Phi_{k,r}$, defined analogously to  $\Phi_{m,n}$, is the list of all possible patterns of $k$ measurements where a total of $r$ photons were measured, i.e., the set of all $k$-tuples of integers $(n_1, \cdots, n_k)$ such that $\sum_i n_i = r$. Recall that $|\Phi_{k,r}| = \binom{k+r-1}{r-1}$. Given an $n \times n$ matrix $A$, Gurvits' algorithm computes an estimate of $\per{A}$ to  within additive precision $\pm \varepsilon \|A\|^n$ in time $O(n^2 \varepsilon^{-2})$. Therefore, as $\tilde U$ is unitary, one can compute an estimate of $\pr[t]{s}$ in polynomial time up to inverse polynomial precision, provided that $|\Phi_{k,r}| = O(poly(m))$. 
On the contrary, one will always be able to compute an estimate (again, up to inverse polynomial precision) of the output probability of a certain state by sampling the linear interferometer.

\begin{table}[h!t]
% \centering
\begin{minipage}[b]{.45\textwidth}
\centering
\begin{tabular}{cc|ccc|}
\cline{3-5}
 &  & \multicolumn{3}{c|}{$k$} \\ \cline{3-5} 
 &  & \multicolumn{1}{c|}{$\ \mathcal{O}(1)\ $} & \multicolumn{1}{c|}{$\mathcal{O}(\log m)$} & $\mathcal{O}(m)$ \\ \hline
\multicolumn{1}{|c|}{} & $\mathcal{O}(1)$ & \multicolumn{1}{c|}{\cmark} & \multicolumn{1}{c|}{\cmark} & \cmark \\ \cline{2-5} 
\multicolumn{1}{|c|}{} & $\mathcal{O}(\log m)$ & \multicolumn{1}{c|}{\cmark} & \multicolumn{1}{c|}{\cmark} & \xmark \\ \cline{2-5} 
\multicolumn{1}{|c|}{} & $\mathcal{O}(m)$ & \multicolumn{1}{c|}{\cmark} & \multicolumn{1}{c|}{\xmark} & \xmark \\ \cline{2-5} 
\multicolumn{1}{|c|}{} & $\mathcal{O}(m\log m)$ & \multicolumn{1}{c|}{\cellcolor[HTML]{C0C0C0}{\color[HTML]{C0C0C0} }} & \multicolumn{1}{c|}{\xmark} & \xmark \\ \cline{2-5} 
\multicolumn{1}{|c|}{\multirow{-5}{*}{$r$}} & $\mathcal{O}(m^2)$ & \multicolumn{1}{c|}{\cellcolor[HTML]{C0C0C0}{\color[HTML]{C0C0C0} }} & \multicolumn{1}{c|}{\cellcolor[HTML]{C0C0C0}{\color[HTML]{C0C0C0} }} & \xmark \\ \hline
\end{tabular}
\vspace{.4em}
\caption{Efficient classical output probability estimation regimes of the SI scheme. The green checkmarks indicate that the output probabilities can be estimated efficiently classically while the red crosses indicate no efficient classical algorithm is known. The gray cells indicate non reachable regimes assuming no concentration of the photons.}
\label{tbl:simulationRegimes}
\end{minipage}
\hspace{.08\textwidth}
\begin{minipage}[b]{.45\textwidth}
\centering
\begin{tabular}{cc|ccc|}
\cline{3-5}
 &  & \multicolumn{3}{c|}{$k$} \\ \cline{3-5} 
 &  & \multicolumn{1}{c|}{$\mathcal{O}(1)$} & \multicolumn{1}{c|}{$\mathcal{O}(\log m)$} & $\mathcal{O}(m)$ \\ \hline
\multicolumn{1}{|c|}{\multirow{3}{*}{$r$}} & $\mathcal{O}(1)$ & \multicolumn{1}{c|}{\cmark} & \multicolumn{1}{c|}{\cmark} & \cmark \\ \cline{2-5} 
\multicolumn{1}{|c|}{} & $\mathcal{O}(\log m)$ & \multicolumn{1}{c|}{\cmark} & \multicolumn{1}{c|}{\cmark} & \xmark \\ \cline{2-5} 
\multicolumn{1}{|c|}{} & $\mathcal{O}(m)$ & \multicolumn{1}{c|}{\cmark} & \multicolumn{1}{c|}{\xmark} & \xmark \\ \hline
\end{tabular}
\vspace{.4em}
\caption{Duplicate of \cite[Table 2]{chabaud_quantum_2021}. It describes the simulability regimes for probability estimation in the setting of ALO.  The green check-marks indicate that the output probabilities can be estimated efficiently classically while the red crosses indicate no efficient classical algorithm is known.}
\label{tbl:simulationRegimesALO}
\end{minipage}
\end{table}

The regimes where this technique can be efficiently simulated classically are summarized in \autoref{tbl:simulationRegimes}. We recall in \autoref{tbl:simulationRegimesALO} the regimes in which probability estimation can be efficiently done classically in the ALO picture introduced in \cite{chabaud_quantum_2021}. We remark that our scheme presents the same potential of quantum speedup in the settings where $r$, the total number of photon measured throughout the computation, is smaller than $\mathcal{O}(m)$. But as we offer the possibility to inject photon states anywhere in the computation, the total number of photons to be considered in our scheme goes up to $\mathcal{O}(m^2)$. Since we start with input states with $n$ photons, some regimes (grayed in \autoref{tbl:simulationRegimes}) are out of reach. Nonetheless, our proposal gives access to new computing regimes where classical simulation is not expected to be efficient. Indeed, in order to estimate the output probabilities efficiently, either the number of possible outcomes is polynomially large, or the distribution is concentrated \cite{raussendorf_one-way_2001}.
In addition, using the SI scheme allows us to start with a lower number of initial photons $n_0$. In the case of ALO, $n_0$ must be greater or equal to $r$, which is not the case for SI. As explained in \autoref{subsubsec:general_exp_framework}, this could help to reach experimentally a regime where probability estimation is not believed to be efficiently classically estimated.\\

However, it is important to note that, as explained in \autoref{subsec:properties_SI},  the use of adaptivity does affect the purity of the state and its distinguishability.  This may lead to an increase in the number of shots needed to estimate the probability  of a quantum state with a quantum processor. While this represents a trade-off, it is  worth noting that, first, quantum advantage can arise when the number of injection  layers $k$ is logarithmic, which can ensure a polynomial decay of the state purity,  as stated in \autoref{cor:PurityEvolutionBound}. More importantly, adaptivity can be  used without reducing the purity significantly. One canonical example is the KLM scheme~\cite{knill_scheme_2001}, where adaptive linear optical circuits are used to  create qubit-based computation. Nevertheless, establishing theoretical guarantees on the purity of the state when using SI remains an open question.

%% file: chapters/06_Subspace_Preserving_Algorithms/Subspace_Preserving_Algorithms.tex
\let\textcircled=\pgftextcircled
\chapter[Subspace Preserving Algorithms]{Subspace Preserving Algorithms}
\label{chap:Subspace_Preserving_Algorithms}
\begin{textblock}{5.3}(0,-4)
	%\textit{`Un scientifique dans son laboratoire est un enfant placé devant des phénomènes naturels qui l'impressionnent comme des contes de fées.'\\}

%\hspace{0.5cm}--- Marie Skłodowska-Curie.
\end{textblock}

\initial{P}\textit{reviously, theoretical guarantees of subspace preserving quantum circuits were discussed. To pursue the idea of using subspaces of polynomial size to reach polynomial advantage while avoiding Barren Plateaus, linear optical circuits that preserve the number of particles in initial Fock states were studied. Using the State Injection scheme proposed in \autoref{chap:Photonic_Suboptimal}, more complex subspace preserving QML architectures can now be developed to achieve quantum utility. In this Chapter, two studies that were made in parallel are merged. The first one is the design of a HW-preserving Quantum Convolutional Neural Network (QCNN) based on RBS gates and a measurement-based technique inspired by state injection. The second is its photonic version, which is not based on the qubit picture and relies on state injection, where an experimental proof of concept for the architecture was led, showing the near-term potential for such methods. For both architectures, numerical experiments were performed to illustrate the capacity of those methods to learn on real world problems.}

    \section{Related Work on QCNNs}\label{sec:Related_Work_QCNNs}

These proposals differ from previous CNN architectures such as \cite{cong_quantum_2019, kerenidis_quantum_2020, wei_quantum_2022} in several ways. First, they mimic classical convolutional layers, and pooling by using a specific subspace preserving encoding. This strategy ensures that these algorithms will be useful for large problems and not only for toy models as explained in \cite{bowles_better_2024}. Therefore, we believe that CNNs could be replaced by these quantum equivalents even for large architectures. Secondly, these proposal offer polynomial speed-ups and are therefore classically simulable, in the sense that a classical algorithm can perform the same computation in polynomial time due to the use of quantum circuits that are subspace preserving. This choice is motivated by the theoretical guarantees of such circuits in the training and expressivity of the model \cite{monbroussou_trainability_2025,fontana_characterizing_2024,ragone_lie_2024}. It is strongly believed that achieving a polynomial advantage in machine learning applications is a promising avenue for demonstrating quantum utility. Indeed, these methods could be used to offer a polynomial speed-up of high degree that could eventually achieve a quantum utility in comparison with CNNs, especially considering that these methods seem to achieve similar performances with fewer parameters. Finally, the classical simulation of the layers is tackled by offering a HW preserving \href{https://github.com/ptitbroussou/HW_QCNN}{simulation library} and a \href{https://github.com/ptitbroussou/Photonic_Subspace_QML_Toolkit}{photonic subspace QML toolkit} that allows to test the new proposals on larger learning problems than usually presented in the QML community. A GPU oriented library is offered that allows one to test the proposed HW-preserving architecture on very complex learning tasks, such as the 10-classes image classification tasks. To our knowledge, no other quantum neural network has been demonstrated to successfully tackle 10-label classification as our approach does in \autoref{sec:Results_and_Simulations}. In addition, recent work \cite{bermejo_quantum_2024} on classical simulation of one specific type of Quantum Convolutional Neural Networks (QCNN) has shown, using the LOWESA algorithm \cite{rudolph_classical_2023,fontana_classical_2025}, that the Iris Cong et al. \cite{cong_quantum_2019} proposal of QCNN is effectively classically simulate-friendly by considering the subspace of the low-weight measurement operators that are sufficient for the classification of “locally-easy” datasets. This method of classical simulation is different from the one used in the proposed simulation libraries, and could be applied to algorithms that are not subspace preserving. In particular, the architectures contain correlated parameters and measurement-controlled operations. Future work may determine if LOWESA or similar algorithms based on Pauli propagation \cite{rudolph_classical_2023} can be adapted to achieve speed-ups paralleling those of our specialized method, especially for challenging classification tasks such as described in this paper, and photonic platforms with large repetition rates. The authors of \cite{bermejo_quantum_2024} demonstrate how to classically simulate a specific instance of a QCNN \cite{cong_quantum_2019} on locally easy datasets. In this Chapter, similar classical datasets as \cite{bermejo_quantum_2024} are used for the HW-preserving QCNN, but the analysis is not restricted to only two classes, which explains why the tasks are not locally easy. By designing the architecture to mimic classical CNNs, both motivation and concrete examples are provided illustrating why the proposed algorithms can address complex learning problems—unlike previous quantum neural network proposals.

    \section{Hamming Weight Preserving Quantum Convolutional Neural Network}\label{sec:HW_QCNN}

A new QML algorithm that behaves as a Convolutional Neural Network (CNN) architecture \cite{oshea_introduction_2015} is proposed. This type of neural network is particularly useful in classical Machine Learning for many use cases including, for example, computer vision tasks \cite{lecun_gradient-based_1998, krizhevsky_imagenet_2017}, and Time Series analysis \cite{younis_multivariate_2022}. The proposed algorithm is illustrated on image classification. This analogy allows to ensure that the new algorithm will still be useful on a larger scale, as CNN is widely used. To design such an algorithm, Hamming weight (HW) preserving quantum circuits \cite{monbroussou_trainability_2025} were used, a particular type of subspace preserving framework that allows one to avoid BP \cite{larocca_diagnosing_2022, ragone_lie_2024, fontana_characterizing_2024, monbroussou_trainability_2025} by considering a Hilbert space of polynomial size at the cost of having only a polynomial advantage, that could be of high degree. While recent works \cite{cerezo_does_2024, goh_lie-algebraic_2025} have shown that the absence of BP in this framework leads to the existence of efficient simulation under certain conditions, i.e., no exponential running time complexity advantage, we play within this framework and measurement based techniques to offer significant advantages to our method. In addition, large simulations using GPU clusters show impressive results for the method in comparison with the classical one, including a reduction of the number of parameters which could lead to an even greater advantage.

\begin{figure}[h]
    \centering
    \captionsetup{justification=centering}
    \includegraphics[width=0.85\textwidth]{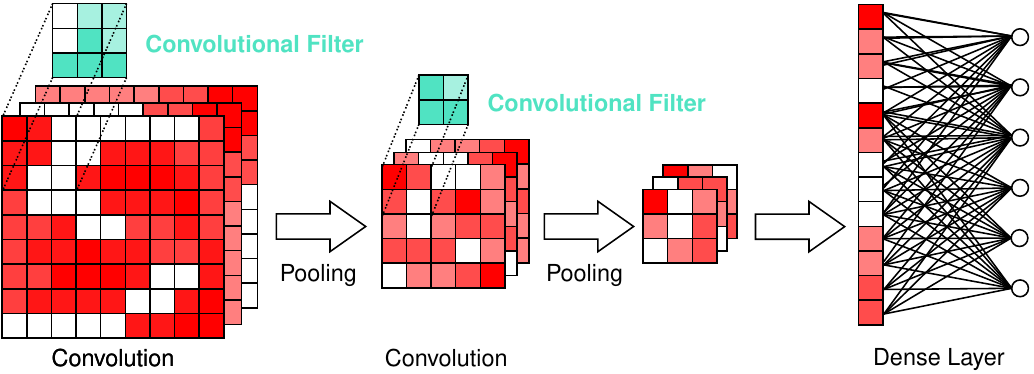}
    \caption{A Convolutional Neural Network architecture. In this example, the input is a batch of 2-dimensional images and is thus a 3-dimensional tensor.}
    \label{fig:figure_intro_CNN}
\end{figure}

%%%%%

        \subsection{Quantum and Classical Convolutional Neural Network Architecture}\label{subsec:QCNN_Architecture}

In this Section, the original HW preserving convolutional architecture is presented. There exist many types of CNN architectures, and the very first one introduced by LeCun \cite{lecun_gradient-based_1998} which is the original version of LeNet is recalled in \autoref{fig:figure_intro_CNN}. This neural network is composed of successive convolutional and pooling layers, and it ends with a dense layer.

This structure is quite simple: the convolution parts extract features from the initial images, the pooling parts reduce the dimension of the images, and the final dense layer mixes the features and performs the classification task. The convolution layers are very appropriate for the feature extraction, as they perform a translation invariant operation on the initial image by applying a convolution filter that is optimized through the training. Usually \cite{lecun_gradient-based_1998, krizhevsky_imagenet_2017, he_deep_2016}, each layer can be followed by the application of a nonlinear function.

It is well known that the success of machine learning algorithms largely depends on data representation \cite{bengio_representation_2013}, as well as the use of non-linearity and complex architectures \cite{montufar_number_2014, raghu_expressive_2017}. Therefore, it is highly unlikely that a quantum neural network could effectively tackle complex learning tasks without incorporating appropriate data representations, sophisticated architectures, and non-linearities that preserve the data structure throughout the computation.

            \subsubsection{Quantum Convolutional Layer}\label{subsubsec:Conv_layer}

In this Section the convolutional layer based on tensor encoding is explained. In the following, the classical convolutional layer is presented, and the Hamming weight preserving quantum convolutional is introduced. It is shown how the quantum version performs a convolution operation that is analog to the classical one, and what their differences are. Both operations are illustrated in \autoref{fig:Classical_Quantum_Convolutional_Layers}.

\subsubsection*{Classical Convolutional Layer:}

\begin{figure}[h!t]
\centering
\begin{subfigure}[t]{.49\textwidth}
    \centering
    \includegraphics[height=0.6\textwidth]{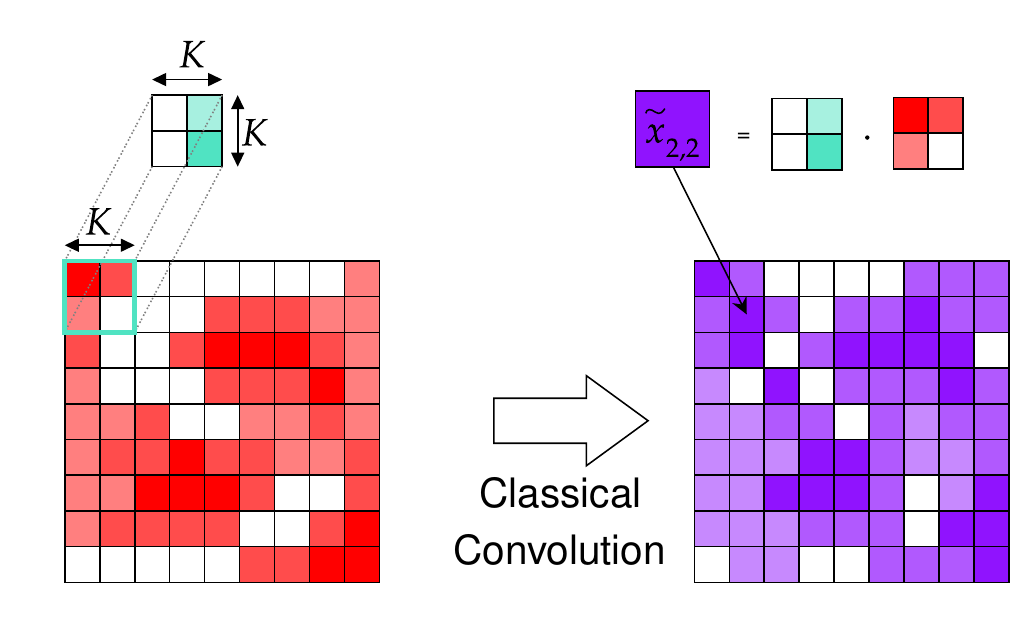}
    \caption{}
    \label{fig:Classical_Conv}
\end{subfigure}
\begin{subfigure}[t]{.49\textwidth}
    \centering
    \includegraphics[height=0.6\textwidth]{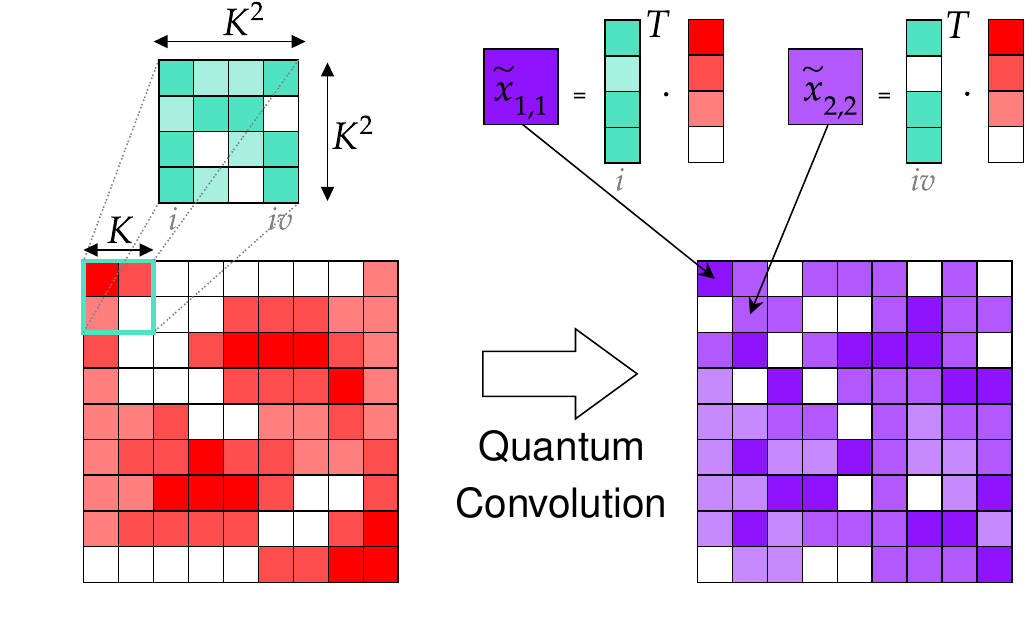}
    \caption{}
    \label{fig:Quantum_Conv}
\end{subfigure}
\caption{Classical (a) and Quantum (b) Convolutional layers. The convolutional filter is represented in blue. }
\label{fig:Classical_Quantum_Convolutional_Layers}
\end{figure}

Let's recall what mathematical operation a classical convolutional layer performs. Consider a 2-dimensional tensor $x = \left(x_{i,j}\right)_{(i,j) \in [d_1]\times[d_2]}$, a convolution filter $W = \left(w_{i,j}\right)_{i,j \in [K]}$ and final image $\Tilde{x} = \left(\Tilde{x}_{i,j}\right)_{(i,j) \in [d_1]\times[d_2]}$. We have:
\begin{equation}\label{eq:classical_2Dconv}
    \forall (i,j) \in [d_1] \times [d_2], \quad \Tilde{x}_{i,j} = \sum_{a, b \in [K]}  w_{a,b} \; x_{i - \lfloor \frac{K}{2} \rfloor + a,  j - \lfloor \frac{K}{2} \rfloor + b} 
\end{equation}
which corresponds to a convolution operation between the Filter tensor and the Filter window around the pixel. In \autoref{fig:Classical_Conv}, this 2-dimensional example is illustrated with the Filter window in green and the Filter in blue. This definition can be extended to any $k$-dimensional tensor and for any convolutional layer of dimension less or equal to $k$. Notice that in the case of a 2-dimensional convolution for a 3-dimensional input such as a batch of square images (see \autoref{fig:figure_intro_CNN}), each image is affected by the same 2-dimensional convolutional operation with the filter, such as described in \autoref{eq:classical_2Dconv}.

\subsubsection*{Tensor Encoding}

To perform the quantum convolutional layer and the encoding, Reconfigurable Beam Splitter (RBS) gates, introduced in \autoref{chap:HW_Preserving_Methods}, are used. It is proposed to encode classical data in such a way that allows to apply a convolutional layer by using HW preserving circuits. More precisely, it is proposed to load any tensor of dimension $k$ by using amplitude encoding on a Tensor Basis of HW $k$. 

\begin{definition}[HW Preserving Tensor encoding]\label{def:TensorEncoding}
    Consider a classical tensor of dimension $k$ such that $x = (x_{1, \dots, 1}, \dots,  x_{d_1, \dots, d_k}) \in \mathbb{R}^{d_1 \times \dots \times d_k}$. An amplitude tensor encoding data loader is a parametrized $n$-qubit quantum circuit (with $n = \sum_{i \in [k]} d_i$) that prepares the quantum states:
    \begin{equation}
        \ket{x} = \frac{1}{||x||} \sum_{i_1 \in [d_1]} \dots \sum_{i_k \in [d_k]}  x_{i_1, \dots, i_k} \ket{e^{d_1}_{i_1}} \otimes \dots \otimes \ket{e^{d_k}_{i_k}} ,
    \end{equation}
    where $\ket{e_{i_l}^{d_l}} = \ket{0 \dots 0 1 0 \dots 0}$ is a state corresponding to a bit-string with $d_l$ bits and only the bit $i_l$ is equal to $1$. Therefore, for any $j \in [k]$, $\left\{\ket{e_{i}^{d_l}} \mid i \in [d_l] \right\}$ is a fixed family of $d$ orthonormal quantum states, and $||\cdot||$ denotes the $2$-norm of $\mathbb{R}^d$.
\end{definition}

For example a $2 \times 2$ matrix image $x$ can be mapped to a state $\ket{x}$ using this encoding:
\begin{equation}
    X = \begin{pmatrix} x_{1,1} & x_{1,2} \\
                        x_{2,1} & x_{2,2} 
        \end{pmatrix} \; \longrightarrow \; \ket{x} = \frac{1}{||x||} \left( x_{1,1}\ket{1010} + x_{1,2}\ket{1001} + x_{2,1}\ket{0110} + x_{2,2}\ket{0101}\right) \, \textrm{.}
\end{equation}

This choice of this encoding gives a structure to the state that allows to apply the Convolutional layer and the Pooling layers described in the following. It can be considered as amplitude encoding on a specific basis, and can be realized thanks to quantum data loaders that perform amplitude encoding on the basis of fixed HW \cite{farias_quantum_2025, johri_nearest_2021, monbroussou_trainability_2025}. The tensor encoding offers the opportunity to use measurement based operation to apply a Pooling operation as described in \autoref{subsubsec:pooling_layer} that reduces the dimension of the state and apply non-linearities while preserving the tensor encoding structure of the final states which has never been done before to our knowledge. This is the key ingredient for the global convolutional architecture. Notice that the final state can be decomposed on $k$ registers, and that all the registers can be considered alone as HW preserving circuits of HW $1$.

\subsubsection*{Hamming-Weight Preserving Convolutional Layer:}

Considering a tensor encoding of dimension $k$, applying a RBS-based quantum circuit on $K$ qubits of one register performs rotations between the states corresponding to each pixel linked with those qubits. As an example, performing a RBS-based quantum circuit on the $K$ first qubits of the line register for a 3 dimension image tensor encoded will affect all the pixels in the $K$ first lines of all the images. By applying the same circuit to each $K$ consecutive qubits of each register, a $k$-dimensional convolution can be performed. On each register, the HW is equal to $1$ (or unary). 

For example, with $k=2$, a 2-dimensional tensor $x = \left(x_{i,j}\right)_{(i,j) \in [d_1]\times[d_2]}$ which is tensor encoded such as in \autoref{def:TensorEncoding} is considered. If a RBS based circuit between all the qubits of indexes $I, \dots, (I+K) \in [d_1]$ of the line register is applied, and another one between all the qubits of indexes $J, \dots, (J+K) \in [d_2]$ of the column register, then it can be considered that the corresponding $K \times K$ pixels form a filter window affected by a unitary matrix $U_\textrm{Filter}(\Theta)$ such that:
\begin{equation}\label{eq:quantum_conv_unitary}
    \sum_{i=I}^{I+K} \sum_{j=J}^{J+K} \Tilde{x}_{i,j} \ket{e_i, e_j} = U_{\textrm{Filter}}(\Theta) \sum_{i=I}^{I+K} \sum_{j=J}^{J+K} x_{i,j} \ket{e_i, e_j} \, \textrm{,}
\end{equation}
with $\Theta$ the RBS parameters, $\ket{e_i}$ is a unary state corresponding to the\autoref{def:TensorEncoding}, $U_{\textrm{Filter}}(\Theta) = (u_{i,j}(\Theta))_{i,j \in [K^2]}$ the quantum convolutional filter, and the final image $\Tilde{x} = \left(\Tilde{x}_{i,j}\right)_{(i,j) \in [d_1]\times[d_2]}$ which is still tensor encoded.

Each pixel in the convolutional window is affected by the quantum filter by a convolutional relation analog to the one given by \autoref{eq:classical_2Dconv}:
\begin{equation}\label{eq:quantum_conv}
    \forall i \in \llbracket I, I+K \rrbracket, \; \forall j \in \llbracket J, J+K \rrbracket, \quad \Tilde{x}_{i,j} = \sum_{a,b \in [K]}  w^{i,j}_{a,b} \; x_{a,b} \quad \textrm{with} \; w^{i,j}_{a,b} = \bra{e_a, e_b} U_{\textrm{Filter}}(\Theta) \ket{e_i,e_j} \, \textrm{.}
\end{equation}
Through \autoref{eq:quantum_conv}, it is shown that applying a same RBS-based circuit to each K consecutive qubits for each register of a tensor encoded state is equivalent to applying a convolution function. The quantum convolution is analog to the classical one in the sense where each pixel in the filter window is affected by a classical convolution operation with a $K \times K$ classical filter corresponding to a part of the $K^2 \times K^2$ quantum filter coefficients. The HW preserving convolutional layer is illustrated in \autoref{fig:Quantum_Conv}. Notice that the same operations are applied to each set of $K \times K$ pixels, and not the same operation to each pixel. This limitation can be bypassed by loading several copies of the initial tensor that are translated in a batch, such a feature is implemented in the offered \href{https://github.com/ptitbroussou/HW_QCNN}{simulation library}. \autoref{eq:quantum_conv_unitary} and \autoref{eq:quantum_conv} can be adapted for any initial tensor dimension and any filter dimension. 

For a $d$-dimensional quantum convolutional layer, the quantum filter unitary is of size $K^d \times K^d$. The quantum convolutional layer hyperparameter is $K$. Even if the quantum filter is bigger than the classical filter, the structure of the quantum convolutional circuit is such that the number of parameters is smaller for the quantum layer, as explained in \autoref{sec:Results_and_Simulations}.

\begin{figure}[h]
    \centering
    \includegraphics[width=0.95\textwidth]{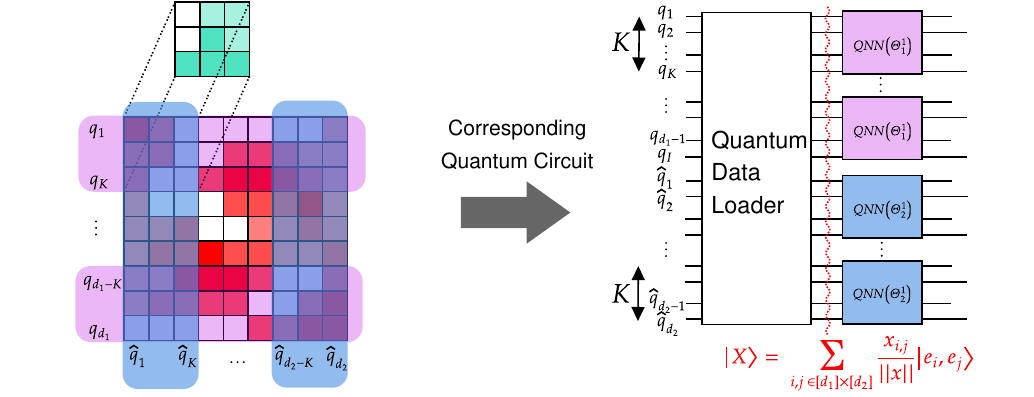}
    \caption{A 2 dimensional convolutional layer using HW preserving quantum circuits and tensor encoding.}
    \label{fig:Convolutional_Layer}
\end{figure}

In this setting, previous work \cite{landman_quantum_2022, cherrat_quantum_2024} have proposed efficient circuits that maximize the controlability with different depth, meaning that those circuits can reach in the unary basis any orthogonal matrices. In the case of the tensor encoded data, each register has a corresponding HW of $1$. In the following, the focus is made on the butterfly circuit \cite{cherrat_quantum_2024}. This choice of circuit is optimal as it allows to create any orthogonal matrix in the subspace of HW $1$ with a minimal logarithmic depth. Therefore, considering a $d$-dimensional quantum filter of size $\prod_{i=1}^d K_i$, the depth of one convolutional layer is $ \max_{i \in [d]} \mathcal{O}\left( \log(K_i) \right)$. The HW preserving convolutional layer is illustrated in \autoref{fig:Convolutional_Layer}. Notice that the $d$-dimensional filter that is applied in the convolutional layer cannot be any matrix of size $K_1 \dots K_d$. As explained in \cite{landman_quantum_2022}, $n$-qubit RBS based circuits can only perform $n \times n$ orthogonal transformations while considering a HW of $1$. The resulting filter, or equivalent unitary on all the registers, is thus a parametrized orthogonal matrix with $\sum_{i=1}^d \frac{K_i (K_i -1)}{2}$ independent parameters.

In terms of time complexity, the classical CNN layer depends on the input size, and its filter size. While considering a batch of $I \times I$ images, the complexity of conventional 2D convolution \cite{wei_rethinking_2021} depends on the size of the input image, the number of channels $C$, and is $\mathcal{O}\left( C^2 \cdot K^2 \cdot I^2\right)$ (we consider that the final are of dimension $I \times I$). For the HW preserving convolutional layer, the complexity only depends on the filter size $K$. As explained previously, considering a butterfly circuit that maximized the expressivity in the subspace of HW $1$, the depth of the quantum circuit is $\mathcal{O}\left(\log(K)\right)$. A comparison of the number of parameters and the time complexity between classical and quantum convolutional architecture layers is presented in \autoref{sec:Results_and_Simulations}. The quantum polynomial advantage increases with the dimension of the tensor, i.e., the HW of the encoding. Therefore, this layer may offer a more interesting advantage for use cases with inputs of large dimensions, such as series classification \cite{ismail_fawaz_deep_2019}.

This convolutional layer avoids Barren Plateaus \cite{mcclean_barren_2018}, because its computation is restricted in a subspace of polynomial size. Considering an input tensor of dimension $k$, such as $x \in \mathbb{R}^{\prod_{i=1}^k d_i}$, the HW of the states is $k$. It is thus fixed with respect to the number of qubits given by $n = \sum_{i=1}^k d_i$. According to recent works \cite{monbroussou_trainability_2025, ragone_lie_2024}, this ensures the absence of vanishing gradient phenomena during the training of this layer. Notice that for very complex datasets, the dimension of the input tensor will not increase, but the size of the architecture (its number of qubits) and its depth would have to increase as for classical CNN architectures.

\subsection{Quantum Pooling Layer}\label{subsubsec:pooling_layer}

In this Section, a Pooling layer that preserves the tensor structure of the quantum state is introduced. This layer allows to reduce the dimension of the image but also to apply some non linearities by using measurements. Applying non linearities in QML architectures is a non trivial task as variational quantum circuits perform linear algebraic operations on the quantum state. Previous works propose to use classical computation between quantum layers to apply non linearities \cite{landman_quantum_2022} or to use specific hardware tools to create non-linearities \cite{steinbrecher_quantum_2019}. Another proposal \cite{cong_quantum_2019} of quantum CNN has considered using measurements and single qubit gates controlled by the outcomes to perform non linearities. However, to our knowledge, there is no existing method to perform a Pooling layer with non linearities that preserves the structure of the state, allowing to keep the subspace preserving properties of the computation. Therefore, this proposal offers the possibility of deep learning architectures that are subspace preserving and thus, that ensure theoretical guarantees on their training. In addition, this method does not require using adaptive measurement techniques, but only to consider CNOT gates and to ignore a part of the qubits in the remaining part of the circuit.

\begin{figure}[h]
    \centering
    \includegraphics[width=0.95\textwidth]{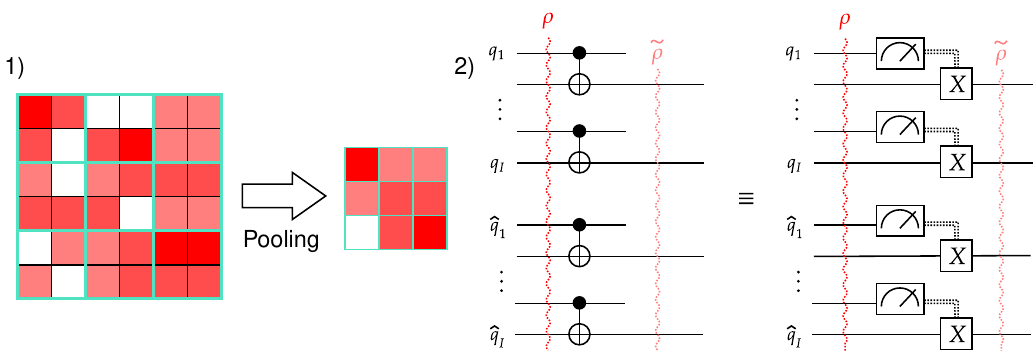}
    \caption{1) Illustration of the Pooling Layer effect on a 2 dimensional image with Pooling windows in blue. 2) Quantum Circuit for the Pooling layer and its equivalent representation using measurement and control X gates. Note that these circuits are only equivalent when considering tensor-encoded 
    states as defined in \autoref{def:TensorEncoding}.}
    \label{fig:Pooling_layer}
\end{figure}

Considering the tensor encoding, the Pooling method consists in applying a CNOT between each pair of qubits in the register corresponding to the dimension that we want to reduce. In the following part of the circuit, only the target qubits will be considered. This method is mathematically equivalent to measuring the control qubits and applying a bit flip operation to the target qubits when measuring the corresponding control qubits in state $\ket{1}$.

This Pooling circuit preserves the tensor encoding structure. Considering an initial state $\rho = \ket{X}\bra{X}$ with $\ket{X}=\sum_{i,j \in [I]}\frac{x_{i,j}}{||x||} \ket{e^I_i} \bigotimes \ket{e^I_j}$, the resulting state considering a Pooling layer for a square image (see \autoref{fig:Pooling_layer}) is:
\begin{equation}
    \Tilde{\rho} = \sum_{i} p_i \ket{\Tilde{X}^i} \bra{\Tilde{X}^i} \quad \text{with} \quad \ket{\Tilde{X}^i} = \sum_{l,k \in [O]}\frac{\Tilde{x}^i_{l,k}}{||x||} \ket{e^O_l} \bigotimes \ket{e^O_k} \, \textrm{,}
\end{equation}

with $O = I/2$. The preservation of the tensor encoding structure allows to implement several convolutional and pooling layers as in most classical deep learning architecture. In addition, this pooling operation is analog to the average pooling operation commonly used in deep learning architecture. Consider the case of a 4 by 4 image in which this pooling operation is applied:
\begin{equation}
    X = \begin{pmatrix} x_{1,1} & x_{1,2} & x_{1,3} & x_{1,4} \\
                        x_{2,1} & x_{2,2} & x_{2,3} & x_{2,4} \\
                        x_{3,1} & x_{3,2} & x_{3,3} & x_{3,4} \\
                        x_{4,1} & x_{4,2} & x_{4,3} & x_{4,4}
        \end{pmatrix} \rightarrow \Tilde{\rho} \, \textrm{.}
\end{equation}
And
\begin{equation}
    \Tilde{\rho} =
        \begin{pmatrix} x^2_{11} + x^2_{12} + x^2_{13} + x^2_{14} & x_{12} x_{14} + x_{22} x_{24} & x_{21} x_{41} + x_{22} x_{42} & x_{22} x_{44} \\
        x_{12} x_{14} + x_{22} x_{24} & x^2_{13} + x^2_{14} + x^2_{23} + x^2_{24} & x_{24} x_{42} & x_{23} x_{43} + x_{24} x_{44} \\
        x_{21} x_{41} + x_{22} x_{42} & x_{24} x_{42} & x^2_{31} + x^2_{32} + x^2_{41} + x^2_{42} & x_{32} x_{34} + x_{42} x_{44} \\
        x_{22} x_{44} & x_{23} x_{43} + x_{24} x_{44} & x_{32} x_{34} + x_{42} x_{44} & x^2_{33} + x^2_{34} + x^2_{43} + x^2_{44}
        \end{pmatrix} \, \textrm{.}
\end{equation}
Notice that the diagonal terms are the sum of the squared values of the pixels in the Pooling windows (see \autoref{fig:Pooling_layer}). Therefore, the probability of measuring the state corresponding to a certain pixel after the Pooling layer is the sum of the probability of measuring the states carrying the pixels in the corresponding Pooling window. Finally, considering a measurement based Pooling operation allows to apply some non linearities to the quantum state which is good as non-linear activation functions are usually used after the Pooling layers. 

This Pooling layer does not require adaptive measurement techniques. While adaptive measurement techniques can be used in QCNN architectures \cite{cong_quantum_2019} and in subspace preserving quantum circuits \cite{monbroussou_toward_2025}, the method does not require to perform in practice an adaptive measurement. As illustrated in \autoref{fig:Pooling_layer}, using CNOTs and discarding the control qubits is sufficient to perform the Pooling. This is the case for the examples given in \autoref{sec:Results_and_Simulations}, but one could imagine more complex architecture using these methods where the control qubits are re-used (for example to increase the depth and the size of the architecture for more complex learning tasks) which would require to measure them during the computation.

\subsection{Quantum Dense Layer}\label{subsubsec:dense_layer}

In this Section, the final part of the subspace preserving deep-learning architecture is discussed. As explained in the introduction of \autoref{subsec:QCNN_Architecture}, an architecture very similar to LeNet (see \autoref{fig:figure_intro_CNN}) is considered. After applying several Convolutional and Pooling Layers, this architecture ends with a vectorization of the image followed by a fully connected layer or dense layer.

\begin{figure}[h]
    \centering
    \includegraphics[width=0.95\textwidth]{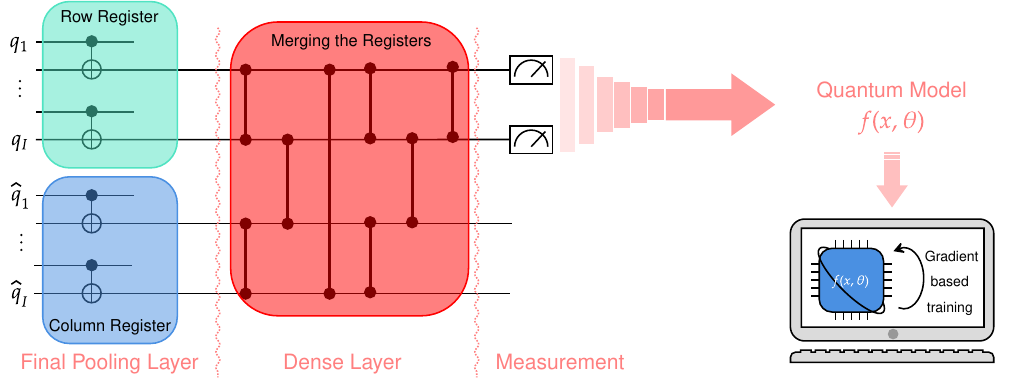}
    \caption{A 2 dimensional convolutional layer using Hamming weight preserving quantum circuits and tensor encoding. Vertical lines in the dense layer represent two-qubit RBS gates, parametrized with independent angles.}
    \label{fig:Dense_Layer}
\end{figure}

In the case of this architecture, no vectorization is required. The dense layer only consists of applying a RBS-based trainable quantum circuit to the remaining qubit while merging all the registers.

Applying such a circuit on a quantum state of fixed HW $k$ corresponds to applying an orthogonal neural network as a dense layer. Previous works \cite{landman_quantum_2022, monbroussou_trainability_2025} have highlighted the fact that using quantum orthogonal neural networks results in powerful neural networks. The number of parameters and the choice of the structure, and more specifically the choice of the connectivity used for this circuit are very important in the maximal controlability of this layer. For example, using a line connectivity for this $n$-qubit layer implies that the layer equivalent unitary is a compound matrix \cite{kerenidis_quantum_2022, monbroussou_trainability_2025}. This reduced the maximal dimension of the Dynamical Lie Algebra of this layer to $n(n-1)/2$, meaning that only a low number of parameters can be useful.

According to the HW of the states in the dense layer, i.e., the dimension $k$ of the tensor considered as the input of the architecture, the dense layer is harder to simulate classically. The impact of a RBS gate parametrized by $\theta$ in a subspace of $n$ qubits and HW $k$ is $\binom{n-2}{k-1}$ $\theta$-planar rotations. Therefore, considering a HW $k$ independent of the number of qubits $n$, the dense layer can be classically simulated. However, the complexity of this simulation could\footnote{to our knowledge, performing the $\binom{n-2}{k-1}$ $\theta$-planar rotations is the best simulation algorithm that exists for RBS-based quantum circuits.} be polynomial of degree $k$.

The quantum dense layer avoid Barren plateaus, as it is a simple HW preserving quantum orthogonal neural network \cite{landman_quantum_2022} with its HW $k$ equal to the dimension of the input tensor. As explained for the Convolutional layers in \autoref{subsubsec:Conv_layer}, the dimension of the tensor $k$ is fixed with respect to the number of qubits that depends on the input size and the number of Pooling layers used. Therefore, it is well known \cite{monbroussou_trainability_2025, ragone_lie_2024} that this layer avoid vanishing gradient phenomena during its training.

    \section{Photonic Quantum Convolutional Neural Network}\label{sec:PQCNN}

\begin{figure}[h!]
    \centering
    \includegraphics[width=1\columnwidth]{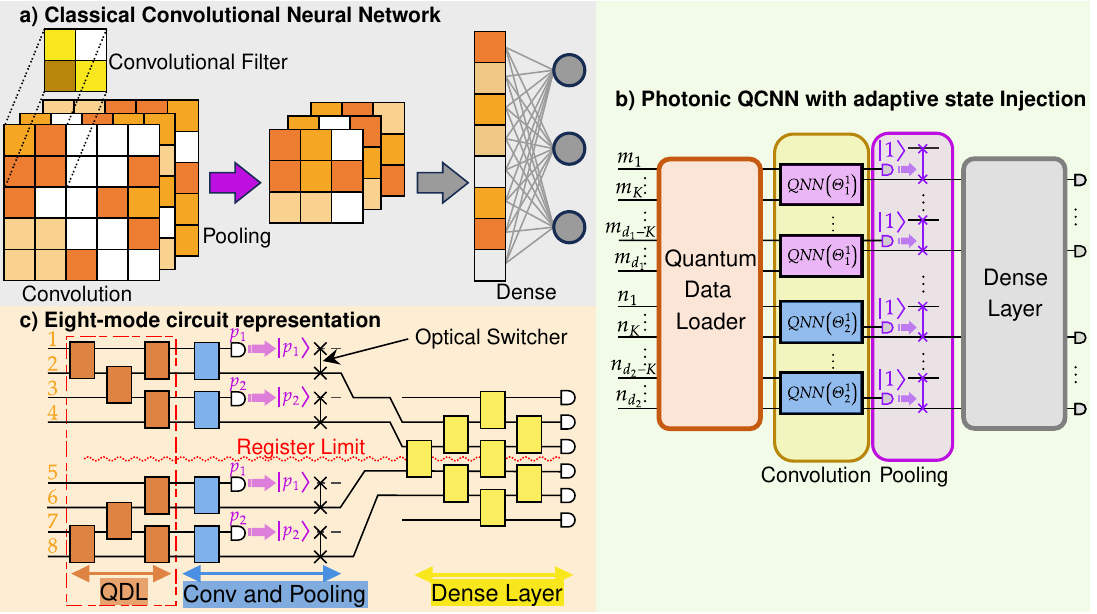}
    \caption{\textbf{Classical Convolutional Neural Networks (CNN) and Photonic Quantum Convolutional Neural Networks (PQCNN) based on adaptive state injection. a)} CNNs alternate convolutional and pooling layers to process the data features and reduce their dimensions. The last layer is a dense layer that connects the remaining features to the output nodes.
    \textbf{b)} PQCNN circuit representation for an input image of size $d_1 \times d_2$. The entire flow is composed of a Quantum Data Loader (QDL), a convolutional layer with a {Quantum Convolutional Filter (QCF)} of size $K \times K$, a pooling layer, and a dense layer before the detection. In the described scenario, the QDL takes as input two separated registers, indicated with $m_i$ and $n_i$, that encode, respectively, the rows and the columns of the input image. In the Pooling Layer, some modes are measured and, upon photon detection, a Fock state with one photon is injected into the adjacent mode. \textbf{c)} In this illustrative PQCNN involving eight modes, the QDL stage is performed by means of four Beam Splitters (BS) acting on the row and column registers respectively, while the convolutional layer is applied through a layer of four BSs on the input image. The distribution at the output of the pooling layer defines the probability of injecting extra photons into the dense layer. The latter is performed through eight BSs over six modes and a final readout layer.} 
    \label{fig:concept}
\end{figure}

The classical Convolutional Neural Network (CNN) is one of the most exploited deep learning architectures \cite{li_survey_2022, alzubaidi_review_2021}, and has been part of significant achievements in many areas, including computer vision, time series analysis, and natural language processing. In its earliest demonstration \cite{lecun_gradient-based_1998}, this architecture is typically composed of convolutional layers that extract important features from the data, pooling layers that reduce the size of the data during the computation, and a final dense neural network that mixes the features and performs the classification task. While CNNs are highly effective in capturing spatial or local patterns in data, their performance can be further improved by introducing mechanisms that provide contextual awareness. 

In \autoref{sec:HW_QCNN}, a quantum counterpart of a CNN is proposed, based on Hamming weight-preserving quantum circuits. Such a scheme aims to construct quantum stages similar to the ones of a typical CNN. As shown in \autoref{fig:concept}b, the architecture of a Photonic Quantum Convolutional Neural Network (PQCNN) comprises of an initial quantum data loading layer that maps the classical image into a quantum register, while the remaining steps reproduce the prototypical CNN operations, i.e. a sequence of convolutional, pooling, and dense layers.

Within the photonic approach proposed here, quantum data loading, convolutional, and dense layers are obtained through linear optics while also exploiting an adaptive state injection scheme in the pooling layer in order to enable dynamic adaptability and nonlinearity, as depicted in \autoref{fig:concept}b.
In the scheme, the pooling layer consists of performing single-photon detection in some modes, and, if a photon is detected, a Fock state with exactly one photon is injected in the next mode.
A particular instance of a PQCNN architecture involving an eight-mode circuit is illustrated in \autoref{fig:concept}c.
In the following subsections, a detailed description of each stage of the proposed PQCNN architecture and of the specific instance that has been implemented in the experiment is provided.

\subsection{Quantum Photonic Data Loader}\label{subsec:Tensor_Encoding}

The first procedure is Quantum Data Loading (QDL), an operative way to encode different data types in quantum states. This algorithm is based on \textit{tensor encoding}, defined in \autoref{def:TensorEncoding}, and adapted for Fock states. Such an encoding maps the data features into the amplitude associated with reference-state vectors. In detail, let us consider a classical tensor of dimension $k$ such that $x = (x_{1, \dots, 1}, \dots,  x_{d_1, \dots, d_k}) \in \mathbb{R}^{d_1 \times \dots \times d_k}$. The corresponding photonic tensor-encoded state is:
    \begin{equation}\label{eq:Tensor_Encoding_modes}
        \ket{x} = \frac{1}{||x||} \sum_{i_1 \in [d_1]} \dots \sum_{i_k \in [d_k]}  x_{i_1, \dots, i_k} \ket{e_{d_1, i_1}} \otimes \dots \otimes \ket{e_{d_k, i_k}}
    \end{equation}
where $\ket{e_{d_l, i_l}} = \ket{0 \dots 0 1 0 \dots 0}$ represents a Fock state over $d_l$ modes, with a single excitation (photon) in mode $i_l$ and vacuum in all other modes. Therefore, the set $\left\{\ket{e_{d_l, i}} \mid i \in [d_l] \right\}$ represents a fixed family of $d_l$ orthonormal quantum states, while $||\cdot||$ denotes the $2$-norm of $\mathbb{R}^d$. Notice that the input state of the algorithm generally requires $m = \sum_{i \in [k]} d_i$ modes dispatched in $k$ different registers with a single particle in each of them. In \autoref{fig:concept}b, the QDL takes two-dimensional images in input. In this case, the first register, with modes $m_{d_1}$, represents rows, and the second, with modes $n_{d_2}$, represents the image columns.

To encode any tensor $x \in \mathbb{R}^{d_1 \times \dots \times d_k}$ within this framework, one needs to use a photonic architecture with $ m=\sum_{i=1}^k d_i$ modes that can freely control the amplitudes of the Fock states used for the tensor encoding in \autoref{eq:Tensor_Encoding_modes}. We note that considering $k$ photons distributed over $k$ registers, each constructed with an independent linear optical circuit spanning $d_j$ modes, constraints the tensor-encoded state of \autoref{eq:Tensor_Encoding_modes} to be a separable state. In order to obtain a generic k-dimensional tensor in the proposed encoding, additional resources in terms of ancilla photons, modes or measurement feedforward would be required \cite{chabaud_quantum_2021,monbroussou_toward_2025}.

\subsection{Photonic Convolutional Layer}\label{subsec:Conv_Layer}

The encoded data is then fed into a convolutional operation.
For $k$-dimensional tensor-encoded inputs, one needs again to consider $k$ separate registers of modes with only a single photon in each of them.
In the quantum circuits proposed in \autoref{subsubsec:Conv_layer}, convolutional layers use the Reconfigurable Beam Splitter (RBS) gate, which applies a planar rotation between the states $\ket{01}$ and $\ket{10}$. When using single photons, this operation can be directly performed with Beam Splitters (BS), another key feature making photonic platforms as the natural candidate for the proposed PQCNN architecture.
In \autoref{subsec:Connection_RBS_BS}, the connection and the differences between BSs and RBS gates are detailed.
The HW preserving convolutional layer can be adapted to photonic platforms as follows: for each register and a {Quantum Convolutional Filter} (QCF) of size $K_1 \times \dots \times K_k$, the convolutional layer consists of applying the same circuit made of BSs to each partition of $K_i$ modes ($i \in \llbracket 1, k \rrbracket$) with the same set of variational parameters. This circuit {is represented in \autoref{fig:concept}b} for a $2$-dimensional tensor input, and a QCF of size $K \times K$. Notice that, as for classical convolutional layers, one can choose to use a convolutional layer of lower dimension with respect to the input dimension by keeping some registers unaffected on the circuit. The depth of the convolutional layer is $\mathcal{O}(K)$ with $K = \max(\{ K_1, \cdots K_k \})$ because only $K(K-1)/2$ parameters are needed to maximize the control of this circuit.

Consider for example a $d_1 \times d_2$ input tensor and a convolutional layer { with a $K \times K$ QCF as represented in \autoref{fig:concept}b}. Then, for any $I,J \in \mathbb{N}$ such that $I K \leq d_1-k$ and $J K \leq d_2-k$, the state $\ket{\Tilde{x}}$ produced after applying the convolution on the initial state $\ket{x}$ is such that:
\begin{equation}\label{eq:Convolution_example_2D}
    \sum_{i=I}^{I+K} \sum_{j=J}^{J+K} \Tilde{x}_{i,j} \ket{e_{d_1,i}, e_{d_2,j}} = U_{\textrm{Filter}}(\Theta) \sum_{i=I}^{I+K} \sum_{j=J}^{J+K} x_{i,j} \ket{e_{d_1,i}, e_{d_2,j}} \, \textrm{,}
\end{equation}
where $\Theta$ indicates the set of variational parameters associated to this layer, $\ket{e_{d_1,i}}$ is a single particle Fock state corresponding to \autoref{eq:Tensor_Encoding_modes}, and $U_{\textrm{Filter}}(\Theta) = (u_{i,j}(\Theta))_{i,j \in [K^2]}$ the QFC. The final state corresponds to a new tensor $\Tilde{x} = \left(\Tilde{x}_{i,j}\right)_{(i,j) \in [d_1]\times[d_2]}$ which is still tensor encoded. 

\subsection{State Injection based Pooling Layer}\label{subsec:SI_Pooling}

Pooling layers play a significant role in the CNN architectures, as they allow one to reduce the dimension of the data through the computation. Usually, such a layer is followed by a nonlinear activation function \cite{montufar_number_2014, raghu_expressive_2017}.
Here, a pooling layer that preserves the structure of the tensor encoded data provided in \autoref{eq:Tensor_Encoding_modes} is introduced, while reducing its size and applying a nonlinearity. The method presented here is based on state injection (see \autoref{fig:concept}b), a measurement-based technique that is suitable for near-term linear optical platforms, introduced in \cite{monbroussou_toward_2025}.
Considering a tensor encoded input state, the pooling method consists of measuring half of the modes for each register. If a photon is measured in one mode, another photon is injected into the following one.
Because of the tensor encoding structure of the state, only one photon could be measured (and injected) per register, which leads to a low number of additional photons needed.
An illustration of this pooling layer is given in \autoref{fig:concept}b. The pooling operation performed is equivalent to the one in \autoref{subsubsec:pooling_layer} when considering tensor encoding on Fock states instead of tensor encoding on states of fixed Hamming weight. The depth of this layer is $\mathcal{O}(1)$, and requires $k$ additional particles, by considering an adaptation of the State Injection. 
This pooling layer allows a reduction by half the size of the input state on each dimension, corresponding to the registers where half of the modes are measured. This operation is structurally similar to the classical average pooling, and one could choose to measure a different number of modes to change the size of the pooling, as long as the structure of the output corresponds to the tensor encoding structure which allows to apply new convolutional layers to create complex neural network architectures.

\subsection{Dense Layer}\label{subsec:Dense_Layer}

As in classical deep-learning architectures, the convolutional and pooling layers introduced previously are used in complex neural networks to extract important features and to reduce the size of the data during the computation. CNN architectures are usually completed with a final dense layer, a linear neural network that concentrates most of the trained parameters and that extracts the key features to perform the learning task. For the present QCNN architecture, it is proposed to use a linear optical layer while merging each register from the tensor encoding structure. This circuit is a linear layer, indeed the following relation holds:
\begin{equation}\label{eq:Photonic_Dense_Linear}
    \rho_{\text{out}} = W(\Theta_{\text{Dense}}) \rho_{\text{in}} W^{\dagger}(\Theta_{\text{Dense}}) \, ,
\end{equation}
with $W$ the $\binom{m+k-1}{k} \times \binom{m+k-1}{k}$ unitary matrix corresponding to the action of the $m$-mode circuit in the subspace of $k$ photons, $\Theta_{\text{Dense}}$ the set of variational parameters, and $\rho_{\text{in}}$ the initial state. As explained in \cite{aaronson_computational_2011}, such Bosonic circuits are limited in their controllability, i.e., in the maximal number of free independent parameters. One could use state injection \cite{monbroussou_toward_2025} layers to increase the Bosonic limit of $m^2-1$ parameters for a $m$ linear optical circuit while preserving the number of particles during the computation to avoid Barren plateaus. {We choose to focus on a linear optical dense layer in the experimental proposal to propose an architecture that can be verified in the very near term with a minimal number of adaptivity layers.}

The output of the QCNN architecture is a probability distribution obtained from the final measurement of the selected optical modes of the dense layer. Depending on the number of classes and the available detector technology, different measurement procedures can be employed to complete the architecture. 
Considering a typical classification task among a restricted number of $d$ classes, each label is assigned according to the probability to detect one or more photons in $d$ distinct modes or in $d$ bins of grouped modes. The training of the variational parameters of the dense layer according to a mean squared error loss ensures that the output label will be the correct one for the classification task. Further training procedure can be performed on the assignment of the labels to the output multi-photon configurations in the dense layer. This last procedure is investigated mainly in the experimental implementation as an additional readout layer that operates as a post-processing stage of the collected data. 

        \subsection{Modular adaptive photonic architecture for PQCNN}\label{subsec:ModularAdaptive}

Here, the overall architecture for the realization of the PQCNN modules described in the previous sections and sketched in \autoref{fig:exp_setup}a through an illustrative example is discussed. The adaptive photonic platform comprises two linear optical circuits interleaved with a feedforward system. The first linear optical circuit includes a QDL, where data is encoded in the amplitude of a quantum state, and then convolutional layer (CONV). These first two parts, named as $A$ and $B$ and identifying the registers employed to encode the images column and rows, can be realized within a single multi-mode linear interferometer. Such an optical circuit can be integrated into a miniaturized chip equipped with full programmability and control over all the internal parameters \cite{carolan_universal_2015,smith_universal_2022,maring_versatile_2024,pentangelo_high-fidelity_2024, giordani_experimental_2023}. The current technology of integrated devices does not allow for rapid on-chip reconfigurability, the key element required to implement the pooling layer based on adaptive state-injections. Hence as an alternative solution one can realize the pooling layer outside the chip. This distributed design could have its own advantages if one wishes to integrate privacy where data encoding is desired to be separated from the computing layer as we discuss later. The adaptive state injection requires the measurements of some output of the convolutional layer and the sequential adaptive injection of photons into the final dense layer (part $C$). Therefore, the feedforward system envisages off-chip delay lines that preserve the coherence of the multi-photon state among the paths to transfer it after the convolutional layer into the dense layer, equipped with fast optical switchers driven by the detectors of the pooling layer which allow the injection of a new single photon if the detector of the pooling clicks. Finally, the dense layer is again a linear optical circuit realized with a second programmable integrated device. The internal parameters of the dense layer are trained to assign labels to the images according to the detection of photons in certain output configurations, the measurement strategy described in the previous section.

\begin{figure*}[htb]
    \centering
    \includegraphics[width=\linewidth]{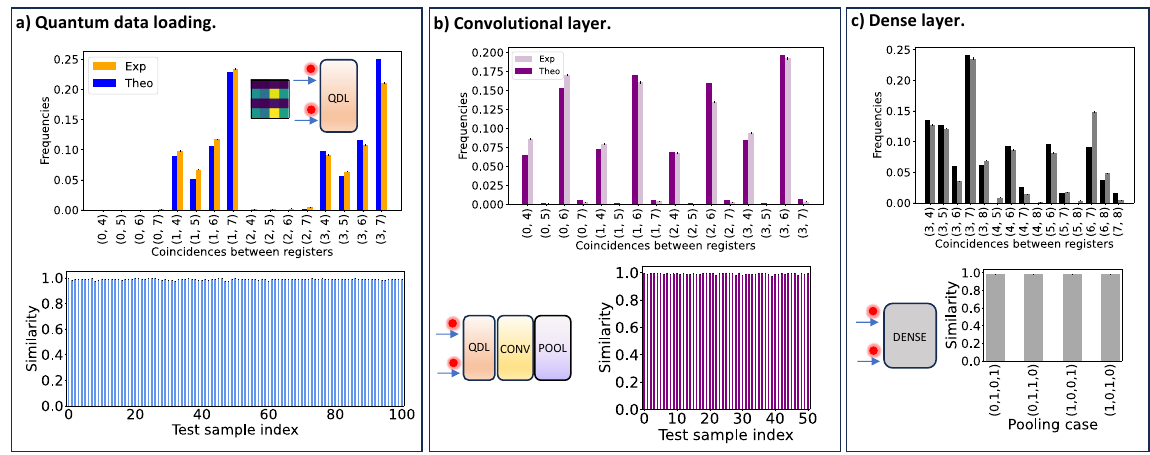}
    \caption{\textbf{Outcomes at each stage of the QCNN. a)} Above, the test image in the inset is encoded in the QDL stage.
    In the orange and blue bar plot, the theoretical and experimental probability distributions obtained at this stage for this particular image are shown. 
    Below, the similarity between the theoretical and experimental distributions for hundred different images.
    \textbf{b)} The convolutional and pooling layers follow after the QDL stage. In the purple and lavender bar plot, the theoretical and experimental probability distributions at the output of the pooling layer for the image encoded in panel a) are reported. The purple bar plot shows the similarity between ideal distributions and experimental data obtained at the output of the pooling layer of fifty different images.  \textbf{c)} Here, the dense layer is tested separately from the previous layers. In the grey and black bar plot, the experimental results obtained for the unitary corresponding to the events where two photons are measured in the second and the last pooling modes, and two photons are injected directly into the dense layers, as shown in the inset, are reported. In the grey histogram, the similarity between theoretical and experimental output distributions for the four unitaries corresponding to such pooling cases is shown.
    All uncertainties are estimated assuming Poissonian statistics.
    }
    \label{fig:results}
\end{figure*}

\section{Experimental Apparatus}\label{sec:Experimental_Apparatus}

\begin{figure*}[h!]
    \centering
    \includegraphics[width=\textwidth]{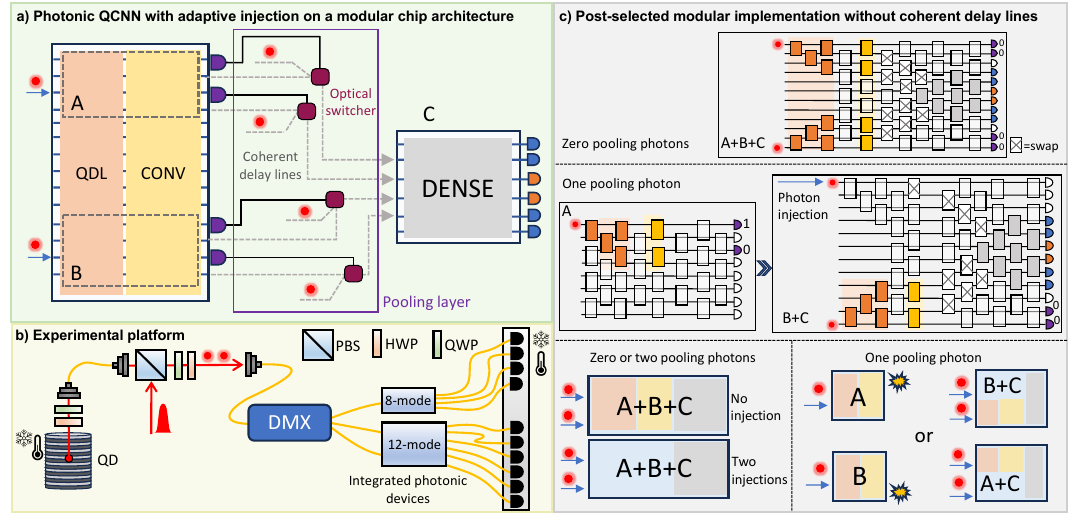}
    \caption{\textbf{QCNN architecture and experimental setup. a)} Modular architecture of a photonic QCNN with linear optics and adaptive state injection. The architecture consists of two linear optical circuits interleaved with a feedforward system. The first linear optical circuit includes the QDL, and the convolutional layer.
    The pooling layer, instead, requires a feedforward system through coherent delay lines to transfer the state after the convolutional into the dense layer, equipped with a fast optical switcher driven by the detectors of the pooling layer. The optical switcher injects (or not) a new single photon, or the state leaving the convolutional layer, according to the measurement outcomes of the detectors of the pooling layer. We experimentally encoded $4 \times 4$ images with rows and columns encoded into two separate registers, indicated here as $A$ and $B$, while  the dense layer is labeled as $C$. \textbf{b)} Scheme of the photonic hardware QOLOSSUS-2. Single photons are generated through a semiconductor quantum dot source (QD). They are then coupled into a temporal-to-spatial demultiplexer (DMX) and manipulated into an 8-mode and a 12-mode universal and programmable integrated interferometers. The output photon statistics are measured through superconducting nanowire single photon detectors (SNSPD). In the figure, only half of the detectors used are shown, as we analyze all output modes for each integrated device. \textbf{c)} To cope with the absence of coherent delay lines and a feedforward system, the current implementation employs post-selection and a different separation of the stages $A,~B,~C$. In detail, above in the panel, the case in which no photons are detected in the pooling modes and, hence, the whole quantum state after the convolutional layer is injected into the dense layer through suitable swap operations is reported. Parts $A,~B,~C$ are therefore realized jointly in the 12-mode circuit.
    In the middle section, the case in which a photon is found in the first pooling mode and, hence, a photon is directly injected into the dense layer for the first register is reported. Here, the first register encoding ($A$) is performed in the 8-mode circuit, while parts $B,~C$ are jointly performed in the 12-mode one. All other cases are summarized in the bottom part of the panel.}
    \label{fig:exp_setup}
\end{figure*}

\textbf{Photonic hardware.} 
The previous model is experimentally tested with a hybrid quantum photonics platform, denominated QOLOSSUS-2, sketched in \autoref{fig:exp_setup}b, including two different integrated photonic devices with 8 and 12 modes, respectively. Single photons are produced through a commercially available (Quandela \emph{e-Delight}) semiconductor quantum dot (QD) single-photon source. It consists of an InGaAs matrix placed in a nanoscale electrically controlled micropillar cavity \cite{somaschi_near-optimal_2016} kept at cryogenic temperature ($\approx 4K$) through an \emph{Attocube-Attodry800} He-closed cycle cryostat. The QD is optically excited with a pulsed laser in resonance with the cavity characteristic wavelength ($928.05$~nm) \cite{somaschi_near-optimal_2016, loredo_generation_2019}. The repetition rate of the laser amounts to 160 MHz. The generated single photons are coupled into a single-mode fiber through a free-space confocal microscope mounted atop the cryostat shroud. Photons are then separated from the residual pumping laser in a cross-polarization scheme \cite{loredo_generation_2019} through the use of a polarizing BS and waveplates. A temporal-to-spatial demultiplexer (DMX) is employed to actively separate the stream of single photons. In particular, the DMX system exploits an acousto-optical modulator programmed to split the train of single photons into three spatially separated modes, which are then temporally synchronized via properly tuned in-fiber delay loops. 

After temporal synchronization, the multi-photon state is then injected into two different programmable universal integrated circuits with 8 and 12 modes respectively \cite{bell_further_2021,giordani_experimental_2023} fabricated through femtosecond laser {waveguide} writing \cite{corrielli_femtosecond_2021, ceccarelli_low_2020, pentangelo_high-fidelity_2024}.
The on-device operations of the 12-mode device are controlled by thermo-optical phase shifters, through the application of external currents over the $132$ heaters on the top of the integrated device. In particular, the optical circuit was developed according to the universal design reported in \cite{bell_further_2021} in which Mach-Zehnder based configurations featuring a pair of reconfigurable internal phases enable the implementation of arbitrary unitary transformations within a more compact physical configuration. The 8-mode chip, designed according to the universal design of \cite{clements_optimal_2016} and encompassing 56 thermo-optic phase shifters, is employed to perform the QDL and convolutional layers only. 
The optical depth of the 12-mode chip is enough to allow for different internal configurations of the required building blocks i.e. QDL, convolutional, and dense layers. We note that the pooling layer is emulated in post selection due to the current unavailability of coherent delay lines and fast optical switches. This emulation procedure is described in the next paragraph. 
After the evolution within the integrated device, photons are detected with superconducting nanowire single-photon detectors (SNSPD) \cite{natarajan_superconducting_2012}.

\textbf{Implementation of the PQCNN via post-selection. }
In what follows, the description of the encoding of the PQCNN architecture is tailored to the experimental photonic platforms described above and reported in \autoref{fig:exp_setup}c.
The goal is to carry out a binary classification of $4 \times 4$ pixel images.
Firstly, the QDL layer is adapted to be encoded in the integrated photonic device that comprises 12 modes, which would in principle encode the full structure $A,~B$ (each comprising QDL and convolutional layer) $+C$ (dense layer). 
The size of the circuit limits the number of layers that can be reserved for the QDL. We opted for an experimental QDL composed of two 4-mode linear-optical circuits that use eight BSs of the device. Each circuit independently encodes a $4$-mode register (see \autoref{fig:exp_setup}c). As a result, the QDL produces tensor-encoded states capable of representing a subset of the possible $4\times 4$ pixel images, like grayscale bars and stripes. 
A suitable dataset was therefore adapted from the publicly available Pennylane Bars-and-Stripes (BAS) dataset, which is here denoted as Custom BAS dataset. Details about this dataset are provided in \autoref{app:Experimental_Details_PQCNN}.
The convolutional stage is encoded in one layer of the circuit with a total number of 4 tunable BSs, highlighted by the yellow area in \autoref{fig:exp_setup}c. The dense layer comprises 8 BSs of the 12-mode device.

The pooling is performed over four modes, two for each register. As previously said, the state injection is emulated via post-selection. Operationally, this means that different experiments are run with different numbers of injected photons and circuit configurations.
Some instances of the pooling configurations and the corresponding circuits are reported in \autoref{fig:exp_setup}c. The current implementation employs different separations of the stages $A,~B,~C$ into the 12- and 8-mode devices. In detail, the top panel of \autoref{fig:exp_setup}c reports the circuit configuration in the case in which no photons are detected in the pooling modes and, hence, the whole quantum state after the convolutional layer is injected into the dense layer through suitable swap operations. Parts $A,~B,~C$ are therefore realized jointly in the 12-mode circuit.
A second scenario is the case in which one photon is found in the first pooling mode in part $A$ and, hence, a photon is directly injected into the dense layer for the first register. Here, the first register encoding ($A$) is performed in the 8-mode circuit, while parts $B,~C$ are jointly performed in the 12-mode one (see middle panel of \autoref{fig:exp_setup}). Similar setup for the other scenario corresponding to one photon detected in the pooling mode of the second QDL register $B$. 

All the post-selection cases and the related circuits and measurement settings are briefly depicted in the last panel of \autoref{fig:exp_setup}c. To summarize, both the scenarios with zero photons in the pooling and no injections, and with two photons in the pooling and two injections, have been fully implemented in the 12-mode device by post-selecting on the output configurations that individuate each of the two configurations. The cases with one photon in the pooling layer are realized by encoding one QDL and one convolutional layer in the 8-mode device (part $A$ or $B$) and the second QDL, convolutional, and the dense layer in the 12-mode device.

    \section{Results and Simulations}\label{sec:Results_and_Simulations}

        \subsection{Hamming-Weight Preserving QCNN}\label{subsec:Results_Sim_HW_QCNN}

The term "model complexity" may refer to different meanings in deep learning, including the expressive capacity and effective model complexity \cite{hu_model_2021}. It may also refer to the time complexity of the different layers \cite{shah_time_2022}. To compare the feed-forward and training running time between the quantum deep learning layers introduced previously and their classical equivalent, two important criteria are particularly significant. First, the number of parameters of the model is a standard metric of the running time, as a low number of parameters reduces the training and the forward pass of a model. In addition, the forward pass running time is very important, and determines the number of basic operations a computer needs to run the model. In \autoref{table:Time_Complexity}, the running time complexity of the forward pass for each convolutional neural network layer is compared with the depth of the analog quantum layers. The depth of the corresponding quantum circuits gives the number of basic quantum operations, i.e., number of parallel gates that should apply.

\begin{table}[h!]
\centering
\begin{tabular}{ |c|c|c|c|c| } 
    \hline
    & Convolutional & Pooling & Orthogonal Dense & Dense \\ 
    \hline
    Classical Complexities & $\mathcal{O}\left((\prod_{i=1}^k K_i) \cdot (\prod_{i=1}^k d_i)\right)$ & $\mathcal{O}(\prod_{i=1}^k d_i)$  & $\mathcal{O}(p \cdot \binom{n}{k})$ & $\mathcal{O}((\sum_{i=1}^k d_i)^2)$ \\ 
    \hline
    Quantum Layer Depth & $\mathcal{O}(\log(K))$ & $\mathcal{O}(1)$ & $\mathcal{O}(\frac{p}{n})$ & - \\
    \hline
\end{tabular}

\caption{Time complexity comparison between classical deep learning layers and Hamming weight preserving quantum analogs. We consider $k$ dimensional convolutional neural network layers with $d_1 \times \cdots \times d_k$ the size of the square input tensor, $\{ K_1, \cdots K_k \}$ the size of the convolutional filter, and $p$ the number of parameters in the orthogonal dense layer. We call $n$ the global number of qubits in the case of the quantum architecture where $n = \sum_{i=1}^k d_i$.}
\label{table:Time_Complexity}
\end{table}

It is important to note that an advantage in time complexity, as reported in this table, does not guarantee an overall running time advantage, as sample complexity must be considered as well. Whether such a gain translates into practice would depend on the type of measurement used in the final architecture. \\ 

In \autoref{subsec:QCNN_Architecture}, the layers were presented in the case of $2$ or $3$-dimensional convolutional architecture. In \autoref{table:Time_Complexity}, the case of $k$-dimensional convolutional layer is considered to consider a general case. The quantum advantage increases with the dimension of the tensor: for the Convolution and Orthogonal Dense layers, it is polynomial of order $k$ the dimension of the input state. However, this dimension corresponds to the global HW, and one should be careful to consider this value independent of the number of qubits $n$ to avoid Barren Plateau \cite{larocca_diagnosing_2022, ragone_lie_2024, monbroussou_trainability_2025}. This should be the case while using QCNN, as the dimension of the tensor depends on the type of data one wants to use. An increase in the complexity of the learning task should result in the increase of the input sizes (not the dimension) and in greater depth of the neural network \cite{raghu_expressive_2017}.

In addition to the running time complexity, the number of parameters of the model and the running time associated with the vectorizations in the model should also be considered. Indeed, preparing the state for each layer in a classical CNN architecture requires to vectorized it, especially when using GPUs for computation \cite{ren_vectorization_2015}. In the case of the quantum models, no adaptation of the state is required for the Convolutional and Pooling layers, as theses layers preserve the structure of the state. The final dense layer only requires to apply RBS between qubits from different registers. 

\begin{table}[h!]
\centering
%\captionsetup{justification=centering}
\begin{tabular}{ |c|c|c|c|c| } 
    \hline
    & Convolutional & Pooling & Orthogonal Dense & Dense \\ 
    \hline
    Classical Layers & $ \prod_{i=1}^k K_i^2$ & $0$  & $p \leq \binom{n}{k}(\binom{n}{k}-1)/2$  & $(\sum_{i=1}^k d_i)^2$ \\ 
    \hline
    Quantum Layers & $\sum_{i=1}^k K_i(K_i - 1)/2$ & $0$ & $p \leq \binom{n}{k}(\binom{n}{k}-1)/2$ & -  \\
    \hline
\end{tabular}
\caption{The number of parameters of classical deep learning layers and of Hamming weight preserving quantum analogs. We consider $k$ dimensional convolutional neural network layers with $d_1 \times \cdots \times d_k$ the size of the square input tensor, $\{ K_1, \cdots K_k \}$ the size of the convolutional filter. We call $n$ the global number of qubits in the case of the quantum architecture where $n = \sum_{i=1}^k d_i$.}
\label{table:Number_Parameters}
\end{table}

\begin{figure}[h!]
\centering
\begin{subfigure}[t]{.45\textwidth}
    \centering
    \includegraphics[height=0.65\textwidth]{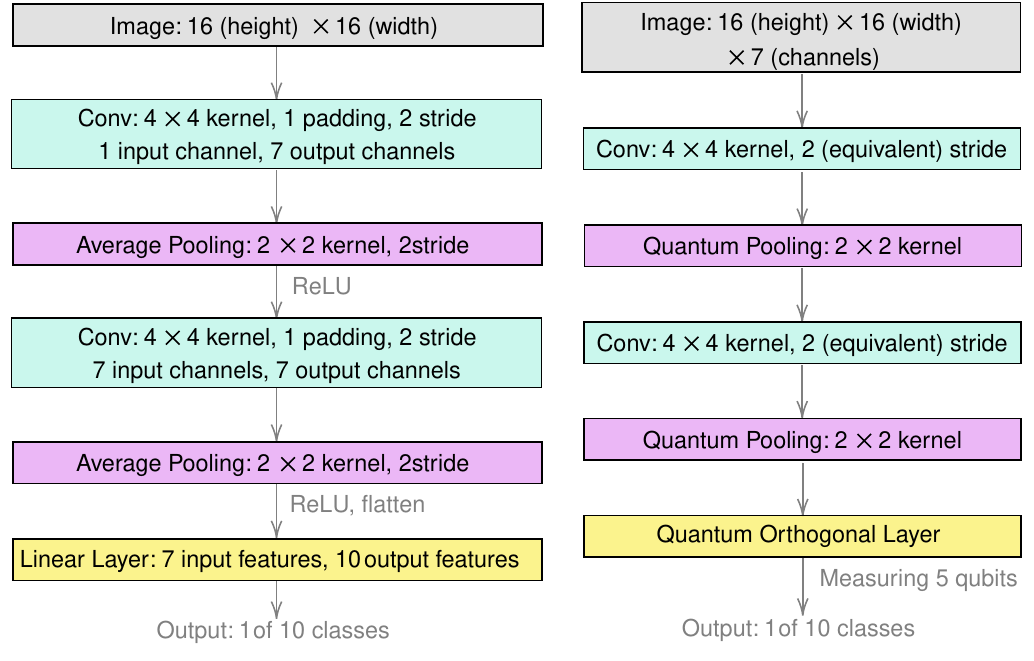}
    \caption{CNN architecture and HW preserving convolutional architecture used for training comparison.}
    \hspace*{0.2in}
    \label{fig:QCNN_CNN_Architecture}
\end{subfigure}
\begin{subfigure}[t]{.45\textwidth}
    \centering
    \includegraphics[height=0.75\textwidth]{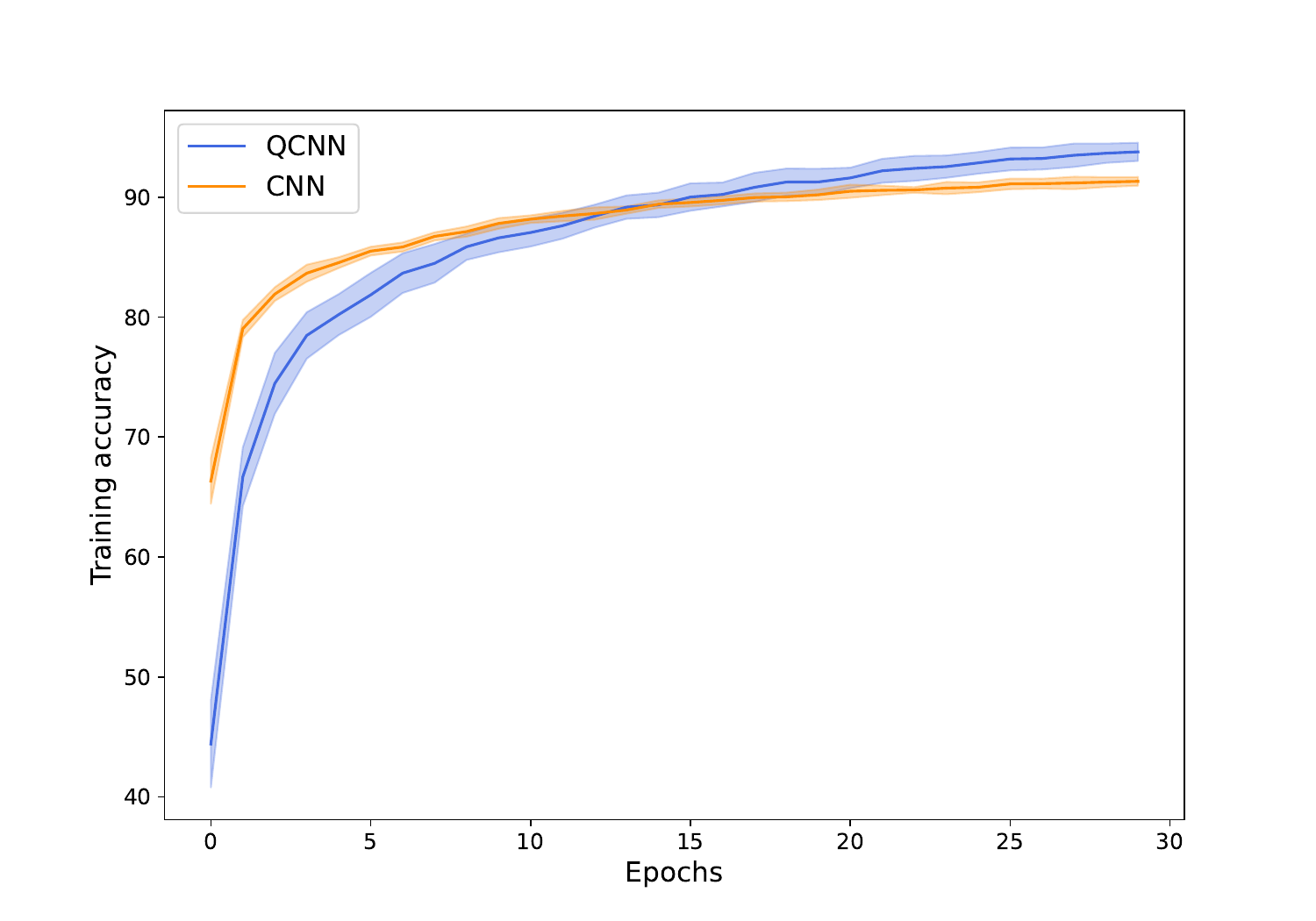}
    \caption{MNIST digit dataset \cite{lecun2010mnist}. 
    }
    \label{fig:QCNN_CNN_MNIST}
\end{subfigure}
\begin{subfigure}[t]{.49\textwidth}
    \centering
    \includegraphics[height=0.75\textwidth]{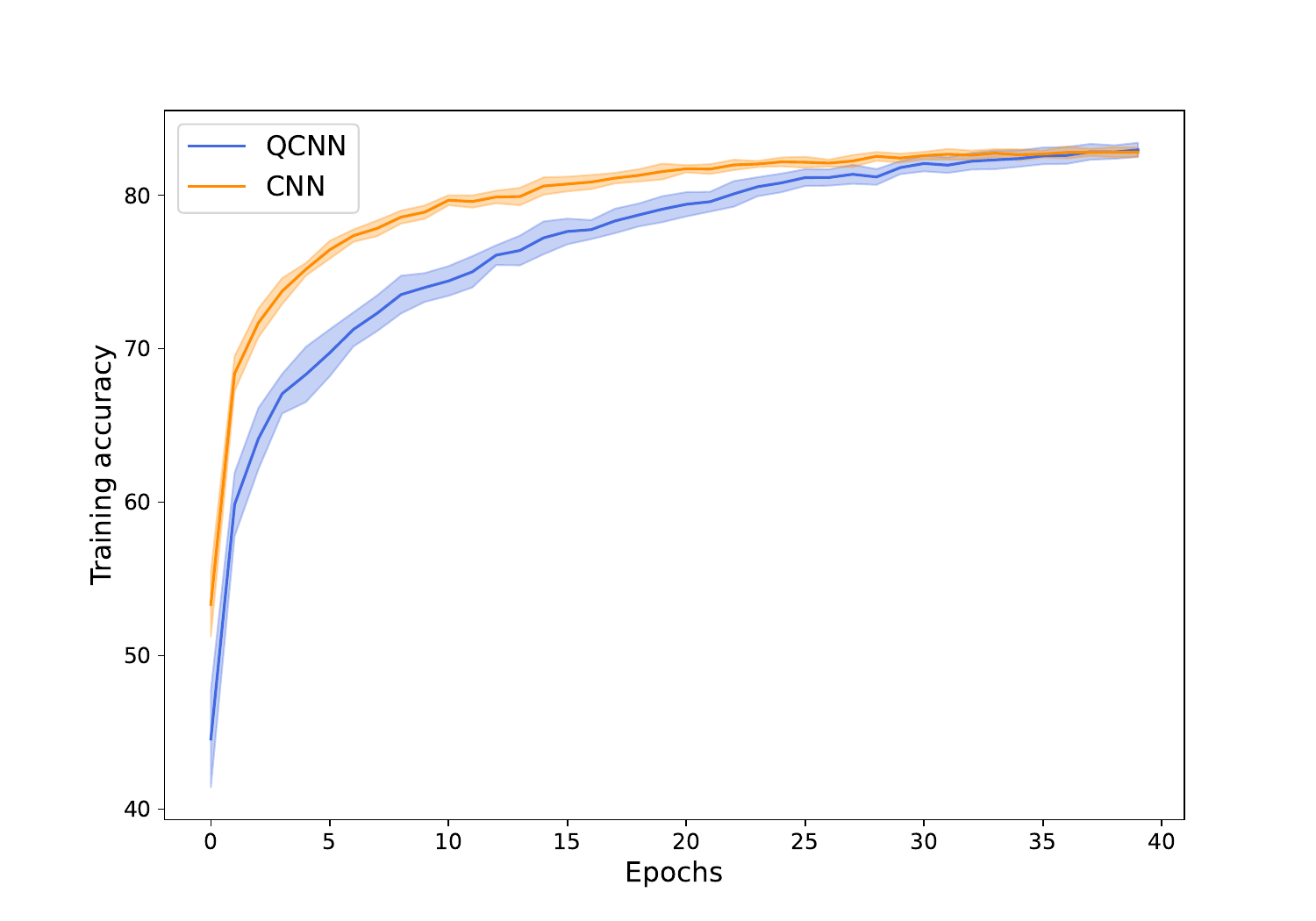}
    \caption{Fashion MNIST dataset \cite{xiao_fashion-mnist_2017}. }
    \hspace*{.2in}
    \label{fig:QCNN_CNN_FashionMNIST}
\end{subfigure}
\begin{subfigure}[t]{.49\textwidth}
    \centering
    \includegraphics[height=0.75\textwidth]{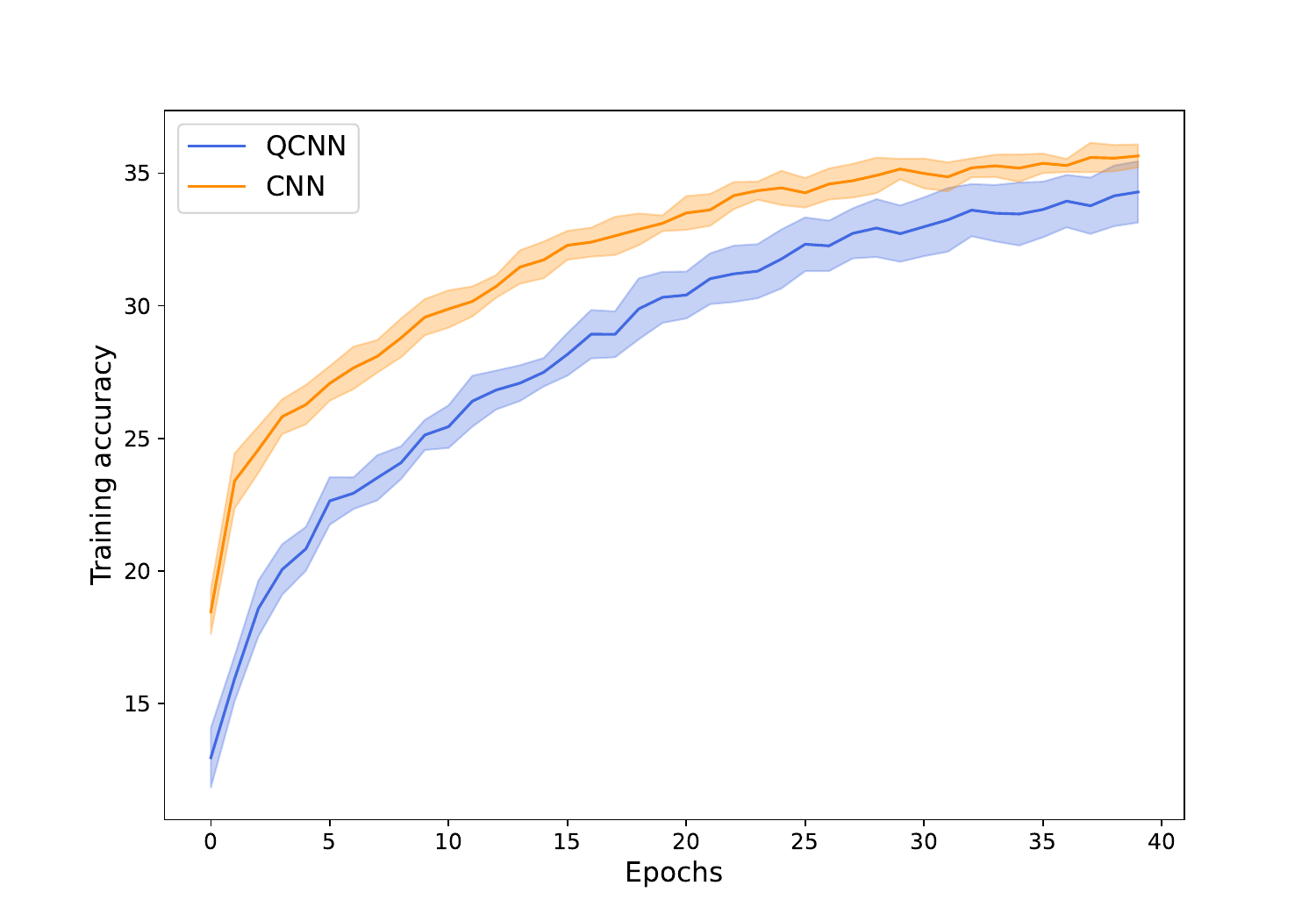}
    \caption{CIFAR-$10$ dataset \cite{Krizhevsky2009LearningML}.}
    \vfill
    \label{fig:QCNN_CNN_CIFAR}
\end{subfigure}
\caption{Average training accuracy and standard deviation comparison between classical CNN architecture and a HW preserving architecture (a) for classification of 10 label datasets (b,c,d), with $2000$ input images. The average values and standard deviation are derived from 10 different trainings. The quantum architecture (QCNN) has $755$ parameters and the classical architecture (CNN) has $990$ parameters. Both architecture (choice of layers and hyper-parameters) are unchanged in all the case, and can be found in our \href{https://github.com/ptitbroussou/HW_QCNN}{GPU-based toolkit}.}
\label{fig:Simulations}
\end{figure}

Thanks to \autoref{table:Time_Complexity} and \autoref{table:Number_Parameters}, it is observed that the Convolutional and Pooling layers offer large polynomial advantages, especially when considering high dimensional input tensors. The quantum filter is less parametrized than the classical one, but simulations presented in \autoref{fig:Simulations} show that those quantum orthogonal filters perform well. Similarly, the quantum orthogonal dense layer performs well with a reduced number of parameters in comparison with classical dense. Previous works \cite{landman_quantum_2022, cherrat_quantum_2024, johri_nearest_2021} have already shown that orthogonal layers perform well in comparison with dense layers. \textbf{The quantum advantage in terms of running time complexity, number of parameters, and lack of vectorization needed open new perspective to design useful subspace preserving QML algorithms.}

In this Section, the proposed method is tested using several very famous datasets used to benchmark classification algorithms. It is proposed in the offered \href{https://github.com/ptitbroussou/HW_QCNN}{GPU-based toolkit} to simulate Hamming-Weight preserving deep-learning architectures. To do so, the code performs linear algebra using the PyTorch \cite{paszke_pytorch_2019} library while only considering the smaller subspace possible. The pooling part of the circuit is simulated using projectors between different subspace bases. Thanks to this new method, it is possible to simulate larger quantum circuits and to perform image classification on $10$-classes datasets and not only binary classification as usually done in QML. This simulation software allows one to mix the subspace preserving simulation with classical layers thanks to its PyTorch module implementation. To our knowledge, this is the most complex image classification task, in the sense of the number of labels, realized with classical data.

\begin{table}[h!]
\centering
\begin{tabular}{ |c|c|c|c|c|c| } 
    \hline
    & Parameters & Dataset & Training Accuracy &  Testing Accuracy & Epochs \\ 
    \hline
    \multirow{3}*{CNN} & \multirow{3}*{990}  & MNIST & $91.33\% \pm 0.36\%$ & $84.59\% \pm 0.91\%$ & 30 \\ 
    & & FashionMNIST & $82.8\% \pm 0.3\%$ & $73.83\% \pm 1.56\%$ & 40 \\ 
    & & CIFAR-10 & $35.65\% \pm 0.43\%$ & $27.79\% \pm 0.85\%$ & 40  \\
    \hline
    \multirow{3}*{QCNN} & \multirow{3}*{755}  & MNIST & $93.79\% \pm 0.76\%$ & $86.79\% \pm 1.45\%$ & 30 \\ 
    & & FashionMNIST & $82.95\% \pm 0.47\%$ & $78.29\% \pm 0.83\%$ & 40\\ 
    & & CIFAR-10 & $34.29\% \pm 1.15\%$ & $28.71 \% \pm 1.05\%$ & 40 \\
    \hline
\end{tabular}
\caption{Simulation Results. We consider $2000$ training samples and $1000$ testing samples. The architectures described in \autoref{fig:Simulations} were trained, with Adam optimizer, and Cross Entropy Loss. All hyper-parameters and computations can be found in \href{https://github.com/ptitbroussou/HW_QCNN}{GPU-based toolkit}.}
\label{table:Simulations}
\end{table}

To benchmark the layers, it is proposed to compare a classical CNN architecture with a quantum one and similar hyper-parameters, for $4$ well known image recognition datasets. Each dataset (\cite{lecun2010mnist,xiao_fashion-mnist_2017,Krizhevsky2009LearningML} has $10$ classes of image, which we prepare by applying a average pooling layer to reduce the size of the input images. Every simulation can be found in the \href{https://github.com/ptitbroussou/HW_QCNN}{GPU-based toolkit}, and an illustration of both architectures is presented on \autoref{fig:QCNN_CNN_Architecture}. The quantum architecture was fixed to be maximal with respect to the GPU capacity: the simulations were ran using a NVIDIA A100 80 GB GPU on a cluster. Then, a classical CNN with the same hyper parameters is chosen: same input, same number of channels, same kernel size, same number of Pooling, equivalent dense layer. The QCNN has less parameters than the classical CNN ($755$ parameters versus $990$) due to the fact that the quantum Convolutional layers and the quantum dense layer limits the filter and the dense equivalent matrix to orthogonal ones, reducing the number of parameters.

To benchmark the method, it is offered to test it with first a Naval Group image classification dataset and then with usual image classification dataset from the Machine Learning community \cite{lecun2010mnist, xiao_fashion-mnist_2017, Krizhevsky2009LearningML}. The Naval Group dataset corresponds to $300$ images of sailboats (first class) and $300$ images of zodiac boats (second class). Each images is a $32 \times 32$ pixels with colors. The images are separated to have $200$ of each class in the training dataset and $100$ of each class in the test dataset.

\begin{figure}[h!]
\centering
\begin{subfigure}[t]{.19\textwidth}
    \centering
    \includegraphics[height=0.99\textwidth]{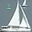}
\end{subfigure}
\begin{subfigure}[t]{.19\textwidth}
    \centering
    \includegraphics[height=0.99\textwidth]{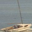}
\end{subfigure}
\begin{subfigure}[t]{.19\textwidth}
    \centering
    \includegraphics[height=0.99\textwidth]{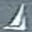}
\end{subfigure}
\begin{subfigure}[t]{.19\textwidth}
    \centering
    \includegraphics[height=0.99\textwidth]{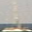}
\end{subfigure}
\begin{subfigure}[t]{.19\textwidth}
    \centering
    \includegraphics[height=0.99\textwidth]{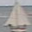}
 \end{subfigure}
\begin{subfigure}[t]{.19\textwidth}
    \centering
    \includegraphics[height=0.99\textwidth]{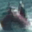}
\end{subfigure}
\begin{subfigure}[t]{.19\textwidth}
    \centering
    \includegraphics[height=0.99\textwidth]{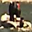}
\end{subfigure}
\begin{subfigure}[t]{.19\textwidth}
    \centering
    \includegraphics[height=0.99\textwidth]{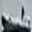}
\end{subfigure}
\begin{subfigure}[t]{.19\textwidth}
    \centering
    \includegraphics[height=0.99\textwidth]{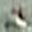}
\end{subfigure}
\begin{subfigure}[t]{.19\textwidth}
    \centering
    \includegraphics[height=0.99\textwidth]{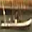}
 \end{subfigure}
\caption{Naval Group dataset: $5$ first images of sailboats class (first row) and $5$ first images of the zodiac boats (second row). Each image is $32 \times 32$ pixels.}
\label{fig:NG_dataset_images}
\end{figure}

\begin{figure}[h!]
  \begin{subfigure}{0.47\columnwidth}
    \includegraphics[width=\linewidth]{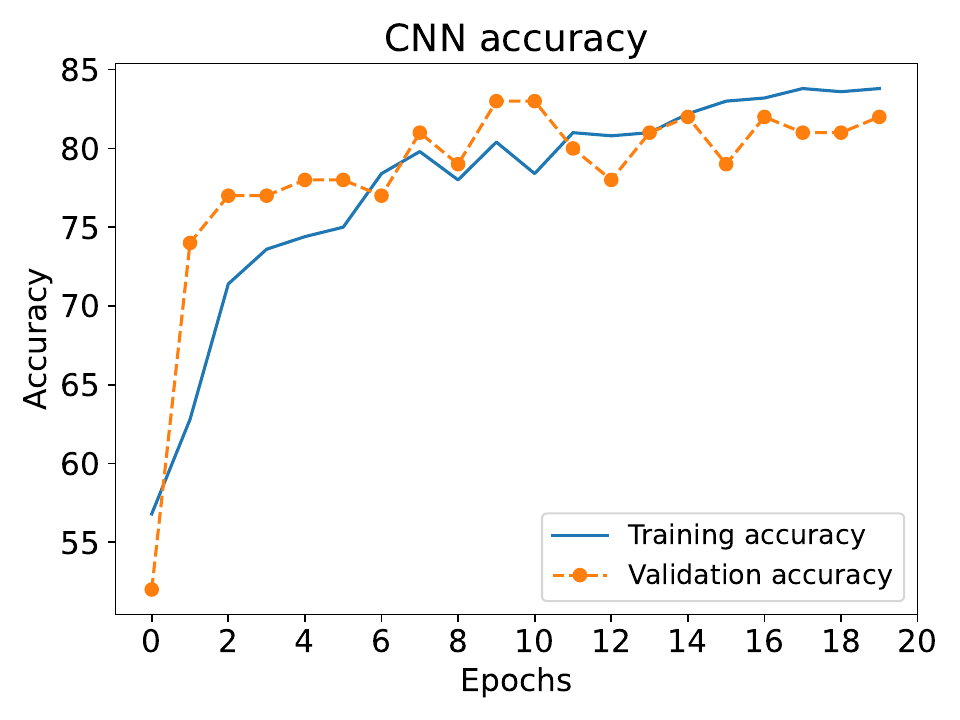}
    \caption{CNN accuracy for 2-dimensional Naval Group dataset.}
    \label{fig:CNN_accuracy_dataset_NG}
  \end{subfigure}
  \hfill
  \begin{subfigure}{0.47\columnwidth}
    \includegraphics[width=\linewidth]{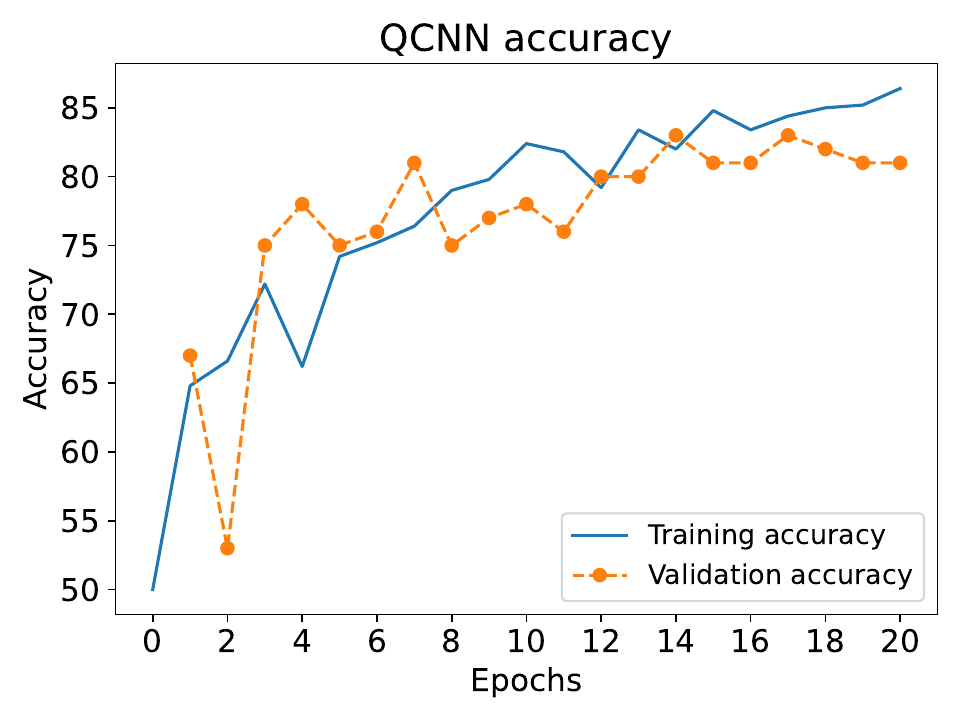}
    \caption{QCNN accuracy for 2-dimensional Naval Group dataset}
    \label{fig:QCNN_accuracy_dataset_NG}
  \end{subfigure}
  \caption{Comparison of the HW preserving QCNN and a classical CNN architecture for the Naval Group dataset. Both architectures are made with two convolutional layer with kernel size of $4$, two (average and quantum) pooling layers, and the CNN is completed by a linear layer while the QCNN is completed with an orthogonal layer made of RBS gates.}
  \label{fig:CNN_QCNN_comparison_dataset_NG}
\end{figure}

\newpage

Results presented in \autoref{fig:Simulations} and \autoref{table:Simulations} show that the architecture offers similar performance than classical CNN architecture. In addition, with the running time complexity advantages, and the lack of vectorization needed, the quantum architecture reaches similar accuracy with fewer parameters due to the orthogonality of its final dense layer, and the structure of its convolutional layers. The model even outperforms the classical architecture for the MNIST and Fashion MNIST dataset classification. In the case of CIFAR-10 dataset, both architectures do not have the complexity to achieve a satisfying result after the training, but we observe similar training behavior and performance. 

\newpage
        \subsection{Photonic QCNN}\label{subsec:Results_Sim_PQCNN}

The PQCNN architecture here proposed offers a polynomial advantage over CNN architectures. In particular, the degree of polynomial advantage depends on the number of input photons, which is given by the dimension $k$ of the input tensor. In \autoref{table:Time_Complexity} we compare layer-by-layer the number of operations, and so the time complexity, required to implement a CNN in terms of the parameters that individuate the size of the network, namely $K_i, \, d_i, \, k, \, m$ defined in Secs. \ref{subsec:Conv_Layer}- \ref{subsec:Dense_Layer},  {with the ones required by a PQCNN.}
The running time of the PQCNN is related to several features of the adopted platform: the characteristics of the optical apparatus, including coupling and {propagation} losses, brightness of the single-photon source{, detection efficiency, the number and reconfiguration speed of adaptive injection layers.}

\begin{table}[h!]
\centering
\begin{tabular}{ |c|c|c|c| } 
    \hline
    & Convolutional layer & Pooling layer & {Dense layer} \\ 
    \hline
    %Classical 
    CNN & $\mathcal{O}\left((\prod_{i=1}^k K_i) \cdot (\prod_{i=1}^k d_i)\right)$ & $\mathcal{O}(\prod_{i=1}^k d_i)$  & $\mathcal{O}(\binom{m+k-1}{k}^2)$  \\ 
    \hline
    PQCNN & $\mathcal{O}(K)$ & $\mathcal{O}(1)$ & $\mathcal{O}(\frac{p}{m})$ \\
    \hline
\end{tabular}
\caption{Comparison between the  CNN and PQCNN required resources. We consider $k$ dimensional convolutional neural network layers with $d_1 \times \cdots \times d_k$ the size of the square input tensor, $d = \max(\{d_i\}_{i=1}^k)$, $\{K_i\}_{i=1}^k$ the size of the convolutional filter, $K = \max(\{K_i\}_{i=1}^k)$, and $p$ the number of parameters in the dense layer. {We call $m$ the total number of modes with $m = \sum_{i=1}^k d_i$.}}
\label{table:Time_Complexity_comp}
\end{table}

\autoref{table:Required_Ressources} summarizes the resources of each layer in the PQCNN, both for the adopted scheme and for the most general architecture which goes beyond the one illustrated in \autoref{fig:concept}, for what concerns the number of modes, input photons, and injected photons. In this Table, for each layer the same input tensor size is considered. In practice, the input tensor size could decrease during the computation because of the pooling layers, but one can choose to use additional modes or photons in a custom architecture. 
\begin{table}[h!]
\centering
\begin{tabular}{ |c|c|c|c|c| } 

    \hline
     \textbf{Adopted} & QDL & Convolutional layer & Pooling layer & Dense layer \\ 
    \hline
    Modes & $m$& $m$& $m$& $m/2+\alpha$ \\
    \hline
    Photons & $k$ & $k$ & $2k$ & $k$ \\
    \hline
    \hline
    \hline
    \textbf{General} & QDL & Convolutional layer & Pooling layer & Dense layer \\ 
    \hline
    Modes & $m'$ & $m'$ & $m'$ & $m'+\alpha'$ \\
    \hline
    Photons & $k'$ & $k'$ & $2k'$ & $k'$ \\
    \hline
\end{tabular}
\caption{Resources required for each layer to build the PQCNN setup. We consider convolutional neural network layers with $d_1 \times \cdots \times d_k$ the size of the square input tensor for each layer (with $m' = \sum_{i=1}^k d_i$ modes and $k'$ photons). We call $\alpha' \in \mathbb{N}$, the number of mode one can add to the dense layer. For the experimental architecture developed in this work, $m=8$, $k=2$, and $\alpha=2$.}
\label{table:Required_Ressources}
\end{table}
\begin{table}[h!]
\centering
\begin{tabular}{ |c|c|c|c|c| } 
    \hline 
    & Input Size & \# Parameters & Train & Test \\
    \hline
    \hline
    BAS & $4 \times 4$ & $10$ & $93.7 \pm 1.6 \%$ & $93.0 \pm 1.2 \%$ \\
    \hline 
    Custom BAS & $4 \times 4$ & $10$ & $91.3 \pm 2.6$ \% & $92.7 \pm 2.1$ \%\\
    \hline
    MNIST & $8 \times 8$ & $30$ & $95.1 \pm2.9 \%$ & $93.1 \pm 3.6 \%$ \\
    \hline
\end{tabular}
\caption{Classification Accuracy for the PQCNN architecture, for different datasets and architectures.}
\label{table:Simulation_Classification}
\end{table}
The number of pooling layers, which are the most challenging part from the hardware point of view, depends on the design choice of the architecture. It usually increases with the input size as pooling layers reduce the size of the input image while convolution layers extract the important features. However, as each pooling layer reduces by half the size of the input, the number of pooling layers will increase logarithmically with the problem size. Notice that one could easily adapt this layer to achieve a reduction of higher or lower order by simply measuring more or fewer modes.

In \autoref{table:Simulation_Classification}, simulation results for datasets of different sizes are presented.
First, the experimental equivalent model is trained for the Custom BAS dataset described in \autoref{app:Experimental_Details_PQCNN}, i.e., the samples that can be encoded with the adopted experimental setups. Then we compare for the Pennylane Bars and Stripes (BAS) dataset with $4 \times 4$ images. 
Finally, the architectures for the MNIST dataset \cite{lecun2010mnist} made of $8 \times 8$ images are compared, by considering a $m=16$ mode-circuit with $k=2$ initial photons, with a single convolutional, a single pooling layer requiring $2k=4$ photons in total, and no additional modes ($\alpha=0$) used in the dense layer. The results for those simulation are close to the results for fault-tolerant quantum architectures of QCNN \cite{li_quantum_2020, kerenidis_quantum_2020}.

All the simulations have been performed using the offered open source library that is tailor-made for QML photonic algorithms by performing the computation in the most suitable subspaces. This Pytorch \cite{paszke_pytorch_2019} based toolkit could be of independent use for photonic simulation and can be found in the \href{https://github.com/ptitbroussou/Photonic_Subspace_QML_Toolkit}{Photonic Subspace QML toolkit}.

%% file: chapters/07_VQC_as_Fourier_Models/VQC_as_Fourier_Models.tex
\let\textcircled=\pgftextcircled
\chapter[VQCs as Fourier Models]{Variational Quantum Circuit as Fourier Models}
\label{chap:VQC_as_Fourier_Models}
\begin{textblock}{5.3}(0,-4)
	%\textit{`Un scientifique dans son laboratoire est un enfant placé devant des phénomènes naturels qui l'impressionnent comme des contes de fées.'\\}

%\hspace{0.5cm}--- Marie Skłodowska-Curie.
\end{textblock}

\initial{M}\textit{any studies have been conducted to understand the potential and limitations of  Quantum models. Multiple works, including the one presented in \autoref{chap:HW_Preserving_Methods}, focus on the trainability of such models and highlight the exponential concentration and vanishing gradient phenomena \cite{mcclean_barren_2018, arrasmith_equivalence_2022,larocca_diagnosing_2022,zoufal_generative_2021}. On the other hand, another fundamental question concerns the expressivity of these models, namely which hypothesis class the quantum model is exploring. Previously, the notion of controllability was discussed, which is an important figure of merit for expressivity. In this chapter, the expressivity of the quantum model is discussed through the lens of its Fourier model, an important notion that can also be used to create surrogate models as explained in \autoref{chap:Fourier_Surrogates}. }

\begin{figure}[h!]
    \centering
    \includegraphics[width=1\linewidth]{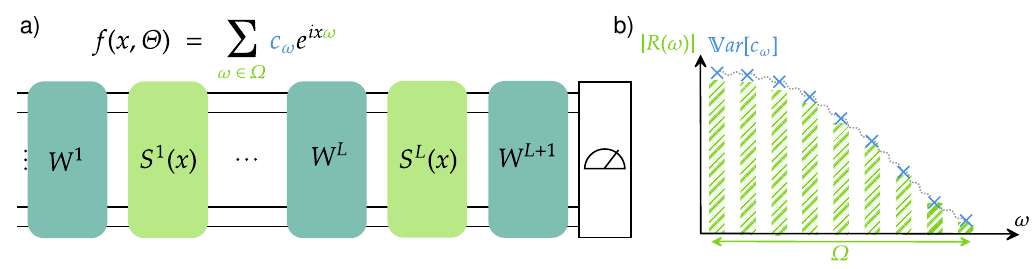}
    \caption{Parameterized quantum models, a) can be seen as Fourier series with frequencies $\omega \in \Omega$. b) illustrates the relation between the frequencies redundancies $|R(\omega)|$, i.e., the number of times a frequency appears in the spectrum, and their Fourier coefficients variance $\Var[c_\omega]$. This connection constrains the expressivity of quantum Fourier models.}
    \label{fig:intro_VQC_Fourier}
\end{figure}
\newpage

From the seminal paper \cite{schuld_effect_2021}, it is known that, when considering an encoding scheme where the classical input is encoded as the time evolution of some Hamiltonian, the quantum model generated by the Variational Quantum Circuit (VQC) can be described as a Fourier series in the classical input. The spectrum is determined by the encoding layers while the Fourier coefficients are mainly controlled by the trainable layers. First, general notions and the framework are recalled in \autoref{sec:Framework}. The main results of this chapter are a new connection highlighted between the Fourier coefficients and the encoding gates as illustrated in \autoref{fig:intro_VQC_Fourier}, and the \emph{vanishing expressivity} phenomenon. This connection is proven in \autoref{sec:Main_Results}, where the connection between vanishing Fourier coefficients and vanishing models is also discussed. Finally, a discussion on the link between Fourier coefficients and controllability is provided, based on simulations and the Fourier Norm Bound \autoref{thm:Mass_Conservation}.

    \section{Framework}\label{sec:Framework}

In this Section, the framework considered throughout this Chapter is presented. First, the considered circuit structure is described, and the method for defining the Fourier representation of the associated quantum model is recalled from \autoref{subsec:Fourier_Model}. In particular, the notion of frequency redundancy is introduced, and how it can be tuned through the choice of the encoding strategy is discussed. Finally, some figure of merits for characterizing the expressivity of Quantum models are provided in \autoref{subsec:Mathematical_Framework}.

\subsection{Quantum Fourier Model}\label{subsec:Fourier_Model}

Considering a standard supervised learning task, where a parameterized function $f$ , called a \textit{model}, must be optimized to match targets in a finite dataset. \textit{Quantum models} on \textit{n} qubits are defined as the family of parameterized functions $f : \mathcal{X} \times \Theta \rightarrow \mathbb{R}$ obtained by measuring the expectation value of some Hermitian observable $O$, such that:
 \begin{equation}\label{eq:quantum_Model}
    f(x,\theta) = \bra{0} U(x,\theta)^\dagger O U(x,\theta) \ket{0} \; , 
 \end{equation}
  where $U(x,\theta)$ is a $2^n$-dimensional unitary , $\theta \in \Theta$ is the vector of trainable parameters and $x= (x_1,\dots,x_D) \in \mathcal{X} \subset \mathbb{R}^D$ is the classical data vector.

A circuit unitary composed of alternating \textit{encoding} and \textit{trainable} layers is considered, as depicted in \autoref{fig:intro_VQC_Fourier}. It has the form: 
\begin{equation}\label{eq:circuit_ansatz_Fourier}
    U(x,\theta) = W^{L+1}(\theta)\left[\prod_{l=1}^L S^l(x)W^l(\theta)\right] \; ,
\end{equation}
where $L$ is the total number of circuit layers (i.e. a circuit layer is made of an encoding layer and a trainable layer), the $W^l(\theta)$s are formed by trainable gates depending on the parameter vector $\theta$, which is optimized during training whereas the $S^l(x)$s only depend on input data values.

In the remainder of this chapter, the \textit{Hamiltonian encoding} strategy is adopted, where the classical input components are encoded as the time evolution of some Hamiltonians $S^l(x)= \prod_{k=1}^D e^{-ix_kH_l^{(k)}}$. From the seminal work \cite{schuld_effect_2021}, it is known that if the Hamiltonian encoding strategy is considered, the quantum model generated by the circuit described in \autoref{eq:circuit_ansatz_Fourier} can be written as a Fourier series. Its spectrum $\Omega$ depends on the eigenvalues of the encoding Hamiltonians, and the associated Fourier coefficients depend mainly on the parameterized unitaries. Under these assumptions, the obtained model is called a \textit{Quantum Fourier Model} (QFM), which is defined as follows:

\begin{equation}\label{eq:quantum_Fourier_Model}
    f(x,\theta) = \sum_{\omega \in \Omega} c_\omega(\theta) e^{i \omega^T x}   \; . 
\end{equation}

The above equation tends to imply that $c_{\omega}(\theta)$ is solely determined by the parameterized unitaries $W^l(\theta)$s. However, this Chapter demonstrates that the dependence of the Fourier coefficients on the encoding gates is more subtle. To highlight the relation between the frequencies and the encoding Hamiltonian's eigenvalues, \autoref{eq:quantum_Model} is expanded in the case of one-dimensional input vectors ($D=1$). The dimension of the Hilbert space is denoted by $d=2^n$ (with $n$ the number of qubits) and it is assumed without loss of generality \footnote{One can simply consider that $S^l(x) = P D P^{-1}$ and \emph{inject} $P$ into the expression of $W^l$ and $P^{-1}$ in $W^{l+1}$.} that $S^l(x) = \text{diag}(\lambda^l_1, \dots, \lambda^l_d)$. The explicit dependence on $\theta$ in $W^{l}(\theta)$ is dropped for simplicity:

\begin{equation}\label{eq:quantum_Fourier_Model_developped}
    f(x,\theta) = \sum_{J, J' \in \llbracket 1,d \rrbracket^L} \sum_{k, k'=1}^d W^{1 *}_{j_1', 0} \dots W^{L+1 *}_{j_L', k'} \cdot W^{1}_{j_1, 0} \dots W^{L+1}_{j_L, k} \cdot O_{k, k'} \cdot e^{-ix (\sum_{l=1}^L (\lambda^l_{j_l} -  \lambda^{l}_{j'_{l}}))}\; . 
\end{equation}

\begin{figure}[h!]
    \centering
    \includegraphics[width=1\linewidth]{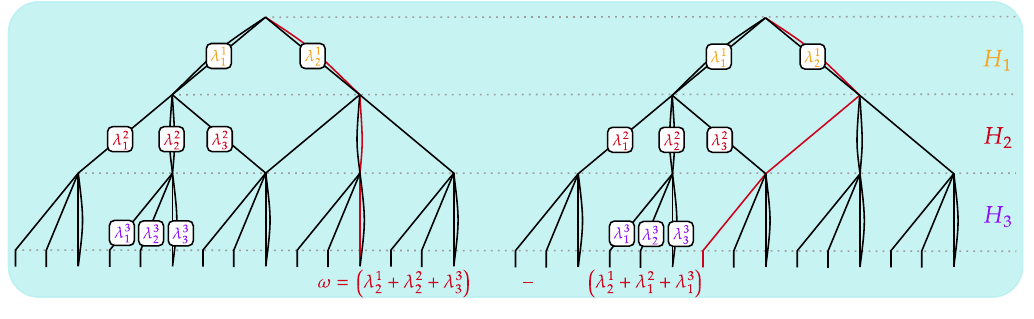}
    \caption{Quantum Spectrum Tree. The frequencies of a Quantum Fourier model are derived from the eigenvalues of the encoding Hamiltonians. Each path in the quantum spectrum tree represents a different choice of eigenvalues $\lambda_{j_\ell}^\ell$ from Hamiltonians $H_\ell$s. Some edges are duplicated, expressing that the some eigenvalue are degenerate for some Hamiltonians (not at scale). Each frequency $\omega$ in the model comes from the  difference of two paths in the tree, as shown in the example in red.
    For a Quantum model acting on $n=2$ qubits with $L=3$ circuit layers and encoding hamiltonians $H_1, H_2 \text{ and } H_3$, we give in red a pair of paths in the tree $J=(4,3,3), J'=(4,1,1)$ generating the frequency $\omega = (\lambda_4^1 + \lambda_3^2+ \lambda_3^3)-(\lambda_4^1 + \lambda_1^2+ \lambda_1^3)$.
    }    \label{fig:Quantum_Spectrum_Trees}
\end{figure}

With $J=(j_1, \dots, j_L)$ a multi-index where each component $j_l$ refers to the choice of the $j^{th}$ eigenvalue of the Hamiltonian $H_l$ ($J$ maps to a path in the tree from \autoref{fig:Quantum_Spectrum_Trees}).

From \autoref{eq:quantum_Fourier_Model_developped}, one can see that the spectrum $\Omega$ can be constructed from the eigenvalues of the encoding Hamiltonians in each layer as follows:
\begin{equation}
\label{Eq:Spectrum_def}
    \Omega = \left\{ \sum_{j_l \in J} \lambda^l_{j_l} - \sum_{j'_l \in J'} \lambda^{l}_{j'_{l}} \middle| (J,J')  \in \llbracket 1,d \rrbracket^L  \right\} \; . 
\end{equation}

The spectrum $\Omega$ contains redundant frequencies by construction but in the remainder of this work, it is considered that $\Omega$ denotes the set of distinct frequencies. 

As shown in \autoref{fig:Quantum_Spectrum_Trees}, the choice of two \emph{paths} $(J,J')$ in the quantum spectrum tree leads to the generation of a frequency $\omega$ by computing the difference of the sum of eigenvalues over each path. One can easily notice that several pairs of paths could lead to the generation of the same frequency. This can happen if an eigenvalue is degenerate, or if several paths of the tree end at the same leaf value (sum of eigenvalues over a path), or eventually if several pairs have the same difference value. The number of these paths evolves with the choice of the different encoding Hamiltonians,  the degeneracy of their eigenvalues, and the number $L$ of circuit layers. 

By grouping the paths $(J,J')$ in \autoref{eq:quantum_Fourier_Model_developped}, that leads to a certain frequency, the \emph{Frequency Generator} $R(\omega)$ is formally defined as the set of all paths leading to the generation of the frequency $\omega$. The cardinality of this set is defined as the frequency's \emph{redundancy} $|R(\omega)|$.  
As demonstrated later, the redundancy of a frequency plays a crucial role in characterizing the expressivity of QFMs.

\begin{definition}[Frequency Generator]\label{def:Redundancy}
    Consider an $L$-layer Quantum Fourier model as described in \autoref{eq:quantum_Model} and \autoref{Eq:Spectrum_def}. For a given frequency $\omega$, its generator $R(\omega)$ is defined as the set of   eigenvalue indices leading to the generation of $\omega$.
    \begin{equation}
        R(\omega) = \left\{ (J,J') \in \llbracket 1,d \rrbracket^L \times \llbracket 1,d \rrbracket^L \middle| \sum_{j_l \in J} \lambda^l_{j_l} - \sum_{j'_l \in J'} \lambda^{l}_{j'_{l}} = \omega  \right\} \; . 
    \end{equation}
    The size of its Generator: $|R(\omega)|$ is called the \textbf{redundancy} of a frequency $\omega$.
\end{definition}

Since $\sum_{\omega \in \Omega} |R(\omega)| = 2^{2n \times L} = d^{2L}$ by construction, the normalized redundancies $\{\frac{|R(\omega)|}{d^{2L}}\}_{\omega \in \Omega}$ define a natural weighted probability distribution over the spectrum $\Omega$.
Therefore, by considering different encoding Hamiltonians, one can obtain different probability distributions over the spectrum that will impact the behavior of the associated Quantum model.

For example, the standard case of Pauli encoding \cite{schuld_effect_2021} is considered, where single-qubit rotation gates are used to encode the classical input $x \in \mathbb{R}$ as the rotation angle. In this case, the encoding Hamiltonian in each layer is a Pauli string. If the Pauli strings do not contain the identity, then the obtained spectrum is simply $\Omega = \llbracket -nL,nL \rrbracket$. Moreover, one can easily show that the spectrum distribution defined by the redundancies follows a standard Gaussian distribution. Hence, this encoding strategy gives rise to a spectrum of linear size (linear in $n$ and $L$) and concentrates the redundancies in the lower values.  

In contrast, the exponential encoding strategy introduced in \cite{shin_exponential_2023}, which uses scaled Pauli rotations for encoding, leads to a spectrum of exponential size consisting of consecutive integer frequencies. Specifically, the obtained spectrum is $\Omega = \left\llbracket -\frac{3^{nL} -1}{2}, \frac{3^{nL} -1}{2} \right\rrbracket$ and some frequencies (not necessarily high frequencies) have redundancies that do not scale exponentially in $n$ and $L$. However, to obtain a fully non-degenerate spectrum (except for the null frequency), a single circuit layer made of a non-local encoding Hamiltonian must be used, as mentioned in \cite{shin_exponential_2023}.
This is the case for the \emph{Golomb} encoding introduced in \cite{peters_generalization_2023}, where the size of the spectrum is exponentially large ($|\Omega| = 2 \binom{d}{2} + 1$) and all non-zero coefficients have a redundancy of one.

\begin{figure}[h!]
    \centering
    \includegraphics[width=1\linewidth]{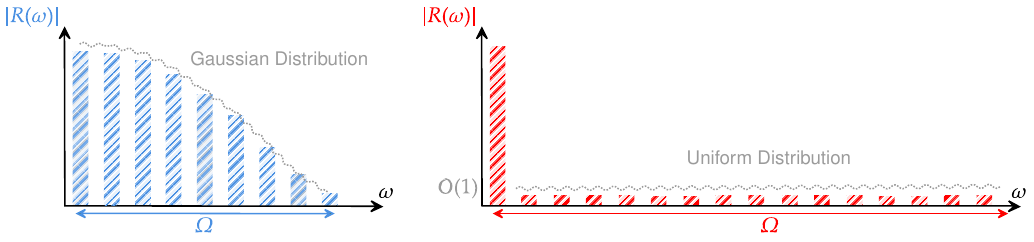}
    \caption{Comparison of two Hamiltonian encoding strategies leading to very different spectrums (x-axis) and distributions (normalized height of the bars). On the left we present an illustration a highly degenerate spectrum (e.g. Pauli encoding spectrum distribution), and on the right an illustration of a weakly degenerate spectrum (e.g.  Golomb encoding spectrum distribution).}
    \label{fig:Pauli_Exp_encoding_distribution}
\end{figure}

As illustrated in \autoref{fig:Pauli_Exp_encoding_distribution}, one could choose a particular set of Hamiltonians to design a quantum model with a specific spectrum  distribution. It is shown later in this Chapter that this choice impacts not only the spectrum of the quantum Fourier model, but also the Fourier coefficients concentration. In addition, previous work \cite{landman_classically_2022} showed the possibility of classically approximating such VQCs for Machine Learning tasks. While having a spectrum of exponential size may be a solution to avoid this classical approximation, it is demonstrated in \autoref{sec:Main_Results} that models with a large spectrum tend to have more constrained Fourier coefficients, hence limiting their expressivity and making their classical approximation potentially more efficient.

\subsection{Expressivity Measures of Quantum Models}\label{subsec:Mathematical_Framework}

In \autoref{chap:Preliminaries} is introduced different metrics of expressivity that are relevant for VQC and this Fourier model study. In the rest of this Chapter, the 2-design hypothesis as defined in \autoref{def:2-design} and \autoref{def:monomial_norm} is used, and its connection between these hypotheses and the concentration phenomena introduced in \autoref{def:Model_Concentration} is discussed.

Along with the expressivity characterization of the parameterized part of a Quantum model by its $\varepsilon$-distance to a 2-design, the expressivity of a QFM should also be examined through its Fourier representation, i.e., the signature of the specific Hamiltonian encoding strategy. In a recent work \cite{xiong_fundamental_2025}, authors proposed defining the \textit{Fourier expressivity} as the smallest set of functions such that the quantum model defined in \autoref{eq:quantum_Model} could be expressed as a linear combination of those functions. According to this definition and the Fourier decomposition of the quantum model (see \autoref{eq:quantum_Fourier_Model}), the Fourier expressivity is bounded by the size of the spectrum, $|\Omega|$.

Similarly, the expressivity of a quantum model is characterized through a Fourier lens in this chapter. Indeed, it is demonstrated that individual Fourier coefficients may suffer from exponential concentration depending on the spectrum distribution. Therefore, a QFM is said to suffer from \textit{vanishing expressivity} if some or all of its Fourier coefficients are exponentially concentrated around their mean.

\begin{definition}[Vanishing Expressivity]\label{def:Vanishing_Expressivity}
    Consider a quantum Fourier model  such as defined in Eq. \eqref{eq:quantum_Fourier_Model} with spectrum $\Omega$. The Fourier model is said to suffer from vanishing expressivity when some Fourier coefficient have an exponentially vanishing variance in the number of qubits $n$:
    \begin{equation}
        \exists \; \Omega_{\text{vanish}} \subset \Omega \quad |  \quad \forall \omega \in \Omega_{\text{vanish}}\ \Var_{\theta}[c_{\omega}(\theta)] = \mathcal{O}\left(\frac{1}{b^n}\right) \; ,
    \end{equation}
    for some constant $b > 1$.
\end{definition}

\section{Main Results}\label{sec:Main_Results}

In this Section,  the main theorems and corollaries on expressivity constraints in quantum Fourier models are presented.
Specifically, the concentration of Fourier coefficients is studied by computing their variance under different assumptions about the distribution of the trainable unitaries. It is then shown that the variance is always constrained by the frequency redundancy and that some Fourier coefficients may exhibit an exponential concentration phenomenon, leading to \textit{vanishing expressivity}. 

To do this, the global 2-design hypothesis is considered for the trainable unitaries for a single layer model in \autoref{thm:single_layer_gloabl_2design} and for a reuploading model in \autoref{Cor:2design_vanishing}. 
Second, the global 2-design assumption is relaxed, and an upper bound on the variance of the Fourier coefficients is given under the $\varepsilon$-approximate 2-design hypothesis in \autoref{thm:bound_approx_2design_informal}. Finally, a brick-wise circuit architecture with local 2-design blocks is considered, and an upper bound on the variance is derived. This circuit architecture falls within the $\varepsilon$-approximate 2-design assumption but has more structure, which allows the locality of the observable to be taken into account.

In the remainder of this Chapter, the focus is on one-dimensional input vectors $(D=1)$. However, the results can be easily extended to the high-dimensional setting under the assumption that the encoding unitaries within a single layer commute.

    \subsection{Trainable Layers as Global 2-design}\label{subsec:Global_2-design}

As described in \autoref{subsec:Mathematical_Framework}, the expressivity of trainable unitaries is often characterized by how uniformly they explore the unitary group; a parametrized unitary is said to be maximally expressive if its distribution approximates the Haar measure. However, it has been shown in \cite{arrasmith_equivalence_2022,mcclean_barren_2018} that the quantum model and its gradient exhibit an exponential concentration phenomena under the 2-design assumption, resulting in an unexpressive model in practice.

The implications of considering maximally expressive trainable unitaries (i.e., each of the trainable layers forms an exact 2-design) on the variance of the Fourier coefficients are explored. First, an exact expression for the variance of the Fourier coefficients of a QFM with a single circuit layer ($L=1$) is presented in \autoref{thm:single_layer_gloabl_2design}. This result is then extended to a reuploading model with $L \geq 1$ in \autoref{Cor:2design_vanishing}.

\begin{restatable}[Fourier coefficients variance with 2-design trainable unitaries, Informal]{thm}{FourierCoefsVar2design}
\label{thm:single_layer_gloabl_2design}
    Consider a quantum model of the form in \autoref{eq:quantum_Model} and a parametrized circuit of the form in \autoref{eq:circuit_ansatz_Fourier} with $L=1$ layers and fixed encoding Hamiltonians resulting in a spectrum $\Omega$. We assume that each of the  trainable layers $W^l(\theta)\;,l \in \{1,2\}$ form independently a 2-design. The expectation and variance of each Fourier coefficient $c_{\omega}(\theta)$ for the frequencies $ \omega \in \Omega$ appearing in the model Fourier decomposition in \autoref{eq:quantum_Fourier_Model} are given by
\begin{equation}\label{eq:coeff_variance_2design_single}
    \begin{aligned}
        \E_{\theta}[c_{\omega}(\theta)]& = \quad\frac{Tr(O)}{d}\delta_{\omega}^0\;,\\
        \Var_{\theta}[c_{\omega}(\theta)] &\in \Theta\left(	\alpha \frac{|\widetilde{R}(\omega)|}{d} - \frac{\alpha}{d^2} \delta_{\omega}^0\right)\;.
    \end{aligned}
\end{equation}
Here, $d=2^n$, $\delta_{i}^j$ is Kronecker function apply on $i$ and $j$, the normalized frequency redundancy $|\widetilde{R}(\omega)| := |R(\omega)|/d^2$ is introduced, and the constant $\alpha:=(d||O||_2^2-Tr(O)^2)/d^2$, which depends on the observable $O$, is defined. 
\end{restatable}

\begin{proof}
    The proof of \autoref{thm:single_layer_gloabl_2design} and its extension to a reuploading model \autoref{Cor:2design_vanishing} is based on Weingarten calculus. For example, the expression of the Fourier coefficient is given by:
    \begin{align}
    c_{\omega} &= \sum_{J,J' \in R(\omega)} \sum_{k,k'} W^{(1)*}_{j'_1 0}  W^{(2)*}_{j'_2 j'_1} \dots W^{(L+1)*}_{k'j'_L} O_{k'k} W^{(L+1)}_{kj_L}\dots W^{(2)}_{j_2 j_1} W^{(1)}_{j_1 0}\; .
    \end{align}
    
    To prove the expression of the expectation value of a Fourier coefficient, it is sufficient to establish this result under the 1-design hypothesis. To do so, we apply the Weingarten formula for the first moment \cite{mele_introduction_2024} and obtain
    \begin{equation}
        \begin{split}
           \E_{W^{(1)},\dots,W^{(L+1)} \sim U(N)} \left[c_{\omega}\right] & =  \smashoperator{\sum_{\substack{k,k' \\ J,J' \in R(\omega)}}}  \frac{\delta_{j_1}^{j'_1} \delta_{j_2}^{j'_2} \ldots \delta_{j_L}^{j'_L} \delta_{k}^{k'} O_{k'k}}{d^{L+1}}  = \smashoperator{\sum_{\substack{k \\ J,J' \in R(\omega)}}} \frac{\delta_{j_1}^{j'_1} \delta_{j_2}^{j'_2} \dots \delta_{j_L}^{j'_L}  O_{kk}}{d^{L+1}}  \\ 
           &= \sum_{J,J' \in R(\omega)} \delta_J^{J'} \frac{\Tr(O)}{d^{L+1}} = \frac{\Tr(O)}{d} \delta_{\omega}^0 \; .
        \end{split}
    \end{equation}

    The variance of a Fourier coefficient for a reuploading VQC (i.e $L>1$) is obtained recursively starting from the variance of a single-layered circuit, using the recursive relation between the partial redundancies, and using Weingarten formula for the first and second moment. The entire proof can be found in \cite{mhiri_constrained_2024}.
\end{proof}

\begin{figure}
    \centering
    \includegraphics[width=1.0\linewidth]{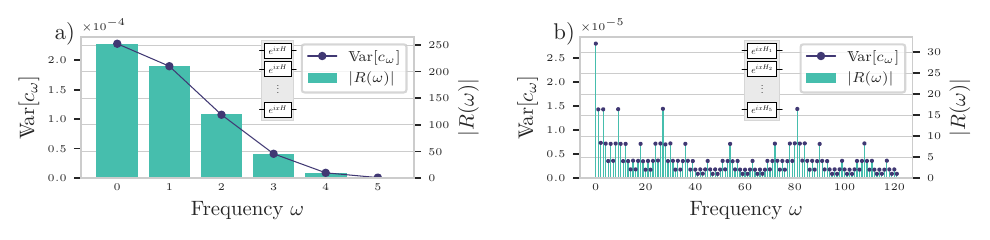}
    \caption{For $n=5$ qubits, one circuit layer $L=1$, five repetitions of the strongly entangling ansatz per trainable layer and global observable; relation between the variance of each Fourier coefficient $\Var\left[c_\omega\right]$ and its redundancy $|R(\omega)|$. Values shown for two different encoding strategies, a) Pauli encoding and b) exponential encoding.}
    \label{fig:variance_redundancies}
\end{figure}
    
\autoref{thm:single_layer_gloabl_2design} establishes that, under the 2-design assumption for the model's trainable unitaries, the variance of a Fourier coefficient depends linearly on its (normalized) frequency redundancy, up to a prefactor. The normalized frequencies sum up to one. Thus, \autoref{thm:single_layer_gloabl_2design} implies that while frequencies with high redundancies exhibit a relatively large variance, those with low redundancies are significantly more constrained.
In other words, the distribution of the Fourier coefficients is dictated by the redundancies and hence by the encoding Hamiltonians. In \autoref{fig:variance_redundancies}, this dependence is illustrated through a numerical study of two models with different encoding Hamiltonians, corresponding to spiked and flat frequency distributions (i.e., defined by normalized redundancies over the model's spectrum). Moreover, the result of \autoref{thm:single_layer_gloabl_2design} is generalized to the setting of reuploading models with $L \geq 1$ alternating layers in \autoref{Cor:2design_vanishing}. Similarly, it is proven that the variance of a Fourier coefficient is linear in its frequency redundancy.

\begin{restatable}[Fourier coefficients variance decay with 2-design trainable unitaries, Informal]{cor}{CorFourierCoefDecay2design}
\label{Cor:2design_vanishing}
 Consider a quantum model of the form in \autoref{eq:quantum_Model} and a parametrized circuit of the form in \autoref{eq:circuit_ansatz_Fourier} with $L\geq 1$ layers and fixed encoding Hamiltonians resulting in a spectrum $\Omega$. We assume that each of the trainable layers $W^l(\theta)$ form independently a 2-design. The variance of each Fourier coefficient $c_{\omega}(\theta) $ for the frequencies $ \omega \in \Omega$ appearing in the model Fourier decomposition in \autoref{eq:quantum_Fourier_Model} is upper bounded by
\begin{align}
   \Var_{\theta}[c_{\omega}(\theta)] &\in \mathcal{O} \left(\alpha \frac{|\widetilde{R}(\omega)|}{d}\right)\;,
\end{align}
where we recall that $d=2^n$ and $\alpha$ is a constant given by $\alpha:=(d||O||_2^2-Tr(O)^2)/d^2$.
\end{restatable}

From Corollary \ref{Cor:2design_vanishing}, it can be seen that, under reasonable assumptions on the observable norm, the variance of all Fourier coefficients decays exponentially in the number of qubits. Specifically, the prefactor $\alpha$ can be bounded by a constant for any observable satisfying $\norm{O}_2^2 \in \mathcal{O}(d)$. Additionally, the normalized redundancy $|\widetilde{R}(\omega)|$ is, by definition, bounded by one. This implies that, irrespective of the frequency redundancy and thus of the encoding strategy, all coefficients concentrate exponentially toward their mean value.
This result can be viewed as an exponential concentration statement for each Fourier coefficient in a reuploading model, aligning with results in \cite{arrasmith_equivalence_2022} on the exponential concentration of the model under the 2-design assumption.

While it is useful to show the connection between the Fourier coefficients and the spectrum redundancies, the consideration of a global 2-design is a strong assumption that leads to the model's exponential concentration. In practice, it is improbable that trainable layers forming a 2-design will be used for learning purposes. Thus, this hypothesis is relaxed in the following by first considering trainable layers that form an approximate 2-design and then considering those made of local 2-design blocks, for models with a single layer $(L=1)$.

    \subsection{Trainable Layers as Global \texorpdfstring{$\varepsilon$}{epsilon}-approximate 2-design}\label{subsec:approx_2-design}

The broader setting is now considered, where the trainable unitaries $W^l(\theta)$ each form an $\varepsilon$-approximate 2-design.   
By moving away from highly expressive trainable unitaries, a question arises regarding whether it is feasible to break free from the constraining redundancy dependence of the Fourier coefficients variance established in \autoref{thm:single_layer_gloabl_2design}, or if such dependency is an inductive bias of the quantum model that still holds even when the trainable unitaries are not maximally expressive (i.e., do not form exact 2-designs).
  
To do this, results from \cite{holmes_connecting_2022} on model concentration for approximate 2-design unitaries are built upon, and encoding-dependent concentration for single Fourier components is explored, giving a finer interpretation of the model's expressivity through the Fourier lens.

In the following theorem, an upper bound on the variance of the Fourier coefficients for a single-layer circuit formed by arbitrary trainable layers is provided.

\begin{restatable}[Fourier coefficients variance decay with approximate 2-design trainable unitaries, Informal]{thm}{FourierCoefdecayApproximate2design}
\label{thm:bound_approx_2design_informal}
       Consider a quantum model of the form in \autoref{eq:quantum_Model} and a parametrized circuit of the form in \autoref{eq:circuit_ansatz_Fourier} with $L=1$ layers and fixed encoding Hamiltonians resulting in a spectrum $\Omega$. We assume that each of the  trainable layer $W^l(\theta)\;,l \in \{1,2\}$ form independently an $\varepsilon$-approximate 2-design. The  variance of each Fourier coefficient $c_{\omega}(\theta)$ for the frequencies $ \omega \in \Omega$ appearing in the model Fourier decomposition in \autoref{eq:quantum_Fourier_Model} is upper bounded as
       \begin{equation}
           \Var[c_\omega] \in \mathcal{O}(Q_\varepsilon(|\widetilde{R}(\omega)|))\;,
       \end{equation}
       where $Q_\varepsilon$ is a polynomial of degree at most 2 in the normalized frequency redundancy  $|\widetilde{R}(\omega)|$ defined for different $\varepsilon$ measures as  

\begin{align}
     Q_{\varepsilon_\diamond}&=  \|O\|_2^2|\widetilde{R}(\omega)|\varepsilon_{\diamond}+\|O\|_1^2 \varepsilon_{\diamond}^2  \;,\label{eq:diamond_norm_bound}\\
      Q_{\varepsilon_\infty}&= \frac{ \|O\|_2^2}{d}\sqrt{|\widetilde{R}(\omega)|}\varepsilon_{\infty} +d^2\|O\|_2^2 |\widetilde{R}(\omega)|\varepsilon_{\infty}^2\;,\label{eq:infty_norm_bound}\\
    Q_{\varepsilon_M}&= \|O\|_2^2 |\widetilde{R}(\omega)| \varepsilon_M+ d^2  \|O\|_2^2|\widetilde{R}(\omega)|^2\varepsilon_M^2\;. \label{eq:monomial_norm_bound}
\end{align}
  
\noindent Here, we use the shorthand $\varepsilon_{\diamond}:= \|\mathcal{A}^{(2)}\|_{\diamond}$ for the diamond norm, $\varepsilon_{\infty} :=\norm{\mathcal{A}^{(2)}}_{\infty}$ for the spectral norm, and $\varepsilon_{M}:= d^2 max_{i,j} |\mathcal{A}^{(2)}|_{i,j}$. We also recall that  $\mathcal{A}^{(2)}$ is a superoperator  defined in \autoref{Eq:Superoperator_A_2-design}.
\end{restatable}

\begin{proof}
    Let us first recall the Fourier coefficient expression for a single encoding layer ($L=1$):
    \begin{equation}\label{eq:Fourier_coeff_L1}
    \begin{aligned}
        c_{\omega} &= \sum_{j_1,j'_1 \in R(\omega)} \sum_{k,k'} W^{(1)*}_{j'_1 0}  W^{(2)*}_{j'_2 j'_1}  O_{k'k}  W^{(2)}_{j_2 j_1} W^{(1)}_{j_1 0}\\
        &=  \sum_{j_1,j'_1 \in R(\omega)} \left( W^{(1)} \ket{0}\bra{0}  W^{(1) \dagger}\right)_{j_1,j'_1} \left( W^{(2) \dagger} O W^{(2)}\right)_{j'_1,j_1}\\
        &= \sum_{j_1,j'_1 \in R(\omega)} Tr \left[W^{(1)} \ket{0}\bra{0}  W^{(1) \dagger} \ket{j'_1}\bra{j_1}\right] Tr \left[ W^{(2) \dagger} O W^{(2)} \ket{j_1}\bra{j'_1}\right]\; .
    \end{aligned}
    \end{equation}
    
    Then, the expectation of the modulus squared of the coefficient $c_{\omega}$ is given by:
    
    \begin{equation}\label{eq:exp_squared_fourier_L1}
    \begin{aligned}
        \mathbb{E}\left[|c_\omega|^2\right]  
        &= \sumrw & \Tr\left[\mathbb{E}_{W^{(1)}}\left[W^{(1)\otimes 2} \ket{00}\bra{00} W^{(1)\dagger \otimes 2} \ket{j'_1 i_1}\bra{j_1 i'_1}\right] \right] \times \\
        && \Tr\left[\mathbb{E}_{W^{(2)}}\left[ W^{(2)\dagger\otimes2}O^{\otimes 2} W^{(2)\otimes2}\ket{j_1 i'_1}\bra{j'_1 i_1}\right]\right]\; .
    \end{aligned}
    \end{equation}
    where we use in the second equality the property $Tr[A] \times Tr[B]=Tr[A \otimes B]$.
    Expectation terms in \autoref{eq:exp_squared_fourier_L1} can be written using the superoperator $\mathcal{A}_{\mathbb{W}}^{(2)}(\cdot):=  \int_{\text{Haar}} d \mu(W) W^{\otimes 2}(\cdot)\left(W^{\dagger}\right)^{\otimes 2}  -\int_{\mathbb{W}} d W W^{\otimes 2}(\cdot)\left(W^{\dagger}\right)^{\otimes 2}$, and the results derive from Weingarten calculus.
\end{proof}

\autoref{thm:bound_approx_2design_informal} shows that
the variance of a Fourier coefficient in the approximate 2-design setting is constrained by the combined action of the normalized frequency redundancy $|\widetilde{R}(\omega)|$ and the $\varepsilon$-distance of the trainable unitaries to a 2-design. 
Specifically, for a fixed choice of the trainable unitaries distribution and thus for a fixed $\varepsilon$ value, the degree to which each Fourier coefficient concentrates around its mean is constrained by its corresponding normalized frequency redundancy. Therefore, it is proven that the vanishing expressivity phenomenon, whereby some Fourier coefficients exhibit exponentially decaying variance, may still hold beyond the 2-design assumption.
It is noted that the bounds in \autoref{eq:diamond_norm_bound} and \autoref{eq:monomial_norm_bound} correspond to different norms used to quantify the distance from a 2-design. These $\varepsilon$-distance definitions are equivalent up to some prefactors \cite{low_pseudo-randomness_2010}. All of them are included because one bound may be tighter than the other depending on the interplay between the observable norm\footnote{We consider Schatten $p$-norms defined as $\norm{O}_p := \left(Tr\left[\left(\sqrt{O^\dagger O}\right)^p\right]\right)^{1/p}$.}, the frequency redundancy, and the $\varepsilon$-distance scalings. Precisely, while $\varepsilon_{\infty}$ saturates at 1 and $\varepsilon_{\diamond}$ at 2, the monomial-based $\varepsilon_{M}$ can take values up to $d^2$. Additionally, it is recalled that the normalized redundancies $|\widetilde{R}(\omega)|$ take values within $[1/d^2,1]$. Hence, these bounds can be used to prove the vanishing expressivity phenomenon introduced in \autoref{def:Vanishing_Expressivity} for frequencies with relatively low redundancies. Namely, a frequency with a normalized redundancy that counterbalances the observable norm will exhibit exponential decay on average over trainable unitaries forming an approximate 2-design. In \autoref{sec:Discussion}, the scalings of the upper bound in \autoref{eq:monomial_norm_bound} and its dependence on the observable and encoding strategy are further discussed.

It is noted here that the upper bounds in \autoref{thm:bound_approx_2design_informal} are generally looser for local observables compared to global ones.
This observation is not surprising, as the obtained bound, being a function of the \textit{global} $\varepsilon$ expressivity measure of the circuit, does not capture the observable-circuit interaction in finer detail. Specifically, the interaction between an $m$-local observable and the remainder of the circuit is captured by the backward light cone of the observable, i.e., the sub-circuit containing all blocks with at least one qubit causally connected to the local observable input qubits. In the next Section, the variance of the Fourier coefficients is explored by taking into account the observable locality.

        \subsection{Trainable Layers as local 2-design Blocks}\label{subsec:local_2-design}
    
\begin{figure*}[h!]
    \centering
    \includegraphics[width=1\linewidth]{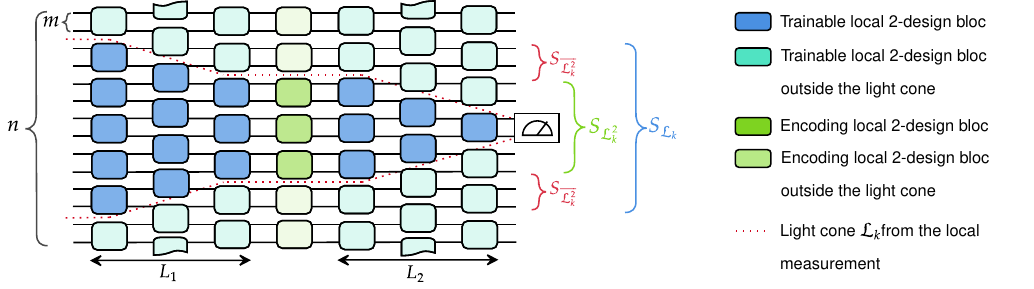}
    \caption{\textbf{Brickwise circuit architecture made of local blocks acting on neighboring qubits.} As shown, $L_1$ is the depth of the pre-encoding trainable block and $L_2$ is the depth of the post-encoding one. 
    We consider an m-local observable acting non-trivially on subsystem $s_k$ and we denote its backward light cone by the subcircuit $\mathcal{L}_k$. We denote by $\mathcal{S}_{\mathcal{L}_k}$ the subsystem on which the backward light cone acts non-trivially and $\mathcal{S}_{E_k}$ the subspace on which the encoding layer (made of the green blocks) acts non trivially inside the light cone. we also define $\mathcal{S}_{\overline{E_k}}$ as the compliment of $\mathcal{S}_{E_k}$ in $\mathcal{S}_{\mathcal{L}_k}$.}
    \label{fig:Figure_Framework_Local2design}
\end{figure*}

In this section, a brickwise circuit architecture formed by trainable local 2-design blocks and local encoding blocks, previously studied in \cite{cerezo_cost_2021}, is considered. As depicted in \autoref{fig:Figure_Framework_Local2design}, the $n$-qubit circuit is made of layers of $m$-qubit unitaries (trainable and encoding unitaries) acting on alternating groups of $m$ neighboring qubits. Each of the trainable blocks is considered to form an exact 2-design on the $m$-qubit subsystem on which it acts non-trivially.
This setting is a special case of the \textit{global} $\varepsilon$-approximate 2-design \cite{harrow_approximate_2023}. However, it yields more accurate results for local observables acting non-trivially on an $m$-qubit subsystem $s_k$ of the form $O= \hat{O}_{s_k} \otimes \mathbb{1}_{\overline{s_k}} $. Indeed, with this circuit architecture, the backward \textit{causal light cone} of such local observables is well defined, and the effective model spectrum can be seen to be reduced. In this setting, the expression for the variance of the Fourier coefficients was derived \cite{mhiri_constrained_2025}. However, since the obtained expression is quite cumbersome, an upper bound on the variance of the Fourier coefficients is presented in the following theorem for two different assumptions on the local observable.

\begin{restatable}[Fourier coefficients variance decay with brickwise local 2-design circuit]{thm}{FourierCoefLocal2Design}
\label{thm:bound_local2design}
 Consider a quantum model of the form in \autoref{eq:quantum_Model} and a parametrized circuit of the form in \autoref{eq:circuit_ansatz_Fourier} using a brickwise architecture with $L=1$ layers and observable $O= \hat{O}_{s_k} \otimes \mathbb{1}_{\overline{s_k}} $ acting non trivially on the $m$-qubit subsystem $s_k$. Assume that each trainable $m$-qubit unitary forms a local 2-design. The  variance of each Fourier coefficient $c_{\omega}(\theta)$ for the frequencies $ \omega \in \Omega$ appearing in the model Fourier decomposition in \autoref{eq:quantum_Fourier_Model} is upper bounded as

\begin{enumerate}
    \item If $||\hat{O}_{s_k}||_2^2 \leq 2^m$ , we have
\begin{equation}\label{eq:cor_bound_local_2design_Pauli}
    \Var[c_{\omega}] \leq \left(  \frac{2^{m+1}}{2^{2m}-1}\right)^{2L_2} |R_{E_k}(\omega)|^2.
\end{equation}
\item If $\hat{O}_{s_k}$ is a projector of rank $r$ , we have
\begin{equation}\label{eq:cor_bound_local_2design_proj}
    \Var[c_{\omega}] \leq \left(  \frac{2^{m+1}}{2^{2m}-1}\right)^{2L_2} \left(\frac{r}{2^m}\right)^2 |R_{E_k}(\omega)|^2.
\end{equation}
\end{enumerate}
Here $R_{E_k}(\omega)$ is the frequency generator obtained from the encoding blocks inside the observable backward light cone $\mathcal{L}_k$ (acting non trivially on $\mathcal{S}_{E_k}$) and $L_2$ is the depth of the post-encoding parameterized block.
\end{restatable}

\autoref{thm:bound_local2design} provides an upper bound on the variance of the Fourier coefficients for circuits made of local 2-design blocks. Once again, this quantity is observed to be constrained by the frequency redundancy. In addition, this result indicates that the vanishing Fourier coefficient phenomenon could depend on the circuit depth for a local observable.
Moreover, the bound in \autoref{thm:bound_local2design} implies that a frequency with relatively low redundancy (i.e., $|R_{E_k}(\omega)|= \mathcal{O}(1)$) will suffer from exponentially vanishing variance for a depth $L_2$ linear in $n$.

\section{Discussing the Quantum Fourier Model Constraints}\label{sec:Discussion}

In this chapter, a connection between the spectrum redundancies and the statistical behaviour of Fourier coefficients for arbitrary trainable unitaries, \textit{on average}, has been established. Namely, an \textit{inductive bias} of the Fourier model has been shown, whereby the variance of a Fourier coefficient is upper bounded by a polynomial in its redundancy. The concept of \textit{vanishing expressivity}, whereby the variance of some Fourier coefficients is exponentially vanishing in the number of qubits, was further introduced.

In this Section, these phenomena are further discussed, and their implications for model design guidelines are studied in \autoref{subsec:Vanishing_Coeff_Cost}.
In addition, a generic bound on the 2-norm of the Fourier coefficients vector is provided, and controllability-related constraints on the Fourier coefficients are briefly discussed in \autoref{subsec:MassConservation_Correlation_Control}.
Finally, the limitations of the framework, the assumptions considered, and thus the limitations of the obtained results are discussed in \autoref{subsec:Limitation_Framework}.

    \subsection{Vanishing Fourier Coefficients and Vanishing Model}\label{subsec:Vanishing_Coeff_Cost}

In this Section, the \textit{vanishing expressivity} phenomenon, whereby the variance of the Fourier coefficients decays exponentially in the system size, is discussed. Specifically, the scaling of the upper bounds established in \autoref{thm:bound_approx_2design_informal} with respect to the different quantities of interest is further discussed. Moreover, this analysis of the Fourier coefficients decay is related to the analysis of the full model decay. Indeed, it was previously stressed in \autoref{sec:Framework} that the vanishing expressivity phenomenon, whereby some Fourier coefficients exhibit exponentially decaying variance on average, is conceptually different from the model's exponential concentration introduced in \autoref{def:Model_Concentration}. Nevertheless, it may still be wondered whether these two phenomena are equivalent or if one implies the other.

Under the 2-design assumption for the trainable unitaries, it was shown in \autoref{thm:single_layer_gloabl_2design} that the model's exponential concentration goes hand in hand with the exponential decay of all Fourier coefficients independently of the encoding strategy, as detailed in Corollary \ref{Cor:2design_vanishing}. However, when using approximate 2-design trainable layers, the link is not trivial.
%%%%%%%%%%%%%%%%
To better understand the relation between the model and the exponential decay of Fourier coefficients beyond the 2-design assumption, an upper bound on the model's variance when using approximate 2-design trainable layers is provided in the following corollary. The ultimate goal is to identify regimes where the model's variance is not exponentially vanishing, whereas all or some of its Fourier coefficients suffer from exponential concentration.

\begin{restatable}[]{cor}{corModelApprox2designInformal}
\label{cor:bound_model_approx_2design_informal}
      Consider a quantum model $f(x,\theta)$ of the form in \autoref{eq:quantum_Model} and a parametrized circuit of the form in \autoref{eq:circuit_ansatz_Fourier} with $L=1$ layers and fixed encoding Hamiltonians. Assume that each of the trainable layers $W^l(\theta), l \in \{1,2\}$ forms independently an $\varepsilon_M$-approximate 2-design according to the monomial definition introduced in \autoref{def:monomial_norm}.
     For a fixed $x \in \mathcal{X}$, the variance of the model $f(x,\theta)$ is upper bounded as
\begin{equation}
      \Var_{\theta}[f(x,\theta)] \in \mathcal{O}\left(\norm{O}_2^2 \varepsilon_M\right)\;.
\end{equation}
\end{restatable}

\begin{proof}
Recall that $f(x,\theta)$ for a fixed data point $x \in \mathbb{R}$ is a real-valued function. Hence, $\forall x \in \mathbb{R}$, its variance is given by $ \Var_{\theta}[f(x,\theta)]= \E_{\theta}[f^2(x,\theta)]-\E_{\theta}[f(x,\theta)]^2$.

Also recall that the model $f$ is given by
\begin{equation}
    f(x,\theta) = Tr[W^{(1) \dagger} S^\dagger(x) W^{(2) \dagger} O W^{(2)} S(x) W^{(1)} \ketbra{0}{0}]\; ,
\end{equation}
where the dependence on trainable parameters $\theta$ is hidden in the trainable unitaries $W^{(1)}$ and $W^{(2)}$.

Hence, the model's second moment with respect to the distributions over $W^{(1)}$ and $W^{(2)}$ can be expressed as
\begin{align}
    \E_{\theta}[f^2(x,\theta)] &:=  \E_{W^{(1)}\sim \mathbb{W},W^{(2)} \sim \mathbb{W}}[f^2(x)]\\
    &= \E_{W^{(1)}\sim \mathbb{W},W^{(2)} \sim \mathbb{W}}[f^2(x)] - \E_{W^{(1)}\sim \text{Haar},W^{(2)} \sim \mathbb{W}}[f^2(x)] + \E_{W^{(1)}\sim \text{Haar},W^{(2)} \sim \mathbb{W}}[f^2(x)]\\
    &= \E_{W^{(2)} \sim \mathbb{W}} \left[ \E_{W^{(1)}\sim  \mathbb{W}}[f^2(x)] - \E_{W^{(1)}\sim  \text{Haar}}[f^2(x)]\right] + \E_{W^{(1)}\sim \text{Haar},W^{(2)} \sim \mathbb{W}}[f^2(x)]\; .
\end{align}

Using the invariance property of the Haar measure, it can be shown that
 \begin{equation}
     (\E_{W^{(1)}\sim \text{Haar},W^{(2)} \sim \text{Haar}}[f(x)])^2 =  (\E_{W^{(1)} \sim \text{Haar}}[Tr[ O W^{(1)}  \ketbra{0}{0} W^{(1) \dagger}]])^2 \; .
 \end{equation}
Similarly, 
 \begin{equation}
     \E_{W^{(1)}\sim \text{Haar},W^{(2)} \sim \mathbb{W}}[f^2(x)] = \E_{W^{(1)} \sim \text{Haar}}[Tr[ O W^{(1)}  \ketbra{0}{0} W^{(1) \dagger}]^2] \; .
 \end{equation}

 %%%
By developing the model's variance expression, it comes:
\begin{align}
    \Var[f(x)] = &\E_{W^{(2)} \sim \mathbb{W}} \left[\Tr\left[ S(x)^{\dagger\otimes 2}W^{(2)\dagger \otimes 2} O^{\otimes 2}W^{(2)\otimes 2} S(x)^{\otimes 2} \mathcal{A}_{\mathbb{W}}(\ketbra{00}{00})   \right]\right] \quad \\
    &+ \Var_{W^{(1)} \sim \text{Haar}}[Tr[ O W^{(1)}  \ketbra{0}{0} W^{(1) \dagger}]] \; .
\end{align}
Using Holder's inequality, the first term can be bounded by:
\begin{equation}
    \Tr\left[ S(x)^{\dagger\otimes 2}W^{(2)\dagger \otimes 2} O^{\otimes 2}W^{(2)\otimes 2} S(x)^{\otimes 2} \mathcal{A}_{\mathbb{W}}(\ketbra{00}{00})   \right] \leq \norm{O}_2^2 \varepsilon \; ,
\end{equation}
The second term can be bounded using Weingarten calculus \cite{mele_introduction_2024}:
\begin{equation}
    \Var_{W^{(1)} \sim \text{Haar}}[Tr[ O W^{(1)}  \ketbra{0}{0} W^{(1) \dagger}]] \leq \norm{O}_2^2 \varepsilon
\end{equation}
which concludes the proof.
\end{proof}

Corollary \ref{cor:bound_model_approx_2design_informal} establishes an upper bound on the full model variance similar to that given for each Fourier coefficient in \autoref{thm:bound_approx_2design_informal}.This result has been established in previous works \cite{larocca_diagnosing_2022,larocca_barren_2025,holmes_connecting_2022} but is adapted here to the monomial distance $\varepsilon_M$ to a 2-design.

By combining \autoref{thm:bound_approx_2design_informal} and Corollary \ref{cor:bound_model_approx_2design_informal}, scenarios can be captured where frequencies with relatively low redundancy are vanishing, whereas there is leeway for the global model not to be.

\begin{figure*}[h!]
    \centering
    \includegraphics[width=1\linewidth]{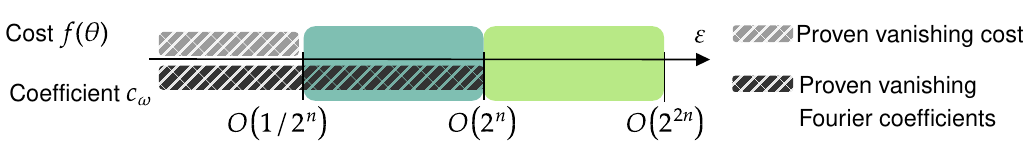}
    \caption{\textbf{Illustration of the vanishing model and vanishing Fourier coefficient phenomena according to the $\varepsilon_M$ distance to a 2-design}. Considering  coefficients $c_{\omega}$ with redundancies $|R(\omega)|= \mathcal{O}(poly(n))$, one can notice that the vanishing expressivity phenomenon can happen outside of the regime with guaranteed exponential concentration of the full model (blue part). Specifically, the dashed gray part corresponds to $\varepsilon_M$ range where the quantum model $f$ is proven to be vanishing while the black one corresponds the the vanishing Fourier coefficients regime. The blue part indicates the regime where the Fourier coefficients are vanishing but not necessarily the case for the corresponding model. Finally, the behavior of the model and its coefficients is unknown in the green part.}
    \label{fig:Figure_Vanishing_Cost_Coeffs_Epsilon}
\end{figure*}

Specifically, for frequencies with redundancies scaling at most polynomially in system size, i.e., $|R(\omega)| \in \mathcal{O}(poly(n))$ (or equivalently $|\widetilde{R}(\omega)| \in \mathcal{O}\left(\frac{poly(n)}{d^2}\right)$), the corresponding upper bound on the coefficient variance in \autoref{eq:monomial_norm_bound} scales as
\begin{equation}
    \Var[c_\omega] \in \mathcal{O} \left(poly(n)\frac{\ \norm{O}_2^2 \varepsilon_M}{d^2}\left(1+\varepsilon_M\right)\right)\;.
\end{equation}
This implies that as long as $\norm{O}_2^2 \varepsilon_M \in \mathcal{O}(poly(n))$, Fourier coefficients with redundancies $|R(\omega)| \in \mathcal{O}(poly(n))$ suffer from exponentially decaying variance. On the other hand, Corollary \ref{cor:bound_model_approx_2design_informal} provides guarantees of exponential decay for the full model only when $\norm{O}_2^2 \varepsilon_M \in \mathcal{O}\left(1/d\right)$. Consequently, there may be leeway for the global model to be non-vanishing while Fourier coefficients with polynomially large redundancies suffer from exponential concentration for a reasonably wide $\varepsilon_M$ range, as depicted in \autoref{fig:Figure_Vanishing_Cost_Coeffs_Epsilon}.

A straightforward construction of an \textit{expressive} model is to use an encoding strategy where the size of the spectrum is exponential in the number of qubits and hence less prone to classical dequantization \cite{landman_classically_2022,sweke_potential_2025} (see \autoref{chap:Fourier_Surrogates}). This implies that the spectrum is weakly degenerate, with many frequencies $\omega$ such that $|R(\omega)|=\Theta(1)$. This is indeed the case for the exponential encoding and the Golomb encoding. 

Consequently, although the quantum model theoretically has access to an exponential number of frequencies, the contribution of each frequency is vanishing. When considering general encoding strategies, \autoref{thm:bound_approx_2design_informal} implies that  frequencies with low redundancies are more likely to suffer from exponential concentration, limiting the expressivity of the quantum model. Specifically, for  fixed trainable unitaries and thus fixed $\varepsilon_M$, the upper bound on the Fourier coefficient variance allows high redundant frequencies to possibly escape exponential concentration while the low redundant ones will exhibit vanishing variance, leading to the \textit{vanishing expressivity} phenomenon.

\subsection{Impact of the Vanishing Expressivity Phenomena on Training and Dequantization}\label{subsec:train_impact}

In the previous Section, the exponential concentration of the Fourier coefficients was compared to that of the full model, with a focus on the settings where these two phenomena can occur independently.
It should be further emphasized that the interpretation of these two behaviors is fundamentally different. Specifically, since the Fourier coefficients are not directly measured to evaluate the model's gradients, the statement that they initially suffer from exponential concentration cannot be directly related to a resource problem (i.e., a finite number of shots) as in other exponential concentration analyses of the whole model. This means that the vanishing Fourier coefficients phenomenon cannot be directly related to trainability issues. 
This observation justifies the choice of the term \textit{vanishing expressivity} for this behavior, based on the intuition that frequencies with vanishing coefficients will have a negligible contribution to the quantum model.

However, a scenario can be envisaged where the signal from each Fourier coefficient is exponentially small. However, it can give rise to a significant signal when merged together. Hence, the exponential concentration of Fourier coefficients does not necessarily imply a constraint on the expressivity of the quantum Fourier model in this case.

Moreover, the analysis of the vanishing expressivity phenomenon holds on average when the parameterized unitaries form approximate 2 design but it gives no guarantees on the model's effective expressivity during the training stage. Consequently, it is possible to start with exponentially small contributions from the Fourier coefficients and still reach all theoretically accessible frequencies, given that the quantum model can be trained efficiently.

\begin{figure*}[htbp]
    \centering

    % Top plot: intersection plot (c)
    \begin{minipage}[t]{0.85\linewidth} % Adjust width as necessary
        \centering
        \includegraphics[width=1\linewidth]{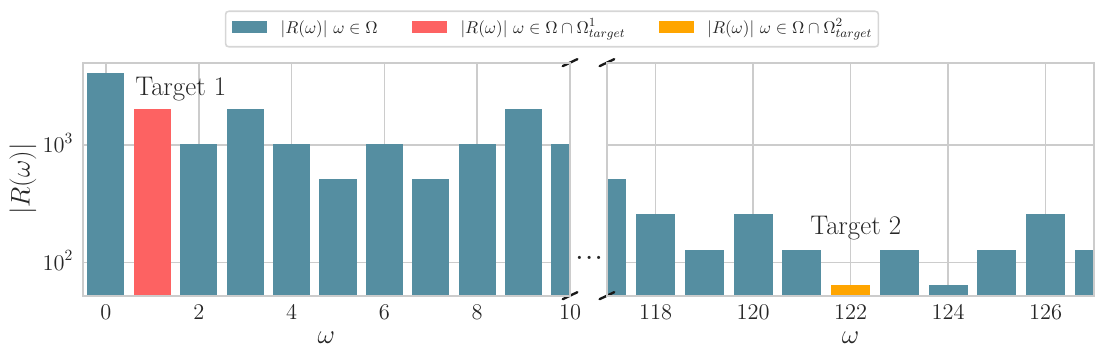}         
        \label{fig:intersection}
    \end{minipage}

    \vspace{-1em} % Space between the top and bottom rows

    \begin{minipage}[t]{0.49\linewidth}
        \centering
        \includegraphics[width=1\linewidth]{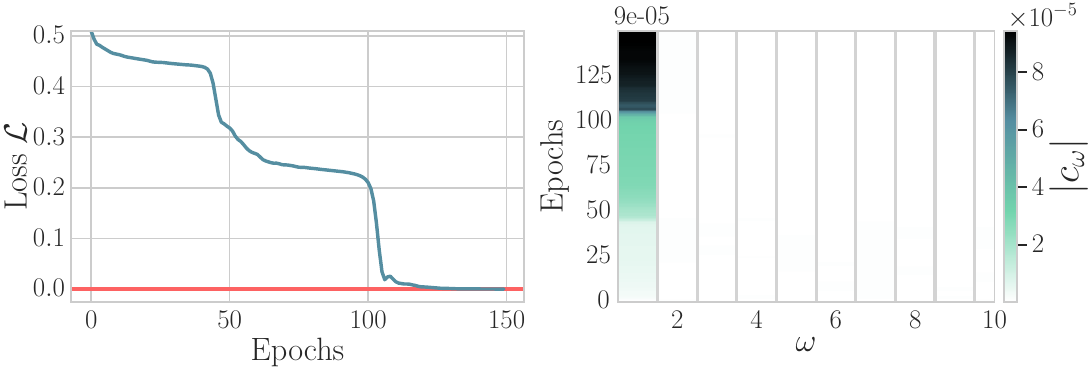}
    \end{minipage}%
    \hfill
    \begin{minipage}[t]{0.49\linewidth}
        \centering
        \includegraphics[width=1\linewidth]{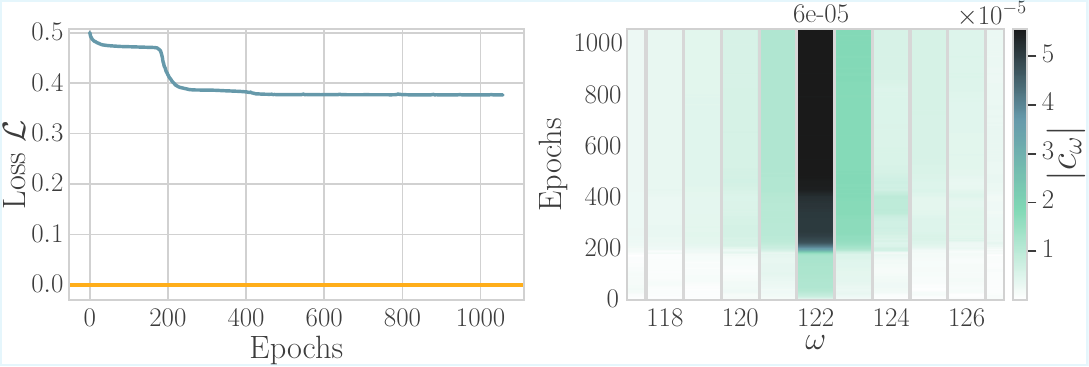}
    \end{minipage}

    \caption{\textbf{Impact of the vanishing expressivity phenomenon on a trained QFM.} The expressive capability of a trained QFM acting on $n=12$ qubits is studied, using the exponential encoding strategy with a) a spectrum $\Omega$ and frequency redundancies $|R(\omega)|$. The model is trained to fit two target frequencies with different redundancies in the model's spectrum: Target 1 corresponds to a high redundant frequency (in red), and Target 2 corresponds to a low redundant frequency (in purple). b) Plot of the loss and the  Fourier coefficients norm evolution fitting Target 1. In c), the plot of the loss and the Fourier coefficients norm evolution fitting Target 2 is presented. The model succeeds in fitting the high redundant frequency but not the low redundant one. }
    \label{fig:training}
\end{figure*}

\emph{Impact of vanishing expressivity on training.} To better understand the consequences of vanishing Fourier coefficients, the analytical results are supplemented with numerical simulations to study the impact of the vanishing expressivity phenomenon on the final trained model. Specifically, the task considered is training a fixed quantum model (fixed trainable unitaries and encoding unitaries) to fit two different sinusoidal functions with two different target frequencies: one highly redundant in the quantum Fourier model spectrum and the other with relatively low redundancy, as depicted in \autoref{fig:training}.
Then the training results are presented, showing that the model manages to reach the highly redundant frequency but not the low redundant one.
This result indeed supports the intuition that frequencies with initially vanishing Fourier coefficients are harder to reach. However, that this behavior could also be due to controllability issues, where no parameter configuration exists that gives the low-redundancy frequency a non-zero weight. In the next Section, the controllability issue is discussed in more detail.

\emph{Impact of vanishing expressivity on Random Fourier Features (RFF, see \autoref{chap:Fourier_Surrogates} and \autoref{thm:rff}) based dequantization.} 
For RFF-based dequantization schemes \cite{sweke_potential_2025} of QFMs, finding the optimal frequency distribution to build the classical surrogate and hence to dequantize the quantum model, requires knowledge about the spectral properties of the final model. Based on these results, the frequency distribution given by the redundancies, as described in \autoref{sec:Framework}, can be proposed as a natural distribution that encodes the bias in the quantum model. However, this choice is based on the assumption that the final model's spectral properties will inherit those of the average-case model (the initial model with random parameter initialization). Although this assumption is not guaranteed to hold in general, the numerics in \autoref{fig:training} show that the decaying Fourier coefficients in ''average'' models persist in the final trained model.

%%%%%%%%%%%%
%%%%%%%%%%

    \subsection{Fourier Norm Bound and Controllability Constraints}\label{subsec:MassConservation_Correlation_Control}

The focus of this Chapter is made on studying variances by considering a uniform distribution over the parameter vector $\theta$.
In this section, it is pointed out that additional constraints can occur due to a lack of model \textit{controllability}, defined as the number of Fourier coefficients that can be independently controlled by tuning the trainable parameter vector.

First, a generic constraint on the quantum model's Fourier coefficients is established in the following theorem. This constraint holds for any Hamiltonian encoding scheme and is independent of the trainable unitaries' distribution.

\begin{restatable}[Fourier Norm Bound]{thm}{FourierNormBound}
\label{thm:Mass_Conservation}
    Consider a quantum model $f(x,\theta)$ of the form in \autoref{eq:quantum_Model} using an observable $O$ and a parametrized circuit of the form in \autoref{eq:circuit_ansatz_Fourier} with $L\geq 1 $ layers. Also assume that the encoding Hamiltonians are fixed, giving rise to a spectrum $\Omega$.
     Then,
    \begin{equation}
        \forall x \in \mathbb{R}^d, \forall \theta \in \Theta, |f(x,\theta)|^2 \leq ||O||^2_{\infty} \; ,
    \end{equation}
    \begin{equation}\label{eq:Slimane}
        \forall \theta \in \Theta, \sum_{\omega \in \Omega}|c_\omega(\theta)|^2 \leq ||O||^2_{\infty} \; .
    \end{equation}
\end{restatable}

The first part of \autoref{thm:Mass_Conservation} is a trivial constraint that holds for any quantum model of the form in \autoref{eq:quantum_Model} even outside of the Fourier framework. This constraint has been mentioned in \cite{sweke_potential_2025} to highlight the fact that a quantum Fourier model cannot achieve any linear function in the Fourier basis given by its spectrum.
The second part of the theorem is more subtle. While very similar to the Parseval identity, the bound in \autoref{eq:Slimane} holds for any real-valued spectrum $\Omega$.
It shows that the 2-norm of the Fourier coefficient vector is upper bounded by the observable largest eigenvalue, introducing another generic constraint on the quantum model Fourier coefficients.

In addition to the previous results, it is important to stress that limitations in the controllability of the trainable unitaries can affect the controllability of the Fourier model. From the expanded expression of the quantum Fourier model given in \autoref{eq:quantum_Fourier_Model_developped}, each Fourier coefficient is defined as a sum and product of coefficients from the trainable unitary matrices.

Depending on the number of parameters and the set of gates chosen, the number of independent matrix coefficients that can be freely controlled through the trainable parameters can vary from one circuit to another. Previous works \cite{larocca_diagnosing_2022, larocca_theory_2023} have studied the controllability of VQCs, by analyzing the corresponding Dynamical Lie Algebra (DLA), or by considering the quantum Fisher Information matrix which characterizes the state controllability. In \cite{fontana_characterizing_2024,ragone_lie_2024}, the authors  highlight a connection between the maximal controllability of a VQC, i.e., the dimension of its DLA, and its capacity to be trained. Hence, this controllability notion will be key in characterizing the controllability of Fourier coefficients.

Namely, from the Fourier coefficient expression given in \autoref{eq:quantum_Fourier_Model_developped}, it can be observed that a pair of paths $(J,J') \in R(\omega)$ from the frequency generator defined in \autoref{def:Redundancy} allocates coefficients of the trainable unitary matrices to the corresponding frequency. In addition, some unitary coefficients are shared among different Fourier coefficients as a consequence of some branches in the generating tree (see \autoref{fig:Quantum_Spectrum_Trees}) being shared between different frequencies. Consequently, this can potentially create correlations between Fourier coefficients. Therefore, if the trainable layers have low controllability, it could lead to the impossibility of independently controlling a large number of Fourier coefficients.This is particularly important because increasing the number of parameters seems to increase the controllability and decrease the distance to a 2-design (see, for example, the evolution of the distance to a 2-design for the Periodic Ansatz in Theorem 1 of \cite{larocca_diagnosing_2022}).

In \autoref{fig:training}, plots show the evolution of the Fourier coefficients when training a QFM to learn two sinusoidal functions. The first one has a target frequency with high redundancy in the quantum Fourier model, and the second one corresponds to a low redundant frequency. It can be noticed that for the second target, the VQC takes more epochs to converge and fails to minimize the loss. In addition, it can be observed that during training, the Fourier coefficients surrounding the target frequency change significantly, due to a lack of controllability over the Fourier coefficients.

    \subsection{Limitations of the Framework}\label{subsec:Limitation_Framework}

In this Section, the limitations of the framework and the assumptions used to derive the main results presented in \autoref{sec:Main_Results} are discussed.

First, the statistical analysis of the Fourier coefficients established in this Chapter holds under the assumption that the trainable parameters are sampled uniformly and independently. Although the case where the trainable unitaries form an approximate 2-design was considered, the obtained constraints on the variance of the Fourier coefficients and their decay hold only on average. Consequently, extrapolating this average case behavior to the final trained model is not systematic. Indeed, this gap between average-case and final-model guarantees constrains the direct applicability of these results for rigorously studying the efficiency of random Fourier features-based dequantization schemes \cite{sweke_potential_2025}. While proving analytically the impact of the frequency redundancies on the \textit{effective expressivity} of the final trained model is a hard task, numerical evidence of this behavior is provided in \autoref{subsec:train_impact}.
Moreover, the upper bounds presented in \autoref{thm:bound_approx_2design_informal} are useful to establish generic theoretical guarantees on the model's expressivity. However, they only apply to quantum models with a single uploading layer. In addition, estimating the $\varepsilon$-distance of the trainable unitaries to 2-designs and its scaling is not efficient in practice. 

Finally, the results could be extended to the case of subspace preserving quantum circuits. In this type of VQC, the computation can be restricted to a particular subspace by using input states that lie in that subspace, reducing the dimension of the effective Hilbert space. These methods can avoid Barren Plateaus while considering subspaces of polynomial size \cite{  larocca_diagnosing_2022,fontana_characterizing_2024,ragone_lie_2024, monbroussou_trainability_2025} (see \autoref{sec:Subspace_Preserving_VQCs}, \autoref{chap:HW_Preserving_Methods}, and \autoref{chap:Photonic_Suboptimal}) but question the quantum advantage of such models \cite{anschuetz_efficient_2023,cerezo_does_2024}. For subspace-preserving unitaries, these results can easily be adapted. The dependency over the frequency distribution will still hold, but the value of $d$ (the dimension of the Hilbert space) will be substituted by the dimension of the subspace. Therefore, models generated by subspace preserving circuits could exhibit a similar \emph{inductive bias} arising from the redundancy constraint on the variance of its Fourier coefficients.

%% file: chapters/08_Fourier_Surrogates/Fourier_Surrogates.tex
\let\textcircled=\pgftextcircled
\chapter[Fourier Surrogates]{Fourier Surrogates}
\label{chap:Fourier_Surrogates}
\begin{textblock}{5.3}(0,-4)
	%\textit{`Un scientifique dans son laboratoire est un enfant placé devant des phénomènes naturels qui l'impressionnent comme des contes de fées.'\\}

%\hspace{0.5cm}--- Marie Skłodowska-Curie.
\end{textblock}

\initial{Q}\textit{uantum Machine Learning algorithms based on Variational Quantum Circuits (VQCs) are important candidates for useful application of quantum computing, that has been discussed throughout this thesis. In \autoref{chap:VQC_as_Fourier_Models}, it is shown that a VQC is a linear model in a feature space determined by its architecture. Such models can be compared to classical ones using various sets of tools, and surrogate models designed to classically approximate their results were proposed. At the same time, quantum advantages for learning tasks have been proven in the case of discrete data distributions and cryptography primitives. This Chapter discusses the notion of quantum advantage defined as the incapacity for a classical model, or surrogate model, to reach the same solution. Using previous results, conditions on the weight vectors of the quantum models that are necessary to avoid dequantization are established. This theory is compatible with previously proven quantum advantages on discrete inputs, and provides examples of advantages for continuous inputs. This separation is connected to large weight vector norm, and it is suggested that this can only happen with a high dimensional feature map. The results demonstrate that it is possible to design quantum models that cannot be classically approximated with good generalization. Finally, a discussion on how concentration issues must be considered to design such instances is presented. This study will aid in the design of near-term quantum models that avoid dequantization methods by ensuring non-classical convergence properties, and to identify existing quantum models that can be classically approximated.}

Machine learning is a heavily explored area in the search of applications for quantum computers \cite{schuld_supervised_2018, cerezo_challenges_2022, biamonte_quantum_2017}. In \autoref{chap:HW_Preserving_Methods}, \autoref{chap:Photonic_Suboptimal}, and \autoref{chap:Subspace_Preserving_Algorithms}, the focus is on using quantum computers as hardware accelerators of classical machine learning routines \cite{monbroussou_subspace_2025, kerenidis_q-means_2019, kerenidis_quantum_2017}, mainly leveraging quantum linear algebra protocols \cite{harrow_quantum_2009, gilyen_quantum_2019}. Another heavily explored area is the use of variational quantum circuits (VQCs) \cite{cerezo_variational_2021} to learn some functions of the data \cite{schuld_circuit-centric_2020}. The initial ideas of variational QML research \cite{schuld_circuit-centric_2020, havlicek_supervised_2019} were that the advantage of quantum computing for machine learning would be to look for models in high dimensional feature spaces, exponentially larger than the initial dimension of the data, and the size of the dataset. %

\begin{figure*}
\includegraphics[width=1\linewidth]{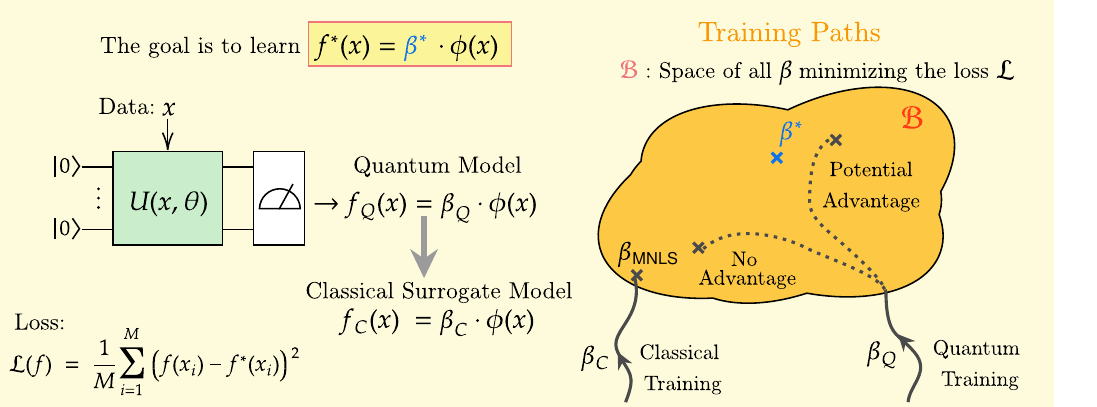}
\caption{A quantum model $f_Q(x) = \beta_Q(\theta)\cdot\phi(x)$ is trained by optimizing its weight vector $\beta_Q(\theta)$. If one can train a surrogate model on a classical computer, using the same (or approximated) feature map $\phi(x)$, it would constitute an obstacle to quantum advantage. It has been shown that during classical linear regression, the weight vector converges towards a specific point $\beta_{\mathrm{MNLS}}$ called the \textit{minimum norm least squares} estimator. Ensuring that the quantum weight vector $\beta_Q$ converges far from $\beta_{\mathrm{MNLS}}$ is therefore a necessary condition to avoid such dequantization.}
\label{fig:Introduction}
\vspace{-2em}
\end{figure*}

It is known that VQCs are linear models in some \textit{feature space} of finite dimension \cite{schuld_effect_2021}.
At first sight, if this feature map can be explicitly computed \cite{schuld_supervised_2018} classically, one may wonder what the interest is in searching for the best parameters of the quantum circuit instead of performing classically a linear regression on the same feature map, using a so called \textit{classical surrogate} model \cite{schreiber_classical_2023}. Even when the feature space is too large to be computed classically, methods exist to reduce its dimension by random sampling, realizing approximated classical models \cite{landman_classically_2022, sweke_potential_2025, sahebi_dequantization_2025}. 

Knowing when these classical models can mimic the quantum ones is crucial to understand potential quantum advantage in such learning tasks. In \cite{you_analyzing_2023}, the authors analyse the optimization dynamics of quantum neural networks and conclude that they are different from the neural tangent kernel. They study in detail the convergence rate of the respective methods, but do not study the actual solutions reached. Using cryptographic primitives, the authors in \cite{jerbi_quantum_2023} study the fact that variational circuits can converge to a different solution than the kernel ridge regression. They point out that there exists functions that are learnable with VQCs but that require exponentially more resources to learn with quantum kernels. They propose several examples of cryptographic inspired VQCs with discrete data that cannot be approximated in their generalization. In this chapter, necessary conditions are presented for a quantum model to avoid such dequantization, that could only be satisfied for high dimensional feature maps. This study can be applied to any quantum circuits with continuous or discrete inputs, and propose conditions that guarantee a quantum model to remain \textit{far} from its equivalent classical model. For that, the important fact is used that classical linear regression causes the optimized weight vector to converge towards a specific solution called \textit{minimum norm least square} (MNLS) estimator. This study focuses on showing when the quantum weight vector does not possess the same bias. \autoref{fig:Introduction} summarizes the methodology. 

Those conditions are then analyzed for several usual frameworks and architectures, showing that the proposed methodology can be seen as a new tool to rule out certain quantum circuits. Using Weingarten calculus, it is demonstrated that some proposed quantum models can be far from the MNLS. It is also shown that cryptographic examples satisfy the proposed condition on the weight vector norm. In addition, the link between these dequantization schemes and concentration, another crucial issue of quantum circuits, is studied. It is proven that a family of models should exist with continuous inputs that avoids both of these problems.

    \section{Results}

\subsection{Setup and Notations}
General forms of quantum machine learning models are considered, or \textit{quantum models} that can be expressed as
\begin{equation}
    f_Q(x) = \text{Tr}(U(x;\: \theta) ^\dagger O U(x;\: \theta) |0^n\rangle\langle 0^n|) \; ,
\end{equation}
where $U(x;\: \theta)$ is a unitary dependent on the input data $x \in \mathbb{R}^d$ and trainable parameters $\theta$.

It is known that most proposed quantum models can be expressed as linear models in a given feature space. That is, there exists a \textit{feature space} $\mathbb{R}^p$ (for some $p\geq1$) and a \textit{feature map} $\phi: \mathbb{R}^d \longrightarrow \mathbb{R}^p$ such that the quantum model can be written as
\begin{equation}
    f_Q(x) = \beta_Q^{\top} \, \phi(x) \,,
\end{equation}
with a $\theta$-dependent \textit{weight vector} $\beta_Q^{\top}$.

Given a \textit{training dataset} of size $M$, consisting of $M$ input points $(x_1, \dots, x_M)$ assumed to have been sampled from some distribution $\mu$ on $\mathbb{R}^d$, and $M$ scalar targets $(y_1, \dots , y_M)$.
This training inputs in feature space form the \textit{data matrix} $\Phi \in \mathbb{R}^{M\times p}$ (as $[\Phi]_{ij}=\phi(x_i)_j$), while the outputs $y_i$ form the vector of targets $y \in \mathbb{R}^{M\times 1}$.

The goal is assumed to be learning a target function linear in the same feature space
\begin{equation}
    f^*(x) = \beta^{*\top} \, \phi(x) \; ,
\end{equation} for some $\beta^* \in \mathbb{R}^p$. For a real world task, there is of course no particular reason for the target function to be expressed in this way. This particular case is however very useful to understand quantum advantage.

During training, the parameters $\theta$ of the quantum circuit are chosen iteratively so as to optimize the \textit{empirical risk} loss
\begin{align}\label{eq:empirical_risk_loss}
    \mathcal{L}(f_Q, f^*) &= \frac{1}{M} \sum_{i=1}^M (f_Q(x_i;\:\theta) - y_i)^2\\
    &= \frac{1}{M}\norm{\Phi\beta_Q(\theta) - y}^2 \,.
\end{align}

On a theoretical level, how well a of particular model $f$ (quantum or classical) \textit{generalizes} to the true solution $f^*$, is captured either by the square of their $L_2$ distance (with respect to distribution $\mu$) 

\begin{equation}\label{eq:true_risk}
    \norm{f - f^*}_\mu^2 := \int_{\mathbb{R}^d} (f(x) - f^*(x))^2 d\mu(x) \, ,
\end{equation}
or by their $\infty$-distance (assuming $f$ and $f^*$ are bounded)
 \begin{equation}
    \norm{f -f^*}_{\infty} := \sup_{x \in \mathcal{X}}\!\big|f(x) - f^*(x)\big|\,.
\end{equation}

\subsection{Bias of Classical Linear Regression}
\label{subsec:mnls}

In this subsection, the known results \cite{bishop2006pattern, hastie_surprises_2022} about the solution of the linear regression problem ared detailed. To minimize the empirical risk loss defined in \autoref{eq:empirical_risk_loss}, one can train the weight vector $\beta$ using a Gradient Descent (GD) method or solve the equivalent Kernel Ridge Regression (KRR) \cite{bishop2006pattern, hofmann2008kernel}. Two regimes can be considered:

\begin{itemize}
    \item The \textbf{underparameterized} regime where the feature space dimension is lower or equal to the number of datapoints: $p \leq M$. In this regime, there is a unique solution, that can be expressed as
\begin{equation}
    \hat{\beta} = (\Phi^\top\Phi)^{-1}\Phi^\top y \,.
\end{equation}
Furthermore, if the data is sampled such that $\Phi^\top\Phi$ is full rank (which is almost always the case) and there is no noise in the observed targets, the estimator $\hat{\beta}$ is equal to the ground truth $\beta^*$.
    \item The \textbf{overparameterized} regime where the feature space dimension is greater than the number of datapoints: $p > M$. In this case, an infinite number of weight vectors can set the empirical risk is zero. However, the algorithms of GD and KRR will converge towards a specific vector $\beta_{\mathrm{MNLS}}$ called \textbf{minimum norm least square} estimator (MNLS). $\beta_{\mathrm{MNLS}}$ is the vector of minimal norm among the minimizers of the empirical loss, and it is provably unique:
\begin{equation}
    \beta_{\mathrm{MNLS}} = \arg \min \norm{\beta}_2 \; \text{with} \;  \mathcal{L}(\beta^\top \phi, f^{\ast}) = 0 \; ,
\end{equation}
which can also be written
\begin{equation}%
    \beta_{\text{MNLS}} = \Phi^\top(\Phi\Phi^\top)^{-1}y \; .
\end{equation}
This behavior is due to the fact that GD and KRR only search a solution in the space spanned by the training datapoints, called the \emph{row space}.
\end{itemize}

\subsection{Classical Models for Dequantization}\label{subsec:rff}

Given a quantum model, one can design a \textit{surrogate} model by considering a classical model with the same feature map $\phi$, defining a new linear model:
\begin{equation}
    f_C(x) = \beta_C^{\top} \, \phi(x) \, ,
\end{equation}
with $\beta_C$ a weight vector that is obtained with a classical computer. %
An obvious obstacle for doing so is the fact that the dimension of the feature map $\phi(x)$ is too big to be stored in memory. However, techniques exist to mitigate this problem.

Authors in \cite{rahimi_random_2007} have introduced the \textit{Random Fourier Features} (RFF) technique to lower computational costs for kernel methods and error bounds were refined in \cite{sutherland_error_2015, li_towards_2021}. This method can be generalized for many other cases in classical linear regression \cite{rahimi_uniform_2008, rahimi_weighted_2008}, and can be applied to the arbitrary basis quantum models defined with a Hamiltonian encoding with any preprocessing function. It has been shown that approximating the target function can be done by learning a function of the form $\hat{f}(x) = \sum_{k=1}^D \beta_i\phi_k(x)$ with $D << p$ where the functions $\phi(\:\cdot\: ;\omega_k)$ are sampled from $\llbracket 1,\: p \rrbracket$. In this case, one only has to learn a vector of dimension $D$.

Studies such as \cite{landman_classically_2022, sweke_potential_2025, sahebi_dequantization_2025} have shown that Random Features Regression can be used to dequantize
quantum models, although limitations exist, particularly for resource-constrained circuits.

In the following, theorem on the RFF technique is presented. This result is very important and will be used in the rest of the chapter.

\begin{restatable}[]{thm}{TheoremRFF}
\label{thm:rff}
    Let  $\phi(x) = [\sqrt{q_1}\:\phi_1(x) \dots \sqrt{q_p}\:\phi_p(x)]^\top$ where $\phi_i(x)$ are basis functions such that $\forall x,\: |\phi_i(x)| \leq 1$ and $q = (q_1, \dots q_p)$ represents a discrete probability distribution, and let $f(x) = \beta^\top \phi(x)$. Let $S$ be a subset of $\llbracket 1, p \rrbracket$ sampled independently with the probability density $q$, with $D = |S|$. The size of the dataset is given by $M$. Then there exists coefficients $c_1, \dots c_D$ such that  $\hat{f}(x) = \sum_{k\in S} c_k\phi_k(x)$ satisfies 
    \begin{equation}
    \label{eqn:approx}
        \Vert \hat{f} - f\Vert _\mu = \mathcal{O}\left( \frac{\max_i |\beta_i|\:\norm{\phi_i}_{\mu}\:/\sqrt{q_i}|}{\sqrt{D}} \right) \, .
    \end{equation}
    Applying the above to $\beta_{\mathrm{MNLS}}$ obtained from a kernel matrix $K$ and target vector $y$ with $\norm{y}_\infty \leq 1$ yields coefficients $c_1, \dots c_D$ such that
    \begin{equation}
    \label{eqn:approx_mnls}
        \Vert \hat{f} - f_{\mathrm{MNLS}}\Vert _\mu = \mathcal{O}\left(\frac{M\:\max_i\norm{\phi_i}_\mu}{\sqrt{D}\:\lambda_{\min}(K)} \right) \, .
    \end{equation}
\end{restatable}

This Theorem is based on the same technique as Theorem 3.1 in \cite{rahimi_uniform_2008}.

\begin{figure*}[t]
\includegraphics[width=1\textwidth]{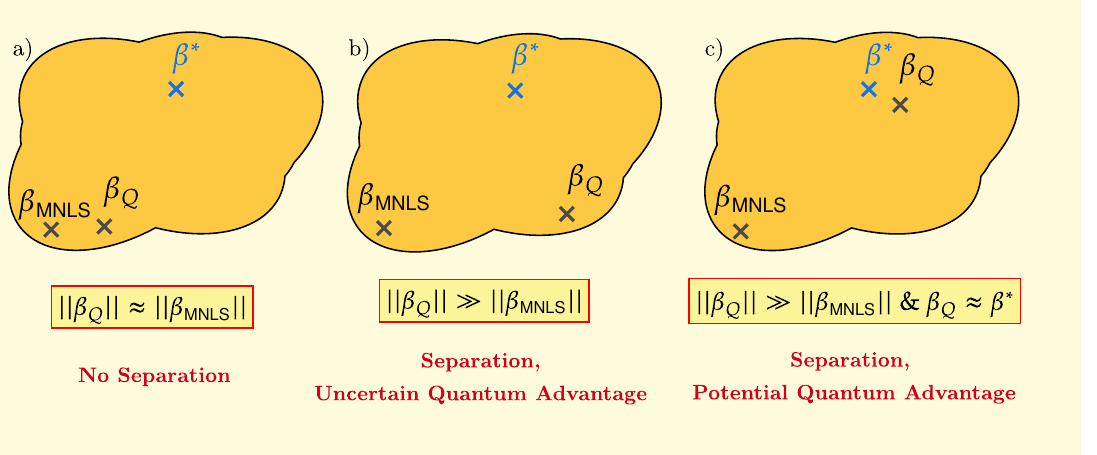}
    \caption{Illustration of the potential quantum advantage. If $\beta_Q$ is close to $\beta_{\mathrm{MNLS}}$ there is no separation between the quantum estimator and the classical one. If $\beta_Q$ and $\beta_{\mathrm{MNLS}}$ are far from each other and far from the ground truth, there is a separation but uncertain quantum advantage. If $\beta_Q$ is closer to the ground truth than $\beta_{\mathrm{MNLS}}$, there is a suggestion of quantum advantage.}
    \label{fig:Quantum_Advantage}
\vspace{-2em}
\end{figure*}

    \section{Bias of Quantum Models and Potential Advantage}\label{sec:Bias_Quantum_Advantage}

In this chapter, a general study of learning through VQCs is proposed, that can be applied to the case where the input variable is continuous or discrete. Most proven quantum advantage results in quantum machine learning come from problems where the input data take discrete values \cite{gyurik_exponential_2024, molteni_exponential_2024, jerbi_quantum_2023, liu_rigorous_2021, jerbi_shadows_2024}, typically $\{0, 1\}^n$. It is convenient because the problems can be linked to cryptography problems which are known or strongly supposed to be hard to solve classically. However many real world use cases utilize continuous vectors, so it is important to have a better understanding in that domain.

\subsection{Underparameterized Regime has Few Advantages}

First of all, it is noted that in the underparameterized regime, there is little potential for solving the linear regression problem more effectively with a quantum computer. The optimal solution to the least square problem has indeed a closed form, and if there is no noise in the data, it is equal to the true weight vector. It means that any other optimization technique will converge towards that optimal solution. Moreover, since the number of data points is assumed to be small enough to be handled with a classical computer, the total number of operations in the procedure is still polynomial in the size of the dataset. An advantage of using a quantum computer to invert the covariance matrix \cite{harrow_quantum_2009} is not excluded, or other more modest polynomial advantages \cite{cerezo_challenges_2022}.%

\subsection{Quantum Models can Differ from Minimum Norm Least Square}

The overparameterized regime is considered. The implicit bias of classical learning algorithms is described in \autoref{subsec:mnls}. A classical linear regression trained with gradient descent, or a kernel ridge regression will output a model $f_{\text{MNLS}}(x) = \beta_{\text{MNLS}}^\top\phi(x)$, and that in lots of cases, if one provides a sampling access to the entries of $\phi(x)$, then $f_{\mathrm{MNLS}}$ could be classically approximated. Contrary to the classical case, one does not have access directly to the coefficients $\beta_Q$ while tuning a quantum model. One instead optimizes a vector of parameters $\theta$ such that $\beta_Q = \beta_Q(\theta)$ and optimizes the loss function $\mathcal{L}(\theta) = \norm{y - X\beta_Q(\theta)}^2$.

{In this case,  during the training, $\beta_Q$ does not remain in the row space (the space spanned by the training datapoints) and does not converge to $\beta_{\text{MNLS}}$. This constitutes a crucial distinction between quantum and classical models. 
If the quantum model would converge to $f_{\mathrm{MNLS}}$, it could be approximated with random feature regression techniques. Therefore, this chapter results suggest that the best usage of quantum computers would not be to reproduce classical linear regressions. The quantum circuit should be used to provide a model $\beta_Q$ such that $\beta_Q \neq \beta_{\mathrm{MNLS}}$. 
It remains to be seen when $\beta_Q$ can converge far from $\beta_{\text{MNLS}}$ or from an approximation of MNLS via random features. 

In practice, $\norm{\beta_Q} \geq \text{poly}(N)$ can be considered in order to have a clear separation. Such examples are developped in \autoref{subsec:Example_quantum_Fourier}. Having a weight vector of large norm will provide a difference with classical models, but a true advantage will be reached if in addition the quantum models is closer to the ground truth than the MNLS. These views are illustrated in \autoref{fig:Quantum_Advantage}.
Since $f_{\mathrm{MNLS}}$ is the interpolating model of minimum norm, any quantum interpolating model $f_Q$ must verify $\norm{\beta_Q} \geq \norm{\beta_{\mathrm{MNLS}}}$. A sufficient condition for separation between $\norm{\beta_Q}$ and $\norm{\beta_{\mathrm{MNLS}}}$ would be that $\norm{\beta_Q} \gg \norm{\beta_{\mathrm{MNLS}}}$. It is stated in the following informal theorem

\begin{restatable}[Informal]{thm}{RelationBetaNormFunctionDist}\label{thm:largeNormBeta}
Let $f_Q$ be an interpolating quantum model, i.e., $\mathcal{L}(f_Q) = 0$. Then, $f_Q$ has a potential quantum advantage if $\norm{\beta_Q} \gg \norm{\beta_{\mathrm{MNLS}}}$.
\end{restatable}

Other works have outlined the differences between quantum and classical linear regression, but none of them mentions the criteria about the norm of the weight vector. The authors in \cite{jerbi_quantum_2023} study the fact that variational circuits express a different solution than the kernel ridge regression (therefore the MNLS). They point out that there exists functions that are learnable with variational quantum circuits but that require exponentially more resources to learn with quantum kernels. In \cite{you_analyzing_2023}, the authors analyse the optimization dynamics of QNNs and conclude that they are different from the neural tangent kernel. They study in detail the convergence rate of the respective methods, but do not study the actual solution reached.

In the following, the weight vector norms are studied for usual VQC framework with continuous input, but also for the cryptographic examples with discrete inputs described in \cite{jerbi_quantum_2023}. The results will emphasize the importance of this criterion to find a quantum advantage.

\section{Examples of Separation from Classical to Quantum Models}
\label{sec:examples}

\subsection{Fourier Model}

Fourier models are defined by the Fourier feature map:%
 
\begin{equation}\label{eq:feat-map}
    \phi(x) = \frac{1}{\sqrt{p}}\begin{bmatrix} \cos(\omega^{\top}x) \\ \sin(\omega^{\top}x) \\ \vdots \end{bmatrix}_{\omega \in \Omega} \; ,
\end{equation}
with $\Omega \subset \mathbb{Z}^d$, and $p=|\Omega|$. The spectrum is assumed to only be composed of vectors of integers, and that $\forall\: \omega \in \Omega, -\omega \notin \Omega$. The input vector $x$ is assumed to be uniformly distributed in $[0, 2\pi]^d$.
This case is very important in the quantum machine learning literature \cite{schuld_effect_2021, schreiber_classical_2023, sweke_potential_2025, peters_generalization_2023}, and could help to understand what quantum circuit design needs to be done in order to do variational circuit learning. 

It has been explained in \autoref{subsec:rff} that the approximability of the MNLS estimator depends on the eigenvalues of the empirical kernel matrix. It can be proven that the smallest eigenvalue of the kernel matrix is a constant with high probability. It enables us to state that the MNLS associated to a Fourier model can be easily approximable, and that the norm of $\norm{\beta_{\mathrm{MNLS}}}$ is bounded by the number of datapoints. This is stated in the following theorem.

\begin{restatable}[]{thm}{WeightVectoretwodesign}\label{thm:WeightVectoretwodesignInformal} (Informal)
Let us consider a Fourier model with a spectrum $\Omega \subset \mathbb{Z}^d$. %
Then, the associated MNLS estimator has a norm scaling like $\norm{\beta_{\mathrm{MNLS}}} = \mathcal{O}(M)$.
\end{restatable}

\begin{proof}
    The proof is an application of Theorem 2.1 from \cite{wolkowicz_bounds_1980}. Let $A$ be a $M\times M$ complex matrix with real eigenvalues. Let $m = \text{tr}(A)/M$  and  $s^2 = \text{tr}(A^2)/M - m^2$. Then, 
    \begin{equation}
        m - s\sqrt{M-1} \leq \lambda_{\text{min}}(A) \leq m - \frac{s}{\sqrt{M-1}} \; .
    \end{equation}

    For the kernel matrix, $m = 1$ and $s^2 = 1 + \frac{1}{M} \sum_{j \neq i}^M k(x_i,x_i)^2 - 1 = \frac{1}{M} \sum_{j \neq i}^M k(x_i,x_j)^2$. As a result, this Theorem can be applied to show that the expectation of $s^2$ is given by
    \begin{itemize}
        \item $\mathbb{E}[s^2] = \displaystyle\frac{M(M-1)}{2Mp} = \displaystyle\frac{(M-1)}{2p}$.
    \end{itemize}
    Furthermore, there exists a constant $C$ such that 
    \begin{itemize}
            \item $\mathrm{Var}[s^2] \leq \displaystyle\frac{C}{p}$.
        \end{itemize}
    Applying those results and Chebyshev's inequality:
        \begin{equation}
        \mathbb{P}(\lambda_{\min}(K)> \frac{1}{2}) \geq 1- \frac{C}{p} \frac{(M-1)^2p^2}{(p-4(M-1)^2)^2} = 1 -\frac{C}{p}\frac{(M-1)^2}{1 - 4\frac{(M-1)^2}{p^2}} \; .
        \end{equation}
\end{proof}

In the case of Fourier models, \autoref{eqn:approx_mnls} from \autoref{thm:rff} can be rewritten, such as: 

\begin{equation}
    \Vert \hat{f} - f_{\mathrm{MNLS}}\Vert _\mu = \mathcal{O}\left( \frac{M}{\sqrt{D}} \right) \, .
\end{equation}

It is the most favorable case to apply random feature regression, it is then enough to have a number of random features polynomial in $M$ to approximate the MNLS estimator.

\subsection{Simple Quantum Fourier Model}\label{subsec:Example_quantum_Fourier}
\vspace{-.5em}

In this section is detailed an example of quantum Fourier model that exhibit the separation mentionned in the previous section, i.e., $\norm{\beta_Q} \gg \norm{\beta_{\mathrm{MNLS}}}$.

Consider a circuit with a diagonal encoding layer $S(x)$ applied to the $|+\rangle^n$ state followed by a trainable unitary $V$ and an observable $O$ such that $\text{Tr}(O) = 0$. This example is illustrated in \autoref{fig:Potential_Advantage_Framework}. 
The quantum model can then be written
\begin{equation}
    f_Q(x) = \text{Tr}(O\: VS(x)(|+\rangle\langle +|^n)S(x)^\dagger V^\dagger) \; .
\end{equation}

\begin{figure}[h!]
    \centering
    \includegraphics[width=0.95\textwidth]{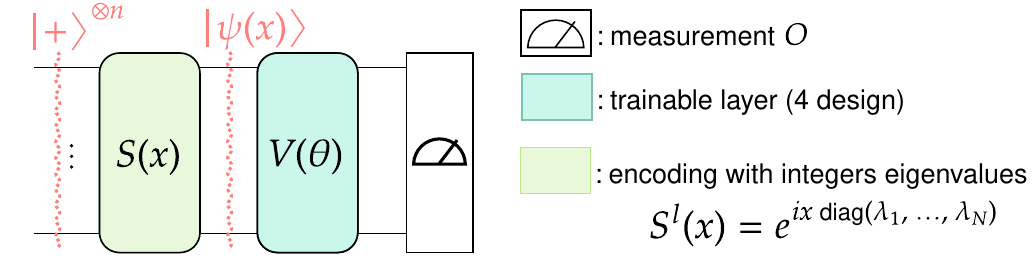}
    \caption{Parameterized quantum models considered: hamiltonian encoding with no integer eigenvalues.}
    \label{fig:Potential_Advantage_Framework}
\end{figure}

The spectrum only depends on the encoding unitary $S(x)$. Two types of encodings are considered among many possibilities that are detailed in \cite{peters_generalization_2023}:

\begin{itemize}
    \item \textbf{The ternary encoding}
    
    $S(x) = \bigotimes_{k=0}^{n-1} RZ_{k}(x\:3^k/2)$
    where $RZ_{k}$ denotes a $Z$ rotation applied to the qubit $k$. The spectrum produced by this encoding is the interval $\llbracket 0, 3^n-1\rrbracket.$ It is the spectrum with the largest size one can produce with one layer of single qubit gates \cite{shin_exponential_2023}.
    
    \item \textbf{The Golomb encoding}
    $S(x) = \exp(-i \displaystyle\frac{x}{2} R_G)$ where $R_G$ is a Golomb ruler \cite{piccard1939ensembles}. The resulting spectrum are all the integers in the set $\llbracket 0, N(N-1)/2\rrbracket.$ Such an encoding is not known to be realizable in polynomial time on a quantum computer, so is of little practical use, but it is an interesting edge case of the results.
\end{itemize}

In this setting, the following result holds:

\begin{restatable}[]{thm}{WeightVectoretwodesign2}\label{thm:WeightVectoretwodesignInformal2} (Informal)
Let us consider a Fourier model.
With high probability,
$\norm{\beta_{\mathrm{MNLS}}}^2 \leq M^2$.
For the Golomb encoding, $\norm{\beta_Q}^2 \sim 2^n$. For the ternary encoding, $\norm{\beta_Q}^2 \sim (3/2)^n$.
\end{restatable}

This result is obtained by integrating order 4 moments of the Haar measure \cite{collins2006integration, weingarten4}. 
This result shows a potential separation between quantum and classical models in the special cases mentioned. Indeed, $\norm{\beta_Q}\gg \norm{\beta_{\mathrm{MNLS}}}$ if it is assumed that $M = O(\text{poly}(n))$.

However, these models are not suitable to be used in a practical case, because considering that the trainable unitary is drawn from a 2-design implies the model concentration and vanishing gradient phenomenon called Barren Plateau \cite{mcclean_barren_2018, holmes_connecting_2022}. This point is discussed \autoref{sec:Expressivity_Measures_Q_Models}.

\subsection{Re-Uploading Fourier Models }\label{subsec:ReUploadingModel}

\vspace{-.5em}

Consider the case of re-uploading model, where the quantum model $f_Q(x,\theta) = \bra{0} U(x,\theta)^\dagger O U(x,\theta) \ket{0}$, is such that the circuit unitary is composed of an \textit{encoding} layer surrounded by two \textit{trainable} layers of the form: 
\begin{equation}\label{eq:circuit_ansatz}
    U(x,\theta) = V^{2}(\theta) S(x) V^{1}(\theta) \, \textrm{,}
\end{equation}
with $V^1(\theta)$ and $V^2(\theta)$ formed by trainable gates depending on the parameter vector $\theta$, which is optimized during training whereas $S(x)$ only depends on input data values. Consider also the Hamiltonian encoding strategy where the classical input components are encoded as the time evolution of some Hamiltonians $S(x)= \prod_{k=1}^D e^{-ix_kH^{(k)}}$. As explained earlier, the quantum model can be written as a Fourier Series where its spectrum $\Omega$ which depends on the eigenvalues of the encoding Hamiltonians. %

In \autoref{chap:VQC_as_Fourier_Models}, it has been shown that the variance of the Fourier coefficients $c_\omega(\Theta)$ depends on the \textit{redundancy} $|R(\omega)|$ of their corresponding frequencies $\omega$. In this chapter, the difference of learning behaviour between the quantum models and the minimum norm least square (MNLS) estimator are investigated. Because the norm of the MNLS estimator is bounded, the proximity to $\beta_Q$ is shown by considering an upper bound on its norm. First, the case where the trainable layers, $V^1(\theta)$ and $V^2(\theta)$, described 2-design over the special unitary group is considered. 

\begin{restatable}[]{thm}{ExpBetaQReuploadingtwodesign}\label{thm:Exp_BetaQ_Reuploading_2design}
  Consider a single layered quantum re-uploading model with an observable $O$ such that $\text{Tr}(O) = 0$, and $\norm{O}^2_2 = N$. Then, $\norm{\beta_Q }_2 \sim \frac{p}{N}$ with $p$ the number of features, and $N$ the number of distinct eigenvalues in the encoding layer.
\end{restatable}

Therefore, the norm of $\beta_Q$ can be very low for low value of $p$, while the case $p \sim N$ may offer a potential advantage. As in the previous example, considering $p \sim N^2$ leads to a clear separation where $\norm{\beta_Q} \gg \norm{\beta_{\mathrm{MNLS}}}$. In \autoref{chap:VQC_as_Fourier_Models}, the authors offer a bound on the variance of Fourier coefficients according to the monomial distance $\varepsilon$ of each trainable layer unitary matrix to a 2 design. Similarly, a bound on the variance of the weight vector norm is provided in \autoref{app:Proof_re_uploading_Fourier_models}, along with more detailed theorems and their corresponding proofs.

Under the hypothesis that the quantum circuit solution minimizes the empirical risk, the $\ell_2$-norm of the quantum circuit weight vector is lower bounded by $\norm{\beta_{\text{MNLS}}}_2$. Those two results can be seen as contradictory, but it simply means that if the trainable layers are close to a 2-design, it could be hard for the quantum circuit to reach a solution that minimizes the empirical risk. %

\subsection{Discrete Inputs VQCs}
\vspace{-.5em}

Several previous works have showed quantum advantage from learning tasks, especially using discrete inputs and cryptography primitives \cite{gyurik_exponential_2024, molteni_exponential_2024, jerbi_quantum_2023, liu_rigorous_2021, jerbi_shadows_2024}. This section details an example of a quantum model that provably cannot be dequantized, and explains how it fits into the general theory.

Let the discrete logarithm unitary be defined as
\begin{equation}
    U_{\text{DLP}}: |i\rangle \longmapsto |\log_g i + 1 \rangle \; ,
\end{equation}
where $g$ is a prime number in $\llbracket 0, N-1\rrbracket$.

Let $|\psi (x)\rangle = \bigotimes_{i=1}^n RY(x_i) |0^n\rangle$ and 
\begin{equation}
    f_{\text{DLP}}(x) = \text{Tr}(U_{\text{DLP}}^{\dagger}Z_n U_{\text{DLP}}|\psi(x)\rangle\langle\psi(x)|) \; .
\end{equation}

$U_{\text{DLP}}^{\dagger}Z_n U_{\text{DLP}}$ is a hermitian diagonal matrix and the coefficients can be written as $(U_{\text{DLP}}^{\dagger}Z_n U_{\text{DLP}})_{ii} = (-1)^{b_n(\log i + 1)}$. $b_n(j)$ is the n-th bit of the binary description of $j$. $f_{\text{DLP}}$ can be rewritten as 

\begin{align}
f_{\text{DLP}}(x) %
&= \sum_{y \in \{0, 1\}^n} \beta_y \phi_y(x) \; ,
\end{align}
where
\begin{equation}
    \phi_y(x) = \frac{1}{2^d} \prod_{i=0}^{d-1} (1 + (-1)^{y_i}\cos (x_i)), \quad y \in \{0, \:1\}^d \; ,
\end{equation}

and $\beta_y = (-1)^{b_n(\log i + 1)}$.

The bounds on the efficiency of Random Feature Regression become exponential in the number of qubits, since for all $y$

\begin{align}
    |b_y \: 2^n\norm{|\phi_y(x)}_{\mu} = \sqrt{2}^n\,.
\end{align}

Therefore it cannot be shown that $f_{\text{DLP}}$ can be learned with Random Features Regression.

It is consistent with the fact that it cannot be efficiently approximated because of the hardness of the discrete logarithm \cite{liu_rigorous_2021}.

\vspace{-.5em}

\section{Discussion}\label{sec:Discussion_Limitations}

\vspace{-.5em}

\subsection{Avoiding Concentration Issues}\label{subsec:LinkConcentration}
\vspace{-.5em}

Concentration phenomenon of parameterized quantum circuits have been studied a lot in the literature. 
A function is said to be concentrated if the variance $\mathrm{Var}_x[f(x)]$ is small. Typically, a quantum model  $f$ is considered concentrated if $\mathrm{Var}[f]\leq 1/\text{poly}(N)$. It is equivalent to the Barren Plateau phenomenon \cite{mcclean_barren_2018}, where the gradient of the loss function is exponentially close to 0.

The quantum model $f_Q$ can only be estimated by taking an average of $N_{\text{shots}}$ measurements with a precision of $ 1/\sqrt{N_{\text{shots}}}$. Thus if $f$ is concentrated, it would take an exponential amount of shots to evaluate it reliably. Therefore it would not be useful as a model. For the  Fourier model, the variance of the function is given by the norm of $\beta$, $\mathrm{Var}[f] = \norm{\beta}^2/p$.
For the weight vectors of the proposed random quantum circuits in \autoref{sec:examples}, $\mathrm{Var}[f]$ is of the order of $1/2^n$ which is concentrated.
A Fourier model with a weight vector norm $\norm{\beta}^2/p \geq \frac{1}{\text{poly}(d)}$ would then be non concentrated and non dequantizable with random features regression.

Constructing such non concentrated quantum models has proven to be a challenge. Recently, the commmunity investigated links between concentration and classical simulability \cite{cerezo_does_2024}. It has been conjectured that quantum models that do not suffer from concentration can be simulated efficiently with classical computers.

Concentration can be avoided in the case of discrete data input where $x \in \{0, 1\}^n$. \cite{cerezo_does_2024} gives some examples, and the discrete log function $f_{\text{DLP}}$ is not concentrated because $\mathbb{E}_{x \in \{0, 1\}^n}[f_{\text{DLP}}(x)^2] - \mathbb{E}_x[f_{\text{DLP}}(x)]^2 = 1/2 - 1/4 = 1/4$ since for half the inputs, $f_{\text{DLP}}(x)=1$ and $0$ for the other half.

The construction of similar examples is of interest, but using data from a continuous distribution, ie $x \sim \mathcal{U}([0, 2\pi]^d)$.
The objective is to find functions such that $\norm{\beta_Q}^2\geq p$ , and it is questioned whether it can be compatible with the fact that $f$ should be bounded independently of $p$, ie $|f(x)|\leq 1$ for all $x$. The fact that $f$ should be bounded comes from the fact that it is the expectation value of an observable. 

In the following, a special family of Fourier models is proposed such that the norm of the weight vector is large (thus far from MNLS), and that is not concentrated. In addition, it is demonstrated that this function is bounded. If a quantum circuit architecture that realizes a function from this family can be found, the conditions for a potential quantum advantage presented previously would be satisfied. An example is found here:

\begin{restatable}[]{thm}{ThmFunctionFarMNLSNoConcentration}
\label{thm:Example_Perfect_Fct}
    Let $\Omega$ a subset of $\llbracket -L, L \rrbracket ^d$, where $L$ is an integer. Consider the following function $f : \mathbb{R}^d \longrightarrow \mathbb{R}$
    \begin{equation}
        f(x) = \frac{1}{\sqrt{p}}\sum_{\omega \in \Omega}  (\beta_{\omega, \cos}\cos(\omega^\top x) + \beta_{\omega, \sin}\sin(\omega^\top x)) \; ,
    \end{equation}
    with $p = |\Omega|$, and $\beta_{\omega,\cos}$, $\beta_{\omega,\sin}$ are all iid uniform random variables in the interval $[-\sigma, \sigma]$ with $\sigma = \Theta(1/(d\:(\log d + \log L)))$
    The following properties hold:
    \begin{enumerate}
        \item $\big|\Vert \beta \Vert^2 - \frac{2}{3}p\sigma^2\big| \leq \sigma^2 \sqrt{p\log (2/\delta)}$ with probability at least $1-\delta$.
     \item $\mathrm{Var}_x[f(x)] \geq \frac{2}{3}\sigma^2 - \frac{\sigma^2}{\sqrt{p}} \sqrt{\log (2/\delta)}$ with probability at least $1-\delta$.
     \item $\forall x \in \mathbb{R}^d, \:|f(x)| \leq 1$ with high probability.
    \end{enumerate}
\end{restatable}

In the above theorem, (1) shows that $\norm{\beta}^2$ is of the order of $p \sigma^2$ therefore of potentially higher norm than $\beta_{\text{MNLS}}$ (which scales like $M$), (2) shows that $f$ is not concentrated, and (3) shows that $f$ is bounded by a constant, which leaves open the amenability to realize it as an quantum expectation value of an observable $O$ with $\norm{O}_{\infty}$ bounded by a constant, which is a property of commonplace quantum observables. This theorem gives a function that is not impossible to achieve from a VQC, far from the corresponding classical model, and not concentrated. However, a quantum circuit capable of implementing such a function needs to be found.

\subsection{Open Questions}\label{subsec:Limitations}

In the previous Sections, it was shown how quantum models can often converge to a solution close to the MNLS estimator, and how to offer potential quantum advantage using arguments on the norm of the weight vector. A VQC could be designed based on these indications to create a model far from its classical counterpart. This section offers open questions and suggestions for future work.

First, the focus is on the analysis of usual classical gradient descent and KRR, giving rise to the bias of converging towards the MNLS estimator. Other classical learning algorithms were not explored that may not converge to the same solution.

This chapter shows how to obtain separation between quantum and classical learning models through the study of their weight vectors. This condition could be applied to VQCs with discrete or continuous input variables. However, finding a VQC with continuous variable input and without concentration issues can be challenging, while possible as explained in \autoref{thm:Example_Perfect_Fct}. The finding of such VQCs, or determining how to use those with discrete input variables for useful learning problems, must be tackled in future work. 

A change of the feature map could be envisioned, as the characteristics of the MNLS solution may then differ. The present study would need to be refined but could be adapted to any feature map. However, an intuitive case where one expects to find a quantum advantage would be when the individual components of the feature map are functions that are easy to compute on a quantum computer, but hard to do so on a classical one. Therefore even trying to train a classical surrogate wouldn't be possible. For instance, one can create a feature map inspired by cryptography \cite{shor_polynomial-time_1997,liu_rigorous_2021}, or create feature maps involving the ground state of data dependent Hamiltonians \cite{umeano_ground_2024}.

Finally, the model was envisioned to be composed of a single quantum circuit, which corresponds to most of the quantum circuits in the literature. One can imagine a succession of quantum circuits interleaved with classical post-processing, such as measurement-based adaptivity as descibed in \autoref{chap:Photonic_Suboptimal} and \autoref{chap:Subspace_Preserving_Algorithms}.

%% file: chapters/09_Conclusion_and_Outlook/Conclusion_and_Outlook.tex
\chapter{Conclusion}
\label{chap:Conclusion_Outlook}

\initial{I}n this thesis, multiple topics of Quantum Machine Learning (QML) were examined in order to identify principles for the design of industrial applications. Based on the Quantum Machine Learning challenges outlined in the introduction, the aspects of \textbf{training}, \textbf{expressivity}, and the capacity to create \textbf{surrogate models} were investigated for near-term methods based on variational quantum circuits.

This thesis focuses on quantum Fourier models in the context of dequantization. While the results presented provide theoretical criteria to identify and avoid dequantizable models, designing quantum models that are provably resistant to dequantization remains a non-trivial task, all the more so when simultaneously avoiding exponential concentration of the loss landscape.

Although many quantum computing projects are driven by the pursuit of exponential speedups over classical systems, recent research has revealed an important limitation: when QML algorithms avoid key issues such as the barren plateau problem (a vanishing-gradient phenomenon that prevents learning), this is often achieved by reducing algorithmic complexity to the point where efficient classical simulation becomes possible. In other words, avoiding failure can entail forfeiting the exponential quantum advantage that was initially sought.

A paradox thus arises at the heart of QML research: exponential advantage typically coincides with a \emph{curse of dimensionality}. Furthermore, the demonstration that a QML model can scale and perform reliably on realistic data, rather than on simplified toy problems, remains extremely challenging. Recent benchmarking studies \cite{bowles_better_2024} have shown that many widely used quantum models fail to outperform well-tuned classical baselines, particularly on tasks more complex than simple binary classification. The promise of quantum learning is therefore often diminished under real-world conditions.

In this thesis, a different approach was adopted by developing a new generation of subspace-preserving QML algorithms. By constraining learning to well-defined regions of the quantum state space, regions in which desirable properties can be rigorously guaranteed, training was maintained efficiently even as the problem scale increased, and vanishing gradients were avoided. These subspace-preserving circuits were shown to offer theoretical guarantees, such as those presented in this manuscript, which are absent from most other QML models. In this way, barren plateaus can be circumvented while still enabling meaningful quantum behaviour.

It is recognised that exponential speedup may not be achievable in the near term. However, it is also argued that a \emph{strong polynomial advantage, combined with appropriate hardware, can be sufficient} to render quantum computing practically useful, especially when efforts are focused on accelerating core building blocks of artificial intelligence. For this reason, particle number preserving architectures were investigated using photonic variational circuits, which were shown to provide significant performance within limited resource budgets.

This thesis did not confine itself to theoretical development; instead, a full-stack platform was realised comprising:
\begin{itemize}
    \item A new mathematical framework for subspace-preserving quantum machine learning;
    \item Open-source simulation tools enabling researchers to run and train these models without access to a quantum device;
    \item A new set of quantum algorithms adapted from classical analogues to both qubit and photonic platforms;
    \item A new computational scheme tailored to photonic architectures, including a state-injection strategy that preserves subspace properties while remaining within the practical limits of current optical technologies.
\end{itemize}

The last point is of particular importance. Many quantum architectures collapse under real-world conditions because they either consume excessive resources or lose the structural properties necessary for stable learning. In contrast, the photonic design proposed here was intentionally kept minimal and scalable: it employs only linear optics, a small number of photons, and introduces just enough nonlinearity to enable deep learning operations, all while maintaining the mathematical subspace that underpins the theoretical guarantees.

%% file: Appendix/Proof_HW_Preserving_Trainability.tex
\let\textcircled=\pgftextcircled
\chapter[Proof on the Trainability of Hamming Weight Preserving Quantum Circuits]{Proof on the Trainability of Hamming Weight Preserving Quantum Circuits}
\label{app:Proof_Trainability_HW_VQCs}

\section{Proof of Lemma ~\ref{lemma:VarianceHWPreserving}}\label{chap:proof_Lemma_Var}

We recall the Lemma~\ref{lemma:VarianceHWPreserving}:

\VarianceHWPreserving*
\begin{proof}

To prove this Lemma, we will first consider the unary case in Section~\ref{chap:proof_lemma_unary}, i.e., the case where the states are in $B^1_n$. In the unary case, RBS and FBS are equivalent. Then, we will show how to extend our result to any HW $k$ for RBS in Section~\ref{chap:proof_lemma_general_RBS}. Finally, we will explain in Section~\ref{chap:proof_lemma_general_FBS} how to adapt this result to the FBS case.

    \subsection{Unary case}\label{chap:proof_lemma_unary}

We consider the case of the squared Euclidean distance cost function. We call $\Delta^L$ the final error: 
\begin{equation}
    \Delta^L = 2(z^L - y) = 2[(w^{\lambda_{\mathrm{max}}} \cdots w^{\lambda + 1}) \cdot w^{\lambda} \cdot \zeta^{\lambda} - y]    
\end{equation}

We consider the case where for each inner layer, there is only one RBS gate considered. The number of parameters $D$ is equal to the number of inner layers $\lambda_{\mathrm{max}}$. We call $\zeta_j$ the amplitude of the $j^{\text{th}}$ state in the state basis considered of $\zeta$. For each inner layer, the action of the gate in the state basis $B_1^n$ results in a rotation of the amplitudes for two states that we call $(l,j)$:
\begin{equation}
    \zeta^{\lambda + 1} = \overrightarrow{cst} + (\cos(\theta_{i}) \cdot \zeta^{\lambda}_{l} + \sin(\theta_{i}) \cdot \zeta^{\lambda}_{j}) \ket{e_l} + (-\sin(\theta_{i}) \cdot \zeta^{\lambda}_{l} + \cos(\theta_{i}) \cdot \zeta^{\lambda}_{j}) \ket{e_j}  
\end{equation}
with $\overrightarrow{cst}^\intercal \cdot \ket{e_j} = \overrightarrow{cst}^\intercal \cdot \ket{e_l} = 0$

For example, the action of a RBS with parameter $\theta_i$ in $B_1^n$ on the two first qubits of a 4-qubit quantum circuit results in the $\theta_i$-planar rotation between the states $\ket{1000}$ and $\ket{0100}$.   

We can define the error according to the final error:
\begin{equation}
    \delta^{\lambda + 1} = (w^{\lambda +1})^{-1} \cdots (w^{\lambda_{\mathrm{max}}})^{-1} \cdot \Delta^l = 2 [w^{\lambda} \cdot \zeta^{\lambda} - (w^{\lambda +1})^{-1} \cdots (w^{\lambda_{\mathrm{max}}})^{-1} \cdot y]
\end{equation}

We have by orthogonality: $\forall \lambda, \quad (w^{\lambda})^{-1} = (w^{\lambda})^{t}$.
    
We call: $\tilde{y} = (w^{\lambda +1})^{t} \cdots (w^{\lambda_{\mathrm{max}}})^{t} \cdot y$ and $\Theta = [0:2\pi]^D$

We use the notation $\tilde{y}_j$ for the amplitude of the $j^{\text{th}}$ state in the state basis considered of $\tilde{y}$.

According to the backpropagation formalism, we have in the unary case:
\begin{equation}
    \frac{\partial \mathcal{C}}{\partial \theta_i} = \delta_l^{\lambda} (-\sin(\theta_i) \zeta_l^{\lambda} + \cos(\theta_i)\zeta_j^{\lambda}) + \delta_j^{\lambda} (-\cos(\theta_i) \zeta_l^{\lambda} - \sin(\theta_i)\zeta_j^{\lambda})
\end{equation}

Therefore, we can express the variance of the cost function gradient as:

\begin{equation}
    \begin{split}
    \mathrm{Var}_{\theta}[\partial_{\theta_i} \mathcal{C}(\theta)]  &= \mathrm{Var}_{\theta} [\delta_l^{\lambda} (-\sin(\theta_i) \zeta_l^{\lambda} + \cos(\theta_i)\zeta_j^{\lambda}) + \delta_j^{\lambda} (-\cos(\theta_i) \zeta_l^{\lambda} - \sin(\theta_i)\zeta_j^{\lambda})] \\
    & = \mathrm{Var}_{\theta}[\delta_l^{\lambda} (-\sin(\theta_i) \zeta_l^{\lambda} + \cos(\theta_i)\zeta_j^{\lambda})] + \mathrm{Var}_{\theta} [\delta_j^{\lambda} (-\cos(\theta_i) \zeta_l^{\lambda} - \sin(\theta_i)\zeta_j^{\lambda})] \\
    & \;+ 2 \cdot \mathbb{C}ov_{\theta} [\delta_l^{\lambda} (-\sin(\theta_i) \zeta_l^{\lambda} + \cos(\theta_i)\zeta_j^{\lambda}) ; \delta_j^{\lambda} (-\cos(\theta_i) \zeta_l^{\lambda} - \sin(\theta_i)\zeta_j^{\lambda})]
    \end{split} 
\end{equation}    

First:
\begin{equation}
    \begin{split}
    \mathrm{Var}_{\theta}&[\delta_l^{\lambda} (-\sin(\theta_i) \zeta_l^{\lambda} + \cos(\theta_i)\zeta_j^{\lambda})] = \mathrm{Var}_{\theta}[2(\cos(\theta_{i}) \cdot \zeta^{\lambda}_{l} + \sin(\theta_{i}) \cdot \zeta^{\lambda}_{j} - \tilde{y}_l)) \cdot (-\sin(\theta_i) \zeta_l^{\lambda} + \cos(\theta_i)\zeta_j^{\lambda})]\\
    & = 4 \int_{\theta \in \Theta} (\frac{1}{2\pi})^D (\delta_l^{\lambda} (-\sin(\theta_i) \zeta_l^{\lambda} + \cos(\theta_i)\zeta_j^{\lambda}) - \mathbb{E}_{\theta}[\delta_l^{\lambda} (-\sin(\theta_i) \zeta_l^{\lambda} + \cos(\theta_i)\zeta_j^{\lambda})])^2 d\theta  
    \end{split} 
\end{equation}

With:
\begin{equation}
    \begin{split}
    \mathbb{E}_{\theta}&[\delta_l^{\lambda} (-\sin(\theta_i) \zeta_l^{\lambda} + \cos(\theta_i)\zeta_j^{\lambda})] =  \int_{\theta \in \Theta} (\frac{1}{2\pi})^D (2(\cos(\theta_{i}) \cdot \zeta^{\lambda}_{l} + \sin(\theta_{i}) \cdot \zeta^{\lambda}_{j} - \tilde{y}_l)) \cdot (-\sin(\theta_i) \zeta_l^{\lambda} + \cos(\theta_i)\zeta_j^{\lambda})) d\theta \\
    & = 2 \int_{\theta \in \Theta} (\frac{1}{2\pi})^D (\cos(\theta_i) \sin(\theta_i) ((\zeta_j^{\lambda})^2 - (\zeta_l^{\lambda})^2) d\theta + 2 \int_{\theta \in \Theta} (\frac{1}{2\pi})^D (\cos^2(\theta_i) - \sin^2(\theta_i)) \zeta_j^{\lambda} \cdot \zeta_l^{\lambda} d\theta \\
    & \quad + 2 \int_{\theta \in \Theta} (\frac{1}{2\pi})^D (\sin(\theta_i) \cdot \zeta_l^{\lambda} - \cos(\theta_i) \cdot \zeta_j^{\lambda}) \tilde{y}_l d\theta
    \end{split}     
\end{equation}

According to our circuit decomposition into inner layers, the previous inner layer $\zeta^{\lambda}$ does not depend on the parameter $\theta_i$ but only on the previous parameters in the circuit. On the other hand, $\tilde{y}$ does not depend on the parameter $\theta_i$ but only on the following parameters in the circuit. Thus, we can take out the integral according to $\theta_i$:

\begin{equation}\label{eq:Expected_value_nul}
    \mathbb{E}_{\theta}[\delta_l^{\lambda} (-\sin(\theta_i) \zeta_l^{\lambda} + \cos(\theta_i)\zeta_j^{\lambda})] = 2 \int_{\theta \in \Theta \backslash \theta_i } (\frac{1}{2\pi})^D (\pi - \pi) \zeta_j^{\lambda} \cdot \zeta_l^{\lambda} d\theta = 0    
\end{equation}

Therefore:
\begin{equation}
    \begin{split}
    \mathrm{Var}_{\theta}&[\delta_l^{\lambda} (-\sin(\theta_i) \zeta_l^{\lambda} + \cos(\theta_i)\zeta_j^{\lambda})] = 4 \int_{\theta \in \Theta} (\frac{1}{2\pi})^D (\delta_l^{\lambda} (-\sin(\theta_i) \zeta_l^{\lambda} + \cos(\theta_i)\zeta_j^{\lambda}))^2 d\theta \\
    & = 4 \int_{\theta \in \Theta} (\frac{1}{2\pi})^D [\cos^2(\theta_i) \sin^2(\theta_i) \cdot ((\zeta_j^{\lambda})^2 - (\zeta_l^{\lambda})^2)^2
    + 2 \cos^3(\theta_i) \sin(\theta_i) \zeta_j^{\lambda} \zeta_l^{\lambda}((\zeta_j^{\lambda})^2 - (\zeta_l^{\lambda})^2) \\
    & - 2 \cos(\theta_i) \sin^3(\theta_i) \zeta_j^{\lambda} \zeta_l^{\lambda}((\zeta_j^{\lambda})^2 - (\zeta_l^{\lambda})^2) + 2 \cos(\theta_i) \sin^3(\theta_i) \zeta_l^{\lambda}((\zeta_j^{\lambda})^2 - (\zeta_l^{\lambda})^2)\tilde{y}_l \\
    & -2 \cos^2(\theta_i) \sin(\theta_i) \zeta_j^{\lambda}((\zeta_j^{\lambda})^2 - (\zeta_l^{\lambda})^2)\tilde{y}_l + [\cos^4(\theta_i) - 2 \cos^2(\theta_i) \sin^2(\theta_i) + \sin^4(\theta_i)](\zeta_l^{\lambda})^2(\zeta_j^{\lambda})^2 \\
    & + 2 \cos^2(\theta_i) \sin(\theta_i)(\zeta_l^{\lambda})^2 \zeta_j^{\lambda} \tilde{y}_l - 2 \sin^3(\theta_i)(\zeta_l^{\lambda})^2 \zeta_j^{\lambda} \tilde{y}_l - 2 \cos^3(\theta_i) \zeta_l^{\lambda} (\zeta_j^{\lambda})^2  \tilde{y}_l + 2 \cos(\theta_i) \sin^2(\theta_i) \zeta_l^{\lambda} (\zeta_j^{\lambda})^2  \tilde{y}_l \\
    & + \sin^2(\theta_i) (\zeta_l^{\lambda})^2 (\tilde{y}_l)^2 - 2 \cos(\theta_i) \sin(\theta_i) \zeta_l^{\lambda} \zeta_j^{\lambda} (\tilde{y}_l)^2 + \cos^2(\theta_i) (\zeta_j^{\lambda})^2 (\tilde{y}_l)^2 d\theta \\
    & = 4 \int_{\theta \in \Theta \backslash \theta_i} (\frac{1}{2\pi})^D [\frac{\pi}{4} (\zeta_l^{\lambda})^4 + \frac{\pi}{2} (\zeta_l^{\lambda})^2 (\zeta_j^{\lambda})^2 + \frac{\pi}{4} (\zeta_j^{\lambda})^4 + \pi (\zeta_l^{\lambda})^2 (\tilde{y}_l)^2 + \pi (\zeta_j^{\lambda})^2 (\tilde{y}_l)^2] d\theta
    \end{split} 
\end{equation}

Finally, we have:
\begin{equation}
    \mathrm{Var}_{\theta}[\delta_l^{\lambda} (-\sin(\theta_i) \zeta_l^{\lambda} + \cos(\theta_i)\zeta_j^{\lambda})] = \int_{\theta \in \Theta} (\frac{1}{2 \pi})^D [\frac{1}{2}(\zeta_l^{\lambda})^4 + (\zeta_l^{\lambda})^2 (\zeta_j^{\lambda})^2 + \frac{1}{2}(\zeta_j^{\lambda})^4 + 2 ((\zeta_l^{\lambda})^2 + (\zeta_j^{\lambda})^2) (\tilde{y}_l)^2] d\theta
\end{equation}
    
With the same methods, it comes:
\begin{equation}
    \mathrm{Var}_{\theta}[\delta_j^{\lambda} (-\cos(\theta_i) \zeta_l^{\lambda} - \sin(\theta_i)\zeta_j^{\lambda})] = \int_{\theta \in \Theta} (\frac{1}{2 \pi})^D [\frac{1}{2}(\zeta_l^{\lambda})^4 + (\zeta_l^{\lambda})^2 (\zeta_j^{\lambda})^2 + \frac{1}{2}(\zeta_j^{\lambda})^4 + 2 ((\zeta_l^{\lambda})^2 + (\zeta_j^{\lambda})^2) (\tilde{y}_l)^2] d\theta
\end{equation}

We can decompose the covariance term:
\begin{equation}
    \begin{split}
    \mathbb{C}ov_{\theta} & [\delta_l^{\lambda} (-\sin(\theta_i) \zeta_l^{\lambda} + \cos(\theta_i)\zeta_j^{\lambda}) ; \delta_j^{\lambda} (-\cos(\theta_i) \zeta_l^{\lambda} - \sin(\theta_i)\zeta_j^{\lambda})]\\
    & = \mathbb{E}_{\theta}[\delta_l^{\lambda} (-\sin(\theta_i) \zeta_l^{\lambda} + \cos(\theta_i)\zeta_j^{\lambda}) \cdot \delta_j^{\lambda} (-\cos(\theta_i) \zeta_l^{\lambda} - \sin(\theta_i)\zeta_j^{\lambda})]\\
    & - \mathbb{E}_{\theta}[\delta_l^{\lambda} (-\sin(\theta_i) \zeta_l^{\lambda} + \cos(\theta_i)\zeta_j^{\lambda})] \cdot \mathbb{E}_{\theta}[\delta_j^{\lambda} (-\cos(\theta_i) \zeta_l^{\lambda} - \sin(\theta_i)\zeta_j^{\lambda})]
    \end{split} 
\end{equation}

As shown with Eq.~\eqref{eq:Expected_value_nul}: 
\begin{equation}
    \mathbb{E}_{\theta}[\delta_l^{\lambda} (-\sin(\theta_i) \zeta_l^{\lambda} + \cos(\theta_i)\zeta_j^{\lambda})] = \mathbb{E}_{\theta}[\delta_j^{\lambda} (-\cos(\theta_i) \zeta_l^{\lambda} - \sin(\theta_i)\zeta_j^{\lambda})] = 0
\end{equation}

\textbf{Therefore, we have that the expectation value of the cost function gradient is null}. In addition, we have:

\begin{equation}
        \begin{split}
            \mathbb{C}ov_{\theta} & [\delta_l^{\lambda} (-\sin(\theta_i) \zeta_l^{\lambda} + \cos(\theta_i)\zeta_j^{\lambda}) ; \delta_j^{\lambda} (-\cos(\theta_i) \zeta_l^{\lambda} - \sin(\theta_i)\zeta_j^{\lambda})]\\
            & = \mathbb{E}_{\theta}[\delta_l^{\lambda} (-\sin(\theta_i) \zeta_l^{\lambda} + \cos(\theta_i)\zeta_j^{\lambda}) \cdot \delta_j^{\lambda} (-\cos(\theta_i) \zeta_l^{\lambda} - \sin(\theta_i)\zeta_j^{\lambda})] \\
            &= \int_{\theta \in \Theta} (\frac{1}{2\pi})^D [\delta_l^{\lambda} (-\sin(\theta_i) \zeta_l^{\lambda} + \cos(\theta_i)\zeta_j^{\lambda}) \cdot \delta_j^{\lambda} (-\cos(\theta_i) \zeta_l^{\lambda} - \sin(\theta_i)\zeta_j^{\lambda})] d\theta \\
            & = \int_{\theta \in \Theta} (\frac{1}{2\pi})^D [\cos^2(\theta_i) \sin^2(\theta_i)(-(\zeta^{\lambda}_l)^4 + 2(\zeta^{\lambda}_l)^2 (\zeta^{\lambda}_j)^2 - (\zeta^{\lambda}_j)^4)\\
            & + \cos(\theta_i)\sin(\theta_i)(\sin^2(\theta_i)-\cos^2(\theta_i))((\zeta^{\lambda}_j)^2 - (\zeta^{\lambda}_l)^2)\zeta^{\lambda}_l\zeta^{\lambda}_j + \cos^2(\theta_i)\sin(\theta_i)((\zeta^{\lambda}_j)^2 - (\zeta^{\lambda}_l)^2)\zeta^{\lambda}_l \tilde{y}^{\lambda}_j \\
            & - \cos(\theta_i)\sin^2(\theta_i)((\zeta^{\lambda}_j)^2 - (\zeta^{\lambda}_l)^2)\zeta^{\lambda}_j \tilde{y}^{\lambda}_j + \cos(\theta_i)\sin(\theta_i)(\cos^2(\theta_i)-\sin^2(\theta_i))((\zeta^{\lambda}_j)^2 - (\zeta^{\lambda}_l)^2)\zeta^{\lambda}_l\zeta^{\lambda}_j \\
            & - (\cos^4(\theta_i) - 2\cos^2(\theta_i)\sin^2(\theta_i) + \sin^4(\theta_i))(\zeta^{\lambda}_l\zeta^{\lambda}_j)^2 + \cos(\theta_i)(\cos^2(\theta_i)-\sin^2(\theta_i))(\zeta^{\lambda}_l)^2\zeta^{\lambda}_j\tilde{y}^{\lambda}_j \\
            & = - \sin(\theta_i)(\cos^2(\theta_i)-\sin^2(\theta_i))\zeta^{\lambda}_l(\zeta^{\lambda}_j)^2\tilde{y}^{\lambda}_j + \cos(\theta_i)\sin^2(\theta_i)((\zeta^{\lambda}_j)^2 - (\zeta^{\lambda}_l)^2)\zeta^{\lambda}_l \tilde{y}^{\lambda}_l \\
            & + \sin(\theta_i)(\sin^2(\theta_i) - \cos^2(\theta_i))(\zeta^{\lambda}_l)^2\zeta^{\lambda}_j \tilde{y}^{\lambda}_l + \cos(\theta_i) \sin(\theta_i) (\zeta^{\lambda}_l)^2 \tilde{y}^{\lambda}_l \tilde{y}^{\lambda}_j - \sin^2(\theta_i) \zeta^{\lambda}_l \zeta^{\lambda}_j \tilde{y}^{\lambda}_l \tilde{y}^{\lambda}_j \\
            & - \cos^2(\theta_i) \sin(\theta_i) ((\zeta^{\lambda}_l)^2 - (\zeta^{\lambda}_j)^2) \zeta^{\lambda}_j \tilde{y}^{\lambda}_l - \cos(\theta_i)(\sin^2(\theta_i) - \cos^2(\theta_i))\zeta^{\lambda}_l(\zeta^{\lambda}_j)^2 \tilde{y}^{\lambda}_l \\
            & + \cos^2(\theta_i) \zeta^{\lambda}_l \zeta^{\lambda}_j \tilde{y}^{\lambda}_l \tilde{y}^{\lambda}_j + \cos(\theta_i) \sin(\theta_i) \zeta^{\lambda}_j \tilde{y}^{\lambda}_l \tilde{y}^{\lambda}_j ] d\theta \\
            & = - \frac{1}{2} \int_{\theta \in \Theta} (\frac{1}{2\pi})^D (\zeta^{\lambda}_l)^4 d\theta - \frac{1}{2} \int_{\theta \in \Theta} (\frac{1}{2\pi})^D (\zeta^{\lambda}_j)^4 d\theta - \int_{\theta \in \Theta} (\frac{1}{2\pi})^D (\zeta^{\lambda}_l)^2 (\zeta^{\lambda}_j)^2 d\theta
        \end{split} 
    \end{equation}

 We can derive the variance of the cost gradient:
    \begin{equation}
        \begin{split}
        \mathrm{Var}_{\theta}[\partial_{\theta_i} \mathcal{C}(\theta)]  =& \mathrm{Var}_{\theta}[\delta_l^{\lambda} (-\sin(\theta_i) \zeta_l^{\lambda} + \cos(\theta_i)\zeta_j^{\lambda})] + \mathrm{Var}_{\theta} [\delta_j^{\lambda} (-\cos(\theta_i) \zeta_l^{\lambda} - \sin(\theta_i)\zeta_j^{\lambda})] \\
        & \;+ 2 \cdot \mathbb{C}ov_{\theta} [\delta_l^{\lambda} (-\sin(\theta_i) \zeta_l^{\lambda} + \cos(\theta_i)\zeta_j^{\lambda}) ; \delta_j^{\lambda} (-\cos(\theta_i) \zeta_l^{\lambda} - \sin(\theta_i)\zeta_j^{\lambda})] \\
        & = \int_{\theta \in \Theta} (\frac{1}{2 \pi})^D [\frac{1}{2}(\zeta_l^{\lambda})^4 + (\zeta_l^{\lambda})^2 (\zeta_j^{\lambda})^2 + \frac{1}{2}(\zeta_j^{\lambda})^4 + 2 ((\zeta_l^{\lambda})^2 + (\zeta_j^{\lambda})^2) (\tilde{y}_l)^2] d\theta \\
        & + \int_{\theta \in \Theta} (\frac{1}{2 \pi})^D [\frac{1}{2}(\zeta_l^{\lambda})^4 + (\zeta_l^{\lambda})^2 (\zeta_j^{\lambda})^2 + \frac{1}{2}(\zeta_j^{\lambda})^4 + 2 ((\zeta_l^{\lambda})^2 + (\zeta_j^{\lambda})^2) (\tilde{y}_l)^2] d\theta \\
        & - \int_{\theta \in \Theta} (\frac{1}{2\pi})^D (\zeta^{\lambda}_l)^4 d\theta - \int_{\theta \in \Theta} (\frac{1}{2\pi})^D (\zeta^{\lambda}_j)^4 d\theta - 2 \int_{\theta \in \Theta} (\frac{1}{2\pi})^D (\zeta^{\lambda}_l)^2 (\zeta^{\lambda}_j)^2 d\theta
        \end{split}
    \end{equation}

We find that:
    \begin{equation}
        \mathrm{Var}_{\theta}[\partial_{\theta_i} \mathcal{C}(\theta)]  = 2 \left(\int_{\theta \in \Theta} (\frac{1}{2\pi})^D (\zeta^{\lambda}_l)^2 + (\zeta^{\lambda}_j)^2 d\theta  \right) \cdot \left(\int_{\theta \in \Theta} (\frac{1}{2\pi})^D (\tilde{y}^{\lambda}_l)^2 + (\tilde{y}^{\lambda}_j)^2 d\theta \right) 
    \end{equation}

    \subsection{Extension to any HW for RBS based VQC}\label{chap:proof_lemma_general_RBS}

We consider once again the case of the squared Euclidean distance cost function. We call $\Delta^L$ the final error: 
    \begin{equation}
        \Delta^L = 2(z^L - y) = 2[(w^{\lambda_{\mathrm{max}}} \cdots w^{\lambda + 1}) \cdot w^{\lambda} \cdot \zeta^{\lambda} - y]    
    \end{equation}

    We still consider the case where for each inner layer, there is only one RBS gate considered. For each inner layer, the action of the gate results in a rotation of the amplitudes between a set of pair of states that we call $\mathcal{R}_{\lambda}$:
    \begin{equation}
        \zeta^{\lambda + 1} = \overrightarrow{cst} + \sum_{(l,j) \in \mathcal{R}_{\lambda}} (\cos(\theta_{i}) \cdot \zeta^{\lambda}_{l} + \sin(\theta_{i}) \cdot \zeta^{\lambda}_{j}) \ket{e_l} + (-\sin(\theta_{i}) \cdot \zeta^{\lambda}_{l} + \cos(\theta_{i}) \cdot \zeta^{\lambda}_{j}) \ket{e_j}  
    \end{equation}
    with $\forall (l,j)\in \mathcal{R}_{\lambda}, \quad \overrightarrow{cst}^\intercal \cdot \ket{e_j} = \overrightarrow{cst}^\intercal \cdot \ket{e_l} = 0$

    For example, the action of a RBS with parameter $\theta_i$ in $B_2^n$ on the two first qubits of a 4-qubit quantum circuit results in the $\theta_i$-planar rotation between the states $\ket{1010}$ and $\ket{0110}$, but also in the same $\theta_i$-planar rotation between the states $\ket{1001}$ and $\ket{0101}$.   

    We can define the error according to the final error:
    \begin{equation}
        \begin{split}
            \delta^{\lambda + 1} &= (w^{\lambda +1})^{-1} \cdots (w^{\lambda_{\mathrm{max}}})^{-1} \cdot \Delta^l = 2 [w^{\lambda} \cdot \zeta^{\lambda} - (w^{\lambda +1})^{-1} \cdots (w^{\lambda_{\mathrm{max}}})^{-1} \cdot y] \\
            &= 2[\overrightarrow{cst} + \sum_{(l,j) \in \mathcal{R}_{\lambda}} (\cos(\theta_{i}) \cdot \zeta^{\lambda}_{l} + \sin(\theta_{i}) \cdot \zeta^{\lambda}_{j}) \ket{e_l} + (-\sin(\theta_{i}) \cdot \zeta^{\lambda}_{l} + \cos(\theta_{i}) \cdot \zeta^{\lambda}_{j}) \ket{e_j} - \tilde{y}^{\lambda}]
        \end{split}
    \end{equation}
    A unique RBS affects $\binom{n-2}{k-1}$ different pairs of states with a rotation, and all the affected states are different. As before, we have:
    \begin{equation}
        \forall (l,j) \in \mathcal{R}_{\lambda}, \quad \delta^{\lambda + 1}_l = 2[\cos(\theta_{i}) \cdot \zeta^{\lambda}_{l} + \sin(\theta_{i}) \cdot \zeta^{\lambda}_{j} - \tilde{y}^{\lambda}_l]
    \end{equation}
    and,
    \begin{equation}
        \forall (l,j) \in \mathcal{R}_{\lambda}, \quad \delta^{\lambda + 1}_j = 2[-\sin(\theta_{i}) \cdot \zeta^{\lambda}_{l} + \cos(\theta_{i}) \cdot \zeta^{\lambda}_{j} - \tilde{y}^{\lambda}_j]
    \end{equation}
    
    We have by orthogonality: $\forall \lambda, \quad (w^{\lambda})^{-1} = (w^{\lambda})^{t}$.
    
    We call: $\tilde{y} = (w^{\lambda +1})^{t} \cdots (w^{\lambda_{\mathrm{max}}})^{t} \cdot y$ and $\Theta = [0:2\pi]^D$
    In the following, we will omit to note the set $\mathcal{R}_{\lambda}$.
    
    \begin{equation}
        \begin{split}
            \mathrm{Var}_{\theta}&[\partial_{\theta_i} \mathcal{C}] = \mathrm{Var}_{\theta}[\sum_{(l,j)} \delta_l^{\lambda} (-\sin(\theta_i) \zeta_l^{\lambda} + \cos(\theta_i)\zeta_j^{\lambda}) + \delta_j^{\lambda} (-\cos(\theta_i) \zeta_l^{\lambda} - \sin(\theta_i)\zeta_j^{\lambda})] \\
            & = \sum_{(l,j)} \mathrm{Var}_{\theta}[\delta_l^{\lambda} (-\sin(\theta_i) \zeta_l^{\lambda} + \cos(\theta_i)\zeta_j^{\lambda}) + \delta_j^{\lambda} (-\cos(\theta_i) \zeta_l^{\lambda} - \sin(\theta_i)\zeta_j^{\lambda})] \\
            & + 2 \sum_{(a,b) \neq (c,d)} \mathbb{C}ov_{\theta} [\delta_a^{\lambda} (-\sin(\theta_i) \zeta_a^{\lambda} + \cos(\theta_i)\zeta_b^{\lambda}) + \delta_b^{\lambda} (-\cos(\theta_i) \zeta_a^{\lambda} - \sin(\theta_i)\zeta_b^{\lambda}), \\
            & \delta_c^{\lambda} (-\sin(\theta_i) \zeta_c^{\lambda} + \cos(\theta_i)\zeta_d^{\lambda}) + \delta_d^{\lambda} (-\cos(\theta_i) \zeta_c^{\lambda} - \sin(\theta_i)\zeta_d^{\lambda})]
        \end{split}        
    \end{equation}

    Using previous result from the previous Section~\ref{chap:proof_lemma_unary}, we have:
    \begin{equation}
        \mathrm{Var}_{\theta}[\partial_{\theta_i} \mathcal{C}(\theta)]  = 2 \left(\int_{\theta \in \Theta} (\frac{1}{2\pi})^D (\zeta^{\lambda}_l)^2 + (\zeta^{\lambda}_j)^2 d\theta  \right) \cdot \left(\int_{\theta \in \Theta} (\frac{1}{2\pi})^D (\tilde{y}^{\lambda}_l)^2 + (\tilde{y}^{\lambda}_j)^2 d\theta \right)
    \end{equation}

    Then,
    \begin{equation}
        \begin{split}
            \mathbb{C}ov_{\theta}&[\delta_a^{\lambda} (-\sin(\theta_i) \zeta_a^{\lambda} + \cos(\theta_i)\zeta_b^{\lambda}) + \delta_b^{\lambda} (-\cos(\theta_i) \zeta_a^{\lambda} - \sin(\theta_i)\zeta_b^{\lambda}), \\
            & \delta_c^{\lambda} (-\sin(\theta_i) \zeta_c^{\lambda} + \cos(\theta_i)\zeta_d^{\lambda}) + \delta_d^{\lambda} (-\cos(\theta_i) \zeta_c^{\lambda} - \sin(\theta_i)\zeta_d^{\lambda})] \\
            & = \mathbb{E}_{\theta}[\left( \delta_a^{\lambda} (-\sin(\theta_i) \zeta_a^{\lambda} + \cos(\theta_i)\zeta_b^{\lambda}) + \delta_b^{\lambda} (-\cos(\theta_i) \zeta_a^{\lambda} - \sin(\theta_i)\zeta_b^{\lambda}) \right) \cdot \\
            & \left( \delta_c^{\lambda} (-\sin(\theta_i) \zeta_c^{\lambda} + \cos(\theta_i)\zeta_d^{\lambda}) + \delta_d^{\lambda} (-\cos(\theta_i) \zeta_c^{\lambda} - \sin(\theta_i)\zeta_d^{\lambda}) \right)] \\
            & - \mathbb{E}_{\theta}[\delta_a^{\lambda} (-\sin(\theta_i) \zeta_a^{\lambda} + \cos(\theta_i)\zeta_b^{\lambda}) + \delta_b^{\lambda} (-\cos(\theta_i) \zeta_a^{\lambda} - \sin(\theta_i)\zeta_b^{\lambda})] \cdot \\
            & \mathbb{E}_{\theta}[\delta_c^{\lambda} (-\sin(\theta_i) \zeta_c^{\lambda} + \cos(\theta_i)\zeta_d^{\lambda}) + \delta_d^{\lambda} (-\cos(\theta_i) \zeta_c^{\lambda} - \sin(\theta_i)\zeta_d^{\lambda})]
        \end{split}        
    \end{equation}

    Using previous result from the previous Section~\ref{chap:proof_lemma_unary} we have:
    \begin{equation}
        \begin{split}
            \mathbb{E}_{\theta}&[\delta_a^{\lambda} (-\sin(\theta_i) \zeta_a^{\lambda} + \cos(\theta_i)\zeta_b^{\lambda}) + \delta_b^{\lambda} (-\cos(\theta_i) \zeta_a^{\lambda} - \sin(\theta_i)\zeta_b^{\lambda})] \\
            & = \mathbb{E}_{\theta}[\delta_c^{\lambda} (-\sin(\theta_i) \zeta_c^{\lambda} + \cos(\theta_i)\zeta_d^{\lambda}) + \delta_d^{\lambda} (-\cos(\theta_i) \zeta_c^{\lambda} - \sin(\theta_i)\zeta_d^{\lambda})] \\
            & = 0
        \end{split}        
    \end{equation}

    We can now derive the value of the covariance term:
    \begin{equation}
        \begin{split}
            \mathbb{C}ov_{\theta}&[\delta_a^{\lambda} (-\sin(\theta_i) \zeta_a^{\lambda} + \cos(\theta_i)\zeta_b^{\lambda}) + \delta_b^{\lambda} (-\cos(\theta_i) \zeta_a^{\lambda} - \sin(\theta_i)\zeta_b^{\lambda}), \\
            & \delta_c^{\lambda} (-\sin(\theta_i) \zeta_c^{\lambda} + \cos(\theta_i)\zeta_d^{\lambda}) + \delta_d^{\lambda} (-\cos(\theta_i) \zeta_c^{\lambda} - \sin(\theta_i)\zeta_d^{\lambda})] \\
            & = \mathbb{E}_{\theta}[\left( \delta_a^{\lambda} (-\sin(\theta_i) \zeta_a^{\lambda} + \cos(\theta_i)\zeta_b^{\lambda}) + \delta_b^{\lambda} (-\cos(\theta_i) \zeta_a^{\lambda} - \sin(\theta_i)\zeta_b^{\lambda}) \right) \cdot \\
            & \left( \delta_c^{\lambda} (-\sin(\theta_i) \zeta_c^{\lambda} + \cos(\theta_i)\zeta_d^{\lambda}) + \delta_d^{\lambda} (-\cos(\theta_i) \zeta_c^{\lambda} - \sin(\theta_i)\zeta_d^{\lambda}) \right)] \\
            & = \mathbb{E}_{\theta}[\delta_a^{\lambda} (-\sin(\theta_i) \zeta_a^{\lambda} + \cos(\theta_i)\zeta_b^{\lambda}) \cdot \delta_c^{\lambda} (-\sin(\theta_i) \zeta_c^{\lambda} + \cos(\theta_i)\zeta_d^{\lambda})] \\
            & + \mathbb{E}_{\theta}[\delta_a^{\lambda} (-\sin(\theta_i) \zeta_a^{\lambda} + \cos(\theta_i)\zeta_b^{\lambda}) \cdot \delta_d^{\lambda} (-\cos(\theta_i) \zeta_c^{\lambda} - \sin(\theta_i)\zeta_d^{\lambda})] \\
            & + \mathbb{E}_{\theta}[\delta_b^{\lambda} (-\cos(\theta_i) \zeta_a^{\lambda} - \sin(\theta_i)\zeta_b^{\lambda}) \cdot \delta_c^{\lambda} (-\sin(\theta_i) \zeta_c^{\lambda} + \cos(\theta_i)\zeta_d^{\lambda})] \\
            & + \mathbb{E}_{\theta}[\delta_b^{\lambda} (-\cos(\theta_i) \zeta_a^{\lambda} - \sin(\theta_i)\zeta_b^{\lambda}) \cdot \delta_d^{\lambda} (-\cos(\theta_i) \zeta_c^{\lambda} - \sin(\theta_i)\zeta_d^{\lambda})] 
        \end{split}        
    \end{equation}

    Using the integral expression of the expectation value:
    \begin{equation}
        \begin{split}
            \mathbb{E}_{\theta}&[\delta_a^{\lambda} (-\sin(\theta_i) \zeta_a^{\lambda} + \cos(\theta_i)\zeta_b^{\lambda}) \cdot \delta_d^{\lambda} (-\cos(\theta_i) \zeta_c^{\lambda} - \sin(\theta_i)\zeta_d^{\lambda})] \\
            & = \int_{\theta \in \Theta} (\frac{1}{2\pi})^D[\frac{1}{2}\left( (\zeta^{\lambda}_b)^2(\zeta^{\lambda}_d)^2 - (\zeta^{\lambda}_b)^2(\zeta^{\lambda}_c)^2 -(\zeta^{\lambda}_a)^2(\zeta^{\lambda}_d)^2 + (\zeta^{\lambda}_a)^2(\zeta^{\lambda}_c)^2 \right) \\
            & + 2 \zeta^{\lambda}_a \zeta^{\lambda}_b \zeta^{\lambda}_c \zeta^{\lambda}_d + 2 \zeta^{\lambda}_a \zeta^{\lambda}_c \tilde{y}^{\lambda}_a \tilde{y}^{\lambda}_c + 2 \zeta^{\lambda}_b \zeta^{\lambda}_d \tilde{y}^{\lambda}_a \tilde{y}^{\lambda}_c] d\theta
        \end{split}        
    \end{equation}
    
    \begin{equation}
        \begin{split}
            \mathbb{E}_{\theta}&[\delta_a^{\lambda} (-\sin(\theta_i) \zeta_a^{\lambda} + \cos(\theta_i)\zeta_b^{\lambda}) \cdot \delta_c^{\lambda} (-\sin(\theta_i) \zeta_c^{\lambda} + \cos(\theta_i)\zeta_d^{\lambda})] \\
            & = \int_{\theta \in \Theta} (\frac{1}{2\pi})^D[\frac{1}{2}\left( (\zeta^{\lambda}_b)^2(\zeta^{\lambda}_c)^2 - (\zeta^{\lambda}_a)^2(\zeta^{\lambda}_c)^2 -(\zeta^{\lambda}_b)^2(\zeta^{\lambda}_d)^2 + (\zeta^{\lambda}_a)^2(\zeta^{\lambda}_d)^2 \right) \\
            & - 2 \zeta^{\lambda}_a \zeta^{\lambda}_b \zeta^{\lambda}_c \zeta^{\lambda}_d + 2 \zeta^{\lambda}_a \zeta^{\lambda}_d \tilde{y}^{\lambda}_a \tilde{y}^{\lambda}_d + 2 \zeta^{\lambda}_b \zeta^{\lambda}_c \tilde{y}^{\lambda}_a \tilde{y}^{\lambda}_d] d\theta
        \end{split}        
    \end{equation}

    \begin{equation}
        \begin{split}
            \mathbb{E}_{\theta}&[\delta_b^{\lambda} (-\cos(\theta_i) \zeta_a^{\lambda} - \sin(\theta_i)\zeta_b^{\lambda}) \cdot \delta_c^{\lambda} (-\sin(\theta_i) \zeta_c^{\lambda} + \cos(\theta_i)\zeta_d^{\lambda})] \\
            & = \int_{\theta \in \Theta} (\frac{1}{2\pi})^D[\frac{1}{2}\left( (\zeta^{\lambda}_a)^2(\zeta^{\lambda}_d)^2 - (\zeta^{\lambda}_b)^2(\zeta^{\lambda}_d)^2 -(\zeta^{\lambda}_a)^2(\zeta^{\lambda}_c)^2 + (\zeta^{\lambda}_b)^2(\zeta^{\lambda}_c)^2 \right) \\
            & - 2 \zeta^{\lambda}_a \zeta^{\lambda}_b \zeta^{\lambda}_c \zeta^{\lambda}_d + 2 \zeta^{\lambda}_a \zeta^{\lambda}_d \tilde{y}^{\lambda}_b \tilde{y}^{\lambda}_c + 2 \zeta^{\lambda}_b \zeta^{\lambda}_c \tilde{y}^{\lambda}_b \tilde{y}^{\lambda}_c] d\theta
        \end{split}        
    \end{equation}

    \begin{equation}
        \begin{split}
            \mathbb{E}_{\theta}&[\delta_b^{\lambda} (-\cos(\theta_i) \zeta_a^{\lambda} - \sin(\theta_i)\zeta_b^{\lambda}) \cdot \delta_d^{\lambda} (-\cos(\theta_i) \zeta_c^{\lambda} - \sin(\theta_i)\zeta_d^{\lambda})]  \\
            & = \int_{\theta \in \Theta} (\frac{1}{2\pi})^D[\frac{1}{2}\left( (\zeta^{\lambda}_a)^2(\zeta^{\lambda}_c)^2 - (\zeta^{\lambda}_a)^2(\zeta^{\lambda}_d)^2 -(\zeta^{\lambda}_b)^2(\zeta^{\lambda}_c)^2 + (\zeta^{\lambda}_b)^2(\zeta^{\lambda}_d)^2 \right) \\
            & - 2 \zeta^{\lambda}_a \zeta^{\lambda}_b \zeta^{\lambda}_c \zeta^{\lambda}_d + 2 \zeta^{\lambda}_a \zeta^{\lambda}_c \tilde{y}^{\lambda}_b \tilde{y}^{\lambda}_d + 2 \zeta^{\lambda}_b \zeta^{\lambda}_d \tilde{y}^{\lambda}_b \tilde{y}^{\lambda}_d] d\theta
        \end{split}        
    \end{equation}

    By summing the terms:

    \begin{equation}
        \begin{split}
            \mathbb{C}ov_{\theta}&[\delta_a^{\lambda} (-\sin(\theta_i) \zeta_a^{\lambda} + \cos(\theta_i)\zeta_b^{\lambda}) + \delta_b^{\lambda} (-\cos(\theta_i) \zeta_a^{\lambda} - \sin(\theta_i)\zeta_b^{\lambda}), \\
            & \delta_c^{\lambda} (-\sin(\theta_i) \zeta_c^{\lambda} + \cos(\theta_i)\zeta_d^{\lambda}) + \delta_d^{\lambda} (-\cos(\theta_i) \zeta_c^{\lambda} - \sin(\theta_i)\zeta_d^{\lambda})] \\
            & = 2 \int_{\theta \in \Theta} (\frac{1}{2\pi})^D [(\zeta_a^{\lambda} \zeta_c^{\lambda} - \zeta_b^{\lambda} \zeta_d^{\lambda})(\tilde{y}^{\lambda}_a \tilde{y}^{\lambda}_c + \tilde{y}^{\lambda}_b \tilde{y}^{\lambda}_d) + (\zeta_a^{\lambda} \zeta_d^{\lambda} - \zeta_b^{\lambda} \zeta_c^{\lambda})(\tilde{y}^{\lambda}_a \tilde{y}^{\lambda}_d + \tilde{y}^{\lambda}_b \tilde{y}^{\lambda}_c)] d\theta
        \end{split}
    \end{equation}

    Finally,

    \begin{equation}
        \begin{split}
            \mathrm{Var}_{\theta}&[\partial_{\theta_i} \mathcal{C}] =  2 \sum_{(l,j)} \left(\int_{\theta \in \Theta} (\frac{1}{2\pi})^D (\zeta^{\lambda}_l)^2 + (\zeta^{\lambda}_j)^2 d\theta  \right) \cdot \left(\int_{\theta \in \Theta} (\frac{1}{2\pi})^D (\tilde{y}^{\lambda}_l)^2 + (\tilde{y}^{\lambda}_j)^2 d\theta \right) \\
            & + 4 \sum_{(a,b) \neq (c,d)} \int_{\theta \in \Theta} (\frac{1}{2\pi})^D [(\zeta_a^{\lambda} \zeta_c^{\lambda} + \zeta_b^{\lambda} \zeta_d^{\lambda})(\tilde{y}^{\lambda}_a \tilde{y}^{\lambda}_c + \tilde{y}^{\lambda}_b \tilde{y}^{\lambda}_d) + (\zeta_a^{\lambda} \zeta_d^{\lambda} - \zeta_b^{\lambda} \zeta_c^{\lambda})(\tilde{y}^{\lambda}_a \tilde{y}^{\lambda}_d - \tilde{y}^{\lambda}_b \tilde{y}^{\lambda}_c)] d\theta
        \end{split}
    \end{equation}

    We now claim that the covariance term is null. To show this, we are about to use an induction proof to prove that $\forall \lambda, \forall a,b \in [d_k], \int_{\theta \in \Theta} (\frac{1}{2\pi})^D \zeta_a^{\lambda} \zeta_b^{\lambda} d\theta = 0 $ while considering our assumption on the input and target output state distribution. First for the basis:
    \begin{equation}
        \forall a,b \in [d_k], \quad \int_{\theta \in \Theta} (\frac{1}{2\pi})^D \zeta_a^{0} \zeta_b^{0} d\theta = \zeta_a^{0} \zeta_b^{0} \int_{\theta \in \Theta} (\frac{1}{2\pi})^D d\theta = \zeta_a^{0} \zeta_b^{0}
    \end{equation}

    Considering the expectation value on the input vectors $\zeta^0$, it comes:
    \begin{equation}
        \forall a,b \in [d_k], \quad \mathbb{E}_{\zeta^0}[\zeta_a^{0} \zeta_b^{0}] = 0
    \end{equation}

    In the following, we consider the expectation value over the input state and output state distribution. For the induction part, we just need to use the recursive formula on the evolution of the inner states. Let us consider the property verified on then layer $\lambda$:
    \begin{equation}
        \forall a,b \in [d_k], \quad \int_{\theta \in \Theta} (\frac{1}{2\pi})^D \zeta_a^{\lambda+1} \zeta_b^{\lambda+1} d\theta = 
        \begin{cases}
            \frac{1}{2}\int_{\theta \in \Theta} (\frac{1}{2\pi})^D \zeta_{a,a'}^{\lambda} \zeta_{b,b'}^{\lambda} d\theta + \frac{1}{2}\int_{\theta \in \Theta} (\frac{1}{2\pi})^D \zeta_{a',a}^{\lambda} \zeta_{b',b}^{\lambda} d\theta = 0 \\
            \int_{\theta \in \Theta} (\frac{1}{2\pi})^D (\pm \cos(\theta_i) \zeta_{a,a'}^{\lambda} \pm \sin(\theta_i) \zeta_{a',a}^{\lambda})  \zeta_b^{\lambda} d\theta = 0  \\
            \int_{\theta \in \Theta} (\frac{1}{2\pi})^D \zeta_a^{\lambda} (\pm \cos(\theta_i) \zeta_{b,b'}^{\lambda} \pm \sin(\theta_i) \zeta_{b',b}^{\lambda}) d\theta = 0  \\
            \int_{\theta \in \Theta} (\frac{1}{2\pi})^D \zeta_a^{\lambda} \zeta_b^{\lambda} d\theta = 0
        \end{cases}
    \end{equation}

    The first case corresponds to the situation where $a$ and $b$ are affected by a rotation in the previous layer $\lambda$. The second and the third case refer to situation where only one state between $a$ and $b$ is affected by a rotation in the previous layer. By integrating over the corresponding parameter, those value are null. Finally, we consider the case where none of them are affected by a rotation in the previous layer. In all the cases if $a \neq b$, the result is zero due to the induction hypothesis. Therefore, we have that:
    \begin{equation}
        \forall \lambda, \forall a,b \in [d_k], \quad a \neq b \implies \quad \int_{\theta \in \Theta} (\frac{1}{2\pi})^D \zeta_a^{\lambda} \zeta_b^{\lambda} d\theta = 0
    \end{equation}

    Which implies that the covariance term is equal to $0$:
    \begin{equation}
        \sum_{(a,b) \neq (c,d)} \int_{\theta \in \Theta} (\frac{1}{2\pi})^D [(\zeta_a^{\lambda} \zeta_c^{\lambda} + \zeta_b^{\lambda} \zeta_d^{\lambda})(\tilde{y}^{\lambda}_a \tilde{y}^{\lambda}_c + \tilde{y}^{\lambda}_b \tilde{y}^{\lambda}_d) + (\zeta_a^{\lambda} \zeta_d^{\lambda} - \zeta_b^{\lambda} \zeta_c^{\lambda})(\tilde{y}^{\lambda}_a \tilde{y}^{\lambda}_d - \tilde{y}^{\lambda}_b \tilde{y}^{\lambda}_c)] d\theta = 0
    \end{equation}

    Finally, we have that:
    
    \begin{equation}
            \mathrm{Var}_{\theta}[\partial_{\theta_i} \mathcal{C}] =  2 \sum_{(l,j)} \left(\int_{\theta \in \Theta} (\frac{1}{2\pi})^D (\zeta^{\lambda}_l)^2 + (\zeta^{\lambda}_j)^2 d\theta  \right) \cdot \left(\int_{\theta \in \Theta} (\frac{1}{2\pi})^D (\tilde{y}^{\lambda}_l)^2 + (\tilde{y}^{\lambda}_j)^2 d\theta \right)
    \end{equation}

We have proved Lemma~\ref{lemma:VarianceHWPreserving} in the specific case of RBS gates. In the following section we show how to adapt the previous proof to FBS gates.

    \subsection{Generalization for FBS gates}\label{chap:proof_lemma_general_FBS}

We consider once again the case of the squared Euclidean cost function. The decomposition of the FBS based quantum circuit for the backpropagation method is very similar to the case of RBS based quantum circuit:
    
    We call $\Delta^L$ the final error: 
    \begin{equation}
        \Delta^L = 2(z^L - y) = 2[(w^{\lambda_{\mathrm{max}}} \cdots w^{\lambda + 1}) \cdot w^{\lambda} \cdot \zeta^{\lambda} - y]    
    \end{equation}

    We consider the case where for each inner layer, there is only one FBS gate considered. For each inner layer, the action of the gate results in a rotation of the amplitudes between a set of pair of states that we call $\mathcal{R}_{\lambda}$:
    \begin{equation}
        \zeta^{\lambda + 1} = \overrightarrow{cst} + \sum_{(l,j) \in \mathcal{R}_{\lambda}} (\cos(\theta_{i}) \cdot \zeta^{\lambda}_{l} + (-1)^{f(l,j, \zeta^{\lambda}_{j})} \sin(\theta_{i}) \cdot \zeta^{\lambda}_{j}) \ket{e_l} + ( (-1)^{f(l,j, \zeta^{\lambda}_{l}) + 1} \sin(\theta_{i}) \cdot \zeta^{\lambda}_{l} + \cos(\theta_{i}) \cdot \zeta^{\lambda}_{j}) \ket{e_j}  
    \end{equation}
    with $\forall (l,j)\in \mathcal{R}_{\lambda}, \quad \overrightarrow{cst}^\intercal \cdot \ket{e_j} = \overrightarrow{cst}^\intercal \cdot \ket{e_l} = 0$

    We can define the error according to the final error:
    \begin{equation}
        \begin{split}
            \delta^{\lambda + 1} &= (w^{\lambda +1})^{-1} \cdots (w^{\lambda_{\mathrm{max}}})^{-1} \cdot \Delta^l = 2 [w^{\lambda} \cdot \zeta^{\lambda} - (w^{\lambda +1})^{-1} \cdots (w^{\lambda_{\mathrm{max}}})^{-1} \cdot y] \\
            &= 2[\overrightarrow{cst} + \sum_{(l,j) \in \mathcal{R}_{\lambda}} (\cos(\theta_{i}) \cdot \zeta^{\lambda}_{l} + (-1)^{f(l,j, \zeta^{\lambda}_{j})} \sin(\theta_{i}) \cdot \zeta^{\lambda}_{j}) \ket{e_l} \\
            & + ((-1)^{f(l,j, \zeta^{\lambda}_{l}) + 1} \sin(\theta_{i}) \cdot \zeta^{\lambda}_{l} + \cos(\theta_{i}) \cdot \zeta^{\lambda}_{j}) \ket{e_j} - \tilde{y}^{\lambda}]
        \end{split}
    \end{equation}
    A unique RBS affects $\binom{n-2}{k-1}$ different pairs of states with a rotation, and all the affected states are different. A unique FBS applied on the same qubits affects the same pairs of states. Therefore:
    \begin{equation}
        \forall (l,j) \in \mathcal{R}_{\lambda}, \quad \delta^{\lambda + 1}_l = 2[\cos(\theta_{i}) \cdot \zeta^{\lambda}_{l} + (-1)^{f(l,j, \zeta^{\lambda}_{j})} \sin(\theta_{i}) \cdot \zeta^{\lambda}_{j} - \tilde{y}^{\lambda}_l]
    \end{equation}
    and,
    \begin{equation}
        \forall (l,j) \in \mathcal{R}_{\lambda}, \quad \delta^{\lambda + 1}_j = 2[(-1)^{f(l,j, \zeta^{\lambda}_{l})+1}\sin(\theta_{i}) \cdot \zeta^{\lambda}_{l} + \cos(\theta_{i}) \cdot \zeta^{\lambda}_{j} - \tilde{y}^{\lambda}_j]
    \end{equation}
    
    Note that by orthogonality: $\forall \lambda, \quad (w^{\lambda})^{-1} = (w^{\lambda})^{t}$.
    
    We call: $\tilde{y} = (w^{\lambda +1})^{t} \cdots (w^{\lambda_{\mathrm{max}}})^{t} \cdot y$ and $\Theta = [0:2\pi]^D$
    In the following, we will omit to note the set $\mathcal{R}_{\lambda}$.
    
    \begin{equation}\label{eq:FBS_developped_expression_var_general_case}
        \begin{split}
            \mathrm{Var}_{\theta}[\partial_{\theta_i} \mathcal{C}] &= \mathrm{Var}_{\theta}[\sum_{(l,j)} \delta_l^{\lambda} (-\sin(\theta_i) \zeta_l^{\lambda} + (-1)^{f(l,j,\zeta_j^{\lambda})} \cos(\theta_i)\zeta_j^{\lambda}) + \delta_j^{\lambda} ((-1)^{f(l,j,\zeta_l^{\lambda})}\cos(\theta_i) \zeta_l^{\lambda} - \sin(\theta_i)\zeta_j^{\lambda})] \\
            & = \sum_{(l,j)} \mathrm{Var}_{\theta}[\delta_l^{\lambda} (-\sin(\theta_i) \zeta_l^{\lambda} + (-1)^{f(l,j,\zeta_j^{\lambda})} \cos(\theta_i)\zeta_j^{\lambda}) + \delta_j^{\lambda} ((-1)^{f(l,j,\zeta_l^{\lambda})}\cos(\theta_i) \zeta_l^{\lambda} - \sin(\theta_i)\zeta_j^{\lambda})] \\
            & + 2 \sum_{(a,b) \neq (c,d)} \mathbb{C}ov_{\theta} [\delta_a^{\lambda} (-\sin(\theta_i) \zeta_a^{\lambda} + (-1)^{f(a,b,\zeta_b^{\lambda})} \cos(\theta_i)\zeta_b^{\lambda}) + \delta_b^{\lambda} ((-1)^{f(a,b,\zeta_a^{\lambda})}\cos(\theta_i) \zeta_a^{\lambda} - \sin(\theta_i)\zeta_b^{\lambda}), \\
            & \delta_c^{\lambda} (-\sin(\theta_i) \zeta_c^{\lambda} + (-1)^{f(c,d,\zeta_d^{\lambda})} \cos(\theta_i)\zeta_d^{\lambda}) + \delta_d^{\lambda} ((-1)^{f(c,d,\zeta_c^{\lambda})}\cos(\theta_i) \zeta_c^{\lambda} - \sin(\theta_i)\zeta_d^{\lambda})]
        \end{split}        
    \end{equation}

    The only change in comparison with the RBS case is with the sign of the sine in the previous equations. However, in the proof of Theorem~\ref{thm:NoBPgeneralcase} given in the previous appendix, we derive each term of Eq.~\eqref{eq:FBS_developped_expression_var_general_case} by integrating over the parameter $\theta_i$. The sign of the sine does not change the integration and as a result we find again that: 
    
    \begin{equation}
        \begin{split}
            \mathrm{Var}_{\theta}&[\partial_{\theta_i} \mathcal{C}] =  2 \sum_{(l,j)} \left(\int_{\theta \in \Theta} (\frac{1}{2\pi})^D (\zeta^{\lambda}_l)^2 + (\zeta^{\lambda}_j)^2 d\theta  \right) \cdot \left(\int_{\theta \in \Theta} (\frac{1}{2\pi})^D (\tilde{y}^{\lambda}_l)^2 + (\tilde{y}^{\lambda}_j)^2 d\theta \right) \\
            & + 4 \sum_{(a,b) \neq (c,d)} \int_{\theta \in \Theta} (\frac{1}{2\pi})^D [(\zeta_a^{\lambda} \zeta_c^{\lambda} + \zeta_b^{\lambda} \zeta_d^{\lambda})(\tilde{y}^{\lambda}_a \tilde{y}^{\lambda}_c + \tilde{y}^{\lambda}_b \tilde{y}^{\lambda}_d) + (\zeta_a^{\lambda} \zeta_d^{\lambda} - \zeta_b^{\lambda} \zeta_c^{\lambda})(\tilde{y}^{\lambda}_a \tilde{y}^{\lambda}_d - \tilde{y}^{\lambda}_b \tilde{y}^{\lambda}_c)] d\theta
        \end{split}
    \end{equation}

We can show in a similar way that the covariance term is null. Finally, we have that:
 \begin{equation}
            \mathrm{Var}_{\theta}[\partial_{\theta_i} \mathcal{C}] =  2 \sum_{(l,j)} \left(\int_{\theta \in \Theta} (\frac{1}{2\pi})^D (\zeta^{\lambda}_l)^2 + (\zeta^{\lambda}_j)^2 d\theta  \right) \cdot \left(\int_{\theta \in \Theta} (\frac{1}{2\pi})^D (\tilde{y}^{\lambda}_l)^2 + (\tilde{y}^{\lambda}_j)^2 d\theta \right)
    \end{equation}
    
\textbf{We have proved Lemma~\ref{lemma:VarianceHWPreserving}.}
\end{proof}

    \section{Formalization and proof of Theorem~\ref{thm:NoBPPSA}}\label{sec:ProofNoBPPSA}
The goal of this section is to make precise and generalize the informal \autoref{thm:NoBPPSA} from the main text, and to prove it.
Let us recall that theorem for convenience:

\NoBPPSA*

In \autoref{subchap:InductiveRelation}, we provide a recurrence relation describing how the squared entries of the intermediate quantum states propagate back and forth throughout the circuit, which will be key in obtaining a precise understanding of how the variance evolves. In \autoref{subchap:StochasticMatrices}, we introduce the concept of stochastic matrices in order to, in \autoref{subchap:variance-formula-recast}, recast the above relation into that language, which will let the final variance quantity be understood through the convergence of certain stochastic matrices powers. In \autoref{subchap:irreducible-and-primitive-stochmats}, we introduce special classes of stochastic matrices, and we show in the following \autoref{subchap:connected-stochmats-props} that the general ansätze we consider in the main text do belong in one such class called primitive stochastic matrices.
Afterwards, we illustrate in \autoref{subchap:convergence-fixed} how the convergence of powers of a fixed stochastic matrix is generally studied, after which we promote, in \autoref{subchap:convergence-sequence-and-conjecture}, the case of single stochastic matrix, to the case of a sequence thereof. For this result to carry through properly, we make there a minor conjecture concerning the structure of the stochastic matrices induced from the general RBS/FBS circuit ansätze considered.
All the previous results are put together in \autoref{subchap:precised-theorem-2}, to obtain \autoref{thm:concluding-back-to-variance}, the precised version of the theorem shown above. In \autoref{subsec:numerical-evidence-spectral-gap}, we provide numerical evidence for the previous conjecture, and lastly in \autoref{subchap:design-discussion} we discuss the differences of our approach with an approach that argues using 
a closeness to unitary $2$-design assumption.

\subsection{Recurrence relation for squared entries of intermediate states}\label{subchap:InductiveRelation}
    
\begin{lemma}[Recurrence relation for squared entries of intermediate states]
\label{lemma:InductiveRelation}
Let us consider a $n$-qubit HW-preserving VQC made of RBS or FBS gates. We consider here a training in the subspace of HW $k$, i.e., corresponding to the basis $B_k^n$.
$\zeta_r^{\lambda}$ denotes the $r^{\mathrm{th}}$ entry (in the basis $B_k^n$) of the intermediate state $\zeta^{\lambda}$.
We have the following recurrence relation between the squared entries of the intermediate state $\zeta^{\lambda+1}$ and those of the previous state $\zeta^\lambda$:
\begin{equation}\label{eq:Inductive_relation}
    \forall r \in [d_k], \quad \int_{\theta \in \Theta} (\frac{1}{2\pi})^D (\zeta_r^{\lambda+1})^2 d\theta = \begin{cases}
        \frac{1}{2}\int_{\theta \in \Theta} (\frac{1}{2\pi})^D (\zeta_{r,r'}^{\lambda})^2 d\theta + \frac{1}{2}\int_{\theta \in \Theta} (\frac{1}{2\pi})^D (\zeta_{r',r}^{\lambda})^2 d\theta \\ \mathrm{or} \quad
        \int_{\theta \in \Theta} (\frac{1}{2\pi})^D (\zeta_r^{\lambda})^2 d\theta\,,
    \end{cases}
\end{equation}
depending on whether or not the state $\ket{e_r}\in B_k^n$ from the computational basis is undergoing a planar rotation with another state $\ket{e_{r'}}$ due to the layer $\lambda$.
\end{lemma}

\begin{proof}
    Eq.~\eqref{eq:Inductive_relation} states that, because the inner states $\zeta^{\lambda+1}$ and $\zeta^{\lambda}$ are connected through the application of one RBS/FBS, and because each RBS/FBS performs $\binom{n-2}{k-1}$ $\theta$-planar rotation between states in the computational basis, a index $r$ of $\zeta^{\lambda+1}$ can take two value:
\begin{itemize}
    \item If the state $\ket{e_r}$ from the computational basis is involved in a planar rotation due to the RBS/FBS, then:
    \begin{equation}\label{eq:Inductive_relation-proof-eq-a}
        \zeta_r^{\lambda+1} = \cos(\theta) \zeta_{r,r'}^{\lambda} \pm \sin(\theta) \zeta_{r',r}^{\lambda}\,,
    \end{equation}
    and thus, we have that:
    \begin{equation}
        \int_{\theta \in \Theta} (\frac{1}{2\pi})^D (\zeta_r^{\lambda+1})^2 d\theta = \frac{1}{2}\int_{\theta \in \Theta} (\frac{1}{2\pi})^D (\zeta_{r,r'}^{\lambda})^2 d\theta + \frac{1}{2}\int_{\theta \in \Theta} (\frac{1}{2\pi})^D (\zeta_{r',r}^{\lambda})^2 d\theta\,.
    \end{equation}
    \item In the opposite case, then we simply have that $\zeta_r^{\lambda+1} = \zeta_{r}^{\lambda}$, and therefore:
    \begin{equation}\label{eq:Inductive_relation-proof-eq-b}
        \int_{\theta \in \Theta} (\frac{1}{2\pi})^D (\zeta_r^{\lambda+1})^2 d\theta = \int_{\theta \in \Theta} (\frac{1}{2\pi})^D (\zeta_r^{\lambda})^2 d\theta\,.
    \end{equation}
\end{itemize}
\end{proof}

This relation can also be stated for the propagation of $\int_{\theta \in \Theta} (\frac{1}{2\pi})^D (\tilde{y}_r^{\lambda})^2 d\theta$, as the backpropagation of the target state only apply planar rotation to this state.

\subsection{Introducing stochastic matrices }\label{subchap:StochasticMatrices}

\paragraph{Stochastic matrices}

Let $T \in \mathbb{R}^{N\times N}$. $T$ is said to be a \emph{column-stochastic} matrix if its columns are probability vectors, i.e. $T_{ij} \geq 0 \ \forall i,j \in [N]$ and $\sum_i T_{ij} =1\ \forall j \in [N]$. Likewise, $T$ is \emph{row-stochastic} if its rows are probability vectors, and is \emph{doubly stochastic} if it is both column-stochastic and row-stochastic.  In the rest of this work, if a matrix is just said to be stochastic (without more precision), we mean to say that is is column-stochastic.

The purpose of defining  column-stochastic matrices is that they are exactly the matrices $M$ for which, when acting on a \textit{probability vector} $\vec{p}$ (i.e. a vector of nonnegative entries that all sum to $1$), the output $M \vec{p}$ remains a probability vector.

Let us note that all three of these subsets of matrices introduced are topologically closed in $\mathbb{R}^{N\times N}$, and therefore if for instance a sequence of doubly stochastic matrices $(T_n)$ converges (with respect to a matrix norm) to a matrix $T_{\infty}$, then $T_{\infty}$ is still doubly stochastic. Furthermore, note that all three of these subsets of matrices are closed under taking a product of two elements.

\paragraph{Eigenvalues, and spectral gap}

Note that $1$ is always an eigenvalue of a column-stochastic matrix $T$, since $T^\intercal$ has the eigenvector $(1,\dots,1)^\intercal$ associated to eigenvalue $1$.

Furthermore, by direct consequence of the Gershgorin circle theorem (\cite[Theorem 6.1.1]{Horn-MatrixAnalysis-2013}), all eigenvalues of a column-stochastic matrix lie in the complex closed unit disk.

Given a column-stochastic matrix $T$, consider studying the behaviour of the sequence of its powers $T^n$ as $n$ increases. A special case of this which comes with the clearest  intuition is when $T$ is diagonalizable. Indeed, since the diagonalization of $T$ provides an expression of the form $T=P \text{diag}(\lambda_1,\dots,\lambda_N) P^{-1}$, so $T^n=P \text{diag}(\lambda_1^n,\dots,\lambda_N^n) P^{-1}$, and thus the behavior of $T^n$ is in that case understood through an exponential vanishing of all its eigenvalues except those of unit modulus. Thus in the case when $1$ is the only unit-modulus eigenvalue, the sequence $(T^n)$ would be converging exponentially fast to a matrix $T_{\infty}$, with the exponential rate being governed by the second largest eigenvalue modulus, denoted $|\lambda_2|$; and the further $|\lambda_2|$ is from $1$, the faster is the rate of the exponential convergence.

In this spirit, we will usually denote by $\Delta:=1-|\lambda_2|$ this gap for the stochastic-matrix $T$, which we will refer to as the \emph{spectral gap} of $T$.

One should keep in mind, though, that in general a stochastic matrix $T$ could have other eigenvalues of unit modulus besides $1$, in which case the sequence $(T^n)$ would not even be necessarily convergent. Besides, a stochastic matrix is not necessarily diagonalizable, either (one such example will be mentioned later in \autoref{subsec:numerical-evidence-spectral-gap}).

\subsection{The variance formula cast in terms of stochastic matrices and probability vectors}\label{subchap:variance-formula-recast}

Let us introduce for all $\lambda \in \llbracket0,\lambda_{\text{max}}\rrbracket$, the vectors $\overrightarrow{Z^\lambda},\overrightarrow{W^\lambda} \in \mathbb{R}^{d_k}$ defined by squaring entry-wise the intermediate states $\zeta^{\lambda}$ and the back-propagated target states $z^{\lambda}$, respectively, i.e.
\begin{align}
    (\overrightarrow{Z^\lambda})_u &:= (\zeta^{\lambda}_u)^2\,,\label{eq:proba-vector-forward-lambda}\\
    (\overrightarrow{W^\lambda})_u &:= (z^{\lambda}_u)^2\,,\label{eq:proba-vector-reverse-lambda}
\end{align}
for all $u\in[d_k]$.

Notice that all these vectors $\overrightarrow{Z^\lambda}$ and $\overrightarrow{W^\lambda}$ are probability vectors (i.e. their entries are nonnegative and sum to $1$).

By careful inspection of Lemma \ref{lemma:InductiveRelation}'s recurrence relation \autoref{eq:Inductive_relation}, one sees that it consists of a linear recurrence relation relating the probability vectors $\overrightarrow{Z^{\lambda+1}}$ to $\overrightarrow{Z^\lambda}$, through multiplication by a stochastic matrix that directly corresponds to the $\lambda^{\text{th}}$ RBS/FBS gate. Namely:
\begin{equation}\label{eq:stochastic-matrix-recurrence-relation}
     \overrightarrow{Z}^{\lambda+1} = T_{\lambda} \cdot \overrightarrow{Z}^{\lambda} \,,
\end{equation}
where $T_{\lambda}$ is defined as the $d_k \times d_k$ column-stochastic matrix that is constructed by taking the VQC's $\lambda^{\text{th}}$ RBS/FBS unitary is (in the subspace $k$) and replacing its $\pm \cos(\theta_i)$ and $\pm \sin(\theta_i)$ entries with $1/2$.
In other words, if $U_{\lambda}(\theta_\lambda)$ denotes the VQC's $\lambda^{\text{th}}$ RBS/FBS unitary in the subspace $k$, then: 
\begin{equation}\label{eq:stochastic-matrix-Tlambda-def}
     \Big(T_\lambda\Big)_{a,b} := \Big(U_{\lambda}(\theta_\lambda:=\pi/4)\Big)_{a,b}^2\,.
\end{equation}
In fact, Lemma \ref{lemma:InductiveRelation}'s recurrence relation \autoref{eq:Inductive_relation} could be shown to hold in full analogy for the back-propagating state vector as well, instead of the forward-propagating one; one would find the "reversed" relation
\begin{equation}\label{eq:stochastic-matrix-recurrence-relation-backward-interm}
     \overrightarrow{Z}^{\lambda-1} = T_{\lambda}^\intercal \cdot \overrightarrow{Z}^{\lambda} \,,
\end{equation}
but since all the stochastic matrices associated to single RBS/FBS gates are symmetric (this follows from their definition in \autoref{eq:stochastic-matrix-Tlambda-def}), the reversed recurrence relation writes as
\begin{equation}\label{eq:stochastic-matrix-recurrence-relation-backward}
     \overrightarrow{Z}^{\lambda-1} = T_{\lambda} \cdot \overrightarrow{Z}^{\lambda} \,.
\end{equation}

Now, suppose the VQC has a CPSA architecture (Definition \ref{def:CPSA}). Its unitary (in the subspace $k$) has a periodic structure of the form
\begin{equation}\label{eq:periodic-structure-appendix}
    U(\theta) = \prod_{l=1}^{L} U_0(\theta_l), \quad U_0(\theta_l) = \prod_{j=1}^J e^{-i \theta_{l,j} H^j_{RBS/FBS}}\,. 
\end{equation}
Let us introduce new indices $(\tilde{l},\tilde{j})$ to specify a given single RBS/FBS gate, similarly to a pair $(l,j)$ but with both indices increasing in a "reversed" way instead (from the end of the circuit to the start).

To summarize, we have presently three "coordinate systems" to specify one of the RBS/FBS gates in this circuit, $\lambda \leftrightarrow (l,j) \leftrightarrow (\tilde{l},\tilde{j})$ (with $\lambda \in \llbracket 1, \lambda_{\text{max}}\rrbracket$, $l \in \llbracket 1, L\rrbracket, j \in \llbracket 1, J\rrbracket$, and $\tilde{l} \in \llbracket 1, L\rrbracket, \tilde{j} \in \llbracket 1, J\rrbracket$), the first two coordinate systems have indices that increase in a "forward" way while the third coordinate system has indices that increase in a "reversed" way; and the three coordinate systems are uniquely related through:
\begin{align}
    \lambda &= (l-1)J + j\,,\label{eq:depth-coordinates-correspondance-1}\\
    \lambda &= (L-(\tilde{l}-1))J - (\tilde{j}-1)\,.\label{eq:depth-coordinates-correspondance-2}
\end{align}
For a given intermediate depth $\lambda$ of the circuit, we will therefore denote by $l(\lambda),j(\lambda)$ and $\tilde{l}(\lambda),\tilde{j}(\lambda)$ the unique values of $l,j$ and $\tilde{l},\tilde{j}$ that correspond to $\lambda$, through Eqs.~\eqref{eq:depth-coordinates-correspondance-1} and \eqref{eq:depth-coordinates-correspondance-2} respectively.

Let us denote by $T$ (without any subscript) the stochastic matrix corresponding to the main pattern $U_0$ of the CPSA architecture (the one that is repeated $L$ times). Namely:
\begin{equation}\label{eq:stochastic-matrix-T-def}
    T := T_J \cdots T_2 \cdot T_1\,. 
\end{equation}
Similarly, let us denote by $\widetilde{T}$ the "reversed" main pattern: 
\begin{equation}\label{eq:stochastic-matrix-Ttilde-def}
    \widetilde{T} := T_1 \cdots T_{J-1} \cdot T_J = T^\intercal\,. 
\end{equation}

By repeated use of the recurrence relation of Eq.~\eqref{eq:stochastic-matrix-recurrence-relation}, the (forward-propagating) probability vector $\overrightarrow{Z}^{\lambda}$ may be related to the initial (left-most) one by:
\begin{equation}\label{eq:repeated-recurrence-relation-forward}
    \overrightarrow{Z}^{\lambda} = T_{\text{rest},\lambda} \cdot T^{l(\lambda)-1} \ \cdot\ \overrightarrow{Z^0}\,,
\end{equation}
where $T_{\text{rest},\lambda} := T_{j(\lambda)} \cdots T_2 \cdot T_1$.
Likewise, the repeated use of the recurrence relation of Eq.~\eqref{eq:stochastic-matrix-recurrence-relation-backward} yields:
\begin{equation}\label{eq:repeated-recurrence-relation-reverse}
    \overrightarrow{W}^{\lambda} = \widetilde{T}_{\text{rest},\lambda} \cdot \widetilde{T}^{\tilde{l}(\lambda)-1} \ \cdot\ \overrightarrow{W^{\lambda_{\text{max}}}}\,,
\end{equation}
where $\widetilde{T}_{\text{rest},\lambda} := T_{\tilde{j}(\lambda)} \cdots T_{J-1} \cdot T_J$.

Notice that Lemma \ref{lemma:VarianceHWPreserving}'s variance formula Eq.~\eqref{eq:VarianceHWPreserving-Var} may now be written more concisely as
\begin{align}\label{eq:variance-concise-1}
    \mathrm{Var}_{\theta}[\partial_{\theta_\lambda} \mathcal{C}] &=  2 \sum_{(u,v)} \left[ \big(\overrightarrow{Z^\lambda}\big)_{\!u} + \big(\overrightarrow{Z^\lambda}\big)_{\!v} \right] \cdot \left[ \big(\overrightarrow{W^\lambda}\big)_{\!u} + \big(\overrightarrow{W^\lambda}\big)_{\!v} \right]\,.
\end{align}
As we will detail in the next section, the four quantities $\big(\overrightarrow{Z^\lambda}\big)_{\!u},\big(\overrightarrow{Z^\lambda}\big)_{\!v},\big(\overrightarrow{W^\lambda}\big)_{\!u}$ and $\big(\overrightarrow{W^\lambda}\big)_{\!v}$ in fact all converge towards the value $1/d_k$ as the depth $\lambda$ goes to infinity. In anticipation of this fact, we suggestively re-write these four terms in Eq.~\eqref{eq:variance-concise-1} as: 
\begin{align}\label{eq:variance-concise-2}
    \mathrm{Var}_{\theta}[\partial_{\theta_\lambda} \mathcal{C}] &=  2 \sum_{(u,v)} \left[ \left(\frac{1}{d_k} + \epsilon^{(l-1,j,u)}\right) + \left(\frac{1}{d_k} + \epsilon^{(l-1,j,v)}\right) \right] \cdot \left[ \left(\frac{1}{d_k} + \tilde{\epsilon}^{(\tilde{l}-1,\tilde{j},u)}\right) + \left(\frac{1}{d_k} + \tilde{\epsilon}^{(\tilde{l}-1,\tilde{j},v)}\right) \right]\,,
\end{align}
where we dropped the dependency on $\lambda$ of $l$ and $\tilde{l}$ to simplify the notation.

We will also obtain, in the next section, upper-bounds on the absolute values of all the $|\epsilon^{(\cdot,\cdot,\cdot)}|$ terms (which we refer to as the error terms) that only depend on the number of (forward or reversed) repetitions ($l$ or $\tilde{l}$). So for now, suppose that we have such bounds, i.e. suppose that for all $j$ and all $u$:
\begin{align}
    |\epsilon^{(l-1,j,u)}| &\leq \mathcal{E}_{l-1}\,,\label{eq:variance-epsilon-commonbound-1}\\
    |\tilde{\epsilon}^{(\tilde{l}-1,\tilde{j},u)}| &\leq \mathcal{E}_{\tilde{l}-1}\,,\label{eq:variance-epsilon-commonbound-2}
\end{align}
 for some $\mathcal{E}_{l-1},\mathcal{E}_{\tilde{l}-1} >0$.
 Injecting the bounds of Eqs.~\eqref{eq:variance-epsilon-commonbound-1} and \eqref{eq:variance-epsilon-commonbound-2} into the variance expression of Eq.~\eqref{eq:variance-concise-2}, simplifying (note that the outer sum is over $\binom{n-2}{k-1}$ terms) and applying some triangle inequalities, leads to:
 \begin{equation}\label{eq:variance-difference-with-idealvariance-general-upper-bound}
     \left| \mathrm{Var}_{\theta}[\partial_{\theta_\lambda} \mathcal{C}] - \frac{1}{d_k} \frac{8 k (n-k)}{n(n-1)} \right| \leq \frac{8 k (n-k)}{n(n-1)} \left( \mathcal{E}_{l-1} + \mathcal{E}_{\tilde{l}-1} + 2 \, d_k \, \mathcal{E}_{l-1} \, \mathcal{E}_{\tilde{l}-1} \right)\,.
 \end{equation}

In the following, we will denote for all $p \in [1,\infty]$, the \textit{entry-wise} $p$-norms and the \textit{Schatten} $p$-norms on square matrices, by $\lVert \cdot \rVert_{\mathrm{ew},p}$ and $\lVert \cdot \rVert_{\mathrm{Sc},p}$ respectively. Recall that these norms are defined for $p \in [1,\infty[$ as $\lVert M \rVert_{\mathrm{ew},p} :=  \big(\sum_{ij}\lvert M_{ij} \rvert^p\big)^{1/p}$ and $\lVert M \rVert_{\mathrm{Sc},p} :=  \big(\sum_{i} (\sigma_{i}(M)^p)\big)^{1/p}$, and for $p=\infty$ as $\lVert M \rVert_{\mathrm{ew},\infty} :=  \max_{ij}\lvert M_{ij} \rvert$ and $\lVert M \rVert_{\mathrm{Sc},\infty} :=  \max_{i} \left(\sigma_{i}(M)\right)$. In the previous expressions, $\sigma_{i}(M)$ denotes the $i^{\mathrm{th}}$ singular value of the matrix $M$.

\begin{lemma}[General bound on variance error terms]
\label{lemma:variance-error-terms-first-general-bound}
For all $j$, all $u$, it holds that for all $l \geq1$:
\begin{align}
    |\epsilon^{(l-1,j,u)}| &\leq d_k \, \lVert T^{l-1} - T_\infty \rVert_{\mathrm{Sc},2} \,,\label{eq:variance-error-terms-first-general-bound-1}\\
    |\tilde{\epsilon}^{(\tilde{l}-1,\tilde{j},u)}| &\leq d_k \, \lVert T^{l-1} - T_\infty \rVert_{\mathrm{Sc},2} \,.\label{eq:variance-error-terms-first-general-bound-2}
\end{align}
Here, $T_\infty$ denotes here the $d_k \times d_k$ matrix with all coefficients equal to $1/d_k$, and $T$ is the stochastic matrix of \autoref{eq:stochastic-matrix-T-def}.
\end{lemma}
\begin{proof}
    We begin by showing the following claim. For $S$ any $d_k \times d_k$ complex matrix, $\overrightarrow{Y}$ any $d_k \times d_k$ probability vector, and for any $u \in [d_k]$, the following holds:
    \begin{equation}\label{eq:variance-error-terms-first-general-bound-subclaim-1}
        \left| (S \cdot \overrightarrow{Y})_u - \frac{1}{d_k}\right| \leq  \lVert S - T_\infty \rVert_{\mathrm{ew},\infty}\,.
    \end{equation}
    Indeed:
    \begin{align*}
        \left| (S \cdot \overrightarrow{Y})_u - \frac{1}{d_k}\right|
        &= \left| \overrightarrow{[S]_{u \bullet}} \cdot \overrightarrow{Y} - \frac{1}{d_k} \ \right|\\
        &= \left| \overrightarrow{[S]_{u \bullet}} \cdot \overrightarrow{Y} - \overrightarrow{[T_\infty]_{u \bullet}} \cdot \overrightarrow{Y} \ \right|\\
        &= \left| \left(\overrightarrow{[S]_{u \bullet}} - \overrightarrow{[T_\infty]_{u \bullet}}\right) \cdot \overrightarrow{Y} \ \right|\\
        &\leq \lVert \overrightarrow{[S]_{u \bullet}} - \overrightarrow{[T_\infty]_{u \bullet}} \rVert_\infty \  \,\lVert \overrightarrow{Y} \rVert_1\\
        &\leq \lVert S - T_\infty \rVert_{\mathrm{ew},\infty}\,.
    \end{align*}
    In the above, we used the bullet to denote a dummy index, so that for instance $\overrightarrow{[S]_{u \bullet}}$ is a vector of size $d_k$, whose entries are the $u^{\mathrm{th}}$ row of the matrix $S$. The second equality holds because $\overrightarrow{Y}$ is a probability vector, the third equality is the $(\infty,1)$-Hölder inequality for vectors, and the last inequality results from the definition of the matrix norm $\lVert \cdot \rVert_{\mathrm{ew},\infty}$ and from $\overrightarrow{Y}$ being a probability vector.

    Choosing for the matrix $S$ the stochastic matrix $T_{\text{rest},\lambda} \cdot T^{l(\lambda)-1}$ (from Eq.~\eqref{eq:repeated-recurrence-relation-forward}), and for the probability vector $\overrightarrow{Y}$ the vector $\overrightarrow{Z^0}$  associated to the initial state (Eq.~\eqref{eq:proba-vector-forward-lambda}), the claim of Eq.~\eqref{eq:variance-error-terms-first-general-bound-subclaim-1} yields:
    \begin{align}
        \left| (T_{\text{rest},\lambda} \cdot T^{l(\lambda)-1} \cdot \overrightarrow{Z^0})_u - \frac{1}{d_k}\right|
        &\leq  \lVert T_{\text{rest},\lambda} \cdot T^{l(\lambda)-1} - T_\infty \rVert_{\mathrm{ew},\infty}\label{eq:variance-error-terms-first-general-bound-nextineqs1}\\
        &\leq  \lVert T_{\text{rest},\lambda} \cdot T^{l(\lambda)-1} - T_\infty \rVert_{\mathrm{Sc},2}\label{eq:variance-error-terms-first-general-bound-nextineqs2}\\
        &\leq  \lVert T_{\text{rest},\lambda} \rVert_{\mathrm{Sc},2} \ \, \lVert T^{l(\lambda)-1} - T_\infty \rVert_{\mathrm{Sc},2}\label{eq:variance-error-terms-first-general-bound-nextineqs3}\\
        &\leq d_k \, \lVert T^{l(\lambda)-1} - T_\infty \rVert_{\mathrm{Sc},2}\,.\label{eq:variance-error-terms-first-general-bound-nextineqs4}
    \end{align}
In the above, the first inequality is the mentioned application of the claim of Eq.~\eqref{eq:variance-error-terms-first-general-bound-subclaim-1}, the second inequality holds because of the relation $\lVert \cdot \rVert_{\mathrm{ew},\infty} \leq \lVert \cdot \rVert_{\mathrm{ew},2} = \lVert \cdot \rVert_{\mathrm{Sc},2}$ between matrix-norms, the third inequality is due to the sub-multiplicativity of Schatten $p$-norms, and the fourth inequality is due to the norm inequalities $\lVert \cdot \rVert_{\mathrm{Sc},2} \leq d_k\lVert \cdot \rVert_{\mathrm{Sc},\infty}$ along with the fact that the moduli of eigenvalues of stochastic matrices are less or equal to one (as mentioned in \autoref{subchap:StochasticMatrices}).

Thus Eq.~\eqref{eq:variance-error-terms-first-general-bound-1} is established. By applying again the claim of Eq.~\eqref{eq:variance-error-terms-first-general-bound-subclaim-1}, but this time choosing for the matrix $S$ the stochastic matrix $\widetilde{T}_{\text{rest},\lambda} \cdot \widetilde{T}^{\tilde{l}(\lambda)-1}$ (from Eq.~\eqref{eq:repeated-recurrence-relation-reverse}), and for the probability vector $\overrightarrow{Y}$ the vector $\overrightarrow{W^{\lambda_{\text{max}}}}$  associated to the target state (Eq.~\eqref{eq:proba-vector-reverse-lambda})
one gets (by following the same reasoning as \autoref{eq:variance-error-terms-first-general-bound-nextineqs1,eq:variance-error-terms-first-general-bound-nextineqs2,eq:variance-error-terms-first-general-bound-nextineqs3,eq:variance-error-terms-first-general-bound-nextineqs4}):
\begin{align}
        \left| (\widetilde{T}_{\text{rest},\lambda} \cdot \widetilde{T}^{\tilde{l}(\lambda)-1} \cdot \overrightarrow{W^{\lambda_{\text{max}}}})_u - \frac{1}{d_k}\right|
        &\leq d_k \, \lVert \widetilde{T}^{l(\lambda)-1} - T_\infty \rVert_{\mathrm{Sc},2}\,.\label{eq:variance-error-terms-first-general-bound-analogous-ineq}
    \end{align}
But in the right-hand-side, $\lVert \widetilde{T}^{l(\lambda)-1} - T_\infty \rVert_{\mathrm{Sc},2} = \lVert T^{l(\lambda)-1} - T_\infty \rVert_{\mathrm{Sc},2}$ (due to the fact that $\widetilde{T}=T^{\intercal}$, that $T_\infty$ is symmetric, and that Schatten-$p$ norms are invariant under transposition), so Eq.~\eqref{eq:variance-error-terms-first-general-bound-2} is established.
\end{proof}

\subsection{\emph{Irreducible} and \emph{primitive} stochastic matrices}
\label{subchap:irreducible-and-primitive-stochmats}

Let us first introduce elementary notions about graphs. In what follows, $N$ denotes any integer such that $N\geq 2$.

By a \emph{directed graph} $\Gamma$, we mean a pair $\Gamma=(V,E)$ where $V$ is any finite set (its elements are the \textit{vertices} of $\Gamma$) and $E$ is any subset of $V^2=V \times V$ (its elements are the \textit{directed edges} of $\Gamma$).
We may denote a directed edge $(i,j)\in E$ by $i \to j$.
Importantly, note that in the above definition of directed graphs, we have allowed them to have \emph{self-loops}, i.e. directed edges $i\to i$ from a vertex to itself.

The adjacency matrix $A(\Gamma)$ of a directed graph $\Gamma=(\llbracket1, N\rrbracket,E)$ over $N$ vertices is the $N \times N$ matrix defined, for all $(i,j) \in \llbracket1, N\rrbracket^2$, by $[A(\Gamma)]_{ij} = 1$ if the graph possesses the directed edge $i \to j$, and $0$ otherwise.

A directed graph $\Gamma=(V,E)$ is said to be \emph{strongly-connected} if for every ordered pair of vertices $(i,j)\in V^2$ there exists a path in the graph from $i$ to $j$ (i.e. on a drawing of the graph one can go from $i$ to $j$ by following arrows). Importantly, one can notice that $\Gamma=(\llbracket1, N\rrbracket,E)$ is strongly-connected if and only if its adjacency matrix $A(\Gamma)$ has the property:
\begin{equation}\label{eq:directed-graph-SC-caract}
\forall (i,j) \in \llbracket1, N\rrbracket^2 \,\ \,\  \exists p_{(i,j)} \geq 1 \,\ \,\  [A(\Gamma)^{p_{(i,j)}}]_{ij}>0\,. 
\end{equation}

Now, Let $T$ be an $N \times N$ stochastic matrix.

The directed graph $\Gamma(T)$ of $T$ is defined as the directed graph $\Gamma=(\llbracket1, N\rrbracket,E)$  over $N$ vertices such that for all $(i,j) \in \llbracket1, N\rrbracket^2$, the graph possesses a directed edge $i \to j$ if and only if $[T]_{ij}>0$.

$T$ is said to be \emph{irreducible} if its directed graph $\Gamma(T)$ is strongly-connected.

Furthermore, $T$ is said to be \emph{primitive} if for a certain power of $p$, all the matrix coefficients of $T^p$ are positive. (It may be checked that if $p$ is such a power, than all subsequent matrix powers $q\geq p$ keep remain with positive coefficients as well.)
Because this property may be written as 
\begin{equation}\label{eq:stoch-mat-primitive}
\exists p \geq 1  \,\ \,\   \forall (i,j) \in \llbracket1, N\rrbracket^2 \,\ \,\  [A(\Gamma(T))^{p}]_{ij}>0\,, 
\end{equation}
notice (by comparing with \autoref{eq:directed-graph-SC-caract}) that $T$ being primitive is in general a stronger property than $T$ being irreducible.

\subsection{\emph{Connected} RBS/FBS patterns, and properties of their associated stochastic matrices}\label{subchap:connected-stochmats-props}

In this section, we first phrase in a precise manner what the assumption of $U_0(\theta)$ being connected (as it was stated in the main text's \autoref{def:CPSA}) means, we then prove that such an assumption indeed implies that the graphs of the associated stochastic matrices $T$ (at all Hamming-weights $k$) are strongly-connected, and we furthermore show that those $T$ are in fact primitive. Lastly, we give a sufficient condition on $U_0$ so that $T$ is symmetric.

Let $U_0=U_0(\theta)$ be a pattern of RBS/FBS gates on $n$ qubits, taken formally as an ordered list of triples $((i_1,j_1,\theta_1),(i_2,j_2,\theta_2),\dots,(i_J,j_J,\theta_J))$, where each entry $(i_k,j_k,\theta_k)$ indicates the presence of an RBS/FBS gate placed from qubit $i_k$ to qubit $j_k$, and set at angle $\theta_k$, and where the ordering of the list corresponds to time.

The \emph{graph $\Gamma(U_0)$} associated to the pattern $U_0$ is defined as the directed graph $\Gamma=(\llbracket1, N\rrbracket,E)$ over $N$ vertices whose directed edges indicate the presence of an RBS/FBS gate in the above ordered list of $U_0$, i.e. $E=\{ (i_1,j_1),(i_2,j_2),\dots,(i_J,j_J) \}$. In words, $\Gamma(U_0)$ may be thought of being the result of taking the $n$-qubit circuit depiction of $U_0$, "flattening out" the time/depth axis, and adding arrow tips on each side of every vertical line that represented a gate.

In this section, $U_0^k$ will denote the unitary matrix of the pattern $U_0(\theta)$ of RBS/FBS gates in the subspace of Hamming weight $k$. We will denote by $T^{(U_0^k)}$ the $d_k \times d_k$ stochastic matrix associated to that pattern of gates $U_0^k$  (\autoref{eq:stochastic-matrix-T-def}, where it was denoted $T$).
Recall from the previous section that $\Gamma(T^{(U_0^k)})$ then denotes the graph over $d_k$ vertices with directed edges $u\to v$ exactly when $[T^{(U_0^k)}]_{uv}>0$.

\begin{definition}[Connected pattern of RBS/FBS gates]
\label{def:connected-pattern}
The pattern $U_0$ of RBS/FBS gates is said to be \emph{connected} if its associated directed graph $\Gamma(U_0)$ is strongly-connected. 
\end{definition}

The following lemma will be useful later in this section:
\begin{lemma}
\label{lem:connected-section-lem1}
Let $T$ be an $N \times N$ stochastic matrix. Let $A,B\geq0$ and let $S_1,\dots,S_A,S'_{1},\dots,S'_B$ be arbitrary $N \times N$ matrices that all have nonnegative entries, and only positive entries on their diagonals.
Then, for all $(u,v) \in \llbracket1, N\rrbracket^2$:
\begin{equation}
    [T]_{uv}>0 \implies [S_1\cdots S_A \,
T S'_{1}\,\cdots S'_B]_{uv}>0\,.
\end{equation}
\end{lemma}
\begin{proof}
The case $A=1,B=0$ is readily shown, since, if $[T]_{uv}>0$, then
\begin{equation}
    [S\,T]_{uv}
    = \sum_{k=1}^N S_{uk} T_{kv}
    = S_{uu} T_{uv} + \sum_{\substack{k=1\\k\neq u}}^N S_{uk} T_{kv}\,,
\end{equation}
and thus this expression is positive, as the first term $S_{uu} T_{uv}$ is positive by assumption and the rest of the summed terms are all nonnegative.

The case $A=0,B=1$ is shown similarly. Then, the cases of general $A$ and $B$ may be shown to follow by induction.
\end{proof}

Let us denote by $\operatorname{Involved}_k(i,j)$ the pairs of indices of basis vectors $B_k^n$ of Hamming-weight $k$ that would be "involved" together in a rotation if an RBS/FBS gate was applied between qubits $i$ and $j$. Explicitly: 
\begin{equation}
\operatorname{Involved}_k(i,j) := \big\{ (u,v)\in \llbracket1, d_k\rrbracket^2 
\ \big| \ (e_u)^i = 1 \mathrm{\;and\;} (e_v)^j = 0\mathrm{\ \;\;or\ \;\;} (e_u)^i = 0 \mathrm{\;and\;} (e_v)^i = 1 \big\}\,,
\end{equation}
where we used notation $(e_u)^i$ for the $i^{\mathrm{th}}$ bit of the $n$-bitstring $e_u$. Note that this set has $2\binom{n-2}{k-1}$ elements.
\begin{lemma}
\label{lem:connected-section-lem2}
If $(i,j)$ is a directed edge in the graph $\Gamma(U_0)$, then for any $k \in \llbracket 1, n-1\rrbracket$,  all the elements of $\operatorname{Involved}_k(i,j)$ are directed edges in the graph $\Gamma(T^{(U_0^k)})$.
\end{lemma}
\begin{proof}
Suppose that the graph $\Gamma(U_0)$ contains the edge $i\to j$. This means that an RBS/FBS gate between qubits $i$ and $j$ is present in the pattern $U_0$, and therefore for every $k$, its associated stochastic matrix for the Hamming-weight $k$, $T^{(U_0^k)}$, is a product (see \autoref{eq:stochastic-matrix-T-def}) of $J$ "elementary" stochastic matrices associated to single RBS/FBS gates (\autoref{eq:stochastic-matrix-Tlambda-def}), and one of them corresponds to a gate between qubits $i$ and $j$. Let us write this as 
\begin{equation}
T^{(U_0^k)} = T_1\cdots T_A \,
\,T^{(i,j)}\, T'_{1}\,\cdots T'_B\,.
\end{equation}
But by the definition \autoref{eq:stochastic-matrix-Tlambda-def} of these elementary stochastic matrices in this product, one has firstly that they all have positive diagonal coefficients everywhere (the $u^{\mathrm{th}}$ diagonal coefficient is either $1/2$ if $u$ is involved in the gate's rotation and $1$ otherwise), and secondly that for all $(u,v) \in \operatorname{Involved}_k(i,j)$, $[T^{(i,j)}]_{ij}=[T^{(i,j)}]_{ji}=1/2$. 
Therefore, the previous \autoref{lem:connected-section-lem1} readily applies, to give, for all $(u,v) \in \operatorname{Involved}_k(i,j)$:
\begin{equation}
[T^{(U_0^k)}]_{uv}>0\,.
\end{equation}
\end{proof}

\begin{lemma}
\label{lem:connected-section-lem3}
If the graph $\Gamma(U_0)$ is strongly-connected, then for all $k \in \llbracket 1, n-1\rrbracket$ the graph $\Gamma(T^{(U_0^k)})$ is strongly-connected as well.
\end{lemma}
\begin{proof}
Fix a $k \in \llbracket 1, n-1\rrbracket$, and let $(u,v)\in \llbracket1, d_k\rrbracket^2$. Let us show that there exists a path $\mathcal{P}=(u \to \cdots \to v)$ of directed edges in the graph $\Gamma(T^{(U_0^k)})$ joining vertex $u$ to vertex $v$.

Let $(h_1^u,\dots,h_k^u)$ and $(h_1^v,\dots,h_k^v)$ be the indices that hold the values $1$ in the $n$-bitstrings $e_u$ and $e_v$, respectively, and  let $(i_1^u,\dots,i_{k'}^u)$ and $(i_1^v,\dots,i_{k'}^v)$ be the respective subsets of those for which none of the remaining indices are shared between the first and second tuples (i.e. all first indices $i_1^u,\dots,i_{k'}^u$ are different from all the second indices $i_1^v,\dots,i_{k'}^v$).

For all $s \in \llbracket 1, k'\rrbracket$, use the assumption that $\Gamma(U_0)$ is strongly-connected to get the existence of a path $\mathcal{P}_s=(i_s^u \to \cdots \to i_s^v)$ of directed edges in $\Gamma(U_0)$.

Now, one constructs the desired path $\mathcal{P}=(u \to \cdots \to v)$, by successively invoking \autoref{lem:connected-section-lem2} along each whole path $\mathcal{P}_s$ in $\Gamma(U_0)$ to get a corresponding path $\mathcal{P}^k_s$ in $\Gamma(T^{(U_0^k)})$, and by concatenating the obtained paths $\mathcal{P}^k_1 , \mathcal{P}^k_2 ,\dots, \mathcal{P}^k_{k'}.$
\end{proof}

\begin{cor}
\label{cor:connected-section-cor1}
If the pattern $U_0$ of RBS/FBS gates is connected (\autoref{def:connected-pattern}), then for every Hamming weight $k \in \llbracket 1, n-1\rrbracket$, the associated stochastic matrix $T^{(U_0^k)}$ is primitive.
\end{cor}
\begin{proof}
Fix a Hamming weight $k \in \llbracket 1, n-1\rrbracket$, and denote $T:=T^{(U_0^k)}$. If $U_0$ is connected, then by \autoref{lem:connected-section-lem3} the stochastic matrix $T$ is irreducible, meaning that the property of \autoref{eq:directed-graph-SC-caract} holds for $\Gamma=\Gamma(T)$. In general, a stochastic matrix $T$ being irreducible does not imply that it is primitive, but in our case this actually follows. Indeed, for each $(u,v)\in \llbracket1, d_k\rrbracket^2$, the characterization of irreducibility of $T$ of \autoref{eq:directed-graph-SC-caract} gives a $p_{(u,v)}\geq1$ such that $[T^{p_{(u,v)}}]_{uv}>0$. Now, for this stochastic matrix $T^{p_{(u,v)}}$, applying \autoref{lem:connected-section-lem1} to it yields (recursively, with $S_1:=T$) that for all subsequent powers $q\geq p_{(u,v)}$, one still has $[T^{q}]_{uv}>0$.
Therefore, taking $p:=\max_{(u,v)\in \llbracket1, d_k\rrbracket^2}(p_{(u,v)})$, it follows that $T^q$ has positive entries whenever $q\geq p$, i.e. we have established that $T$ is primitive.
\end{proof}

\begin{lemma}\label{lem:T-is-doubly-stoch}
For every Hamming weight $k \in \llbracket 1, n-1\rrbracket$, the stochastic matrix $T^{(U_0^k)}$ associated to any pattern of RBS/FBS gates $U_0$ is always doubly-stochastic.
\end{lemma}
\begin{proof}
    Denote $T:=T^{(U_0^k)}$.
    $T$ is constructed as a product (c.f. \autoref{eq:stochastic-matrix-T-def}) of "elementary" stochastic matrices (\autoref{eq:stochastic-matrix-Tlambda-def}) that are doubly-stochastic (since those are in fact symmetric, by definition). Hence, recalling that the set of doubly-stochastic matrices is closed under product, $T$ is doubly-stochastic as well.
\end{proof}

\begin{lemma}
\label{lem:connected-section-palindrome}
If the pattern $U_0$ if RBS/FBS gates is a palindrome, then for every Hamming weight $k \in \llbracket 1, n-1\rrbracket$, the associated stochastic matrix $T^{(U_0^k)}$ is symmetric.

By the pattern being a palindrome, we mean here that the ordered list defining the pattern is of the form
\begin{equation*}
   \Big((i_1,j_1,\;\theta_1),\dots,\;(i_{M-1},j_{M-1},\theta_{M-1}),\;(i_M,j_M,\theta_M),\;(i_{M-1},j_{M-1},\theta_{M+1}),\;\dots,(i_1,j_1,\theta_{2M-1})\Big)\,.
\end{equation*}
\end{lemma}
\begin{proof}
Indeed, in this case, the stochastic matrix $T:=T^{(U_0^k)}$ is of form
\begin{align} 
    T 
    &= T_{1} \, T_{2} \cdots T_{M-1} \, T_{M} \, T_{M-1} \cdots T_{2} \, T_{1}\\
    &= T_{1} \, T_{2} \cdots T_{M-1} \, T_{M} \, T_{M}\, T_{M-1} \cdots T_{2} \, T_{1}\\
    &= T_{(B)} T_{(F)}
\end{align}
where we denoted $T_{(F)}:=T_{M}\, T_{M-1} \cdots T_{2} \, T_{1}$ and $T_{(B)}:=T_{1} \, T_{2} \cdots T_{M-1} \, T_{M}$. In the second inequality, $T_M = T_M^2$ was used, as it can indeed by checked that all stochastic matrices corresponding to single RBS/FBS gates square to themselves (from their definition of \autoref{eq:stochastic-matrix-Tlambda-def}).
But we have
\begin{align} 
    T_{(F)}^\intercal
    &= (T_{M} \, T_{M-1} \cdots T_{2} \, T_{1})^\intercal\\
    &= T_{1}^\intercal \, T_{2}^\intercal \cdots T_{M-1}^\intercal \, T_{M}^\intercal\\
    &= T_{1} \, T_{2} \cdots T_{M-1} \, T_{M}\\
    &= T_{(B)}
\end{align}
where the third equality is because all stochastic matrices corresponding to single RBS/FBS gates are symmetric (this also stems from their definition of \autoref{eq:stochastic-matrix-Tlambda-def}).
Hence,
    \begin{align}
    T = T_{(F)}^\intercal T_{(F)}\,,
    \end{align}
which establishes that $T$ is symmetric.
\end{proof}

\subsection{Convergence of powers of a fixed stochastic matrix}\label{subchap:convergence-fixed}

\begin{thm}[Exponential convergence of $(T^l)_{l \in \mathbb{N}}$]
\label{thm:stoch-mat-convergence}
Let $N\geq1$ be fixed, and let $T \in \mathbb{R}^{N \times N}$ be a column-stochastic matrix. If $T$ is \emph{primitive}, then the following points hold:
\begin{enumerate}[align=left]
    \item[1. (Convergence to rank-one matrix)] The sequence of matrix powers $(T^l)_{l \in \mathbb{N}}$ converges to a certain matrix $T_{\infty}$, whose columns are all identical and equal to some probability vector $\vec{\pi}=(\pi_1,\dots,\pi_N)^\intercal$.

    \item[2. (Upper-bound on rate of convergence)]
    There exists constants $l_0 \in \mathbb{N}$ and $A,B>0$ (depending on $T$) such that for all $l \geq l_0$,
    \begin{equation}\label{eq:stoch-mat-convergence}
    \lVert T^l - T_{\infty} \rVert_{\mathrm{ew},\infty} \leq \frac{A}{\exp(B\, l)}\,,
    \end{equation}

    \item[3. (Double-stochastic case)] If $T$ is furthermore doubly stochastic, then $\vec{\pi}=(1/N,\dots,1/N)^\intercal$, i.e.
    \begin{equation}
    T_\infty = \begin{pmatrix} 
    1/N & \dots  & 1/N\\
    \vdots & \ddots & \vdots\\
    1/N & \dots  & 1/N 
    \end{pmatrix}\,.
    \end{equation}
\end{enumerate}
\end{thm}
\begin{proof}
$ $\newline
\begin{enumerate}[align=left]
    \item[1.] This is part of the content of the Perron–Frobenius theorem (for the special case of primitive stochastic matrices), see for instance \cite[Chapter 8]{Horn-MatrixAnalysis-2013}.

    \item[2.] Recall that all eigenvalues $\lambda$ of $T$ are contained in the complex closed unit disk. One of the other points of the Perron–Frobenius theorem for primitive stochastic matrices $T$ is that, if $\lambda$ is any eigenvalue of $T$ different from $1$, then $|\lambda|<1$. Let $\lambda_2$ denote any eigenvalue of $T$ that achieves the highest modulus value $|\lambda|$, among all eigenvalues $\lambda$ besides $1$. We will now explicitly show how this implies that $(T^l)_{l \in \mathbb{N}}$ converges exponentially with a rate $B$ governed by this largest eigenvalue modulus. We do so in pedagogical detail, notably because later on we will remark how these methods succeed or fail to be conclusive in the more generalized setting of starting with not one but a \emph{sequence} of stochastic matrices.
    
    Firstly, suppose it is the case that $T$ is \emph{normal}, i.e. $T^\dagger T = T T^\dagger$, as it is the most intuitive case. By the spectral theorem, this is equivalent to the existence of a \emph{unitary} matrix $P$ such that $T = P D P^{-1}$, where $D:=\text{diag}(1,\lambda_2,\dots,\lambda_k)$ and $1,\lambda_2,\dots,\lambda_k$ are the eigenvalues of $T$ (repeated with multiplicity). It then follows that $T^l = P D^l P^{-1}$, from which taking the limit $l \to \infty$ on both sides gives (by the previous point of the current theorem) $T_\infty = P E_1 P^{-1}$, with $E_1:=\text{diag}(1,0,\dots,0)$. Therefore, we have
    \begin{equation}\label{eq:stoch-mat-convergence-normal-eq1}
    \lVert T^l - T_{\infty} \rVert_{\mathrm{Sc},2} = \lVert P D^l P^{-1} - P E_1 P^{-1} \rVert_{\mathrm{Sc},2} = \lVert P (D^l - E_1) P^{-1} \rVert_{\mathrm{Sc},2}\,,
    \end{equation}
    But since $P$ is unitary, we have by unitary invariance of the Schatten $p$-norms that
    \begin{equation}\label{eq:stoch-mat-convergence-normal-eq2}
    \lVert P (D^l - E_1) P^{-1} \rVert_{\mathrm{Sc},2} = \lVert D^l - E_1 \rVert_{\mathrm{Sc},2}\,.
    \end{equation}
    Furthermore, we have
    \begin{align}
    \lVert D^l - E_1 \rVert_{\mathrm{Sc},2}
    &= \lVert \text{diag}(0,\lambda_2^l,\lambda_3^l,\dots,\lambda_N^l) \rVert_{\mathrm{Sc},2}\nonumber\\
    &= \sqrt{|\lambda_2|^{2l} + |\lambda_3|^{2l} + \cdots + |\lambda_N|^{2l}}
    \leq \sqrt{N-1}\,|\lambda_2|^{l}\,.\label{eq:stoch-mat-convergence-normal-eq3}
    \end{align}
    Therefore, combining \autoref{eq:stoch-mat-convergence-normal-eq1,eq:stoch-mat-convergence-normal-eq2,eq:stoch-mat-convergence-normal-eq3}, along with the norm inequality $\lVert \cdot \rVert_{\mathrm{ew},\infty} \leq \lVert \cdot \rVert_{\mathrm{ew},2} = \lVert \cdot \rVert_{\mathrm{Sc},2}$, yields the claim of \autoref{eq:stoch-mat-convergence} (assuming $T$ to be normal) with
    \begin{equation}
    l_0=1,\quad A=\sqrt{N-1},\quad \text{and }  B=\ln(1/|\lambda_2|).
    \end{equation}

    Without an assumption of normality of the matrix $T$, it is still possible to show the exponential convergence with $l$, by making use of the Jordan canonical form (which applies to any square matrix) of $T$. Indeed, it provides the existence of an \emph{invertible} matrix $P$ such that $T = P J P^{-1}$, where $J$ is a matrix of the form
    \begin{equation}
    J = 
        \begin{pmatrix}
        1&0&\;&\;&\;\\
        \;&\lambda_{2}&\bullet&\;&\;\\
        \;&\;&\ddots&\ddots&\;\\
        \;&\;&\;&\ddots&\bullet\\
        \;&\;&\;&\;&\lambda_{N}
        \end{pmatrix}\,,
    \end{equation}
    where the elements on the diagonal are the eigenvalues of $T$ (repeated with algebraic multiplicity), where the elements on the superdiagonal take values $0$ or $1$ (in some manner that depends on the geometric multiplicity of the eigenvalue to the left of it), and where all other entries are zero. A precise statement may be found in \cite[Section 3.1]{Horn-MatrixAnalysis-2013}, but the important takeaway is that even if $J$ is not exactly diagonal, its powers $J^l$ will still converge towards $E_1$ in the same asymptotic fashion (i.e. exponentially fast with the rate being governed by $|\lambda_2|$). Indeed, it can be readily shown (e.g. from \cite[Section 3.2.5]{Horn-MatrixAnalysis-2013}) that its powers verify the property that for all $l \geq N$,
    \begin{equation}
    \lVert J^l - E_1 \rVert_{\mathrm{ew},\infty} \leq \frac{N}{s^{N-1}} \ \, l^{N-1} |\lambda_2|^l\,,
    \end{equation}
    where $s(T)>0$ denotes the smallest positive modulus of an eigenvalue of $T$.
    But since
    \begin{equation}
    l^{N-1} |\lambda_2|^l = l^{N-1} \exp\left(-\ln(1/|\lambda_2|) \, l\right) \in \underset{l\to\infty}{\mathcal{O}}\left( \exp\left(-\frac{1}{2}\ln(1/|\lambda_2|) \, l\right) \right)\,,
    \end{equation}
    there exists some $l_0(N)\geq1$ such that for all $l\geq l_0(N)$,
    \begin{equation}\label{eq:stoch-mat-convergence-nonnormal-eq0}
    \lVert J^l - E_1 \rVert_{\mathrm{ew},\infty} \leq \frac{N}{s^{N-1}} \ \exp\left(-\frac{1}{2}\ln(1/|\lambda_2|) \, l\right)\,\,.
    \end{equation}

    Besides, just as in \autoref{eq:stoch-mat-convergence-normal-eq1}, we also have
    \begin{equation}\label{eq:stoch-mat-convergence-nonnormal-eq1}
    \lVert T^l - T_{\infty} \rVert_{\mathrm{Sc},2} = \lVert P J^l P^{-1} - P E_1 P^{-1} \rVert_{\mathrm{Sc},2} = \lVert P (J^l - E_1) P^{-1} \rVert_{\mathrm{Sc},2}\,,
    \end{equation}
    however here $P$ is not a priori unitary, so we may merely invoke sub-multiplicativity of Schatten or entry-wise norms (and not unitary invariance) which gives
    \begin{equation}\label{eq:stoch-mat-convergence-nonnormal-eq2}
    \lVert P (J^l - E_1) P^{-1} \rVert_{\mathrm{Sc},2} \leq \lVert P \rVert_{\mathrm{Sc},2} \, \lVert P ^{-1}\rVert_{\mathrm{Sc},2} \,\,  \lVert J^l - E_1 \rVert_{\mathrm{Sc},2}:=\mathrm{cond}(P)\,\lVert J^l - E_1 \rVert_{\mathrm{Sc},2}\,.
    \end{equation}
    where $\mathrm{cond}(P):=\lVert P \rVert_{\mathrm{Sc},2} \, \lVert P ^{-1}\rVert_{\mathrm{Sc},2}$ is the so-called \emph{condition number} of the matrix $P$ (with respect to $\lVert \cdot \rVert_{\mathrm{Sc},2}$).
    
    Hence, combining \autoref{eq:stoch-mat-convergence-nonnormal-eq0,eq:stoch-mat-convergence-nonnormal-eq1,eq:stoch-mat-convergence-nonnormal-eq2} yields this time the claim of \autoref{eq:stoch-mat-convergence} (without assuming $T$ to be normal) with
    \begin{equation}
    l_0=l_0(N),\quad A=\frac{N \, \mathrm{cond}(P)}{s(T)^{N-1}},\quad \text{and }  B=\frac{1}{2}\ln(1/|\lambda_2|).
    \end{equation}

    \item[3.] If $T$ is doubly stochastic, then so is $T^l$ for all $l\geq1$, and therefore as mentioned above the limit $T_\infty$ is also doubly stochastic. Since $T_\infty$ is a doubly stochastic matrix of rank 1 (as its columns are all equal), it 
    is necessarily equal to
    \begin{equation}
    T_\infty = \begin{pmatrix} 
    1/N & \dots  & 1/N\\
    \vdots & \ddots & \vdots\\
    1/N & \dots  & 1/N 
    \end{pmatrix}\,.
\end{equation}
    Indeed, if a matrix is row-stochastic and rank-1, all of its rows are equal --- due to the fact that because its first column is nonzero (since it is a probability vector) rank-1 implies that all its other columns are scalar multiples of the first, and hence must be equal to the first (since they must all be probability vectors). Likewise, if a matrix is column-stochastic and rank-1, all of its columns are equal. It follows that if a matrix is doubly stochastic and rank-1, all of its entries are equal, and hence equal to $1/N$.
\end{enumerate}
\end{proof}

The next theorem, taken from the literature of mixing times of Markov chains, gives qualitatively the same result (with different constants involved), but it turns out that in the next section, where the constants will become sequences, only this result will be able to be converted into our more general setting of interest (while the previous \autoref{thm:stoch-mat-convergence} won't be usable in general).
\begin{thm}[{Adapted from \cite[Theorem 1.2]{JerisonGeneralMixing2013}}]
\label{thm:fixed-stoch-mat-Jeriso-convergence}
Let $N\geq1$ be fixed, let $T \in \mathbb{R}^{N \times N}$ be a column-stochastic matrix, and suppose that $T$ is \emph{primitive}. Denote by $\Delta:=1-|\lambda_2|$ the \emph{spectral gap} of $T$, and introduce the quantities 
\begin{align}
    A&:=\frac{2}{e}  \exp\Big(\big[ \ln(1/\Delta) + 2(1+\ln(2))\big] N \Big)\,,\\[9pt]
    B&:=\frac{\Delta}{2}\,.
\end{align}
Then, for all $l\in \mathbb{N}$ such that $A/\exp(B \, l) < 1$:
\begin{equation}\label{eq:fixed-stoch-mat-Jeriso-convergence-stoch-mat-convergence}
    \lVert T^l - T_{\infty} \rVert_{\mathrm{ew},\infty} \leq \frac{A}{\exp(B \, l)}\,.
\end{equation}
Here, $T_{\infty}$ denotes a certain  $N \times N$ stochastic matrix of rank $1$.
\end{thm}
\begin{proof}
Since $T$ is primitive, we know it converges to some rank-1 column-stochastic matrix $T_\infty$ (see the proof of \autoref{thm:stoch-mat-convergence}). Therefore, the result of \cite[Theorem 1.2]{JerisonGeneralMixing2013} applies -- and it is straight-forward manipulation of inequalities to recast it into our above statement.
\end{proof}

\subsection{Convergence of powers of a sequence of stochastic matrices, and a spectral gap conjecture}\label{subchap:convergence-sequence-and-conjecture}

\begin{thm}
\label{thm:stoch-mat-sequence-convergence}
Let $(T_n)_{n\geq2}$ be a sequence of $N_n \times N_n$ stochastic matrices that are doubly-stochastic and primitive. Furthermore, suppose that

\begin{equation}\label{eq:stoch-mat-sequence-convergence-poly-assumptions}
    \Delta_n \in \Omega\big(1/\mathrm{poly}(n)\big)\,,
\end{equation}
where $\Delta_n:=1-|\lambda_2(T_n)|$ denotes the spectral gap of the stochastic matrix $T_n$.

Then, for any sequence $l_n \in \Omega( \Delta_n^{-1} \, N_n \, n)$, there exists a constant $c>0$ such that:
\begin{equation}\label{eq:stoch-mat-Jeriso-seuquence-convergence}
    \lVert T_n^{l(n)} - T_{n,\infty} \rVert_{\mathrm{ew},\infty} \in \mathcal{O}\left(\frac{1}{\exp(c\, n)}\right)\,.
\end{equation}
where $T_{n,\infty}$ denotes the $N_n \times N_n$ matrix with all entries equal to $1/N_n$.

\end{thm}
\begin{proof}

Denote
\begin{align}
    A_n&:=\frac{2}{e}  \exp\Big(\big[ \ln(1/\Delta_n) + 2(1+\ln(2))\big] N_n \Big)\,,\\[9pt]
    B_n&:=\frac{\Delta_n}{2}\,.
\end{align}

First, we claim that there exists $c_A >0$ such that
\begin{equation}\label{eq:stoch-mat-sequence-convergence-epsilon-claim1}
    A_n \in \mathcal{O}\left(\exp(c_A\,N_n\, \ln(n))\right)\,.
\end{equation}
Indeed, the assumption $(1/\Delta_n) \in \mathcal{O}(\mathrm{poly}(n)$ implies that $\big[ \ln(1/\Delta_n) + 2(1+\ln(2))\big] \in \mathcal{O}\left(\ln(n)\right)$, and hence $\big[ \ln(1/\Delta_n) + 2(1+\ln(2))\big]N_n \in \mathcal{O}\left(N_n\,\ln(n)\right)$, which indeed implies that there exists a constant $c_A>0$ such that \autoref{eq:stoch-mat-sequence-convergence-epsilon-claim1} holds.

Second, we claim that for any choice of sequence $(l_n) \in \Omega( \Delta_n^{-1} \, N_n \, n)$, there exists a constant $c_B>0$ such that
\begin{equation}\label{eq:stoch-mat-sequence-convergence-epsilon-claim2}
    \frac{1}{\exp(B_n \, l_n)} \in \mathcal{O}\left(\frac{1}{\exp(c_B \, N_n \, n)}\right)\,.
\end{equation}

Indeed, picking any $l_n \in \Omega( \Delta_n^{-1} \, N_n \, n)$ 
implies that $(B_n \, l_n) \in \Omega(N_n \, n)$, which indeed implies that there exists a constant $c_B>0$ such that \autoref{eq:stoch-mat-sequence-convergence-epsilon-claim2} holds.

Therefore, \autoref{eq:stoch-mat-sequence-convergence-epsilon-claim1,eq:stoch-mat-sequence-convergence-epsilon-claim2} together give that
\begin{equation}\label{eq:stoch-mat-sequence-convergence-epsilon-intermeq1}
    \frac{A_n}{\exp(B_n \, l_n)} \in \mathcal{O}\left(\frac{1}{\exp\left( 
N_n\left[ c_B \, n - c_A \, \ln(n) \right] \right)}\right)\,.
\end{equation}
But since
\begin{equation}
    N_n\left[ c_B \, n - c_A \, \ln(n) \right]
    \geq \left[ c_B \, n - c_A \, \ln(n) \right]
    = n\left( c_B - c_A \frac{\ln(n)}{n} \right)
    \in \Omega \left( n \right)\,,
\end{equation}
it holds that
\begin{equation}\label{eq:stoch-mat-sequence-convergence-epsilon-intermeq2}
     N_n\left[ c_B \, n - c_A \, \ln(n) \right] \in \Omega \left( n \right)\,,
\end{equation}
and thus \autoref{eq:stoch-mat-sequence-convergence-epsilon-intermeq1,eq:stoch-mat-sequence-convergence-epsilon-intermeq2} imply that there exists some $c>0$ such that 
\begin{equation}\label{eq:stoch-mat-sequence-convergence-epsilon-request-bound}
    \frac{A_n}{\exp(B_n \, l_n)} \in \mathcal{O}\left(\frac{1}{\exp\left(c \, n\right)}\right)\,.
\end{equation}

We have thus shown that there for any sequence $(l_n) \in \Omega( \Delta_n^{-1} \, N_n \, n)$ there exists a $c>0$ such that \autoref{eq:stoch-mat-sequence-convergence-epsilon-request-bound} holds.
Now, take any such $(l_n)$ and apply, for each $n\geq2$, \autoref{thm:fixed-stoch-mat-Jeriso-convergence} to the stochastic matrix $T:=T_n$, and to the power $l:=l_n$. Because \autoref{eq:stoch-mat-sequence-convergence-epsilon-request-bound} implies that there exists an $n_0\geq1$ such that ${A_n}/{\exp(B_n \, l_n)} \leq 1$, these applications of \autoref{thm:fixed-stoch-mat-Jeriso-convergence} give us, for all $n\geq n_0$, the result that
\begin{equation}\label{eq:stoch-mat-Jeriso-seuquence-convergence-applied-to-seq}
    \lVert T_n^{l(n)} - T_{n,\infty} \rVert_{\mathrm{ew},\infty} \leq \frac{A_n}{\exp(B_n\, l_n)}\,.
\end{equation}
Combining \autoref{eq:stoch-mat-sequence-convergence-epsilon-request-bound,eq:stoch-mat-Jeriso-seuquence-convergence-applied-to-seq} yields
\begin{equation}\label{eq:stoch-mat-Jeriso-seuquence-convergence-applied-proved}
    \lVert T_n^{l(n)} - T_{n,\infty} \rVert_{\mathrm{ew},\infty} \in \mathcal{O}\left(\frac{1}{\exp\left(c \, n\right)}\right)\,.
\end{equation}
\end{proof}

Note that in the case where the matrices $T_n$ are normal (this is for instance the case when the pattern $U_0$ is a palindrome, due to \autoref{lem:connected-section-palindrome}), it is possible to prove the result of \autoref{thm:stoch-mat-sequence-convergence} more simply, by relying on the previous spectral theorem argument detailed in the proof of \autoref{thm:stoch-mat-convergence} instead of on the result of \autoref{thm:fixed-stoch-mat-Jeriso-convergence}, and by additionally assuming that $N_n \in \mathcal{O}\left(\mathrm{poly}(n)\right)$. 
Indeed, it yields for any sequence $(l_n)$ that for all $n\geq2$,
\begin{equation}
\lVert T_n^{l_n} - T_{n,\infty} \rVert_{\mathrm{ew},\infty} \leq \frac{\sqrt{N_n -1}}{\exp(\,\ln(1/\lambda_2(T_n))\, l_n\,)}\,,
\end{equation}
and thus, since $N_n \in \mathcal{O}\left(\mathrm{poly}(n)\right)$, taking $l_n \in \Omega\left(n/\ln(1/\lambda_2(T_n))\right)$ suffices to obtain $\lVert T_n^{l_n} - T_{n,\infty} \rVert_{\mathrm{ew},\infty} \in \mathcal{O}(1/c\,\exp(n))$ for some $c>0$; and because $\ln(1/\lambda_2(T_n)) \geq 1/\Delta_n$ (for all $n\geq2$) and $1/\Delta_n \in \Omega\left(1/\mathrm{poly}(n)\right)$, such an $l_n$ can be chosen to be in $\mathcal{O}\left(\mathrm{poly}(n)\right)$ as well.  

However, in the case where the matrices $T_n$ are not normal, the Jordan canonical form argument that was given as well in the proof of \autoref{thm:stoch-mat-convergence} cannot be successfully employed to prove the result of \autoref{thm:stoch-mat-sequence-convergence}, as the obtained bound would involve a condition number $\mathrm{cond}(P_n)$ (with $P_n$ the change of basis matrix that converts $T_n$ into its Jordan canonical form), over which we do not have any control of its scaling behavior with $n$.

\paragraph{A spectral gap conjecture}

As per the assumption \autoref{eq:stoch-mat-sequence-convergence-poly-assumptions} in the previous theorem, we will need, in order to arrive at our conclusion of absence of Barren Plateaus, to make a conjecture on the size of the spectral gaps of the relevant stochastic matrices. We state this in \autoref{conj:spectral-gap}, and we we provide numerical evidence that this conjecture holds, which we defer to \autoref{subsec:numerical-evidence-spectral-gap}. The actual proof of \autoref{conj:spectral-gap} is left for future work.

\begin{conj}[Spectral gaps of connected RBS/FBS patterns are inverse-polynomially large]
\label{conj:spectral-gap}

Let $(U_{0,n})_{n\geq2}$ be a sequence of connected patterns of RBS/FBS gates (\autoref{def:connected-pattern}), where each $U_{0,n}$ is such a pattern over $n$ qubits.
If the number $J_n$ of gates in the pattern $U_{0,n}$ satisfies $J_n \in \mathcal{O}\left(\mathrm{poly}(n)\right)$, then for any fixed Hamming weight $k \in \mathcal{O}(1)$, the associated $d_{k,n} \times d_{k,n}$ stochastic matrix $T_n:=T^{(U_{0,n}^k)}$ (\autoref{eq:stochastic-matrix-T-def}) satisfies
\begin{equation}
    \Delta_n \in \Omega\left(1/\mathrm{poly}(n)\right)\,,
\end{equation}
where $\Delta_n:=1-|\lambda_2(T_n)|$ denotes the spectral gap of the stochastic matrix $T_n$.
\end{conj}

\subsection{Precised version of \autoref{thm:NoBPPSA}, and proof}\label{subchap:precised-theorem-2}

Putting it all together, we finally obtain:

\begin{thm}[Absence of Barren Plateaus]
\label{thm:concluding-back-to-variance}
Let $(U_{0,n})_{n\geq2}$ be a sequence of connected patterns of RBS/FBS gates (\autoref{def:connected-pattern}), where each $U_{0,n}$ is such a pattern over $n$ qubits.
Assume that the number of gates $J_n \geq 1$ in the pattern $U_{0,n}$ satisfies $J_n \in \mathcal{O}\left(\mathrm{poly}(n)\right)$, and assume any fixed Hamming weight $k \in \mathcal{O}(1)$.

Then, for any integer sequences $(L_n)_{n\geq2}$ and $(l_n)_{n\geq2}$ satisfying $1 \leq l_n \leq L_n$ (for all $n\geq2$) as well as
\begin{equation}\label{eq:concluding-back-to-variance-hyp-Ln}
L_n\,,\;\;l_n\,,\;\;(L_n - l_n) \;\;\;\in\;\;\; \Omega\left( \Delta_n^{-1} \, d_{k,n} \, n\right)\,,
\end{equation}
the quantum circuit comprised of $L_n$ repetitions of the RBS/FBS patten $U_{0,n}$ has -- for any initial and target states of Hamming weight $k$ -- a cost function whose gradient for the parameter of the $j^{\mathrm{th}}$ gate in the $l_{n}^{\,\mathrm{th}}$ repetition (c.f. \autoref{eq:periodic-structure-appendix}) has, for any $j\in\llbracket 1, J_n \rrbracket$, a variance of inverse-polynomial order, i.e.:
\begin{equation}
    \mathrm{Var}_{\theta}[\partial_{\theta_{\lambda(l_n,j)}} \mathcal{C}]
    \in
    \Theta\left( 1/\mathrm{poly}(n) \right)\,.
\end{equation}
Here, $\Delta_n:=1-|\lambda_2(T_n)|$ denotes the spectral gap of the stochastic matrix $T_n:=T^{(U_{0,n}^k)}$ (\autoref{eq:stochastic-matrix-T-def}), and $d_{k,n}:=\binom{n}{k}$.

Assuming that \autoref{conj:spectral-gap} holds, there exists sequences $(L_n)$ that simultaneously satisfy $L_n \in \Omega\left( \Delta_n^{-1} \, d_{k,n} \, n\right)$ and $L_n \in \mathcal{O}\left(\mathrm{poly}(n)\right)$.

In particular, letting $q_\Delta\in\mathbb{N}$ be the lowest integer such that $(1/\Delta_n) \in \mathcal{O}(n^{q_\Delta})$, and letting $q_k := \min(k,n-k)$ (the lowest integer such that $d_{k,n} \in \mathcal{O}(n^{q_k})$), the choices
\begin{equation}
    L_n := n^{q_\Delta + q_k + 1}
    \qquad\mathrm{and}\qquad\;\;
    l_n := \lfloor \alpha \, L_n \rfloor  \,,
\end{equation}
for any fixed constant $\alpha$ such that $0<\alpha<1$,
satisfy the assumptions of \autoref{eq:concluding-back-to-variance-hyp-Ln}, and thus one can say that there is an absence of Barren Plateaus for CPSA ansätze (\autoref{def:CPSA}) with $L_n = n^{q_\Delta + q_k + 1}$ repetitions, for angles located at any constant fraction of the depth.
\end{thm}

\begin{proof}
Given choices of sequences $(L_n)$ and $(l_n)$ that satisfy the assumptions of the theorem, define the third sequence $(\tilde{l}_n)$ in accordance to the different "coordinates systems" $(l,j)\leftrightarrow(\tilde{l},\tilde{j})$ discussed around \autoref{eq:depth-coordinates-correspondance-1,eq:depth-coordinates-correspondance-2}, i.e. by:
\begin{equation}
    \tilde{l}_n := L_n - l_n +1\,.
\end{equation}
Due to \autoref{eq:concluding-back-to-variance-hyp-Ln}, the sequences $(l_n)$ and $(\tilde{l}_n)$ are both in $\Omega( \Delta_n^{-1} \, d_{k,n} \, n)$, and hence so are the sequences $(l_n -1)_{n\geq2}$ and $(\tilde{l}_n -1)_{n\geq2}$.
Furthermore, since for each $n$ the pattern $U_{0,n}$ is assumed to be connected, the associated stochastic matrices $T_n:=T^{(U_{0,n}^k)}$ are all primitive by \autoref{cor:connected-section-cor1}, and they are all doubly-stochastic as well by \autoref{lem:T-is-doubly-stoch}.

Therefore, one can apply \autoref{thm:stoch-mat-sequence-convergence} to $(T_n)_{n\geq2}$, and with either sequences of powers $(l_n -1)_{n\geq2}$ or $(\tilde{l}_n -1)_{n\geq2}$. Doing so separately, using both of them, yields respectively constants $c,\tilde{c}>0$ such that
\begin{align}
    \lVert T_n^{l(n)-1} - T_{n,\infty} \rVert_{\mathrm{ew},\infty} 
    &\in \mathcal{O}\left(1/\exp(c\, n)\right)\,,\\
    \lVert T_n^{\tilde{l}(n)-1} - T_{n,\infty} \rVert_{\mathrm{ew},\infty} 
    &\in \mathcal{O}\left(1/\exp(\tilde{c}\, n)\right)\,.
\end{align}  
Thus, letting $c':=\min(c,\tilde{c})$:
\begin{equation}\label{eq:concluding-back-to-variance-proof-eq1}
    \lVert T_n^{l(n)-1} - T_{n,\infty} \rVert_{\mathrm{ew},\infty},\;\;\lVert T_n^{\tilde{l}(n)-1} - T_{n,\infty} \rVert_{\mathrm{ew},\infty}
    \;\in\; \mathcal{O}\left(1/\exp(c'\, n)\right)\,.
\end{equation}

Combining \autoref{eq:concluding-back-to-variance-proof-eq1} with \autoref{eq:variance-difference-with-idealvariance-general-upper-bound,lemma:variance-error-terms-first-general-bound}, and with $\lVert \cdot \rVert_{\mathrm{Sc},2} \leq d_{k,n}\lVert \cdot \rVert_{\mathrm{ew},\infty}$ , one obtains:
\begin{equation}\label{eq:variance-difference-with-idealvariance-general-upper-bound--re}
     \left| \mathrm{Var}_{\theta}[\partial_{\theta_{\lambda(l_n,j)}} \mathcal{C}] - \frac{1}{d_{k,n}} \frac{8 k (n-k)}{n(n-1)} \right| 
     \in
     \mathcal{O}\left( \frac{8 k (n-k)}{n(n-1)} \left[ \frac{1}{\exp(c'\,n)} + \frac{1}{\exp(c'\,n)} + \frac{2 \, d_{k,n}}{\exp(2\,c'\,n)} \right] \right)\,,
 \end{equation}
and hence, since 
\begin{equation}
    \frac{8 k (n-k)}{n(n-1)}  \in \mathcal{O}(1)
\end{equation}
and
\begin{equation}
    \left[ \frac{1}{\exp(c'\,n)} + \frac{1}{\exp(c'\,n)} + \frac{2 \, d_{k,n}}{\exp(2\,c'\,n)} \right]  \in \mathcal{O}\left(\frac{1}{\exp(c'\,n)}\right)\,,
\end{equation}
one gets 
\begin{equation}\label{eq:variance-difference-with-idealvariance-general-upper-bound--last-proof-eq}
     \left| \mathrm{Var}_{\theta}[\partial_{\theta_{\lambda(l_n,j)}} \mathcal{C}] - \frac{1}{d_{k,n}} \frac{8 k (n-k)}{n(n-1)} \right| 
     \in
     \mathcal{O}\left(\frac{1}{\exp(c'\,n)}\right)\,.
 \end{equation}
And thus, since
 \begin{equation}
     \frac{1}{d_{k,n}} \frac{8 k (n-k)}{n(n-1)} \;\in\;
     \Theta\left(\frac{1}{\mathrm{poly}(n)}\right)\,,
 \end{equation}
 \autoref{eq:variance-difference-with-idealvariance-general-upper-bound--last-proof-eq} implies that
 \begin{equation}
     \mathrm{Var}_{\theta}[\partial_{\theta_{\lambda(l_n,j)}} \mathcal{C}]
     \;\in\;
     \Theta\left(\frac{1}{\mathrm{poly}(n)}\right)\,.
 \end{equation}

Lastly, if \autoref{conj:spectral-gap} holds, it implies (since $d_{k,n}\in \mathcal{O}\left(\mathrm{poly}(n)\right)$ that $\left( \Delta_n^{-1} \, d_{k,n} \, n\right) \in \mathcal{O}\left(\mathrm{poly}(n)\right)$, implying that indeed $\Omega\left( \Delta_n^{-1} \, d_{k,n} \, n\right) \,\cap\, \mathcal{O}\left(\mathrm{poly}(n)\right)$ is non-empty, thereby justifying the existence of sequences $(L_n)$ being in both $\Omega\left( \Delta_n^{-1} \, d_{k,n} \, n\right)$ and $\mathcal{O}\left(\mathrm{poly}(n)\right)$.
\end{proof}

\subsection{Numerical evidence supporting \autoref{conj:spectral-gap}}\label{subsec:numerical-evidence-spectral-gap}

We consider three sequences of connected patterns $(U_{0,n})_{n\geq2}$ here, labeled \texttt{line-down}, \texttt{line-up}, and \texttt{pyramid}.
The circuit of \texttt{line-down} consists of a cascading line of $J=n-1$ RBS gates, going downwards and rightwards (the first gate RBS gate connects qubit n$^{\circ}$1 to qubit n$^{\circ}$2, and so on).
The circuit of \texttt{line-downup} consists of the previous cascading line of $n-1$ RBS gates, followed by a second cascading line of $n-2$ RBS gates, this time going back upwards and rightwards, for a total of $J=2n -3$ gates.
The circuit of \texttt{pyramid} is an arrangement of $J=n(n-1)/2$ RBS gates into a "triangle" (see e.g. \cite[Section 2.3.2]{landman_quantum_2022}).

These three RBS patterns are all connected (\autoref{def:connected-pattern}). Note that there is no need to consider FBS gates at all because they have identical stochastic matrices than those of RBS gates.

For all three of these RBS patterns $(U_{0,n})_{n\geq2}$, for the Hamming weight values $k=2,3$, and for qubit counts $n\in\llbracket4,50\rrbracket$, we numerically construct their associated stochastic matrices $T_n:=T^{(U_{0,n}^k)}$ (\autoref{eq:stochastic-matrix-T-def}), and we numerically evaluate the eigenvalues of $T_n$, from which we deduce their spectral gap values $\Delta_n=1-|\lambda_2|$. (All of this is performed using \textit{Numpy}, in double precision.)

As a side remark, note that the patterns \texttt{line-downup} and \texttt{pyramid} are both palindromes (\autoref{lem:connected-section-palindrome}), however the pattern  \texttt{line-down} is not.
In fact we checked with the symbolic computation software \textit{Mathematica} that in general the stochastic matrices $T_n$ associated to the \texttt{line-down} pattern are neither symmetric, nor normal, and not even diagonalizable. (Eigenvalues are still well-defined, of course, even for non-diagonalizable matrices.)

We then plot the obtained $\Delta_n$ values as a function of the number of qubits $n$. Our \autoref{conj:spectral-gap} is claiming that $\Delta_n$ should be decaying at most polynomially fast (and not faster), which corresponds to the claim that graphically, on a "loglog" plot (where both the $x$ and $y$ axes have logarithmic scaling), $\Delta_n$ should vanish "at most in a straight descending line" (and not faster).
If however, on a "semilog" plot (where the $x$ axis has a regular scaling and the $y$ axis has a logarithmic scaling), we were to observe that $\Delta_n$ vanishes "in a straight descending line" (or faster), then it would indicate an exponentially-vanishing trend.
Hence, we perform two types of linear regressions, corresponding to both the semilog and loglog plot types just described, to respectively assess how good does a polynomial decay fit to the data, and how bad does an exponential decay fit to the data.
Since we are trying to evaluate the \emph{asymptotic} nature of the decays anyways, we offset the start of the fitted region to $n=20$ (in hopes of better matching the asymptotic regime of the data, but while still keeping a good amount of data points).
The $r^2$ value quantifies how well the respective model fits the data (the closer to $1$, the better of a fit).

\begin{figure}[h!]
    \centering
    \includegraphics[width=1\textwidth]{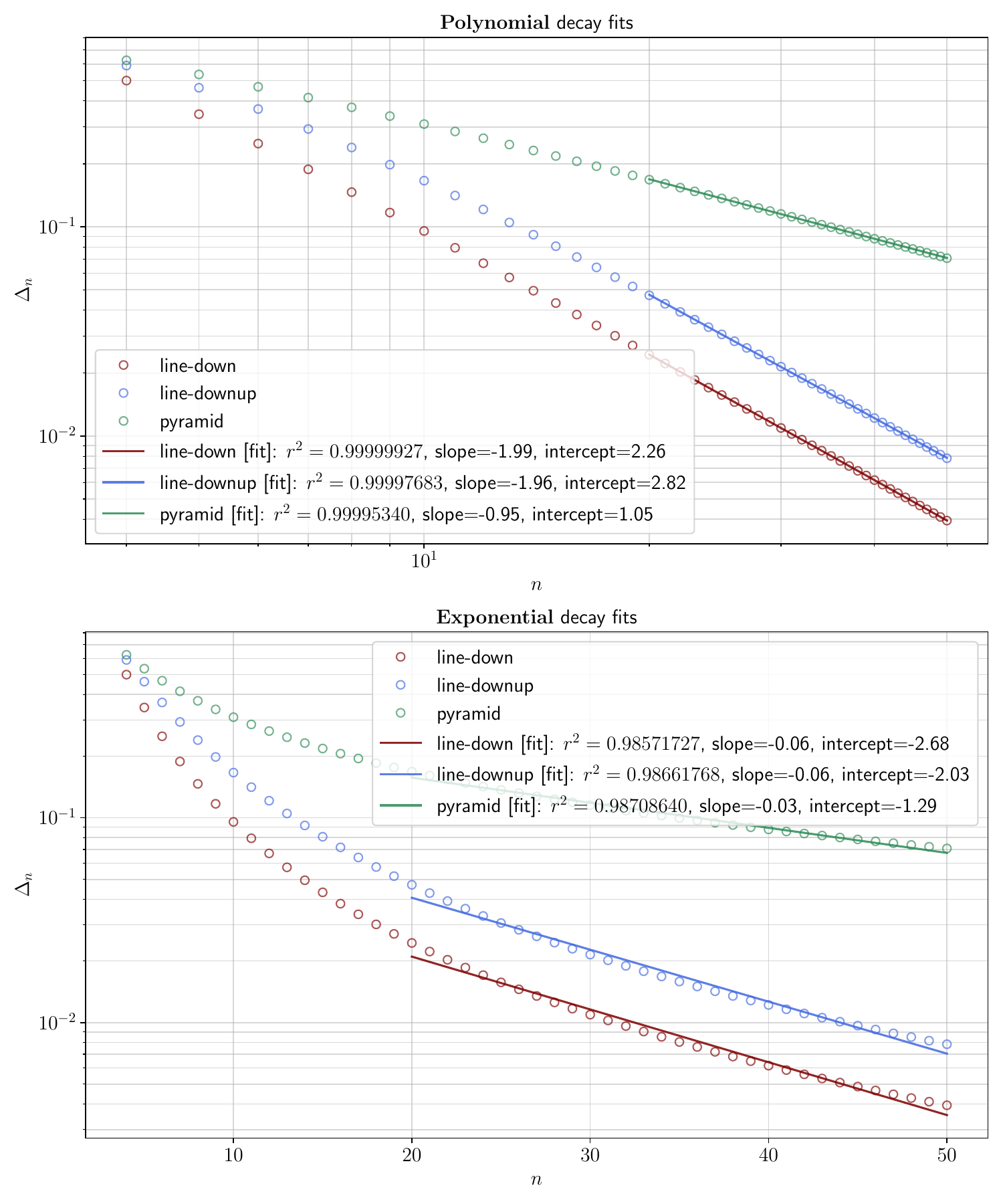}
    \caption{Numerical evidence for the inverse-polynomial largeness of the spectral gap $\Delta_n$; Hamming-weights of $k=1$ and $k=2$ both produce this exact figure.}
    \label{fig:numerical-evidence-spectral-gap-k=2}
\end{figure}

The results are presented in \autoref{fig:numerical-evidence-spectral-gap-k=2}. In fact, the plots we obtain for $k=1$ or $k=2$ have no perceivable difference at all -- numerically, we observe differences in data point values of order $10^{-13}$. Therefore, we only plot one of them to avoid an unnecessary "duplicate" figure, but the whole \autoref{fig:numerical-evidence-spectral-gap-k=2} (including the shown parameters of the fitting results) is to be taken for both cases of $k=1$ and $k=2$.\footnote{Perhaps we could hence add to \autoref{conj:spectral-gap} that either the spectral gaps values $\Delta^{(k)}_n$ are strictly independent of $k$, or that their differences for different values of $k$ vanishes exponentially with $n$.}
As a side note, we observe numerically for these two cases that, even though the second largest eigenvalue moduli $|\lambda_2|$ have practically identical values (up to some $10^{-13}$) throughout all the values $n$, the third largest eigenvalue moduli $|\lambda_3|$ still differ significantly for the first few values of $n$.

For each of the three patterns (and for both $k=1,2$), the $r^2$ values obtained indicate that a polynomial decay trend fits the spectral gap data much better than an exponential decay trend (as the polynomial $r^2$ values are closer to $1$ than the exponential $r^2$ values are, by multiple orders of magnitude), thereby supporting our \autoref{conj:spectral-gap}. The better fitting of the polynomial decay regressions may also be appreciated visually on these plots.

\subsection{A priori differences between $2$-design arguments and \autoref{thm:concluding-back-to-variance}}\label{subchap:design-discussion}
For the purpose of discussion, let us recall some definitions (c.f. e.g. \cite{larocca_diagnosing_2022}).
A VQC $U(\theta)$ (with some given probability distribution $\mu_\Theta$ on the angles $\theta$), defined over a $d$-dimensional quantum system, is said to be an \emph{$\epsilon$-approximate $2$-design} if
\begin{equation}
\lVert \mathcal{A}_{U} \rVert_{\mathrm{Sc},\infty}\leq \epsilon\,,
\end{equation}
where we introduced the linear operator
\begin{equation}
\mathcal{A}_{U} := 
\mathbb{E}_{U \sim \mu_{\mathrm{Haar}}}[ U^{\otimes 2} \otimes  \overline{U}\,^{\otimes 2} ] - \mathbb{E}_{\theta \sim \mu_\Theta}[ U(\theta)^{\otimes 2} \otimes  \overline{U(\theta)}\,^{\otimes 2} ]\,,
\end{equation}
with $\overline{U}$ denoting the entry-wise complex-conjugation of the matrix $U$, and $\mu_{\mathrm{Haar}}$ the Haar measure on $\mathrm{U}(d)$.

Since our result \autoref{thm:concluding-back-to-variance} of absence of Barren Plateaus requires a number of repetitions $L_n$ that is polynomially large in the qubit number $n$ (at least of order $\Delta_n^{-1} \, d_{k,n} \, n$) to hold, one may wonder if this polynomial repetition number $L_n$ is already enough to guarantee the overall circuit to be an \emph{$\epsilon_n$-approximate $2$-design} with $\epsilon_n\in\mathcal{O}\left( 1/\exp(\alpha\,n) \right)$ (for some $\alpha>0$). As, if that is the case, then (by definition of being an $\epsilon_n$-approximate $2$-design), the variance quantity we study, being the variance of an observable expectation cost, can be approximated to order $\mathcal{O}(\epsilon_n)$ by the corresponding variance taken over the Haar ensemble of $d_k \times d_k$ unitaries (see e.g. \cite[Appendix D]{holmes_connecting_2022}). Since the latter exact Haar variance, which can be calculated using formulas derived from Weingarten calculus of the unitary Haar measure (see e.g. \cite[Appendix E.1]{holmes_connecting_2022}), ought to coincide with our asymptotic (polynomially-vanishing) variance value of
\begin{equation}
\frac{1}{d_{k,n}} \frac{8 k (n-k)}{n(n-1)}\,,
\end{equation}
one would be able to derive that our variance quantity of study lies in $\Theta( \frac{1}{d_{k,n}} \frac{8 k (n-k)}{n(n-1)} )$ just from the fact that the polynomial repetition number of the pattern leads to an $\mathcal{O}( 1/\exp(\alpha\,n))$-approximate $2$-design -- without a need to resort to our \autoref{thm:concluding-back-to-variance}.

However, it is not at all immediate if $L_n \in \Omega(\Delta_n^{-1} \, d_{k,n} \, n)$ repetitions of a connected RBS/FBS ansatz $U_{0,n}$ suffices to guarantee that the total unitary is an $\mathcal{O}( 1/\exp(\alpha\,n))$-approximate $2$-design (for some $\alpha>0$).
In fact, the existence of a repetition number $L_n \in O(\mathrm{poly}(n))$ such that the total unitary is an $\mathcal{O}( 1/\exp(\alpha\,n))$-approximate $2$-design (for some $\alpha>0$) is equivalent (due to \cite[Section 4.1]{larocca_diagnosing_2022}) to:
\begin{equation}\label{eq:design-discussion-exp-2-design-underlying-hyp}
\ln\left( \frac{1}{\lVert \mathcal{A}_{U_{0,n}} \rVert_{\mathrm{Sc},\infty}} \right) \in \Omega\left( \frac{1}{\mathrm{poly}(n)} \right)\,.
\end{equation}
It is not immediate to theoretically prove that \autoref{eq:design-discussion-exp-2-design-underlying-hyp} holds in our setting, and even assessing its numerical validity may not be so straightforward, as constructing the operator $\mathcal{A}_{U_{0,n}}$ numerically could be costly. In fact, we are not aware of any existing literature exploring numerically the validity of \autoref{eq:design-discussion-exp-2-design-underlying-hyp} for any setting of VQC ansätze.

In contrast, in this work's \autoref{thm:concluding-back-to-variance}, it is the assumption
\begin{equation}\label{eq:design-discussion-our-hyp}
\Delta_n \in \Omega\left( \frac{1}{\mathrm{poly}(n)} \right)\,
\end{equation}
(i.e. our spectral gap \autoref{conj:spectral-gap}) that guarantees that the variance lies in $\Theta( \frac{1}{d_{k,n}} \frac{8 k (n-k)}{n(n-1)} )$.
This spectral gap quantity $\Delta_n$ is conceptually simpler (for instance, it does not directly involve any probability measures, like $\mathcal{A}_{U_{0,n}}$ does), it is simpler to evaluate numerically (up to the difficulty of numerically evaluating eigenvalues of $d_{k,n}$-sized matrices), and doing so we were able to offer direct numerical evidence for the validity of \autoref{eq:design-discussion-our-hyp} in the previous \autoref{subsec:numerical-evidence-spectral-gap}, for several RBS/FBS ansätze.

Exploring whether the assumption responsible for the fast-enough convergence of second moments to that of the Haar measure (\autoref{eq:design-discussion-exp-2-design-underlying-hyp}), and the assumption responsible for the fast-enough convergence of stochastic matrix powers (\autoref{eq:design-discussion-our-hyp}), are actually equivalent for given RBS/FBS patterns, would be an interesting future direction of work.

Lastly, let us note that our \autoref{thm:NoBPgeneralcase} makes no assumptions on the number of repetitions $L$, and so its result may not be obtained in any way using closeness to unitary $2$-design assumptions.

\section{Proof of Theorem~\ref{thm:NoBPgeneralcase}}\label{chap:proof_BP_general}

We recall the Theorem~\ref{thm:NoBPgeneralcase}:

\NoBPgeneralcase*
\begin{proof}
    
According to Lemma~\ref{lemma:VarianceHWPreserving}, we have:
\begin{equation}
    \mathbb{E}_{\theta}[\partial_{\theta_i} \mathcal{C}] = 0 \, \textrm{,}
\end{equation}
\begin{equation}
    \mathrm{Var}_{\theta}[\partial_{\theta_i} \mathcal{C}] =  2 \sum_{(l,j)} \left(\int_{\theta \in \Theta} (\frac{1}{2\pi})^D (\zeta^{\lambda}_l)^2 + (\zeta^{\lambda}_j)^2 d\theta  \right) \cdot \left(\int_{\theta \in \Theta} (\frac{1}{2\pi})^D (\tilde{y}^{\lambda}_l)^2 + (\tilde{y}^{\lambda}_j)^2 d\theta \right) \, \textrm{.}
\end{equation}

By the assumption on the input state $\zeta^0$'s distribution, we have:
    \begin{equation}
        \forall r \in [d_k], \quad \mathbb{E}_{\zeta^0,y} \left[\int_{\theta \in \Theta} (\frac{1}{2\pi})^D (\zeta_r^{0})^2 d\theta\right] = \frac{1}{d_k} \, \textrm{.}
    \end{equation}
From the recurrence relation given by Eq.~\eqref{eq:Inductive_relation} of Lemma~\ref{lemma:InductiveRelation}, it follows that:
    \begin{equation}\label{eq:thmNoBPgeneralcase-proof-eq-1}
        \forall \lambda \in \llbracket 0, \lambda_{\max} \rrbracket, \;\forall r \in [d_k], \quad \mathbb{E}_{\zeta^0,y} \left[\int_{\theta \in \Theta} (\frac{1}{2\pi})^D (\zeta_r^{\lambda})^2 d\theta \right] = \frac{1}{d_k} \, \textrm{.}
    \end{equation}

Indeed, to be more explicit, we have using the notations of \autoref{subchap:variance-formula-recast} (\autoref{eq:repeated-recurrence-relation-forward}):
\begin{align}
    \overrightarrow{Z}^{\lambda} &= T_{\text{rest},\lambda} \cdot T^{l(\lambda)-1} \ \cdot\ \overrightarrow{Z^0}\,,\\
\intertext{and hence}
    \mathbb{E}_{\zeta^0,y}\big[\overrightarrow{Z}^{\lambda}\big]
    &= \big(T_{\text{rest},\lambda} \cdot T^{l(\lambda)-1}\big) \ \cdot\ \mathbb{E}_{\zeta^0,y}\big[\overrightarrow{Z^0}]\\
     &= \big(T_{\text{rest},\lambda} \cdot T^{l(\lambda)-1}\big) \ \cdot\ (1/d_k,\dots,1/d_k)^\intercal\\
     &= (1/d_k,\dots,1/d_k)^\intercal\,,
\end{align}
where the last equality follows from the fact that $\big(T_{\text{rest},\lambda} \cdot T^{l(\lambda)-1}\big)$ is row-stochastic (c.f. \autoref{lem:T-is-doubly-stoch}).

As explained in Section \ref{subchap:InductiveRelation}, the recurrence relation given by Eq.~\eqref{eq:Inductive_relation} can also be applied for the backpropagation of the target state $\tilde{y}$, and so analogously as above, we find, due to the assumption on the target state $y$'s distribution, that:
    \begin{equation}\label{eq:thmNoBPgeneralcase-proof-eq-2}
        \forall \lambda \in \llbracket 0, \lambda_{\max} \rrbracket, \;\forall r \in [d_k], \quad \mathbb{E}_{\zeta^0,y} \left[\int_{\theta \in \Theta} (\frac{1}{2\pi})^D (\tilde{y}_r^{\lambda})^2 d\theta \right] = \frac{1}{d_k} \, \textrm{.}
    \end{equation}

Using Eqs.~\eqref{eq:thmNoBPgeneralcase-proof-eq-1} and \eqref{eq:thmNoBPgeneralcase-proof-eq-2}, Lemma \ref{lemma:VarianceHWPreserving} yields, for all $\lambda$:
\begin{equation}
    \mathrm{Var}_{\theta}[\partial_{\theta_{\lambda}} \mathcal{C}(\theta)] = \mathbb{E}_{\zeta^0,y}  \left[ 2 \sum_{l,j} \left( \frac{1}{\left(2 \pi\right)^D} \int_{\theta} (\zeta_l^{\lambda})^2 + (\zeta_j^{\lambda})^2 d\theta \right)
    \cdot \left( \frac{1}{\left(2 \pi\right)^D} \int_{\theta} (\tilde{y}_l^{\lambda})^2 + (\tilde{y}_j^{\lambda})^2 d\theta \right) \right] = 2 \sum_{l,j} \frac{4}{d_k^2}\,.
\end{equation}

Each $(l,j)$ represents the indices of two basis states that are involved in the rotations created by the RBS/FBS gate corresponding to the inner layer $\lambda$. Considering $n$ qubits and a Hamming weight of $k$, there are $\binom{n-2}{k-1}$ such different pairs which are involved in rotations. And since
    \begin{equation}
        \binom{n-2}{k-1} = \frac{(n-2)!}{(k-1)! (n-1-k)!} = \frac{k(n-k)}{n(n-1)} \binom{n}{k} = \frac{k(n-k)}{n(n-1)} d_k \,,
    \end{equation}
we can conclude that for any  $\lambda \in \llbracket0, \lambda_{\mathrm{max}}\rrbracket$:
    \begin{equation}
        \begin{aligned}
            \mathbb{E}_{\zeta^0,y} \mathrm{Var}_{\theta}[\partial_{\theta_{\lambda}} \mathcal{C}(\theta)] &= \frac{k(n-k)}{n(n-1)}\frac{8}{d_k}\,.
        \end{aligned}
    \end{equation}
\end{proof}

%% file: Appendix/Experimental_Details_PQCNN.tex
\let\textcircled=\pgftextcircled
\chapter[Experimental Details on the PQCNN]{Experimental Details on the PQCNN}
\label{app:Experimental_Details_PQCNN}

In this Section, we discuss the Quantum Data Loading part of the architecture. As explained in \autoref{subsec:Tensor_Encoding}, the Photonic QCCN is based on the tensor encoding. We recall the expression of the corresponding state for a classical tensor of dimension $k$ such that $x = (x_{1, \dots, 1}, \dots,  x_{d_1, \dots, d_k}) \in \mathbb{R}^{d_1 \times \dots \times d_k}$. The corresponding photonic tensor encoded state is described by \autoref{eq:Tensor_Encoding_modes}:
    \begin{equation*}
        \ket{x} = \frac{1}{||x||} \sum_{i_1 \in [d_1]} \dots \sum_{i_k \in [d_k]}  x_{i_1, \dots, i_k} \ket{e_{d_1, i_1}} \otimes \dots \otimes \ket{e_{d_k, i_k}} ,
    \end{equation*}

where $\ket{e_{d_l, i_l}} = \ket{0 \dots 0 1 0 \dots 0}$ represents a Fock state over $d_l$ modes, with a single excitation (photon) in mode $i_l$ and vacuum in all other modes. To encode a normalized tensor of size $d_1 \times d_2 \times \dots \times d_k$ from an input Fock state of $k$ particles, one need to use a quantum circuit with at least $\prod_{i=1}^k d_i -1$ degrees of freedom, i.e., a quantum circuit that can freely control the amplitudes of the output state in the following basis:
\begin{equation}\label{eq:Basis_Tensor_Encoding}
    B = \left\{ \ket{e_{d_1, i_1}} \otimes \dots \otimes \ket{e_{d_k, i_k}} \right\}_{(i_1, \dots, i_k) \in [d_1] \times \dots \times [d_k]}
\end{equation}

However, a linear optical circuit is limited in its controlability for input states with several particles as explained in \cite{aaronson_computational_2011} due to the photonic homomorphism illustrated in \autoref{fig:Unitary_Bosonic_Circuits}. In the following, we first explain how we encode our data for the experiment introduced in \autoref{sec:Experimental_Apparatus}. Then we propose possible way to encode larger data points on a larger photonic architecture.

    \section{Quantum Data Loading in the Experiment Considered}

\begin{figure}[h!]
    \centering
    \includegraphics[width=0.95\linewidth]{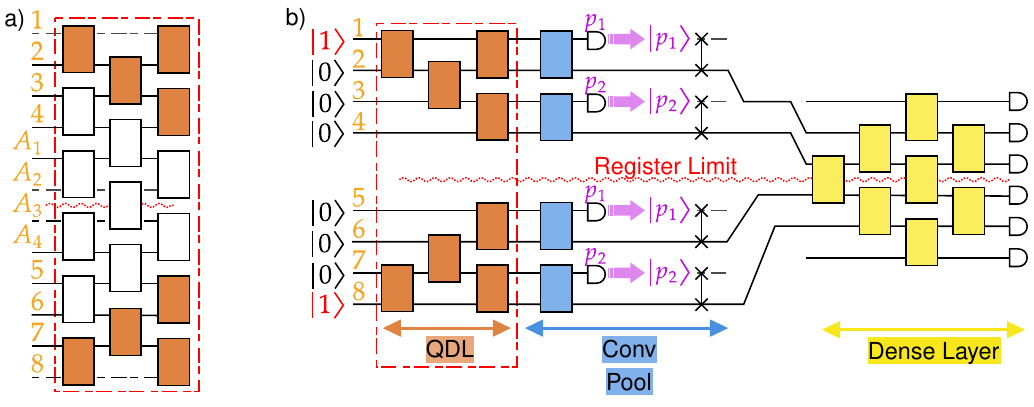}
    \caption{\textbf{Photonic QCNN architecture for experimentation. a)} Part of the $12 \times 12$ mode circuit used to encode the classical data. \textbf{b)} Photonic QCNN architecture experimentally tested. The red dotted curvy line represent the separation between the line and column registers.}
    \label{fig:QCNN_eq_QDL}
\end{figure}

The Photonic QCNN algorithm is experimentally tested with a hybrid quantum photonic platform sketched in Fig.~\ref{fig:exp_setup}a. A quantum data-loader, a quantum convolutional layer, a pooling layer, and a final dense layer are created using the $12$-mode programmable integrated interferometer as described in Fig.~\ref{fig:exp_setup}c. The Photonic QCNN architecture chosen is represented in Fig.~\ref{fig:QCNN_eq_QDL}b, and the 

\begin{figure}
    \centering
    \includegraphics[width=0.25\linewidth]{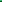}
    \hspace*{0.2in}
    \includegraphics[width=0.25\linewidth]{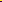}
    \hspace*{0.2in}
    \includegraphics[width=0.25\linewidth]{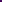}
    \caption{From left to right: sample from BAS dataset; plain sample with all pixels with the same value; sample from the Custom dataset, performed by applying a Gaussian noise on the plain sample corresponding QDL parameters.}
    \label{fig:dataset_sample_examples}
\end{figure}

Due to the limitation of the chip size, we encode our data only using $3$ parameters per register, without any parameter that link the registers. As a result, we are constrained in the amount of classical data that can be encoded. For example, the Pennylane Bars and Stripes (BAS) dataset required to encode any sample $x \in \mathbb{R}^{4 \times 4}$ which is not possible with our architecture. Therefore, we design the Custom BAS dataset in order to only consider samples of size $4 \times 4$ pixels but with more structure in order to allow our experimental linear optical data loading circuit to work. To do so, we simply choose to first design a test dataset made of bars and plot image, but with all the bright pixels to be equal in value, and all the dark pixels to be equal to $0$. Such image is easy to load considering our experimental QDL (see \autoref{fig:QCNN_eq_QDL}): for an image with lines, one just needs to tune the BS on the line register to have the photon in a uniform superposition on the corresponding modes, and the second photon on the column register to be uniformly distributed. Then we design a Custom dataset by applying a Gaussian noise on the corresponding QDL parameters for each possible set of lines or bars. An illustration is given in \autoref{fig:dataset_sample_examples}.

    \section{Quantum Data Loading for Large Data}

For complex learning problem, the QCNN architecture must use a QDL that can perform the tensor encoding of large tensors. To do so, a QDL needs enough "controllability", meaning that it must be able to freely control the amplitudes of the orthogonal state basis described in \autoref{eq:Basis_Tensor_Encoding}. We refer to this number of controllable orthogonal states as \textit{degrees of freedom}. The limitation of $m$-mode linear-optical circuit controlability is explained in \cite{aaronson_computational_2011} where the photonic homomorphism (see \autoref{fig:Unitary_Bosonic_Circuits}) limits the number of degrees of freedom to $m^2-1$, or $m(m-1)/2$ if we are not considering the phases of the state as in the tensor encoding. We can suggest several solutions in order to increase the controllability of a photonic QDL. First, one can consider additional ancillas mode to increase the controlability. On can go beyond the photonic homomorphism limitations by considering non-linearities, post-processing, or adaptivity scheme \cite{chabaud_quantum_2021, monbroussou_toward_2025}.

%% file: Appendix/Proof_Fourier_Surrogate.tex
\let\textcircled=\pgftextcircled
\chapter[Proof on the Re-Uploading Fourier Models]{Proof on the Re-Uploading Fourier Models}
\label{app:Proof_re_uploading_Fourier_models}

In this Section, we offer more formal versions of the Theorems presented in \autoref{subsec:ReUploadingModel} with their corresponding proofs. We start by \autoref{thm:Exp_BetaQ_Reuploading_2design}:

\begin{restatable}[Formal version of \autoref{thm:Exp_BetaQ_Reuploading_2design}]{thm}{ExpBetaQReuploadingtwodesignFormal}\label{thm:Exp_BetaQ_Reuploading_2design_Formal}
    Consider a single layered quantum re-uploading model with Fourier coefficients $c_{\omega}(\theta)$, with spectrum $\Omega$. We assume that each of the two parameterized unitaries are drawn form a 2-design. The variance of $\norm{\beta_Q}_2$ is given by:
    \begin{equation}
        \begin{split}
            \mathbb{E}_{\text{Haar}}[\norm{\beta_Q}^2_2] =  &\left(\frac{N\norm{O}_2^2-\text{Tr}(O)^2}{N(N^2-1)}\right) \frac{N^2p}{N(N+1)}\\
            &+ \frac{\text{Tr}(O)^2}{N^2} \, \textrm{,}
        \end{split}
    \end{equation}
    with $O$ the measurement observable.
\end{restatable}

We observe that, assuming $\text{Tr}(O)=0$ and $\norm{O}_2^2=N$, the expected values $\mathbb{E}_{\text{Haar}}[\norm{\beta_Q}^2_2]$ is scaling as $\frac{p}{N}$ such as in previous examples from \autoref{subsec:Limitations}. Using Jensen's inequality, we have that $\mathbb{E}_{\text{Haar}}[\norm{\beta_Q}_2] \leq \sqrt{\mathbb{E}_{\text{Haar}}[\norm{\beta_Q}^2_2]}$. Therefore, we have that the norm of $\beta_Q$ can be very low for low value of $p$, while the case $p \sim N$ may offer a potential advantage. Note that one could obtain an expression for $\mathbb{V}[\norm{\beta_Q}^2_2]$ using integration of Haar momentum, but under the hypothesis that the trainable layer unitaries form an $8$-design; this would require using Weingarten calculus of order $8$.

\begin{proof}
The expression of the quantum model weight vector l2-norm is $\norm{\beta_Q}_2 = \sqrt{\sum_{i=1}^{|\Omega|} |c_{\omega}|^2}$, thus we have:
    \begin{equation}
        \mathbb{E}[\norm{\beta_Q}_2^2] = \sum_{\omega \in \Omega} \mathbb{E}[|c_{\omega}|^2] \, \textrm{.}
    \end{equation}

\begin{restatable}[from \cite{mhiri_constrained_2024}]{thm}{ThmMhiri1}
\label{thm:single_layer_global_2design}
     Consider a single layered Quantum Fourier model with Fourier coefficients $c_{\omega}(\theta)$, with spectrum $\Omega$, and redundancies $|R(\omega)|$. We assume that  each of the two parameterized layers form independently a 2-design (under the uniform distribution over their parameters). The expectation and variance of each Fourier coefficient in the spectrum $\Omega$ is given by:
\begin{equation}%\label{eq:coeff_variance_2design_single}
    \begin{aligned}
        \mathbb{E}_{\theta}[c_{\omega}(\theta)]\quad&= \quad\frac{Tr(O)}{N}\delta_{\omega}^0 \, \textrm{,}\\
        \text{Var}_{\theta}[c_{\omega}(\theta)] &=	 \left(\frac{N\norm{O}_2^2-Tr(O)^2}{N(N^2-1)}\right) \frac{|R(\omega)|}{N(N+1)}    +   \frac{Tr(O)^2-N\norm{O}^2}{N^2(N^2-1)}\delta_{\omega}^0 \, \textrm{.}
    \end{aligned}
\end{equation}
\end{restatable}

According to the fact that $\sum_{\omega \in \Omega} |R(\omega)| = N^2 = 2^{2n}$, we now that each Fourier coefficient variance is vanishing when the trainable layers describe a 2-design. In addition, we observe that:
\begin{equation}
    \mathbb{E}_{\text{Haar}}[\norm{\beta_Q}^2_2] = \mathbb{E}_{\text{Haar}}[p\sum_{\omega \in \Omega} |c_{\omega}(\theta)|^2] = \sum_{\omega \in \Omega}p \mathbb{E}_{\text{Haar}}[|c_{\omega}(\theta)|^2] =  \sum_{\omega \in \Omega}p (\text{Var}_{\text{Haar}}[c_{\omega}(\theta)] + \mathbb{E}_{\text{Haar}}[c_{\omega}(\theta)]^2)
\end{equation}
And thus:
\begin{equation}
    \mathbb{E}_{\text{Haar}}[\norm{\beta_Q}^2_2] = \left(\frac{N\norm{O}_2^2-Tr(O)^2}{N(N^2-1)}\right) \frac{pN^2}{N(N+1)} + \frac{Tr(O)^2}{N^2}
\end{equation}
\end{proof}

In \cite{mhiri_constrained_2024} and in \autoref{chap:VQC_as_Fourier_Models}, we offered a bound on the variance of Fourier coefficients according to the monomial distance $\varepsilon$ of each trainable layer unitary matrix to a 2 design. Similarly, we offer a bound on the variance of the weight vector:

\begin{restatable}[]{thm}{bornbetaqhamiltonianencoding}
\label{thm:born_betaq_hamiltonian_encoding}
     Consider a single layered quantum re-uploading model with Fourier coefficients $c_{\omega}(\theta)$, with spectrum $\Omega$. We assume that each of the two parameterized unitaries form an $\varepsilon$-approximate 2-design according to the monomial definition. The expectation and variance of $\norm{\beta_Q}_2$ are given by:
\begin{equation}
    \begin{split}
        \mathbb{E}_{\theta}[\norm{\beta_Q}^2_2] \leq \mathbb{E}_{\text{Haar}}[\norm{\beta_Q}^2_2]&+\left( \frac{C_1 \varepsilon}{N^2}+\frac{C_2\varepsilon}{N(N+1)}\right)N^2 \\
        &+C_2\frac{\varepsilon^2}{N^2}N^4
    \end{split}
\end{equation}
where $C_1 = \frac{N\norm{O}_2^2-Tr(O)^2}{N(N^2-1)}$,$C_2=\sum_{l,k}\frac{|O^{\bigotimes{2}}_{l,k}|}{N^2}$, and $\mathbb{E}_{\text{Haar}}[\norm{\beta_Q}^2_2]$ as defined in \autoref{thm:Exp_BetaQ_Reuploading_2design}.
\end{restatable}

Once again, one can use Jensen's inequality to derive a similar bound on $\mathbb{E}_{\theta}[\norm{\beta_Q}_2]$.  According to the trainable layers monomial distance $\varepsilon$ to a 2-design, the choice of the observable, and the dimension of the feature map $p$, the l2-norm of the quantum weight vector can be sufficiently low and close to $\norm{\beta_{\mathrm{MNLS}}}_2$ or very large (when $p \sim N^2$).

\begin{proof}
In a more general setting where trainable layers are $\varepsilon$-approximate 2-design according to the monomial distance, we use the results from \autoref{chap:VQC_as_Fourier_Models} and \cite{mhiri_constrained_2024} that provide a bound on the Fourier coefficients variance:

\begin{restatable}[from \cite{mhiri_constrained_2024}]{thm}{ThmMhiri2}
\label{thm:bound_approx_2design}
        Consider a single layered Quantum Fourier model with spectrum $\Omega$, Fourier coefficients $c_{\omega}(\theta)$   and redundancies $|R(\omega)|$. We assume that each of the two parameterized layers forms an $\varepsilon$-approximate 2-design according to the monomial definition. The variance of the model's Fourier coefficients obeys the following bound:
    \begin{equation}
     \text{Var}_{\theta}[c_{\omega}(\theta)] \leq \text{Var}_{\text{Haar}}[c_{\omega}(\theta)]+\left( \frac{C_1 \varepsilon}{d^2}+\frac{C_2\varepsilon}{d(d+1)}\right)|R(\omega)|+C_2\frac{\varepsilon^2}{d^2}|R(\omega)|^2 \, \textrm{,}
   \end{equation}
where $C_1 = \frac{d\norm{O}^2-Tr(O)^2}{d(d^2-1)},C_2=\sum_{l,k} \frac{|O^{\bigotimes{2}}_{l,k}|}{d^2} $ and $\text{Var}_{\text{Haar}}[c_{\omega}]$ is the variance of a Fourier coefficient under the 2-design assumption given in \autoref{thm:single_layer_global_2design}.
\end{restatable}
    
Considering a single layered Quantum Fourier model, and assuming that each of the two parametrized layers form independently a 2-design (under the uniform distribution over their parameters), we can use the results from \autoref{thm:single_layer_global_2design} and \autoref{thm:bound_approx_2design}. By applying \autoref{thm:bound_approx_2design}, it comes directly:
    \begin{equation}
        \mathbb{E}_{\theta}[\norm{\beta_Q}^2_2] \leq \mathbb{E}_{\text{Haar}}[\norm{\beta_Q}^2_2]+\left( \frac{C_1 \varepsilon}{d^2}+\frac{C_2\varepsilon}{d(d+1)}\right)|R(\omega)|+C_2\frac{\varepsilon^2}{d^2}|R(\omega)|^2 \, \textrm{.}
    \end{equation}
    By using Jensen inequality, and by considering the concavity of the square root function:
    \begin{equation}
        \mathbb{E}_{\theta}[\norm{\beta_Q}_2] \leq \sqrt{\mathbb{E}_{\text{Haar}}[\norm{\beta_Q}^2_2]+\left( \frac{C_1 \varepsilon}{d^2}+\frac{C_2\varepsilon}{d(d+1)}\right)|R(\omega)|+C_2\frac{\varepsilon^2}{d^2}|R(\omega)|^2}
    \end{equation}
\end{proof}

%% file: References.bib
@article{jerbi_quantum_2023,
	title = {Quantum machine learning beyond kernel methods},
	volume = {14},
	copyright = {2023 The Author(s)},
	issn = {2041-1723},
	url = {https://www.nature.com/articles/s41467-023-36159-y},
	doi = {10.1038/s41467-023-36159-y},
	language = {en},
	number = {1},
	urldate = {2025-07-01},
	journal = {Nature Communications},
	author = {Jerbi, Sofiene and Fiderer, Lukas J. and Poulsen Nautrup, Hendrik and Kübler, Jonas M. and Briegel, Hans J. and Dunjko, Vedran},
	month = jan,
	year = {2023},
	note = {Publisher: Nature Publishing Group},
	pages = {517},
}

@article{fontana_characterizing_2024,
	title = {Characterizing barren plateaus in quantum ansätze with the adjoint representation},
	volume = {15},
	copyright = {2024 The Author(s)},
	issn = {2041-1723},
	url = {https://www.nature.com/articles/s41467-024-49910-w},
	doi = {10.1038/s41467-024-49910-w},
	language = {en},
	number = {1},
	urldate = {2025-07-01},
	journal = {Nature Communications},
	author = {Fontana, Enrico and Herman, Dylan and Chakrabarti, Shouvanik and Kumar, Niraj and Yalovetzky, Romina and Heredge, Jamie and Sureshbabu, Shree Hari and Pistoia, Marco},
	month = aug,
	year = {2024},
	note = {Publisher: Nature Publishing Group},
	pages = {7171},
}

@article{larocca_diagnosing_2022,
    title = {Diagnosing {Barren} {Plateaus} with {Tools} from {Quantum} {Optimal} {Control}},
    volume = {6},
    url = {https://quantum-journal.org/papers/q-2022-09-29-824/},
    doi = {10.22331/q-2022-09-29-824},
    language = {en-GB},
    urldate = {2025-07-01},
    journal = {Quantum},
    author = {Larocca, Martin and Czarnik, Piotr and Sharma, Kunal and Muraleedharan, Gopikrishnan and Coles, Patrick J. and Cerezo, M.},
    month = sep,
    year = {2022},
    note = {Publisher: Verein zur Förderung des Open Access Publizierens in den Quantenwissenschaften},
    pages = {824},
}

@article{holmes_connecting_2022,
    title = {Connecting {Ansatz} {Expressibility} to {Gradient} {Magnitudes} and {Barren} {Plateaus}},
    volume = {3},
    url = {https://link.aps.org/doi/10.1103/PRXQuantum.3.010313},
    doi = {10.1103/PRXQuantum.3.010313},
    number = {1},
    urldate = {2025-07-01},
    journal = {PRX Quantum},
    author = {Holmes, Zoë and Sharma, Kunal and Cerezo, M. and Coles, Patrick J.},
    month = jan,
    year = {2022},
    note = {Publisher: American Physical Society},
    pages = {010313},
}

@article{monbroussou_trainability_2025,
    title = {Trainability and {Expressivity} of {Hamming}-{Weight} {Preserving} {Quantum} {Circuits} for {Machine} {Learning}},
    volume = {9},
    url = {https://quantum-journal.org/papers/q-2025-05-15-1745/},
    doi = {10.22331/q-2025-05-15-1745},
    language = {en-GB},
    urldate = {2025-07-01},
    journal = {Quantum},
    author = {Monbroussou, Léo and Mamon, Eliott Z. and Landman, Jonas and Grilo, Alex B. and Kukla, Romain and Kashefi, Elham},
    month = may,
    year = {2025},
    note = {Publisher: Verein zur Förderung des Open Access Publizierens in den Quantenwissenschaften},
    pages = {1745},
}

@misc{kerenidis_quantum_2022,
    title = {Quantum machine learning with subspace states},
    url = {http://arxiv.org/abs/2202.00054},
    doi = {10.48550/arXiv.2202.00054},
    urldate = {2025-07-02},
    publisher = {arXiv},
    author = {Kerenidis, Iordanis and Prakash, Anupam},
    month = feb,
    year = {2022},
    note = {arXiv:2202.00054 [quant-ph]},
}

@article{cherrat_quantum_2024,
    title = {Quantum {Vision} {Transformers}},
    volume = {8},
    url = {https://quantum-journal.org/papers/q-2024-02-22-1265/},
    doi = {10.22331/q-2024-02-22-1265},
    language = {en-GB},
    urldate = {2025-07-02},
    journal = {Quantum},
    author = {Cherrat, El Amine and Kerenidis, Iordanis and Mathur, Natansh and Landman, Jonas and Strahm, Martin and Li, Yun Yvonna},
    month = feb,
    year = {2024},
    note = {Publisher: Verein zur Förderung des Open Access Publizierens in den Quantenwissenschaften},
    pages = {1265},
}

@article{johri_nearest_2021,
    title = {Nearest centroid classification on a trapped ion quantum computer},
    volume = {7},
    copyright = {2021 The Author(s)},
    issn = {2056-6387},
    url = {https://www.nature.com/articles/s41534-021-00456-5},
    doi = {10.1038/s41534-021-00456-5},
    language = {en},
    number = {1},
    urldate = {2025-07-02},
    journal = {npj Quantum Information},
    author = {Johri, Sonika and Debnath, Shantanu and Mocherla, Avinash and Singk, Alexandros and Prakash, Anupam and Kim, Jungsang and Kerenidis, Iordanis},
    month = aug,
    year = {2021},
    note = {Publisher: Nature Publishing Group},
    pages = {122},
}

@article{jain_quantum_2024,
    title = {Quantum {Fourier} networks for solving parametric {PDEs}},
    volume = {9},
    issn = {2058-9565},
    url = {https://dx.doi.org/10.1088/2058-9565/ad42ce},
    doi = {10.1088/2058-9565/ad42ce},
    language = {en},
    number = {3},
    urldate = {2025-07-02},
    journal = {Quantum Science and Technology},
    author = {Jain, Nishant and Landman, Jonas and Mathur, Natansh and Kerenidis, Iordanis},
    month = may,
    year = {2024},
    note = {Publisher: IOP Publishing},
    pages = {035026},
}

@misc{coyle_training-efficient_2025,
    title = {Training-efficient density quantum machine learning},
    url = {http://arxiv.org/abs/2405.20237},
    doi = {10.48550/arXiv.2405.20237},
    urldate = {2025-07-02},
    publisher = {arXiv},
    author = {Coyle, Brian and Raj, Snehal and Mathur, Natansh and Cherrat, El Amine and Jain, Nishant and Kazdaghli, Skander and Kerenidis, Iordanis},
    month = may,
    year = {2025},
    note = {arXiv:2405.20237 [quant-ph]},
}

@misc{mathur_bayesian_2025,
    title = {Bayesian {Quantum} {Orthogonal} {Neural} {Networks} for {Anomaly} {Detection}},
    url = {http://arxiv.org/abs/2504.18103},
    doi = {10.48550/arXiv.2504.18103},
    urldate = {2025-07-02},
    publisher = {arXiv},
    author = {Mathur, Natansh and Coyle, Brian and Jain, Nishant and Raj, Snehal and Tandon, Akshat and Krauser, Jasper Simon and Stoessel, Rainer},
    month = apr,
    year = {2025},
    note = {arXiv:2504.18103 [quant-ph]},
}

@misc{raj_hyper_2025,
    title = {Hyper {Compressed} {Fine}-{Tuning} of {Large} {Foundation} {Models} with {Quantum} {Inspired} {Adapters}},
    url = {http://arxiv.org/abs/2502.06916},
    doi = {10.48550/arXiv.2502.06916},
    urldate = {2025-07-02},
    publisher = {arXiv},
    author = {Raj, Snehal and Coyle, Brian},
    month = feb,
    year = {2025},
    note = {arXiv:2502.06916 [cs]},
}

@misc{goh_lie-algebraic_2025,
    title = {Lie-algebraic classical simulations for quantum computing},
    url = {http://arxiv.org/abs/2308.01432},
    doi = {10.48550/arXiv.2308.01432},
    urldate = {2025-07-04},
    publisher = {arXiv},
    author = {Goh, Matthew L. and Larocca, Martin and Cincio, Lukasz and Cerezo, M. and Sauvage, Frédéric},
    month = mar,
    year = {2025},
    note = {arXiv:2308.01432 [quant-ph]},
}

@article{mcclean_barren_2018,
    title = {Barren plateaus in quantum neural network training landscapes},
    volume = {9},
    copyright = {2018 The Author(s)},
    issn = {2041-1723},
    url = {https://www.nature.com/articles/s41467-018-07090-4},
    doi = {10.1038/s41467-018-07090-4},
    language = {en},
    number = {1},
    urldate = {2025-07-04},
    journal = {Nature Communications},
    author = {McClean, Jarrod R. and Boixo, Sergio and Smelyanskiy, Vadim N. and Babbush, Ryan and Neven, Hartmut},
    month = nov,
    year = {2018},
    note = {Publisher: Nature Publishing Group},
    pages = {4812},
}

@article{landman_quantum_2022,
    title = {Quantum {Methods} for {Neural} {Networks} and {Application} to {Medical} {Image} {Classification}},
    volume = {6},
    url = {https://quantum-journal.org/papers/q-2022-12-22-881/},
    doi = {10.22331/q-2022-12-22-881},
    language = {en-GB},
    urldate = {2025-07-02},
    journal = {Quantum},
    author = {Landman, Jonas and Mathur, Natansh and Li, Yun Yvonna and Strahm, Martin and Kazdaghli, Skander and Prakash, Anupam and Kerenidis, Iordanis},
    month = dec,
    year = {2022},
    note = {Publisher: Verein zur Förderung des Open Access Publizierens in den Quantenwissenschaften},
    pages = {881},
}

@article{miszczak_symbolic_2017,
    title = {Symbolic integration with respect to the {Haar} measure on the unitary groups},
    issn = {2300-1917},
    url = {https://journals.pan.pl/dlibra/publication/121307/edition/105697},
    urldate = {2025-07-04},
    journal = {Bulletin of the Polish Academy of Sciences: Technical Sciences; 2017; 65; No 1; 21-27},
    author = {Miszczak, J. A. and Puchała, Z.},
    year = {2017},
}

@article{knill_scheme_2001,
    title = {A scheme for efficient quantum computation with linear optics},
    volume = {409},
    copyright = {2001 Macmillan Magazines Ltd.},
    issn = {1476-4687},
    url = {https://www.nature.com/articles/35051009},
    doi = {10.1038/35051009},
    language = {en},
    number = {6816},
    urldate = {2025-07-05},
    journal = {Nature},
    author = {Knill, E. and Laflamme, R. and Milburn, G. J.},
    month = jan,
    year = {2001},
    note = {Publisher: Nature Publishing Group},
    pages = {46--52},
}

@article{knill_quantum_2002,
    title = {Quantum gates using linear optics and postselection},
    volume = {66},
    url = {https://link.aps.org/doi/10.1103/PhysRevA.66.052306},
    doi = {10.1103/PhysRevA.66.052306},
    number = {5},
    urldate = {2025-07-05},
    journal = {Physical Review A},
    author = {Knill, E.},
    month = nov,
    year = {2002},
    note = {Publisher: American Physical Society},
    pages = {052306},
}

@article{zhong_quantum_2020,
    title = {Quantum computational advantage using photons},
    volume = {370},
    url = {https://www.science.org/doi/10.1126/science.abe8770},
    doi = {10.1126/science.abe8770},
    number = {6523},
    urldate = {2025-07-05},
    journal = {Science},
    author = {Zhong, Han-Sen and Wang, Hui and Deng, Yu-Hao and Chen, Ming-Cheng and Peng, Li-Chao and Luo, Yi-Han and Qin, Jian and Wu, Dian and Ding, Xing and Hu, Yi and Hu, Peng and Yang, Xiao-Yan and Zhang, Wei-Jun and Li, Hao and Li, Yuxuan and Jiang, Xiao and Gan, Lin and Yang, Guangwen and You, Lixing and Wang, Zhen and Li, Li and Liu, Nai-Le and Lu, Chao-Yang and Pan, Jian-Wei},
    month = dec,
    year = {2020},
    note = {Publisher: American Association for the Advancement of Science},
    pages = {1460--1463},
}

@article{bremner_classical_2010,
    title = {Classical simulation of commuting quantum computations implies collapse of the polynomial hierarchy},
    volume = {467},
    url = {https://royalsocietypublishing.org/doi/10.1098/rspa.2010.0301},
    doi = {10.1098/rspa.2010.0301},
    number = {2126},
    urldate = {2025-07-05},
    journal = {Proceedings of the Royal Society A: Mathematical, Physical and Engineering Sciences},
    author = {Bremner, Michael J. and Jozsa, Richard and Shepherd, Dan J.},
    month = aug,
    year = {2010},
    note = {Publisher: Royal Society},
    pages = {459--472},
}

@incollection{gard_introduction_2015,
    title = {An {Introduction} to {Boson}-{Sampling}},
    isbn = {978-981-4678-69-8},
    url = {https://www.worldscientific.com/doi/abs/10.1142/9789814678704_0008},
    urldate = {2025-07-05},
    booktitle = {From {Atomic} to {Mesoscale}},
    publisher = {WORLD SCIENTIFIC},
    author = {Gard, Bryan T. and Motes, Keith R. and Olson, Jonathan P. and Rohde, Peter P. and Dowling, Jonathan P.},
    month = mar,
    year = {2015},
    doi = {10.1142/9789814678704_0008},
    pages = {167--192},
}

@article{preskill_quantum_2018,
    title = {Quantum {Computing} in the {NISQ} era and beyond},
    volume = {2},
    url = {https://quantum-journal.org/papers/q-2018-08-06-79/},
    doi = {10.22331/q-2018-08-06-79},
    language = {en-GB},
    urldate = {2025-07-05},
    journal = {Quantum},
    author = {Preskill, John},
    month = aug,
    year = {2018},
    note = {Publisher: Verein zur Förderung des Open Access Publizierens in den Quantenwissenschaften},
    pages = {79},
}

@article{polino_photonic_2024,
    title = {Photonic implementation of quantum gravity simulator},
    volume = {3},
    issn = {2791-1519, 2791-1519},
    url = {https://www.spiedigitallibrary.org/journals/advanced-photonics-nexus/volume-3/issue-3/036011/Photonic-implementation-of-quantum-gravity-simulator/10.1117/1.APN.3.3.036011.full},
    doi = {10.1117/1.APN.3.3.036011},
    number = {3},
    urldate = {2025-07-05},
    journal = {Advanced Photonics Nexus},
    author = {Polino, Emanuele and Polacchi, Beatrice and Poderini, Davide and Agresti, Iris and Carvacho, Gonzalo and Sciarrino, Fabio and Biagio, Andrea Di and Rovelli, Carlo and Christodoulou, Marios},
    month = may,
    year = {2024},
    note = {Publisher: SPIE},
    pages = {036011},
}

@article{steinbrecher_quantum_2019,
    title = {Quantum optical neural networks},
    volume = {5},
    copyright = {2019 The Author(s)},
    issn = {2056-6387},
    url = {https://www.nature.com/articles/s41534-019-0174-7},
    doi = {10.1038/s41534-019-0174-7},
    language = {en},
    number = {1},
    urldate = {2025-07-05},
    journal = {npj Quantum Information},
    author = {Steinbrecher, Gregory R. and Olson, Jonathan P. and Englund, Dirk and Carolan, Jacques},
    month = jul,
    year = {2019},
    note = {Publisher: Nature Publishing Group},
    pages = {60},
}

@article{fu_photonic_2023,
    title = {Photonic machine learning with on-chip diffractive optics},
    volume = {14},
    copyright = {2023 The Author(s)},
    issn = {2041-1723},
    url = {https://www.nature.com/articles/s41467-022-35772-7},
    doi = {10.1038/s41467-022-35772-7},
    language = {en},
    number = {1},
    urldate = {2025-07-05},
    journal = {Nature Communications},
    author = {Fu, Tingzhao and Zang, Yubin and Huang, Yuyao and Du, Zhenmin and Huang, Honghao and Hu, Chengyang and Chen, Minghua and Yang, Sigang and Chen, Hongwei},
    month = jan,
    year = {2023},
    note = {Publisher: Nature Publishing Group},
    pages = {70},
}

@article{chabaud_quantum_2021,
    title = {Quantum machine learning with adaptive linear optics},
    volume = {5},
    url = {https://quantum-journal.org/papers/q-2021-07-05-496/},
    doi = {10.22331/q-2021-07-05-496},
    language = {en-GB},
    urldate = {2025-07-05},
    journal = {Quantum},
    author = {Chabaud, Ulysse and Markham, Damian and Sohbi, Adel},
    month = jul,
    year = {2021},
    note = {Publisher: Verein zur Förderung des Open Access Publizierens in den Quantenwissenschaften},
    pages = {496},
}

@article{marcus_permanents_1965,
    title = {Permanents},
    volume = {72},
    issn = {0002-9890},
    url = {https://www.jstor.org/stable/2313846},
    doi = {10.2307/2313846},
    number = {6},
    urldate = {2025-07-05},
    journal = {The American Mathematical Monthly},
    author = {Marcus, Marvin and Minc, Henryk},
    year = {1965},
    note = {Publisher: [Taylor \& Francis, Ltd., Mathematical Association of America]},
    pages = {577--591},
}

@inproceedings{aaronson_computational_2011,
    address = {New York, NY, USA},
    series = {{STOC} '11},
    title = {The computational complexity of linear optics},
    isbn = {978-1-4503-0691-1},
    url = {https://dl.acm.org/doi/10.1145/1993636.1993682},
    doi = {10.1145/1993636.1993682},
    urldate = {2025-07-05},
    booktitle = {Proceedings of the forty-third annual {ACM} symposium on {Theory} of computing},
    publisher = {Association for Computing Machinery},
    author = {Aaronson, Scott and Arkhipov, Alex},
    year = {2011},
    pages = {333--342},
}

@article{parellada_no-go_2023,
    title = {No-go theorems for photon state transformations in quantum linear optics},
    volume = {54},
    issn = {2211-3797},
    url = {https://www.sciencedirect.com/science/article/pii/S2211379723009014},
    doi = {10.1016/j.rinp.2023.107108},
    urldate = {2025-07-05},
    journal = {Results in Physics},
    author = {Parellada, Pablo V. and Gimeno i Garcia, Vicent and Moyano-Fernández, Julio José and Garcia-Escartin, Juan Carlos},
    month = nov,
    year = {2023},
    pages = {107108},
}

@article{xiong_fundamental_2025,
    title = {On fundamental aspects of quantum extreme learning machines},
    volume = {7},
    issn = {2524-4914},
    url = {https://doi.org/10.1007/s42484-025-00239-7},
    doi = {10.1007/s42484-025-00239-7},
    language = {en},
    number = {1},
    urldate = {2025-07-06},
    journal = {Quantum Machine Intelligence},
    author = {Xiong, Weijie and Facelli, Giorgio and Sahebi, Mehrad and Agnel, Owen and Chotibut, Thiparat and Thanasilp, Supanut and Holmes, Zoë},
    month = feb,
    year = {2025},
    pages = {20},
}

@article{larocca_theory_2023,
    title = {Theory of overparametrization in quantum neural networks},
    volume = {3},
    copyright = {2023 The Author(s), under exclusive licence to Springer Nature America, Inc.},
    issn = {2662-8457},
    url = {https://www.nature.com/articles/s43588-023-00467-6},
    doi = {10.1038/s43588-023-00467-6},
    language = {en},
    number = {6},
    urldate = {2025-07-06},
    journal = {Nature Computational Science},
    author = {Larocca, Martín and Ju, Nathan and García-Martín, Diego and Coles, Patrick J. and Cerezo, Marco},
    month = jun,
    year = {2023},
    note = {Publisher: Nature Publishing Group},
    pages = {542--551},
}

@article{bamber_how_1985,
    title = {How many parameters can a model have and still be testable?},
    volume = {29},
    issn = {0022-2496},
    url = {https://www.sciencedirect.com/science/article/pii/0022249685900057},
    doi = {10.1016/0022-2496(85)90005-7},
    number = {4},
    urldate = {2025-07-06},
    journal = {Journal of Mathematical Psychology},
    author = {Bamber, Donald and van Santen, Jan P. H.},
    month = dec,
    year = {1985},
    pages = {443--473},
}

@incollection{paszke_pytorch_2019,
    address = {Red Hook, NY, USA},
    title = {{PyTorch}: an imperative style, high-performance deep learning library},
    shorttitle = {{PyTorch}},
    number = {721},
    urldate = {2025-07-06},
    booktitle = {Proceedings of the 33rd {International} {Conference} on {Neural} {Information} {Processing} {Systems}},
    publisher = {Curran Associates Inc.},
    author = {Paszke, Adam and Gross, Sam and Massa, Francisco and Lerer, Adam and Bradbury, James and Chanan, Gregory and Killeen, Trevor and Lin, Zeming and Gimelshein, Natalia and Antiga, Luca and Desmaison, Alban and Köpf, Andreas and Yang, Edward and DeVito, Zach and Raison, Martin and Tejani, Alykhan and Chilamkurthy, Sasank and Steiner, Benoit and Fang, Lu and Bai, Junjie and Chintala, Soumith},
    year = {2019},
    pages = {8026--8037},
}

@misc{cerezo_does_2024,
    title = {Does provable absence of barren plateaus imply classical simulability? {Or}, why we need to rethink variational quantum computing},
    shorttitle = {Does provable absence of barren plateaus imply classical simulability?},
    url = {http://arxiv.org/abs/2312.09121},
    doi = {10.48550/arXiv.2312.09121},
    urldate = {2025-07-06},
    publisher = {arXiv},
    author = {Cerezo, M. and Larocca, Martin and García-Martín, Diego and Diaz, N. L. and Braccia, Paolo and Fontana, Enrico and Rudolph, Manuel S. and Bermejo, Pablo and Ijaz, Aroosa and Thanasilp, Supanut and Anschuetz, Eric R. and Holmes, Zoë},
    month = mar,
    year = {2024},
    note = {arXiv:2312.09121 [quant-ph]},
}

@article{monbroussou_subspace_2025,
    title = {Subspace preserving quantum convolutional neural network architectures},
    volume = {10},
    issn = {2058-9565},
    url = {https://dx.doi.org/10.1088/2058-9565/adbf43},
    doi = {10.1088/2058-9565/adbf43},
    language = {en},
    number = {2},
    urldate = {2025-07-01},
    journal = {Quantum Science and Technology},
    author = {Monbroussou, Léo and Landman, Jonas and Wang, Letao and Grilo, Alex B and Kashefi, Elham},
    month = mar,
    year = {2025},
    note = {Publisher: IOP Publishing},
    pages = {025050},
}

@article{makarov_theory_2022,
    title = {Theory for the {Beam} {Splitter} in {Quantum} {Optics}: {Quantum} {Entanglement} of {Photons} and {Their} {Statistics}, {HOM} {Effect}},
    copyright = {© 2022 by the author.  Licensee MDPI, Basel, Switzerland. This article is an open access article distributed under the terms and conditions of the Creative Commons Attribution (CC BY) license (https://creativecommons.org/licenses/by/4.0/).  Notwithstanding the ProQuest Terms and Conditions, you may use this content in accordance with the terms of the License.},
    shorttitle = {Theory for the {Beam} {Splitter} in {Quantum} {Optics}},
    url = {https://www.proquest.com/docview/2756756548?pq-origsite=wos&sourcetype=Scholarly%20Journals},
    doi = {10.3390/math10244794},
    language = {English},
    urldate = {2025-07-06},
    author = {Makarov, Dmitry},
    year = {2022},
    note = {Num Pages: 4794
Publisher: MDPI AG},
    pages = {4794},
}

@article{aaronson_generalizing_2014,
    title = {Generalizing and derandomizing {Gurvits}'s approximation algorithm for the permanent},
    volume = {14},
    issn = {1533-7146},
    number = {7\&8},
    journal = {Quantum Info. Comput.},
    author = {Aaronson, Scott and Hance, Travis},
    year = {2014},
    pages = {541--559},
}

@article{bartolucci_fusion-based_2023,
    title = {Fusion-based quantum computation},
    volume = {14},
    copyright = {2023 The Author(s)},
    issn = {2041-1723},
    url = {https://www.nature.com/articles/s41467-023-36493-1},
    doi = {10.1038/s41467-023-36493-1},
    language = {en},
    number = {1},
    urldate = {2025-07-09},
    journal = {Nature Communications},
    author = {Bartolucci, Sara and Birchall, Patrick and Bombín, Hector and Cable, Hugo and Dawson, Chris and Gimeno-Segovia, Mercedes and Johnston, Eric and Kieling, Konrad and Nickerson, Naomi and Pant, Mihir and Pastawski, Fernando and Rudolph, Terry and Sparrow, Chris},
    month = feb,
    year = {2023},
    note = {Publisher: Nature Publishing Group},
    pages = {912},
}

@article{raussendorf_one-way_2001,
    title = {A {One}-{Way} {Quantum} {Computer}},
    volume = {86},
    url = {https://link.aps.org/doi/10.1103/PhysRevLett.86.5188},
    doi = {10.1103/PhysRevLett.86.5188},
    number = {22},
    urldate = {2025-07-09},
    journal = {Physical Review Letters},
    author = {Raussendorf, Robert and Briegel, Hans J.},
    month = may,
    year = {2001},
    note = {Publisher: American Physical Society},
    pages = {5188--5191},
}

@article{cong_quantum_2019,
    title = {Quantum convolutional neural networks},
    volume = {15},
    copyright = {2019 The Author(s), under exclusive licence to Springer Nature Limited},
    issn = {1745-2481},
    url = {https://www.nature.com/articles/s41567-019-0648-8},
    doi = {10.1038/s41567-019-0648-8},
    language = {en},
    number = {12},
    urldate = {2025-07-09},
    journal = {Nature Physics},
    author = {Cong, Iris and Choi, Soonwon and Lukin, Mikhail D.},
    month = dec,
    year = {2019},
    note = {Publisher: Nature Publishing Group},
    pages = {1273--1278},
}

@article{maring_versatile_2024,
    title = {A versatile single-photon-based quantum computing platform},
    volume = {18},
    copyright = {2024 The Author(s)},
    issn = {1749-4893},
    url = {https://www.nature.com/articles/s41566-024-01403-4},
    doi = {10.1038/s41566-024-01403-4},
    language = {en},
    number = {6},
    urldate = {2025-07-09},
    journal = {Nature Photonics},
    author = {Maring, Nicolas and Fyrillas, Andreas and Pont, Mathias and Ivanov, Edouard and Stepanov, Petr and Margaria, Nico and Hease, William and Pishchagin, Anton and Lemaître, Aristide and Sagnes, Isabelle and Au, Thi Huong and Boissier, Sébastien and Bertasi, Eric and Baert, Aurélien and Valdivia, Mario and Billard, Marie and Acar, Ozan and Brieussel, Alexandre and Mezher, Rawad and Wein, Stephen C. and Salavrakos, Alexia and Sinnott, Patrick and Fioretto, Dario A. and Emeriau, Pierre-Emmanuel and Belabas, Nadia and Mansfield, Shane and Senellart, Pascale and Senellart, Jean and Somaschi, Niccolo},
    month = jun,
    year = {2024},
    note = {Publisher: Nature Publishing Group},
    pages = {603--609},
}

@article{wang_high-efficiency_2017,
    title = {High-efficiency multiphoton boson sampling},
    volume = {11},
    copyright = {2017 Springer Nature Limited},
    issn = {1749-4893},
    url = {https://www.nature.com/articles/nphoton.2017.63},
    doi = {10.1038/nphoton.2017.63},
    language = {en},
    number = {6},
    urldate = {2025-07-09},
    journal = {Nature Photonics},
    author = {Wang, Hui and He, Yu and Li, Yu-Huai and Su, Zu-En and Li, Bo and Huang, He-Liang and Ding, Xing and Chen, Ming-Cheng and Liu, Chang and Qin, Jian and Li, Jin-Peng and He, Yu-Ming and Schneider, Christian and Kamp, Martin and Peng, Cheng-Zhi and Höfling, Sven and Lu, Chao-Yang and Pan, Jian-Wei},
    month = jun,
    year = {2017},
    note = {Publisher: Nature Publishing Group},
    pages = {361--365},
}

@article{zhong_12-photon_2018,
    title = {12-{Photon} {Entanglement} and {Scalable} {Scattershot} {Boson} {Sampling} with {Optimal} {Entangled}-{Photon} {Pairs} from {Parametric} {Down}-{Conversion}},
    volume = {121},
    url = {https://link.aps.org/doi/10.1103/PhysRevLett.121.250505},
    doi = {10.1103/PhysRevLett.121.250505},
    number = {25},
    urldate = {2025-07-09},
    journal = {Physical Review Letters},
    author = {Zhong, Han-Sen and Li, Yuan and Li, Wei and Peng, Li-Chao and Su, Zu-En and Hu, Yi and He, Yu-Ming and Ding, Xing and Zhang, Weijun and Li, Hao and Zhang, Lu and Wang, Zhen and You, Lixing and Wang, Xi-Lin and Jiang, Xiao and Li, Li and Chen, Yu-Ao and Liu, Nai-Le and Lu, Chao-Yang and Pan, Jian-Wei},
    month = dec,
    year = {2018},
    note = {Publisher: American Physical Society},
    pages = {250505},
}

@article{calvarese_strategies_2022,
    title = {Strategies for improved temporal response of glass-based optical switches},
    volume = {12},
    copyright = {2022 The Author(s)},
    issn = {2045-2322},
    url = {https://www.nature.com/articles/s41598-021-04218-3},
    doi = {10.1038/s41598-021-04218-3},
    language = {en},
    number = {1},
    urldate = {2025-07-09},
    journal = {Scientific Reports},
    author = {Calvarese, Matteo and Paiè, Petra and Ceccarelli, Francesco and Sala, Federico and Bassi, Andrea and Osellame, Roberto and Bragheri, Francesca},
    month = jan,
    year = {2022},
    note = {Publisher: Nature Publishing Group},
    pages = {239},
}

@inproceedings{smith_universal_2022,
    title = {A {Universal} 20-mode {Quantum} {Photonic} {Processor} in {Silicon} {Nitride}},
    copyright = {© 2022 The Author(s)},
    url = {https://opg.optica.org/abstract.cfm?uri=QUANTUM-2022-QW4B.2},
    doi = {10.1364/QUANTUM.2022.QW4B.2},
    language = {EN},
    urldate = {2025-07-09},
    booktitle = {Quantum 2.0 {Conference} and {Exhibition} (2022), paper {QW4B}.2},
    publisher = {Optica Publishing Group},
    author = {Smith, Devin and Taballione, Caterina and Anguita, Malaquias Correa and Goede, Michiel De and Venderbosch, Pim and Kassenberg, Ben and Snijders, Henk and Epping, Jörn P. and Meer, Reinier van der and Pinkse, Pepijn W. H. and Vlekkert, Hans van den and Renema, Jelmer J.},
    month = jun,
    year = {2022},
    pages = {QW4B.2},
}

@article{tian_piezoelectric_2024,
    title = {Piezoelectric actuation for integrated photonics},
    volume = {16},
    copyright = {© 2024 Optica Publishing Group},
    issn = {1943-8206},
    url = {https://opg.optica.org/aop/abstract.cfm?uri=aop-16-4-749},
    doi = {10.1364/AOP.529288},
    language = {EN},
    number = {4},
    urldate = {2025-07-09},
    journal = {Advances in Optics and Photonics},
    author = {Tian, Hao and Liu, Junqiu and Attanasio, Alaina and Siddharth, Anat and Blésin, Terence and Wang, Rui Ning and Voloshin, Andrey and Lihachev, Grigory and Riemensberger, Johann and Kenning, Scott E. and Tian, Yu and Chang, Tzu Han and Bancora, Andrea and Snigirev, Viacheslav and Shadymov, Vladimir and Kippenberg, Tobias J. and Bhave, Sunil A.},
    month = dec,
    year = {2024},
    note = {Publisher: Optica Publishing Group},
    pages = {749--867},
}

@article{hong_measurement_1987,
    title = {Measurement of subpicosecond time intervals between two photons by interference},
    volume = {59},
    url = {https://link.aps.org/doi/10.1103/PhysRevLett.59.2044},
    doi = {10.1103/PhysRevLett.59.2044},
    number = {18},
    urldate = {2025-07-09},
    journal = {Physical Review Letters},
    author = {Hong, C. K. and Ou, Z. Y. and Mandel, L.},
    month = nov,
    year = {1987},
    note = {Publisher: American Physical Society},
    pages = {2044--2046},
}

@article{thomas_noise_2025,
    title = {Noise {Performance} of {On}-{Chip} {Nano}-{Mechanical} {Switches} for {Quantum} {Photonics} {Applications}},
    volume = {8},
    copyright = {© 2024 The Authors. Advanced Quantum Technologies published by Wiley-VCH GmbH},
    issn = {2511-9044},
    url = {https://onlinelibrary.wiley.com/doi/abs/10.1002/qute.202400012},
    doi = {10.1002/qute.202400012},
    language = {en},
    number = {2},
    urldate = {2025-07-09},
    journal = {Advanced Quantum Technologies},
    author = {Thomas, Rodrigo A. and Qvotrup, Celeste and Liu, Zhe and Midolo, Leonardo},
    year = {2025},
    note = {\_eprint: https://advanced.onlinelibrary.wiley.com/doi/pdf/10.1002/qute.202400012},
    pages = {2400012},
}

@article{memeo_micro-opto-mechanical_2024,
    title = {Micro-opto-mechanical glass interferometer for megahertz modulation of optical signals},
    volume = {11},
    copyright = {© 2024 Optica Publishing Group},
    issn = {2334-2536},
    url = {https://opg.optica.org/optica/abstract.cfm?uri=optica-11-2-178},
    doi = {10.1364/OPTICA.506669},
    language = {EN},
    number = {2},
    urldate = {2025-07-09},
    journal = {Optica},
    author = {Memeo, Roberto and Crespi, Andrea and Osellame, Roberto},
    month = feb,
    year = {2024},
    note = {Publisher: Optica Publishing Group},
    pages = {178--183},
}

@inproceedings{kerenidis_quantum_2020,
    title = {Quantum {Algorithms} for {Deep} {Convolutional} {Neural} {Networks}},
    url = {https://iclr.cc/virtual_2020/poster_Hygab1rKDS.html},
    language = {en},
    urldate = {2025-07-11},
    author = {Kerenidis, Iordanis and Landman, Jonas and Prakash, Anupam},
    month = apr,
    year = {2020},
}

@article{wei_quantum_2022,
    title = {A quantum convolutional neural network on {NISQ} devices},
    volume = {32},
    issn = {2309-4710},
    url = {https://doi.org/10.1007/s43673-021-00030-3},
    doi = {10.1007/s43673-021-00030-3},
    language = {en},
    number = {1},
    urldate = {2025-07-11},
    journal = {AAPPS Bulletin},
    author = {Wei, ShiJie and Chen, YanHu and Zhou, ZengRong and Long, GuiLu},
    month = jan,
    year = {2022},
    pages = {2},
}

@misc{bowles_better_2024,
    title = {Better than classical? {The} subtle art of benchmarking quantum machine learning models},
    shorttitle = {Better than classical?},
    url = {http://arxiv.org/abs/2403.07059},
    doi = {10.48550/arXiv.2403.07059},
    urldate = {2025-07-11},
    publisher = {arXiv},
    author = {Bowles, Joseph and Ahmed, Shahnawaz and Schuld, Maria},
    month = mar,
    year = {2024},
    note = {arXiv:2403.07059 [quant-ph]},
}

@misc{bermejo_quantum_2024,
    title = {Quantum {Convolutional} {Neural} {Networks} are ({Effectively}) {Classically} {Simulable}},
    url = {http://arxiv.org/abs/2408.12739},
    doi = {10.48550/arXiv.2408.12739},
    urldate = {2025-07-11},
    publisher = {arXiv},
    author = {Bermejo, Pablo and Braccia, Paolo and Rudolph, Manuel S. and Holmes, Zoë and Cincio, Lukasz and Cerezo, M.},
    month = aug,
    year = {2024},
    note = {arXiv:2408.12739 [quant-ph]},
}

@article{fontana_classical_2025,
    title = {Classical simulations of noisy variational quantum circuits},
    volume = {11},
    copyright = {2025 The Author(s)},
    issn = {2056-6387},
    url = {https://www.nature.com/articles/s41534-024-00955-1},
    doi = {10.1038/s41534-024-00955-1},
    language = {en},
    number = {1},
    urldate = {2025-07-11},
    journal = {npj Quantum Information},
    author = {Fontana, Enrico and Rudolph, Manuel S. and Duncan, Ross and Rungger, Ivan and Cîrstoiu, Cristina},
    month = may,
    year = {2025},
    note = {Publisher: Nature Publishing Group},
    pages = {84},
}

@misc{rudolph_classical_2023,
    title = {Classical surrogate simulation of quantum systems with {LOWESA}},
    url = {http://arxiv.org/abs/2308.09109},
    doi = {10.48550/arXiv.2308.09109},
    urldate = {2025-07-11},
    publisher = {arXiv},
    author = {Rudolph, Manuel S. and Fontana, Enrico and Holmes, Zoë and Cincio, Lukasz},
    month = aug,
    year = {2023},
    note = {arXiv:2308.09109 [quant-ph]},
}

@article{larocca_barren_2025,
    title = {Barren plateaus in variational quantum computing},
    volume = {7},
    copyright = {2025 Springer Nature Limited},
    issn = {2522-5820},
    url = {https://www.nature.com/articles/s42254-025-00813-9},
    doi = {10.1038/s42254-025-00813-9},
    language = {en},
    number = {4},
    urldate = {2025-07-11},
    journal = {Nature Reviews Physics},
    author = {Larocca, Martín and Thanasilp, Supanut and Wang, Samson and Sharma, Kunal and Biamonte, Jacob and Coles, Patrick J. and Cincio, Lukasz and McClean, Jarrod R. and Holmes, Zoë and Cerezo, M.},
    month = apr,
    year = {2025},
    note = {Publisher: Nature Publishing Group},
    pages = {174--189},
}

@misc{oshea_introduction_2015,
    title = {An {Introduction} to {Convolutional} {Neural} {Networks}},
    url = {http://arxiv.org/abs/1511.08458},
    doi = {10.48550/arXiv.1511.08458},
    urldate = {2025-07-11},
    publisher = {arXiv},
    author = {O'Shea, Keiron and Nash, Ryan},
    month = dec,
    year = {2015},
    note = {arXiv:1511.08458 [cs]},
}

@article{lecun_gradient-based_1998,
    title = {Gradient-based learning applied to document recognition},
    volume = {86},
    issn = {1558-2256},
    url = {https://ieeexplore.ieee.org/document/726791},
    doi = {10.1109/5.726791},
    number = {11},
    urldate = {2025-07-11},
    journal = {Proceedings of the IEEE},
    author = {Lecun, Y. and Bottou, L. and Bengio, Y. and Haffner, P.},
    month = nov,
    year = {1998},
    pages = {2278--2324},
}

@article{krizhevsky_imagenet_2017,
    title = {{ImageNet} classification with deep convolutional neural networks},
    volume = {60},
    issn = {0001-0782},
    url = {https://dl.acm.org/doi/10.1145/3065386},
    doi = {10.1145/3065386},
    number = {6},
    urldate = {2025-07-11},
    journal = {Commun. ACM},
    author = {Krizhevsky, Alex and Sutskever, Ilya and Hinton, Geoffrey E.},
    year = {2017},
    pages = {84--90},
}

@article{cerezo_variational_2021,
    title = {Variational quantum algorithms},
    volume = {3},
    copyright = {2021 Springer Nature Limited},
    issn = {2522-5820},
    url = {https://www.nature.com/articles/s42254-021-00348-9},
    doi = {10.1038/s42254-021-00348-9},
    language = {en},
    number = {9},
    urldate = {2025-07-11},
    journal = {Nature Reviews Physics},
    author = {Cerezo, M. and Arrasmith, Andrew and Babbush, Ryan and Benjamin, Simon C. and Endo, Suguru and Fujii, Keisuke and McClean, Jarrod R. and Mitarai, Kosuke and Yuan, Xiao and Cincio, Lukasz and Coles, Patrick J.},
    month = sep,
    year = {2021},
    note = {Publisher: Nature Publishing Group},
    pages = {625--644},
}

@inproceedings{younis_multivariate_2022,
    title = {Multivariate {Time} {Series} {Analysis}: {An} {Interpretable} {CNN}-based {Model}},
    shorttitle = {Multivariate {Time} {Series} {Analysis}},
    url = {https://ieeexplore.ieee.org/document/10032335},
    doi = {10.1109/DSAA54385.2022.10032335},
    urldate = {2025-07-11},
    booktitle = {2022 {IEEE} 9th {International} {Conference} on {Data} {Science} and {Advanced} {Analytics} ({DSAA})},
    author = {Younis, Raneen and Zerr, Sergej and Ahmadi, Zahra},
    month = oct,
    year = {2022},
    pages = {1--10},
}

@inproceedings{he_deep_2016,
    title = {Deep {Residual} {Learning} for {Image} {Recognition}},
    url = {https://ieeexplore.ieee.org/document/7780459},
    doi = {10.1109/CVPR.2016.90},
    urldate = {2025-07-11},
    booktitle = {2016 {IEEE} {Conference} on {Computer} {Vision} and {Pattern} {Recognition} ({CVPR})},
    author = {He, Kaiming and Zhang, Xiangyu and Ren, Shaoqing and Sun, Jian},
    month = jun,
    year = {2016},
    note = {ISSN: 1063-6919},
    pages = {770--778},
}

@article{bengio_representation_2013,
    title = {Representation {Learning}: {A} {Review} and {New} {Perspectives}},
    volume = {35},
    issn = {0162-8828},
    shorttitle = {Representation {Learning}},
    url = {https://doi.org/10.1109/TPAMI.2013.50},
    doi = {10.1109/TPAMI.2013.50},
    number = {8},
    urldate = {2025-07-11},
    journal = {IEEE Trans. Pattern Anal. Mach. Intell.},
    author = {Bengio, Yoshua and Courville, Aaron and Vincent, Pascal},
    year = {2013},
    pages = {1798--1828},
}

@incollection{montufar_number_2014,
    series = {Guide {Proceedings}},
    title = {On the number of linear regions of deep neural networks},
    url = {https://dlnext.acm.org/doi/10.5555/2969033.2969153},
    urldate = {2025-07-11},
    booktitle = {Proceedings of the 28th {International} {Conference} on {Neural} {Information} {Processing} {Systems} - {Volume} 2},
    author = {Montúfar, Guido and {View Profile} and Pascanu, Razvan and {View Profile} and Cho, Kyunghyun and {View Profile} and Bengio, Yoshua and {View Profile}},
    month = dec,
    year = {2014},
    doi = {10.5555/2969033.2969153},
    pages = {2924--2932},
}

@inproceedings{raghu_expressive_2017,
    address = {Sydney, NSW, Australia},
    series = {{ICML}'17},
    title = {On the expressive power of deep neural networks},
    urldate = {2025-07-11},
    booktitle = {Proceedings of the 34th {International} {Conference} on {Machine} {Learning} - {Volume} 70},
    publisher = {JMLR.org},
    author = {Raghu, Maithra and Poole, Ben and Kleinberg, Jon and Ganguli, Surya and Dickstein, Jascha Sohl},
    year = {2017},
    pages = {2847--2854},
}

@article{farias_quantum_2025,
    title = {Quantum encoder for fixed-{Hamming}-weight subspaces},
    volume = {23},
    url = {https://link.aps.org/doi/10.1103/PhysRevApplied.23.044014},
    doi = {10.1103/PhysRevApplied.23.044014},
    number = {4},
    urldate = {2025-07-11},
    journal = {Physical Review Applied},
    author = {Farias, Renato M.S. and Maciel, Thiago O. and Camilo, Giancarlo and Lin, Ruge and Ramos-Calderer, Sergi and Aolita, Leandro},
    month = apr,
    year = {2025},
    note = {Publisher: American Physical Society},
    pages = {044014},
}

@inproceedings{wei_rethinking_2021,
    title = {Rethinking {Convolution}: {Towards} an {Optimal} {Efficiency}},
    shorttitle = {Rethinking {Convolution}},
    url = {https://www.semanticscholar.org/paper/Rethinking-Convolution%3A-Towards-an-Optimal-Wei-Tian/9894aaae697283a7a880bf6b7b3aa81f0ab88b82},
    urldate = {2025-07-11},
    author = {Wei, Tao and Tian, Yonghong and Chen, C.},
    month = may,
    year = {2021},
}

@article{ismail_fawaz_deep_2019,
    title = {Deep learning for time series classification: a review},
    volume = {33},
    issn = {1573-756X},
    shorttitle = {Deep learning for time series classification},
    url = {https://doi.org/10.1007/s10618-019-00619-1},
    doi = {10.1007/s10618-019-00619-1},
    language = {en},
    number = {4},
    urldate = {2025-07-11},
    journal = {Data Mining and Knowledge Discovery},
    author = {Ismail Fawaz, Hassan and Forestier, Germain and Weber, Jonathan and Idoumghar, Lhassane and Muller, Pierre-Alain},
    month = jul,
    year = {2019},
    pages = {917--963},
}

@article{monbroussou_toward_2025,
    title = {Toward quantum advantage with photonic state injection},
    volume = {7},
    url = {https://link.aps.org/doi/10.1103/PhysRevResearch.7.033051},
    doi = {10.1103/PhysRevResearch.7.033051},
    number = {3},
    urldate = {2025-07-15},
    journal = {Physical Review Research},
    author = {Monbroussou, Léo and Mamon, Eliott Z. and Thomas, Hugo and Yacoub, Verena and Chabaud, Ulysse and Kashefi, Elham},
    month = jul,
    year = {2025},
    note = {Publisher: American Physical Society},
    pages = {033051},
}

@article{li_survey_2022,
    title = {A {Survey} of {Convolutional} {Neural} {Networks}: {Analysis}, {Applications}, and {Prospects}},
    volume = {33},
    issn = {2162-2388},
    shorttitle = {A {Survey} of {Convolutional} {Neural} {Networks}},
    url = {https://ieeexplore.ieee.org/document/9451544},
    doi = {10.1109/TNNLS.2021.3084827},
    number = {12},
    urldate = {2025-07-15},
    journal = {IEEE Transactions on Neural Networks and Learning Systems},
    author = {Li, Zewen and Liu, Fan and Yang, Wenjie and Peng, Shouheng and Zhou, Jun},
    month = dec,
    year = {2022},
    pages = {6999--7019},
}

@article{alzubaidi_review_2021,
    title = {Review of deep learning: concepts, {CNN} architectures, challenges, applications, future directions},
    volume = {8},
    issn = {2196-1115},
    shorttitle = {Review of deep learning},
    url = {https://doi.org/10.1186/s40537-021-00444-8},
    doi = {10.1186/s40537-021-00444-8},
    number = {1},
    urldate = {2025-07-15},
    journal = {Journal of Big Data},
    author = {Alzubaidi, Laith and Zhang, Jinglan and Humaidi, Amjad J. and Al-Dujaili, Ayad and Duan, Ye and Al-Shamma, Omran and Santamaría, J. and Fadhel, Mohammed A. and Al-Amidie, Muthana and Farhan, Laith},
    month = mar,
    year = {2021},
    pages = {53},
}

@article{somaschi_near-optimal_2016,
    title = {Near-optimal single-photon sources in the solid state},
    volume = {10},
    copyright = {2016 Springer Nature Limited},
    issn = {1749-4893},
    url = {https://www.nature.com/articles/nphoton.2016.23},
    doi = {10.1038/nphoton.2016.23},
    language = {en},
    number = {5},
    urldate = {2025-07-17},
    journal = {Nature Photonics},
    author = {Somaschi, N. and Giesz, V. and De Santis, L. and Loredo, J. C. and Almeida, M. P. and Hornecker, G. and Portalupi, S. L. and Grange, T. and Antón, C. and Demory, J. and Gómez, C. and Sagnes, I. and Lanzillotti-Kimura, N. D. and Lemaítre, A. and Auffeves, A. and White, A. G. and Lanco, L. and Senellart, P.},
    month = may,
    year = {2016},
    note = {Publisher: Nature Publishing Group},
    pages = {340--345},
}

@article{loredo_generation_2019,
    title = {Generation of non-classical light in a photon-number superposition},
    volume = {13},
    copyright = {2019 The Author(s), under exclusive licence to Springer Nature Limited},
    issn = {1749-4893},
    url = {https://www.nature.com/articles/s41566-019-0506-3},
    doi = {10.1038/s41566-019-0506-3},
    language = {en},
    number = {11},
    urldate = {2025-07-17},
    journal = {Nature Photonics},
    author = {Loredo, J. C. and Antón, C. and Reznychenko, B. and Hilaire, P. and Harouri, A. and Millet, C. and Ollivier, H. and Somaschi, N. and De Santis, L. and Lemaître, A. and Sagnes, I. and Lanco, L. and Auffèves, A. and Krebs, O. and Senellart, P.},
    month = nov,
    year = {2019},
    note = {Publisher: Nature Publishing Group},
    pages = {803--808},
}

@article{bell_further_2021,
    title = {Further compactifying linear optical unitaries},
    volume = {6},
    issn = {2378-0967},
    url = {https://doi.org/10.1063/5.0053421},
    doi = {10.1063/5.0053421},
    number = {7},
    urldate = {2025-07-17},
    journal = {APL Photonics},
    author = {Bell, B. A. and Walmsley, I. A.},
    month = jul,
    year = {2021},
    pages = {070804},
}

@article{giordani_experimental_2023,
    title = {Experimental certification of contextuality, coherence, and dimension in a programmable universal photonic processor},
    volume = {9},
    url = {https://www.science.org/doi/10.1126/sciadv.adj4249},
    doi = {10.1126/sciadv.adj4249},
    number = {44},
    urldate = {2025-07-17},
    journal = {Science Advances},
    author = {Giordani, Taira and Wagner, Rafael and Esposito, Chiara and Camillini, Anita and Hoch, Francesco and Carvacho, Gonzalo and Pentangelo, Ciro and Ceccarelli, Francesco and Piacentini, Simone and Crespi, Andrea and Spagnolo, Nicolò and Osellame, Roberto and Galvão, Ernesto F. and Sciarrino, Fabio},
    month = nov,
    year = {2023},
    note = {Publisher: American Association for the Advancement of Science},
    pages = {eadj4249},
}

@article{corrielli_femtosecond_2021,
    title = {Femtosecond laser micromachining for integrated quantum photonics},
    volume = {10},
    copyright = {De Gruyter expressly reserves the right to use all content for commercial text and data mining within the meaning of Section 44b of the German Copyright Act.},
    issn = {2192-8614},
    url = {https://www.degruyterbrill.com/document/doi/10.1515/nanoph-2021-0419/html},
    doi = {10.1515/nanoph-2021-0419},
    language = {en},
    number = {15},
    urldate = {2025-07-17},
    journal = {Nanophotonics},
    author = {Corrielli, Giacomo and Crespi, Andrea and Osellame, Roberto},
    month = nov,
    year = {2021},
    note = {Publisher: De Gruyter},
    pages = {3789--3812},
}

@article{ceccarelli_low_2020,
    title = {Low {Power} {Reconfigurability} and {Reduced} {Crosstalk} in {Integrated} {Photonic} {Circuits} {Fabricated} by {Femtosecond} {Laser} {Micromachining}},
    volume = {14},
    copyright = {© 2020 The Authors. Published by Wiley-VCH GmbH},
    issn = {1863-8899},
    url = {https://onlinelibrary.wiley.com/doi/abs/10.1002/lpor.202000024},
    doi = {10.1002/lpor.202000024},
    language = {en},
    number = {10},
    urldate = {2025-07-17},
    journal = {Laser \& Photonics Reviews},
    author = {Ceccarelli, Francesco and Atzeni, Simone and Pentangelo, Ciro and Pellegatta, Francesco and Crespi, Andrea and Osellame, Roberto},
    year = {2020},
    note = {\_eprint: https://onlinelibrary.wiley.com/doi/pdf/10.1002/lpor.202000024},
    pages = {2000024},
}

@article{pentangelo_high-fidelity_2024,
    title = {High-fidelity and polarization-insensitive universal photonic processors fabricated by femtosecond laser writing},
    volume = {13},
    copyright = {De Gruyter expressly reserves the right to use all content for commercial text and data mining within the meaning of Section 44b of the German Copyright Act.},
    issn = {2192-8614},
    url = {https://www.degruyterbrill.com/document/doi/10.1515/nanoph-2023-0636/html},
    doi = {10.1515/nanoph-2023-0636},
    language = {en},
    number = {12},
    urldate = {2025-07-17},
    journal = {Nanophotonics},
    author = {Pentangelo, Ciro and Giano, Niki Di and Piacentini, Simone and Arpe, Riccardo and Ceccarelli, Francesco and Crespi, Andrea and Osellame, Roberto},
    month = may,
    year = {2024},
    note = {Publisher: De Gruyter},
    pages = {2259--2270},
}

@article{natarajan_superconducting_2012,
    title = {Superconducting nanowire single-photon detectors: physics and applications},
    volume = {25},
    issn = {0953-2048},
    shorttitle = {Superconducting nanowire single-photon detectors},
    url = {https://dx.doi.org/10.1088/0953-2048/25/6/063001},
    doi = {10.1088/0953-2048/25/6/063001},
    language = {en},
    number = {6},
    urldate = {2025-07-17},
    journal = {Superconductor Science and Technology},
    author = {Natarajan, Chandra M and Tanner, Michael G and Hadfield, Robert H},
    month = apr,
    year = {2012},
    note = {Publisher: IOP Publishing},
    pages = {063001},
}

@article{carolan_universal_2015,
    title = {Universal linear optics},
    volume = {349},
    url = {https://www.science.org/doi/10.1126/science.aab3642},
    doi = {10.1126/science.aab3642},
    number = {6249},
    urldate = {2025-07-17},
    journal = {Science},
    author = {Carolan, Jacques and Harrold, Christopher and Sparrow, Chris and Martín-López, Enrique and Russell, Nicholas J. and Silverstone, Joshua W. and Shadbolt, Peter J. and Matsuda, Nobuyuki and Oguma, Manabu and Itoh, Mikitaka and Marshall, Graham D. and Thompson, Mark G. and Matthews, Jonathan C. F. and Hashimoto, Toshikazu and O’Brien, Jeremy L. and Laing, Anthony},
    month = aug,
    year = {2015},
    note = {Publisher: American Association for the Advancement of Science},
    pages = {711--716},
}

@article{clements_optimal_2016,
    title = {Optimal design for universal multiport interferometers},
    volume = {3},
    issn = {2334-2536},
    url = {https://opg.optica.org/optica/abstract.cfm?uri=optica-3-12-1460},
    doi = {10.1364/OPTICA.3.001460},
    language = {EN},
    number = {12},
    urldate = {2025-07-17},
    journal = {Optica},
    author = {Clements, William R. and Humphreys, Peter C. and Metcalf, Benjamin J. and Kolthammer, W. Steven and Walmsley, Ian A.},
    month = dec,
    year = {2016},
    note = {Publisher: Optica Publishing Group},
    pages = {1460--1465},
}

@article{hu_model_2021,
    title = {Model complexity of deep learning: a survey},
    volume = {63},
    issn = {0219-3116},
    shorttitle = {Model complexity of deep learning},
    url = {https://doi.org/10.1007/s10115-021-01605-0},
    doi = {10.1007/s10115-021-01605-0},
    language = {en},
    number = {10},
    urldate = {2025-07-17},
    journal = {Knowledge and Information Systems},
    author = {Hu, Xia and Chu, Lingyang and Pei, Jian and Liu, Weiqing and Bian, Jiang},
    month = oct,
    year = {2021},
    pages = {2585--2619},
}

@article{shah_time_2022,
    series = {4th {International} {Conference} on {Innovative} {Data} {Communication} {Technology} and {Application}},
    title = {Time {Complexity} in {Deep} {Learning} {Models}},
    volume = {215},
    issn = {1877-0509},
    url = {https://www.sciencedirect.com/science/article/pii/S1877050922020944},
    doi = {10.1016/j.procs.2022.12.023},
    urldate = {2025-07-17},
    journal = {Procedia Computer Science},
    author = {Shah, Bhoomi and Bhavsar, Hetal},
    month = jan,
    year = {2022},
    pages = {202--210},
}

@article{lecun2010mnist,
  title={MNIST handwritten digit database},
  author={LeCun, Yann and Cortes, Corinna and Burges, CJ},
  journal={ATT Labs [Online]. Available: http://yann.lecun.com/exdb/mnist},
  volume={2},
  year={2010}
}

@misc{xiao_fashion-mnist_2017,
    title = {Fashion-{MNIST}: a {Novel} {Image} {Dataset} for {Benchmarking} {Machine} {Learning} {Algorithms}},
    shorttitle = {Fashion-{MNIST}},
    url = {http://arxiv.org/abs/1708.07747},
    doi = {10.48550/arXiv.1708.07747},
    urldate = {2025-07-18},
    publisher = {arXiv},
    author = {Xiao, Han and Rasul, Kashif and Vollgraf, Roland},
    month = sep,
    year = {2017},
    note = {arXiv:1708.07747 [cs]},
}

@inproceedings{Krizhevsky2009LearningML,
  title={Learning Multiple Layers of Features from Tiny Images},
  author={Alex Krizhevsky},
  year={2009},
  url={https://api.semanticscholar.org/CorpusID:18268744}
}

@article{li_quantum_2020,
    title = {A quantum deep convolutional neural network for image recognition},
    volume = {5},
    issn = {2058-9565},
    url = {https://dx.doi.org/10.1088/2058-9565/ab9f93},
    doi = {10.1088/2058-9565/ab9f93},
    language = {en},
    number = {4},
    urldate = {2025-07-18},
    journal = {Quantum Science and Technology},
    author = {Li, YaoChong and Zhou, Ri-Gui and Xu, RuQing and Luo, Jia and Hu, WenWen},
    month = jul,
    year = {2020},
    note = {Publisher: IOP Publishing},
    pages = {044003},
}

@article{arrasmith_equivalence_2022,
    title = {Equivalence of quantum barren plateaus to cost concentration and narrow gorges},
    volume = {7},
    issn = {2058-9565},
    url = {https://dx.doi.org/10.1088/2058-9565/ac7d06},
    doi = {10.1088/2058-9565/ac7d06},
    language = {en},
    number = {4},
    urldate = {2025-07-19},
    journal = {Quantum Science and Technology},
    author = {Arrasmith, Andrew and Holmes, Zoë and Cerezo, M and Coles, Patrick J},
    month = aug,
    year = {2022},
    note = {Publisher: IOP Publishing},
    pages = {045015},
}

@misc{zoufal_generative_2021,
    title = {Generative {Quantum} {Machine} {Learning}},
    url = {http://arxiv.org/abs/2111.12738},
    doi = {10.48550/arXiv.2111.12738},
    urldate = {2025-07-19},
    publisher = {arXiv},
    author = {Zoufal, Christa},
    month = nov,
    year = {2021},
    note = {arXiv:2111.12738 [quant-ph]},
}

@article{schuld_effect_2021,
    title = {Effect of data encoding on the expressive power of variational quantum-machine-learning models},
    volume = {103},
    url = {https://link.aps.org/doi/10.1103/PhysRevA.103.032430},
    doi = {10.1103/PhysRevA.103.032430},
    number = {3},
    urldate = {2025-07-19},
    journal = {Physical Review A},
    author = {Schuld, Maria and Sweke, Ryan and Meyer, Johannes Jakob},
    month = mar,
    year = {2021},
    note = {Publisher: American Physical Society},
    pages = {032430},
}

@article{shin_exponential_2023,
    title = {Exponential data encoding for quantum supervised learning},
    volume = {107},
    url = {https://link.aps.org/doi/10.1103/PhysRevA.107.012422},
    doi = {10.1103/PhysRevA.107.012422},
    number = {1},
    urldate = {2025-07-19},
    journal = {Physical Review A},
    author = {Shin, S. and Teo, Y. S. and Jeong, H.},
    month = jan,
    year = {2023},
    note = {Publisher: American Physical Society},
    pages = {012422},
}

@article{peters_generalization_2023,
    title = {Generalization despite overfitting in quantum machine learning models},
    volume = {7},
    url = {https://quantum-journal.org/papers/q-2023-12-20-1210/},
    doi = {10.22331/q-2023-12-20-1210},
    language = {en-GB},
    urldate = {2025-07-19},
    journal = {Quantum},
    author = {Peters, Evan and Schuld, Maria},
    month = dec,
    year = {2023},
    note = {Publisher: Verein zur Förderung des Open Access Publizierens in den Quantenwissenschaften},
    pages = {1210},
}

@inproceedings{landman_classically_2022,
    title = {Classically {Approximating} {Variational} {Quantum} {Machine} {Learning} with {Random} {Fourier} {Features}},
    url = {https://openreview.net/forum?id=ymFhZxw70uz},
    language = {en},
    urldate = {2025-07-19},
    author = {Landman, Jonas and Thabet, Slimane and Dalyac, Constantin and Mhiri, Hela and Kashefi, Elham},
    month = sep,
    year = {2022},
}

@article{sim_expressibility_2019,
    title = {Expressibility and {Entangling} {Capability} of {Parameterized} {Quantum} {Circuits} for {Hybrid} {Quantum}-{Classical} {Algorithms}},
    volume = {2},
    copyright = {© 2019 WILEY-VCH Verlag GmbH \& Co. KGaA, Weinheim},
    issn = {2511-9044},
    url = {https://onlinelibrary.wiley.com/doi/abs/10.1002/qute.201900070},
    doi = {10.1002/qute.201900070},
    language = {en},
    number = {12},
    urldate = {2025-07-19},
    journal = {Advanced Quantum Technologies},
    author = {Sim, Sukin and Johnson, Peter D. and Aspuru-Guzik, Alán},
    year = {2019},
    note = {\_eprint: https://advanced.onlinelibrary.wiley.com/doi/pdf/10.1002/qute.201900070},
    pages = {1900070},
}

@misc{low_pseudo-randomness_2010,
    title = {Pseudo-randomness and {Learning} in {Quantum} {Computation}},
    url = {http://arxiv.org/abs/1006.5227},
    doi = {10.48550/arXiv.1006.5227},
    urldate = {2025-07-19},
    publisher = {arXiv},
    author = {Low, Richard A.},
    month = jun,
    year = {2010},
    note = {arXiv:1006.5227 [quant-ph]},
}

@article{monbroussou_photonic_2025,
    title = {Photonic quantum convolutional neural networks with adaptive state injection},
    volume = {7},
    issn = {2577-5421, 2577-5421},
    url = {https://www.spiedigitallibrary.org/journals/advanced-photonics/volume-7/issue-6/066012/Photonic-quantum-convolutional-neural-networks-with-adaptive-state-injection/10.1117/1.AP.7.6.066012.full},
    doi = {10.1117/1.AP.7.6.066012},
    number = {6},
    urldate = {2026-01-08},
    journal = {Advanced Photonics},
    publisher = {SPIE},
    author = {Monbroussou, Léo and Polacchi, Beatrice and Yacoub, Verena and Caruccio, Eugenio and Rodari, Giovanni and Hoch, Francesco and Carvacho, Gonzalo and Spagnolo, Nicolò and Giordani, Taira and Bossi, Mattia and Rajan, Abhiram and Giano, Niki Di and Albiero, Riccardo and Ceccarelli, Francesco and Osellame, Roberto and Kashefi, Elham and Sciarrino, Fabio},
    month = nov,
    year = {2025},
    pages = {066012},
}

@article{heurtel_perceval_2023,
    title = {Perceval: {A} {Software} {Platform} for {Discrete} {Variable} {Photonic} {Quantum} {Computing}},
    volume = {7},
    shorttitle = {Perceval},
    url = {https://quantum-journal.org/papers/q-2023-02-21-931/},
    doi = {10.22331/q-2023-02-21-931},
    language = {en-GB},
    urldate = {2025-07-19},
    journal = {Quantum},
    author = {Heurtel, Nicolas and Fyrillas, Andreas and Gliniasty, Grégoire de and Bihan, Raphaël Le and Malherbe, Sébastien and Pailhas, Marceau and Bertasi, Eric and Bourdoncle, Boris and Emeriau, Pierre-Emmanuel and Mezher, Rawad and Music, Luka and Belabas, Nadia and Valiron, Benoît and Senellart, Pascale and Mansfield, Shane and Senellart, Jean},
    month = feb,
    year = {2023},
    note = {Publisher: Verein zur Förderung des Open Access Publizierens in den Quantenwissenschaften},
    pages = {931},
}

@article{thabet_when_2025,
    title = {When quantum and classical models disagree: learning beyond minimum norm least square},
    copyright = {2026 The Author(s)},
    issn = {2056-6387},
    shorttitle = {When quantum and classical models disagree},
    url = {https://www.nature.com/articles/s41534-026-01217-y},
    doi = {10.1038/s41534-026-01217-y},
    language = {en},
    urldate = {2026-04-14},
    journal = {npj Quantum Information},
    publisher = {Nature Publishing Group},
    author = {Thabet, Slimane and Monbroussou, Léo and Mamon, Eliott Z. and Landman, Jonas},
    month = mar,
    year = {2026},
}

@book{schuld_supervised_2018,
    address = {Cham},
    series = {Quantum {Science} and {Technology}},
    title = {Supervised {Learning} with {Quantum} {Computers}},
    copyright = {https://www.springer.com/tdm},
    isbn = {978-3-319-96423-2 978-3-319-96424-9},
    url = {https://link.springer.com/10.1007/978-3-319-96424-9},
    language = {en},
    urldate = {2025-07-01},
    publisher = {Springer International Publishing},
    author = {Schuld, Maria and Petruccione, Francesco},
    year = {2018},
    doi = {10.1007/978-3-319-96424-9},
}

@article{cerezo_challenges_2022,
    title = {Challenges and opportunities in quantum machine learning},
    volume = {2},
    copyright = {2022 Springer Nature America, Inc.},
    issn = {2662-8457},
    url = {https://www.nature.com/articles/s43588-022-00311-3},
    doi = {10.1038/s43588-022-00311-3},
    language = {en},
    number = {9},
    urldate = {2025-07-01},
    journal = {Nature Computational Science},
    author = {Cerezo, M. and Verdon, Guillaume and Huang, Hsin-Yuan and Cincio, Lukasz and Coles, Patrick J.},
    month = sep,
    year = {2022},
    note = {Publisher: Nature Publishing Group},
    pages = {567--576},
}

@article{biamonte_quantum_2017,
    title = {Quantum machine learning},
    volume = {549},
    copyright = {2017 Macmillan Publishers Limited, part of Springer Nature. All rights reserved.},
    issn = {1476-4687},
    url = {https://www.nature.com/articles/nature23474},
    doi = {10.1038/nature23474},
    language = {en},
    number = {7671},
    urldate = {2025-07-22},
    journal = {Nature},
    author = {Biamonte, Jacob and Wittek, Peter and Pancotti, Nicola and Rebentrost, Patrick and Wiebe, Nathan and Lloyd, Seth},
    month = sep,
    year = {2017},
    note = {Publisher: Nature Publishing Group},
    pages = {195--202},
}

@incollection{kerenidis_q-means_2019,
    address = {Red Hook, NY, USA},
    title = {q-means: a quantum algorithm for unsupervised machine learning},
    shorttitle = {q-means},
    number = {372},
    urldate = {2025-07-22},
    booktitle = {Proceedings of the 33rd {International} {Conference} on {Neural} {Information} {Processing} {Systems}},
    publisher = {Curran Associates Inc.},
    author = {Kerenidis, Iordanis and Landman, Jonas and Luongo, Alessandro and Prakash, Anupam},
    year = {2019},
    pages = {4134--4144},
}

@inproceedings{kerenidis_quantum_2017,
    address = {Dagstuhl, Germany},
    series = {Leibniz {International} {Proceedings} in {Informatics} ({LIPIcs})},
    title = {Quantum {Recommendation} {Systems}},
    volume = {67},
    isbn = {978-3-95977-029-3},
    url = {https://drops.dagstuhl.de/entities/document/10.4230/LIPIcs.ITCS.2017.49},
    doi = {10.4230/LIPIcs.ITCS.2017.49},
    urldate = {2025-07-22},
    booktitle = {8th {Innovations} in {Theoretical} {Computer} {Science} {Conference} ({ITCS} 2017)},
    publisher = {Schloss Dagstuhl – Leibniz-Zentrum für Informatik},
    author = {Kerenidis, Iordanis and Prakash, Anupam},
    editor = {Papadimitriou, Christos H.},
    year = {2017},
    note = {ISSN: 1868-8969},
    pages = {49:1--49:21},
}

@article{harrow_quantum_2009,
    title = {Quantum {Algorithm} for {Linear} {Systems} of {Equations}},
    volume = {103},
    url = {https://link.aps.org/doi/10.1103/PhysRevLett.103.150502},
    doi = {10.1103/PhysRevLett.103.150502},
    number = {15},
    urldate = {2025-07-22},
    journal = {Physical Review Letters},
    author = {Harrow, Aram W. and Hassidim, Avinatan and Lloyd, Seth},
    month = oct,
    year = {2009},
    note = {Publisher: American Physical Society},
    pages = {150502},
}

@inproceedings{gilyen_quantum_2019,
    address = {New York, NY, USA},
    series = {{STOC} 2019},
    title = {Quantum singular value transformation and beyond: exponential improvements for quantum matrix arithmetics},
    isbn = {978-1-4503-6705-9},
    shorttitle = {Quantum singular value transformation and beyond},
    url = {https://dl.acm.org/doi/10.1145/3313276.3316366},
    doi = {10.1145/3313276.3316366},
    urldate = {2025-07-22},
    booktitle = {Proceedings of the 51st {Annual} {ACM} {SIGACT} {Symposium} on {Theory} of {Computing}},
    publisher = {Association for Computing Machinery},
    author = {Gilyén, András and Su, Yuan and Low, Guang Hao and Wiebe, Nathan},
    year = {2019},
    pages = {193--204},
}

@article{schuld_circuit-centric_2020,
    title = {Circuit-centric quantum classifiers},
    volume = {101},
    url = {https://link.aps.org/doi/10.1103/PhysRevA.101.032308},
    doi = {10.1103/PhysRevA.101.032308},
    number = {3},
    urldate = {2025-07-22},
    journal = {Physical Review A},
    author = {Schuld, Maria and Bocharov, Alex and Svore, Krysta M. and Wiebe, Nathan},
    month = mar,
    year = {2020},
    note = {Publisher: American Physical Society},
    pages = {032308},
}

@article{havlicek_supervised_2019,
    title = {Supervised learning with quantum-enhanced feature spaces},
    volume = {567},
    copyright = {2019 The Author(s), under exclusive licence to Springer Nature Limited},
    issn = {1476-4687},
    url = {https://www.nature.com/articles/s41586-019-0980-2},
    doi = {10.1038/s41586-019-0980-2},
    language = {en},
    number = {7747},
    urldate = {2025-07-22},
    journal = {Nature},
    author = {Havlíček, Vojtěch and Córcoles, Antonio D. and Temme, Kristan and Harrow, Aram W. and Kandala, Abhinav and Chow, Jerry M. and Gambetta, Jay M.},
    month = mar,
    year = {2019},
    note = {Publisher: Nature Publishing Group},
    pages = {209--212},
}

@article{schreiber_classical_2023,
    title = {Classical {Surrogates} for {Quantum} {Learning} {Models}},
    volume = {131},
    url = {https://link.aps.org/doi/10.1103/PhysRevLett.131.100803},
    doi = {10.1103/PhysRevLett.131.100803},
    number = {10},
    urldate = {2025-07-22},
    journal = {Physical Review Letters},
    author = {Schreiber, Franz J. and Eisert, Jens and Meyer, Johannes Jakob},
    month = sep,
    year = {2023},
    note = {Publisher: American Physical Society},
    pages = {100803},
}

@article{sweke_potential_2025,
    title = {Potential and limitations of random {Fourier} features for dequantizing quantum machine learning},
    volume = {9},
    url = {https://quantum-journal.org/papers/q-2025-02-20-1640/},
    doi = {10.22331/q-2025-02-20-1640},
    language = {en-GB},
    urldate = {2025-07-01},
    journal = {Quantum},
    author = {Sweke, Ryan and Recio-Armengol, Erik and Jerbi, Sofiene and Gil-Fuster, Elies and Fuller, Bryce and Eisert, Jens and Meyer, Johannes Jakob},
    month = feb,
    year = {2025},
    note = {Publisher: Verein zur Förderung des Open Access Publizierens in den Quantenwissenschaften},
    pages = {1640},
}

@misc{sahebi_dequantization_2025,
    title = {On {Dequantization} of {Supervised} {Quantum} {Machine} {Learning} via {Random} {Fourier} {Features}},
    url = {http://arxiv.org/abs/2505.15902},
    doi = {10.48550/arXiv.2505.15902},
    urldate = {2025-07-22},
    publisher = {arXiv},
    author = {Sahebi, Mehrad and Barthe, Alice and Suzuki, Yudai and Holmes, Zoë and Grossi, Michele},
    month = may,
    year = {2025},
    note = {arXiv:2505.15902 [quant-ph]},
}

@inproceedings{you_analyzing_2023,
    title = {Analyzing {Convergence} in {Quantum} {Neural} {Networks}: {Deviations} from {Neural} {Tangent} {Kernels}},
    shorttitle = {Analyzing {Convergence} in {Quantum} {Neural} {Networks}},
    url = {https://proceedings.mlr.press/v202/you23a.html},
    language = {en},
    urldate = {2025-07-22},
    booktitle = {Proceedings of the 40th {International} {Conference} on {Machine} {Learning}},
    publisher = {PMLR},
    author = {You, Xuchen and Chakrabarti, Shouvanik and Chen, Boyang and Wu, Xiaodi},
    month = jul,
    year = {2023},
    note = {ISSN: 2640-3498},
    pages = {40199--40224},
}

@book{bishop2006pattern,
	title        = {Pattern recognition and machine learning},
	author       = {Bishop, Christopher M},
	publisher    = {Springer},
	volume       = 4,
	number       = 4
}

@article{hastie_surprises_2022,
    title = {{SURPRISES} {IN} {HIGH}-{DIMENSIONAL} {RIDGELESS} {LEAST} {SQUARES} {INTERPOLATION}},
    volume = {50},
    issn = {0090-5364},
    url = {https://www.ncbi.nlm.nih.gov/pmc/articles/PMC9481183/},
    doi = {10.1214/21-aos2133},
    number = {2},
    urldate = {2025-07-22},
    journal = {Annals of statistics},
    author = {Hastie, Trevor and Montanari, Andrea and Rosset, Saharon and Tibshirani, Ryan J.},
    month = apr,
    year = {2022},
    pmid = {36120512},
    pmcid = {PMC9481183},
    pages = {949--986},
}

@article{hofmann2008kernel,
	title        = {Kernel methods in machine learning},
	author       = {Hofmann, Thomas and Sch{\"o}lkopf, Bernhard and Smola, Alexander J},
	year         = 2008
}

@inproceedings{rahimi_random_2007,
    title = {Random {Features} for {Large}-{Scale} {Kernel} {Machines}},
    volume = {20},
    url = {https://proceedings.neurips.cc/paper/2007/hash/013a006f03dbc5392effeb8f18fda755-Abstract.html},
    urldate = {2025-07-22},
    booktitle = {Advances in {Neural} {Information} {Processing} {Systems}},
    publisher = {Curran Associates, Inc.},
    author = {Rahimi, Ali and Recht, Benjamin},
    year = {2007},
}

@inproceedings{sutherland_error_2015,
    address = {Arlington, Virginia, USA},
    series = {{UAI}'15},
    title = {On the error of random fourier features},
    isbn = {978-0-9966431-0-8},
    urldate = {2025-07-22},
    booktitle = {Proceedings of the {Thirty}-{First} {Conference} on {Uncertainty} in {Artificial} {Intelligence}},
    publisher = {AUAI Press},
    author = {Sutherland, Danica J. and Schneider, Jeff},
    year = {2015},
    pages = {862--871},
}

@article{li_towards_2021,
    title = {Towards a {Unified} {Analysis} of {Random} {Fourier} {Features}},
    volume = {22},
    issn = {1533-7928},
    url = {http://jmlr.org/papers/v22/20-1369.html},
    number = {108},
    urldate = {2025-07-22},
    journal = {Journal of Machine Learning Research},
    author = {Li, Zhu and Ton, Jean-Francois and Oglic, Dino and Sejdinovic, Dino},
    year = {2021},
    pages = {1--51},
}

@inproceedings{rahimi_weighted_2008,
    title = {Weighted {Sums} of {Random} {Kitchen} {Sinks}: {Replacing} minimization with randomization in learning},
    volume = {21},
    shorttitle = {Weighted {Sums} of {Random} {Kitchen} {Sinks}},
    url = {https://proceedings.neurips.cc/paper/2008/hash/0efe32849d230d7f53049ddc4a4b0c60-Abstract.html},
    urldate = {2025-07-22},
    booktitle = {Advances in {Neural} {Information} {Processing} {Systems}},
    publisher = {Curran Associates, Inc.},
    author = {Rahimi, Ali and Recht, Benjamin},
    year = {2008},
}

@inproceedings{rahimi_uniform_2008,
    title = {Uniform approximation of functions with random bases},
    url = {https://ieeexplore.ieee.org/document/4797607},
    doi = {10.1109/ALLERTON.2008.4797607},
    urldate = {2025-07-22},
    booktitle = {2008 46th {Annual} {Allerton} {Conference} on {Communication}, {Control}, and {Computing}},
    author = {Rahimi, Ali and Recht, Benjamin},
    month = sep,
    year = {2008},
    pages = {555--561},
}

@misc{gyurik_exponential_2024,
    title = {Exponential separations between classical and quantum learners},
    url = {http://arxiv.org/abs/2306.16028},
    doi = {10.48550/arXiv.2306.16028},
    urldate = {2025-07-22},
    publisher = {arXiv},
    author = {Gyurik, Casper and Dunjko, Vedran},
    month = nov,
    year = {2024},
    note = {arXiv:2306.16028 [quant-ph]},
}

@misc{molteni_exponential_2024,
    title = {Exponential quantum advantages in learning quantum observables from classical data},
    url = {http://arxiv.org/abs/2405.02027},
    doi = {10.48550/arXiv.2405.02027},
    urldate = {2025-07-22},
    publisher = {arXiv},
    author = {Molteni, Riccardo and Gyurik, Casper and Dunjko, Vedran},
    month = dec,
    year = {2024},
    note = {arXiv:2405.02027 [quant-ph]},
}

@article{liu_rigorous_2021,
    title = {A rigorous and robust quantum speed-up in supervised machine learning},
    volume = {17},
    copyright = {2021 The Author(s), under exclusive licence to Springer Nature Limited},
    issn = {1745-2481},
    url = {https://www.nature.com/articles/s41567-021-01287-z},
    doi = {10.1038/s41567-021-01287-z},
    language = {en},
    number = {9},
    urldate = {2025-07-22},
    journal = {Nature Physics},
    author = {Liu, Yunchao and Arunachalam, Srinivasan and Temme, Kristan},
    month = sep,
    year = {2021},
    note = {Publisher: Nature Publishing Group},
    pages = {1013--1017},
}

@article{jerbi_shadows_2024,
    title = {Shadows of quantum machine learning},
    volume = {15},
    copyright = {2024 The Author(s)},
    issn = {2041-1723},
    url = {https://www.nature.com/articles/s41467-024-49877-8},
    doi = {10.1038/s41467-024-49877-8},
    language = {en},
    number = {1},
    urldate = {2025-07-22},
    journal = {Nature Communications},
    author = {Jerbi, Sofiene and Gyurik, Casper and Marshall, Simon C. and Molteni, Riccardo and Dunjko, Vedran},
    month = jul,
    year = {2024},
    note = {Publisher: Nature Publishing Group},
    pages = {5676},
}

@article{shor_polynomial-time_1997,
    title = {Polynomial-{Time} {Algorithms} for {Prime} {Factorization} and {Discrete} {Logarithms} on a {Quantum} {Computer}},
    volume = {26},
    issn = {0097-5397},
    url = {https://epubs.siam.org/doi/10.1137/S0097539795293172},
    doi = {10.1137/S0097539795293172},
    number = {5},
    urldate = {2025-07-22},
    journal = {SIAM Journal on Computing},
    author = {Shor, Peter W.},
    month = oct,
    year = {1997},
    note = {Publisher: Society for Industrial and Applied Mathematics},
    pages = {1484--1509},
}

@misc{umeano_ground_2024,
    title = {Ground state-based quantum feature maps},
    url = {http://arxiv.org/abs/2404.07174},
    doi = {10.48550/arXiv.2404.07174},
    urldate = {2025-07-22},
    publisher = {arXiv},
    author = {Umeano, Chukwudubem and Kyriienko, Oleksandr},
    month = dec,
    year = {2024},
    note = {arXiv:2404.07174 [quant-ph]},
}

@article{cerezo_cost_2021,
    title = {Cost function dependent barren plateaus in shallow parametrized quantum circuits},
    volume = {12},
    copyright = {2021 The Author(s)},
    issn = {2041-1723},
    url = {https://www.nature.com/articles/s41467-021-21728-w},
    doi = {10.1038/s41467-021-21728-w},
    language = {en},
    number = {1},
    urldate = {2025-07-28},
    journal = {Nature Communications},
    author = {Cerezo, M. and Sone, Akira and Volkoff, Tyler and Cincio, Lukasz and Coles, Patrick J.},
    month = mar,
    year = {2021},
    note = {Publisher: Nature Publishing Group},
    pages = {1791},
}

@article{anschuetz_efficient_2023,
    title = {Efficient classical algorithms for simulating symmetric quantum systems},
    volume = {7},
    url = {https://quantum-journal.org/papers/q-2023-11-28-1189/},
    doi = {10.22331/q-2023-11-28-1189},
    language = {en-GB},
    urldate = {2025-07-28},
    journal = {Quantum},
    author = {Anschuetz, Eric R. and Bauer, Andreas and Kiani, Bobak T. and Lloyd, Seth},
    month = nov,
    year = {2023},
    note = {Publisher: Verein zur Förderung des Open Access Publizierens in den Quantenwissenschaften},
    pages = {1189},
}

@article{liu_quantum_2019,
    title = {Quantum {Fisher} information matrix and multiparameter estimation},
    volume = {53},
    issn = {1751-8121},
    url = {https://dx.doi.org/10.1088/1751-8121/ab5d4d},
    doi = {10.1088/1751-8121/ab5d4d},
    language = {en},
    number = {2},
    urldate = {2025-07-28},
    journal = {Journal of Physics A: Mathematical and Theoretical},
    author = {Liu, Jing and Yuan, Haidong and Lu, Xiao-Ming and Wang, Xiaoguang},
    month = dec,
    year = {2019},
    note = {Publisher: IOP Publishing},
    pages = {023001},
}

@article{haug_capacity_2021,
    title = {Capacity and {Quantum} {Geometry} of {Parametrized} {Quantum} {Circuits}},
    volume = {2},
    url = {https://link.aps.org/doi/10.1103/PRXQuantum.2.040309},
    doi = {10.1103/PRXQuantum.2.040309},
    number = {4},
    urldate = {2025-07-28},
    journal = {PRX Quantum},
    author = {Haug, Tobias and Bharti, Kishor and Kim, M.S.},
    month = oct,
    year = {2021},
    note = {Publisher: American Physical Society},
    pages = {040309},
}

@article{ragone_lie_2024,
    title = {A {Lie} algebraic theory of barren plateaus for deep parameterized quantum circuits},
    volume = {15},
    copyright = {2024 The Author(s)},
    issn = {2041-1723},
    url = {https://www.nature.com/articles/s41467-024-49909-3},
    doi = {10.1038/s41467-024-49909-3},
    language = {en},
    number = {1},
    urldate = {2025-07-16},
    journal = {Nature Communications},
    author = {Ragone, Michael and Bakalov, Bojko N. and Sauvage, Frédéric and Kemper, Alexander F. and Ortiz Marrero, Carlos and Larocca, Martín and Cerezo, M.},
    month = aug,
    year = {2024},
    note = {Publisher: Nature Publishing Group},
    pages = {7172},
}

@article{ren_vectorization_2015,
    title = {On {Vectorization} of {Deep} {Convolutional} {Neural} {Networks} for {Vision} {Tasks}},
    volume = {29},
    copyright = {Copyright (c)},
    issn = {2374-3468},
    url = {https://ojs.aaai.org/index.php/AAAI/article/view/9488},
    doi = {10.1609/aaai.v29i1.9488},
    language = {en},
    number = {1},
    urldate = {2025-08-18},
    journal = {Proceedings of the AAAI Conference on Artificial Intelligence},
    author = {Ren, Jimmy and Xu, Li},
    month = feb,
    year = {2015},
}

@article{mele_introduction_2024,
    title = {Introduction to {Haar} {Measure} {Tools} in {Quantum} {Information}: {A} {Beginner}'s {Tutorial}},
    volume = {8},
    shorttitle = {Introduction to {Haar} {Measure} {Tools} in {Quantum} {Information}},
    url = {https://quantum-journal.org/papers/q-2024-05-08-1340/},
    doi = {10.22331/q-2024-05-08-1340},
    language = {en-GB},
    urldate = {2025-08-18},
    journal = {Quantum},
    author = {Mele, Antonio Anna},
    month = may,
    year = {2024},
    note = {Publisher: Verein zur Förderung des Open Access Publizierens in den Quantenwissenschaften},
    pages = {1340},
}

@article{harrow_approximate_2023,
    title = {Approximate {Unitary} t-{Designs} by {Short} {Random} {Quantum} {Circuits} {Using} {Nearest}-{Neighbor} and {Long}-{Range} {Gates}},
    volume = {401},
    issn = {1432-0916},
    url = {https://doi.org/10.1007/s00220-023-04675-z},
    doi = {10.1007/s00220-023-04675-z},
    language = {en},
    number = {2},
    urldate = {2025-08-18},
    journal = {Communications in Mathematical Physics},
    author = {Harrow, Aram W. and Mehraban, Saeed},
    month = jul,
    year = {2023},
    pages = {1531--1626},
}

@article{wolkowicz_bounds_1980,
    series = {Special {Volume} {Dedicated} to {Alson} {S}. {Householder}},
    title = {Bounds for eigenvalues using traces},
    volume = {29},
    issn = {0024-3795},
    url = {https://www.sciencedirect.com/science/article/pii/002437958090258X},
    doi = {10.1016/0024-3795(80)90258-X},
    urldate = {2025-08-18},
    journal = {Linear Algebra and its Applications},
    author = {Wolkowicz, Henry and Styan, George P. H.},
    month = feb,
    year = {1980},
    pages = {471--506},
}

@article{piccard1939ensembles,
	title        = {Sur les ensembles de distances des emsembles de points d'un espace euclidien},
	author       = {Piccard, Sophie},
	year         = 1939,
	journal      = {}
}

@article{collins2006integration,
	title        = {Integration with respect to the Haar measure on unitary, orthogonal and symplectic group},
	author       = {Collins, Beno{\^\i}t and {\'S}niady, Piotr},
	year         = 2006,
	journal      = {Communications in Mathematical Physics},
	publisher    = {Springer},
	volume       = 264,
	number       = 3,
	pages        = {773--795}
}

@online{weingarten4,
	title        = {Weingarten functions for p = 4},
	author       = {Motohisa Fukuda},
	year         = 1999,
	url          = {https://motohisafukuda.github.io/RTNI/PYTHON/Weingarten/functions4.html},
	urldate      = {2024-08-30}
}

@article{aaronson_read_2015,
    title = {Read the fine print},
    volume = {11},
    copyright = {2015 Springer Nature Limited},
    issn = {1745-2481},
    url = {https://www.nature.com/articles/nphys3272},
    doi = {10.1038/nphys3272},
    language = {en},
    number = {4},
    urldate = {2025-09-07},
    journal = {Nature Physics},
    author = {Aaronson, Scott},
    month = apr,
    year = {2015},
    note = {Publisher: Nature Publishing Group},
    pages = {291--293},
}

@article{lloyd_quantum_2014,
    title = {Quantum principal component analysis},
    volume = {10},
    copyright = {2014 Springer Nature Limited},
    issn = {1745-2481},
    url = {https://www.nature.com/articles/nphys3029},
    doi = {10.1038/nphys3029},
    language = {en},
    number = {9},
    urldate = {2025-09-07},
    journal = {Nature Physics},
    author = {Lloyd, Seth and Mohseni, Masoud and Rebentrost, Patrick},
    month = sep,
    year = {2014},
    note = {Publisher: Nature Publishing Group},
    pages = {631--633},
}

@article{mhiri_constrained_2024,
    title = {Constrained and {Vanishing} {Expressivity} of {Quantum} {Fourier} {Models}},
    volume = {9},
    url = {https://quantum-journal.org/papers/q-2025-09-03-1847/},
    doi = {10.22331/q-2025-09-03-1847},
    language = {en-GB},
    urldate = {2025-09-10},
    journal = {Quantum},
    author = {Mhiri, Hela and Monbroussou, Leo and Herrero-Gonzalez, Mario and Thabet, Slimane and Kashefi, Elham and Landman, Jonas},
    month = sep,
    year = {2025},
    note = {Publisher: Verein zur Förderung des Open Access Publizierens in den Quantenwissenschaften},
    pages = {1847},
}

@article{reck_experimental_1994,
    title = {Experimental realization of any discrete unitary operator},
    volume = {73},
    url = {https://link.aps.org/doi/10.1103/PhysRevLett.73.58},
    doi = {10.1103/PhysRevLett.73.58},
    number = {1},
    urldate = {2025-10-01},
    journal = {Physical Review Letters},
    author = {Reck, Michael and Zeilinger, Anton and Bernstein, Herbert J. and Bertani, Philip},
    month = jul,
    year = {1994},
    note = {Publisher: American Physical Society},
    pages = {58--61},
}

@article{mhiri_constrained_2025,
    title = {Constrained and {Vanishing} {Expressivity} of {Quantum} {Fourier} {Models}},
    volume = {9},
    url = {https://quantum-journal.org/papers/q-2025-09-03-1847/},
    doi = {10.22331/q-2025-09-03-1847},
    language = {en-GB},
    urldate = {2025-09-10},
    journal = {Quantum},
    author = {Mhiri, Hela and Monbroussou, Leo and Herrero-Gonzalez, Mario and Thabet, Slimane and Kashefi, Elham and Landman, Jonas},
    month = sep,
    year = {2025},
    note = {Publisher: Verein zur Förderung des Open Access Publizierens in den Quantenwissenschaften},
    pages = {1847},
}

@article{tang_quantum_2021,
    title = {Quantum {Principal} {Component} {Analysis} {Only} {Achieves} an {Exponential} {Speedup} {Because} of {Its} {State} {Preparation} {Assumptions}},
    volume = {127},
    url = {https://link.aps.org/doi/10.1103/PhysRevLett.127.060503},
    doi = {10.1103/PhysRevLett.127.060503},
    number = {6},
    urldate = {2026-06-08},
    journal = {Physical Review Letters},
    publisher = {American Physical Society},
    author = {Tang, Ewin},
    month = aug,
    year = {2021},
    pages = {060503},
}

@misc{lloyd_quantum_2013,
    title = {Quantum algorithms for supervised and unsupervised machine learning},
    url = {http://arxiv.org/abs/1307.0411},
    doi = {10.48550/arXiv.1307.0411},
    urldate = {2026-06-08},
    publisher = {arXiv},
    author = {Lloyd, Seth and Mohseni, Masoud and Rebentrost, Patrick},
    month = nov,
    year = {2013},
    note = {arXiv:1307.0411 [quant-ph]},
}

@book{Horn-MatrixAnalysis-2013,
	title        = {Matrix Analysis},
	author       = {Horn, Roger A. and Johnson, Charles R.},
	year         = 2013,
	publisher    = {Cambridge University Press},
	doi          = {https://doi.org/10.1017/CBO9780511810817},
	isbn         = {978-0-521-83940-2 978-0-521-54823-6},
	edition      = {Second edition}
}

@misc{JerisonGeneralMixing2013,
	title        = {General Mixing Time Bounds for Finite Markov Chains via the Absolute Spectral Gap},
	author       = {Jerison, Daniel},
	year         = 2013,
	month        = oct,
	publisher    = {arXiv},
	number       = {arXiv:1310.8021},
	doi          = {10.48550/arXiv.1310.8021},
	url          = {https://doi.org/10.48550/arXiv.1310.8021},
	eprint       = {1310.8021},
	primaryclass = {math},
	archiveprefix = {arXiv},
}
